\documentclass[12pt, a4paper, twoside]{report}
\usepackage{cite}

\usepackage[bottom]{footmisc}

\usepackage[hidelinks,colorlinks=true,linkcolor=blue,citecolor=blue]{hyperref}

\pdfoutput=1

\usepackage{blindtext}
\usepackage[margin=2cm] {geometry}
\usepackage{setspace}
\usepackage{fancyhdr}
\usepackage[toc,page]{appendix}
\usepackage[nottoc]{tocbibind}

\usepackage{enumerate}
\usepackage{bm}
\usepackage{bbm}
\usepackage{enumitem}
\usepackage[usenames,dvipsnames]{xcolor}
\usepackage{amsmath}
\usepackage{amsfonts}
\usepackage{amssymb}
\usepackage{graphicx}
\usepackage{hhline}
\usepackage{subfig}
\usepackage{hyperref}
\usepackage{color}
\usepackage{verbatim}
\usepackage{amsmath,amsthm,amssymb}
\usepackage{lscape}
\usepackage{float}
\usepackage{caption}
\usepackage{lscape}
\usepackage{breqn}
\usepackage{multicol}
\usepackage{graphics}
\usepackage{tikz,todonotes}
\usepackage[utf8]{inputenc}
\usepackage{autobreak}
\usepackage{verbatim}
\usetikzlibrary{arrows,positioning,decorations.markings,decorations.pathmorphing,calc}
\pgfdeclarelayer{edgelayer}
\pgfdeclarelayer{nodelayer}
\usetikzlibrary{decorations.pathreplacing}
\pgfsetlayers{edgelayer,nodelayer,main}
\tikzset{none/.style={draw=none}}
\tikzset{new edge style 2/.style={black}}
\tikzset{new style 0/.style={black}}
\tikzset{rednode/.style={draw=none, scale=0.3pt,fill=red,circle, draw}}
\tikzset{redline/.style={line width=0.3mm,red}}
\tikzset{greyE/.style={line width=0.1mm,gray}}
\usepackage[utf8]{inputenc}
\usetikzlibrary{arrows,positioning,decorations.markings,decorations.pathmorphing,calc}
\usepackage{multirow}

\definecolor{hyperref}{RGB}{026,028,087}

\newcommand{\beq}{\begin{equation}}
\newcommand{\eeq}{\end{equation}}
\newcommand{\bea}{\begin{eqnarray}}
\newcommand{\eea}{\end{eqnarray}}
\def\be{\begin{equation}}
\def\ee{\end{equation}}
\def\ba{\begin{eqnarray}}
\def\ea{\end{eqnarray}}
\newcommand{\bal}{\begin{aligned}}
\newcommand{\eal}{\end{aligned}}

\def\beq{\begin{equation}}
\def\eeq{\end{equation}}

\newcommand{\mpl}{M_{\rm Pl}}

\newcommand{\K}{\mathcal K}
\newcommand{\Q}{\mathcal Q}

\renewcommand{\[}{\left[}
\renewcommand{\]}{\right]}
\renewcommand{\L}{\mathcal L}

\def\nn{\nonumber}

\def\d{\mathrm{d}}

\usepackage[normalem]{ulem}

\def\H{\mathcal{H}}

\def\L{\mathcal{L}}
\def\K{\mathcal{K}}

\def\stu{St\"uckelberg }

\newcommand{\Ostro}{{Ostrogradsky }}

\def\d{\mathrm{d}}
\def\mn{_{\mu \nu}}
\def\mnup{^{\mu \nu}}
\def\ab{_{\alpha \beta}}
\def\abup{^{\alpha \beta}}
\def\mupn{^\mu_{\phantom{\mu}\nu}}
\def\({\left(}
\def\){\right)}

\def\mpl{M_{\rm Pl}}
\def\p{\partial}
\def\ie{{\em i.e. }}

\def\Pro{Proca-Nuevo}
\def\Pros{Proca-Nuevo }
\def\X{\mathcal{X}}
\def\Z{\mathcal{Z}}

\newcommand{\ud}[2]{^{#1}_{\phantom{#1}#2}}
\newcommand{\du}[2]{_{#1}^{\phantom{#1}#2}}

\begin{document}

\includegraphics[width=10cm]{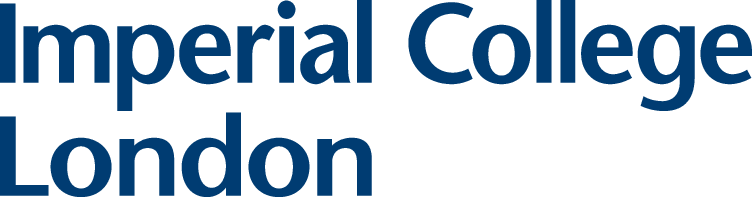}
\vfill
\begin{center}
{\huge Consistency of Scalar and Vector \\ Effective Field Theories}
\rule{15cm}{1pt}

\vspace{2cm}
\textbf{\Large{Victor M\'{a}ty\'{a}s Pozsgay}}\\
Supervisor: Claudia de Rham \\
Examiners: Arttu Rajantie (Internal), Gianmassimo Tasinato (Swansea University) \\
May 2023
\vspace{2cm}

Department of Physics \\
Imperial College London
\vspace{2cm}

Submitted in part fulfilment of the requirements for the degree of\\
Doctor of Philosophy in Theoretical Physics of Imperial College London\\
and the Diploma of Imperial College London
\end{center}
\vfill
\clearpage

\chapter*{Abstract}

In the absence of a theory of everything, modern physicists need to rely on other predictive tools and turned to Effective Field Theories (EFTs) in a number of fields, including but not limited to statistical mechanics, condensed matter, particle physics, cosmology and gravity. The coefficients of an EFT can be constrained with high precision by experiments, which can involve high-energy particle colliders for instance but are generally left free from the theoretical point of view. The focus of this thesis is to use various consistency criteria to get theoretical constraints on the low-energy coefficients of EFTs. In particular, we construct a new model of massive spin-1 field by requiring that the theory is free of any ghostly degree of freedom. We then study its cosmological perturbations and ask that all propagating modes are stable and subluminal, reducing the space of viable cosmological solutions. Finally, we implement a method to get `causality bounds', which are obtained by requiring infrared causality. This is imposed by forbidding any resolvable time advance in the EFT. We derive such `causality bounds' for shift-symmetric and Galileon scalar EFTs, before turning to gauge-symmetric vector fields. We prove that our causality bounds can be competitive with positivity bounds and can even be used in scenarios that are out of reach of the positivity approach. The result of this thesis, by exploring several consistency criteria, is to provide compact causality bounds for low-energy EFT coefficients, in addition to constraints coming from the absence of ghosts, stability and cosmological viability.

\chapter*{Acknowledgements}

I would like to thank my supervisor, Claudia de Rham, who was always available, incredibly helpful and supportive, in all possible ways. I would also like to thank all my collaborators but most importantly the inspiring Andrew Tolley and Mariana Carrillo Gonz\'{a}lez. Thank you to Arttu Rajantie and Gianmassimo Tasinato who kindly agreed to be my examiners.

\chapter*{Copyright Statement}

The copyright of this thesis rests with the author. Unless otherwise indicated, its contents are licensed under a Creative Commons Attribution-Non
Commercial 4.0 International Licence (CC BY-NC). Under this licence, you may copy and redistribute the material in any medium or format. You may also create and distribute modified versions of the work. This is on the condition that: you credit the author and do not use it, or any derivative works, for a commercial purpose. When reusing or sharing this work, ensure you make the licence terms clear to others by naming the licence and linking to the licence text. Where a work has been adapted, you should indicate that the work has been changed and describe those changes. Please seek permission from the copyright holder for uses of this work that are not included in this licence or permitted under UK Copyright Law.

\chapter*{Statement of Originality}

I confirm that this work is my own, and that contributions from others have been appropriately referenced. A significant portion of the material in this thesis has appeared in the publications listed below, but this document as a whole has not been submitted for publication or for degree assessment elsewhere.

\chapter*{List of Publications}

This thesis is built on the publications \cite{deRham:2020yet, deRham:2021efp, CarrilloGonzalez:2022fwg, deRham:2023brw} and on the work \cite{CarrilloGonzalez:2023cbf} to appear shortly. I have been the only PhD student involved on the projects \cite{deRham:2020yet, deRham:2021efp, CarrilloGonzalez:2022fwg, deRham:2023brw} and have derived all the results there. I have also derived all the results on the `causality bounds' side of the most recent project \cite{CarrilloGonzalez:2023cbf}, but haven't been involved on the derivation of the positivity bounds that briefly appear in Chapter \ref{chap:causalVector} solely as a comparison with my bounds. Authors are listed alphabetically as is customary in our field.

\begin{enumerate}
	\item \cite{deRham:2020yet} C. de Rham and \textbf{V. Pozsgay}, \textit{New class of Proca interactions}, \href{https://journals.aps.org/prd/abstract/10.1103/PhysRevD.102.083508}{\textit{Phys. Rev. D} \textbf{102} (2020) 083508} [\href{https://arxiv.org/abs/2003.13773}{2003.13773}]. \\
	\vspace{-1cm}
	\item \cite{deRham:2021efp} C. de Rham, S. Garcia-Saenz, L. Heisenberg and \textbf{V. Pozsgay}, \textit{Cosmology of Extended Proca-Nuevo}, \href{https://iopscience.iop.org/article/10.1088/1475-7516/2022/03/053}{\textit{JCAP} \textbf{03} (2022) 053} [\href{https://arxiv.org/abs/2110.14327}{2110.14327}]. \\
	\vspace{-1cm}
	\item \cite{CarrilloGonzalez:2022fwg} M. Carrillo Gonzalez, C. de Rham, \textbf{V. Pozsgay} and A. J. Tolley, \textit{Causal effective field theories}, \href{https://journals.aps.org/prd/abstract/10.1103/PhysRevD.106.105018}{\textit{Phys. Rev. D} \textbf{106} (2022) 105018} [\href{https://arxiv.org/abs/2207.03491}{2207.03491}]. \\
	\vspace{-1cm}
	\item \cite{deRham:2023brw} C. de Rham, S. Garcia-Saenz, L. Heisenberg, \textbf{V. Pozsgay} and X. Wang, \textit{To Half--Be or Not To Be?}, \href{https://link.springer.com/article/10.1007/JHEP06(2023)088}{\textit{JHEP} \textbf{06} (2023) 088} [\href{https://arxiv.org/abs/2303.05354}{2303.05354}]. \\
	\vspace{-1cm}
	\item \cite{CarrilloGonzalez:2023cbf} M. Carrillo Gonz\'alez,  C. de Rham, S. Jaitly, \textbf{V. Pozsgay} and A. Tokareva, \textit{Positivity-causality competition: a road to ultimate EFT consistency constraints}, \href{https://arxiv.org/abs/2307.04784}{2307.04784}. \\
\end{enumerate}

\tableofcontents
\newpage

\phantomsection
\listoffigures 

\newpage

\phantomsection
\listoftables 
\newpage

%

\chapter*{Conventions}
\addcontentsline{toc}{chapter}{Conventions}

\paragraph{Metric.} Throughout this thesis, we will work in $D$ spacetime dimensions, unless stated otherwise. When specifying to certain proofs, we might work in $2d$ or $4d$. The flat metric (or Minkowski metric) is written in the mostly positive convention

\begin{equation}
	\eta\mn = {\rm diag}(-1, +1, \cdots, + 1) = \eta\mnup \,.
\end{equation}
Unless stated otherwise, indices are raised and lowered using this metric. Especially, we will use square brackets to denote the trace of a given tensor and will use the Minkowski metric to contract its indices

\begin{equation}
	\left[ \Q \right] = \Q\ud{\mu}{\mu} = \eta\mnup \Q\mn = \eta\mn \Q\mnup \,.
\end{equation}
We also use symmetrization and anti-symmetrization of tensor indices with weight $1$,

\begin{equation}
	\Q_{[\mu \nu]} = \Q\mn - \Q_{\nu \mu} \,, \qquad \Q_{(\mu \nu)} = \Q\mn + \Q_{\nu \mu} \,.
\end{equation}
Finally, the totally antisymmetric Levi-Civita tensor is normalized such that

\begin{equation}
	\varepsilon^{\mu_1 \cdots \mu_n} \varepsilon_{\mu_1 \cdots \mu_n} = - n!  \,.
\end{equation}

\paragraph{Kinematics.} We will consider $2 \rightarrow 2$ scattering processes where ingoing particles $1$ and $2$ interact to produce outgoing particles $3$ and $4$. Each particle is described by its mass $m_i$ and momentum $p_i$, such that

\begin{equation}
	p_i^2 = - m_i^2 \,, \qquad p_1 + p_2 = p_3 + p_4 \,.
\end{equation}
The Mandelstam variables are defined as follows

\begin{equation}
	s = - (p_1+p_2)^2 \,, \qquad t = - (p_1 -p_3)^2 \,, \qquad u = (p_1-p_4)^2 \,,
\end{equation}
such that

\begin{equation}
	s+t+u = \sum_{i=1}^4 m_i^2 \,.
\end{equation}
For more details about the kinematics, please refer to Appendix \ref{sec:appKin}.

\chapter*{List of abbreviations}
\addcontentsline{toc}{chapter}{List of abbreviations}

\begin{table}[h!]
\begin{center}
\begin{tabular}{ l  l }
  \textbf{ADM} & \text{Arnowitt-Deser-Misner} \\
  \textbf{AdS} & \text{Anti-de Sitter} \\
  \textbf{DGP} & \text{Dvali-Gabadadze-Porrati} \\
  \textbf{BD} & \text{Boulware-Deser} \\
  \textbf{DHOST} & \text{Degenerate Higher-Order Scalar-Tensor} \\
  \textbf{DL} & \text{Decoupling Limit} \\
  \textbf{dRGT} & \text{de Rham-Gabadadze-Tolley} \\
  \textbf{dS} & \text{de Sitter} \\
  \textbf{EFT} & \text{Effective Field Theory} \\
  \textbf{EPN} & \text{Extended Proca-Nuevo} \\
  \textbf{FLRW} & \text{Friedmann–Lemaître–Robertson–Walker} \\
  \textbf{GP} & \text{Generalized Proca} \\
  \textbf{GR} & \text{General Relativity} \\
  \textbf{IR} & \text{Infrared} \\
  \textbf{LO} & \text{Leading Order} \\
  \textbf{NEV} & \text{Null Eigenvector} \\
  \textbf{NLO} & \text{Next-to-Leading Order} \\
  \textbf{PN} & \text{Proca-Nuevo} \\
  \textbf{QFT} & \text{Quantum Field Theory} \\
  \textbf{UV} & \text{Ultraviolet} \\
  \textbf{WKB} & \text{Wentzel–Kramers–Brillouin} \\
  
\end{tabular}
\end{center}
\end{table}

\chapter{Introduction}
\label{chap:intro}

This thesis sits at the crossroads of gravity, cosmology, quantum field theory (QFT) and more mathematical considerations, and I believe it perfectly illustrates the versatility and the universality of effective field theories (EFTs), which will be the main tool used to study physical phenomena throughout this manuscript.

The aspiration of physicists around the world is to understand the laws of nature. This is done through experiments, data collection, observation, reproduction, and testing, a scientific procedure with rigorous methods that can eventually lead to the recognition of a given pattern. This pattern in turn needs to be modelled by means of mathematical equations. These equations then give predictive power to the scientist which can suddenly go from a passive position to one where outcomes can be predicted before they even happen. Finally, if these innovative predictions are confirmed by further experiments, the theory or model can be consecrated as a consistent one and suddenly the door is open for technology to turn this abstract idea into concrete progress, hopefully benefiting mankind, though not always.

In the case of gravity for example, the discovery of the laws of gravitation by Newton was revolutionary as it allowed us to better understand the motion of celestial objects and our place in the universe. However, the moment where the theory hits its range of validity always comes sooner or later, and when this happens, one has to understand what went wrong in the first place and how to accommodate for strange new observations that don't fit the previous model. Newtonian gravity works perfectly for everyday life but when gravitational forces become too extreme, the deviation between the predictions and the experiments is too significant, signalling the breakdown of the model. The story is well known, Einstein came in with his theory of General Relativity, which recovers Newtonian gravity at low energy and exactly provides the expected corrections in regimes where gravitational effects are important. This is simply one example among many. Establishing a model is just a step along the way. Eventually, this model will need to be refined, and the new one too, until we possibly reach the holy grail of the Theory of Everything.

Adopting the point of view of EFTs might not be as exciting as jumping on the quest to discover such a Theory of Everything. But it's a pragmatic choice that has allowed the physics community to be able to make a lot of progress in a variety of fields ranging from statistical mechanics to particle physics, hydrodynamics and subjects closer to the focus of this thesis such as gravity, cosmology and QFT. Working with EFTs is making the choice of humility: we might not know the deep physical processes happening at high energies, but we don't need them strictly speaking to make predictions. The theory at hand will inevitably possess a lot of coefficients whose exact values are inherited by high-energy processes (momentarily) beyond our reach. These values, rather than being calculated with the equations of the model are instead fixed by experiments, but in the end, it still results in a fully predictive theory (at least within a given range). This thesis aims to explore diverse theoretical rather than experimental tools to fix or at least constrain these EFT coefficients, using fundamental physical principles.

\pagebreak

\section{(Extended) Proca-Nuevo}
\label{sec:introEPN}

Ever since its original formulation, General Relativity (GR) has been tirelessly tested and even though there exists discrepancies in cosmological measurements (most well-known are the Hubble and the $\sigma_8$ tensions), gravitational wave experiments and predictions are in agreement with an unexpected degree of precision. GR is one of the most successful physical theories but it leaves some cosmological questions unanswered. Indeed, the universe's expansion can be explained by the introduction of dark matter in addition to a cosmological constant but its value is not technically natural. Despite decades of efforts, no fully satisfying argument has been proposed to tackle the cosmological constant problem \cite{Weinberg:1988cp}. This motivates the study of modified theories of gravity as well as  theories endowed  with additional degrees of freedom. A scalar field can indeed lead to an accelerated expansion while preserving a homogeneous and isotropic matter distribution. In this context, the Galileon was introduced in \cite{Nicolis:2008in} and the Generalized Galileon in \cite{Deffayet:2009mn} as the most general interactions for a scalar field that remain free from \Ostro instabilities. It turns out that Galileon-like interactions date back to much earlier and first appeared thanks to Horndeski in the context of scalar--tensor theories \cite{Horndeski:1974wa}. These Horndeski and Galileon theories are ubiquitous to many models of modified gravity at large distances \cite{Dvali:2000hr,Luty:2003vm,Nicolis:2004qq,Nicolis:2008in,deRham:2009rm,deRham:2010gu,deRham:2010eu,deRham:2010ik}. 

Following this idea, modifications of General Relativity were then extended to Galileon--like theories of massive Proca fields, where it becomes possible to construct derivative self--interactions for such a massive spin--1 field without \Ostro instabilities and thus propagating only three physical degrees of freedom. Such theories, classified under the name of  Generalized Proca (GP), or sometimes vector--Galileons, were thoroughly investigated in \cite{Tasinato:2014eka, Heisenberg:2014rta, Allys:2015sht, Allys:2016jaq, Jimenez:2016isa,Heisenberg:2018vsk,GallegoCadavid:2019zke}. 

Upon constructing the GP set of interactions \cite{Tasinato:2014eka,Heisenberg:2014rta}, an important implicit ingredient is that the equations of motion for both the helicity--0 and --1 modes of the massive spin--1 field remain at most second order in derivatives. This assumption appears to be related to the requirement that the constraint is uniquely determined by the equation of motion with respect to the component  $A_0$ of the vector field\footnote{This specific assumption is not explicitly formulated as such in the generic formalism of \cite{ErrastiDiez:2019trb} but other implicit assumptions on how the constraint ought to manifest itself effectively reduce the formalism to the same type of GP interactions.}.  Under this assumption, the theory is indeed unique as shown. Phrased in this way, however, it is natural to explore whether the constraint could manifest itself differently while preserving the correct number of degrees of freedom. The analogue of this possibility was successfully explored within the context of massive gravity \cite{deRham:2010kj}, first considered in \cite{deRham:2010gu} and implemented in \cite{deRham:2011rn}. The possibility was then also later implemented within the context of scalar--tensor theories, coming under the name of `Beyond--Horndeski'
\cite{Gleyzes:2014dya,Zumalacarregui:2013pma,Langlois:2015cwa,Langlois:2015skt} and further degenerate higher--order theories (DHOSTs) were considered in \cite{Langlois:2015cwa,BenAchour:2016cay,Crisostomi:2016tcp,Crisostomi:2016czh,Ezquiaga:2016nqo,Motohashi:2016ftl}. Implementations of constraints can indeed be subtle in theories with multiple fields as highlighted in \cite{deRham:2011qq,deRham:2016wji}. With this perspective in mind, we shall consider a new type of Proca interactions dubbed Proca-Nuevo (PN) (and later Extended Proca-Nuevo (EPN)) \cite{deRham:2020yet} in Chapter \ref{chap:PN} which manifest a constraint and hence only propagate three dynamical degrees of freedom in four spacetime dimensions, but differ from the standard GP interactions. Since massive gravity has provided an original framework for exploring non--trivial implementations of constraints, it shall serve as a guiding tool in constructing consistent fully non--linear Proca interactions and will allow us to prove the existence of a new type of massive spin--1 field theory  that is free of \Ostro instabilities, and propagates the required number of degrees of freedom.

Let us turn now to the issue of the constraint algebra of vector field theories. Indeed, the proof of the existence of a constraint was given in \cite{deRham:2020yet}, but the full constraint analysis was yet to be done and was then the focus of \cite{deRham:2023brw}, which will be summarized in Chapter \ref{chap:Constraint}. In order to understand why such an analysis was needed in the first place, consider electrodynamics as a simple example. In this case, the $U(1)$ gauge symmetry ensures the existence of a first class constraint associated with the local symmetry. This constraint removes a pair of conjugate variables in phase space (equivalently two degrees of freedom in field space) leading to $D-2$ propagating degrees of freedom in $D$ dimensions. The addition of a mass term (and more generally of non-$U(1)$-invariant self-interactions\footnote{The breaking of $U(1)$ invariance has to occur at the linear level about the vacuum to avoid infinitely strong coupling issues.}) breaks this symmetry and the first class constraint is downgraded to a pair of second class constraints, each removing half a degree of freedom (see \cite{Heisenberg:2014rta,BeltranJimenez:2019wrd,ErrastiDiez:2020dux} for constraint analyses of GP). The resulting theories then propagate $D-1$ physical degrees of freedom in $D$ dimensions. The claim of the absence of ghosts in (E)PN is based on the existence of a second class constraint realized in the form of a null equation for the Hessian matrix. The existence of a second class constraint removes one degree of freedom in the $2D$-dimensional Hamiltonian phase space, potentially leaving $D-1/2$ degrees of freedom in field space and not fully exorcising the Ostrogradsky ghost. It was then argued in \cite{deRham:2020yet,deRham:2021efp} that there should exist a secondary constraint based on the fact that the existence of half degrees of freedom should not arise in Lorentz and parity invariant theories (see Section \ref{sec:HalfDegree}). Conversely, it was claimed in \cite{ErrastiDiez:2022qvd} that (E)PN might be the first counterexample to this expectation\footnote{A similar claim was made in \cite{Kluson:2011qe} that ghost-free massive gravity failed to have a secondary constraint. This constraint was later explicitly derived in \cite{Hassan:2011ea}. In the words of \cite{Hassan:2011ea}, ``{\it an odd dimensional phase space --- an odd situation indeed}''.}. Motivated by this, we present in Chapter \ref{chap:Constraint} (based on \cite{deRham:2023brw}) a complete analysis of the constraint algebra of EPN, proving that (as expected), the primary second class constraint is followed by a secondary constraint which fully takes care of eradicating the whole would-be ghostly degree of freedom. 

After having successfully proposed a new massive and ghost-free vector effective field theory, one could wonder if this only advances the classification of consistent EFTs or if EPN can be used to model interesting physical phenomena. With astrophysical and cosmological applications in mind, the embedding of these effective field theories in a fully gravitational framework is an exciting  problem connecting with  the ongoing program of classifying viable extensions of GR. Similarly to their scalar-tensor counterparts, generalized vector-tensor theories have been shown to exhibit intriguing phenomenological properties in astrophysical systems \cite{Heisenberg:2017xda,Heisenberg:2017hwb,Kase:2017egk,Kase:2018owh,Kase:2020yhw,Garcia-Saenz:2021uyv,Brihaye:2021qvc} and cosmology \cite{Jimenez:2013qsa,DeFelice:2016yws,DeFelice:2016uil,Heisenberg:2016wtr,Nakamura:2017dnf,deFelice:2017paw,Nakamura:2018oyy,DeFelice:2020sdq,Heisenberg:2020xak,Garnica:2021fuu}. In the latter case, of particular interest, is the fact that a time-dependent vector condensate could behave as a dark energy fluid, driving the observed accelerated cosmic expansion in the present-day universe, with a technically natural vector mass and dark energy scale \cite{Heisenberg:2020jtr,deRham:2021yhr}. An important milestone in this program was the discovery of GP, featuring some unique properties in relation to the screening mechanisms and the coupling to alternative theories of gravity \cite{DeFelice:2016cri,Heisenberg:2018acv,Garcia-Saenz:2021acj}.

(E)PN theory successfully exploits the fact that multi-field systems may in principle evade the Ostrogradsky theorem if the equations happen to be degenerate \cite{deRham:2016wji,BeltranJimenez:2019wrd} (see Chapters \ref{chap:PN} and \ref{chap:Constraint}), and does it through a non-trivial realization of the primary constraint, motivated by the decoupling limit of massive gravity \cite{deRham:2010kj,Ondo:2013wka}. 

Going beyond the simple flat geometry, one could wonder whether EPN can be coupled to gravity in a consistent way. This question will be the focus of Chapter \ref{chap:CovEPN} where we explore alternative covariantization schemes for EPN. We explicitly show that no ghosts appear below the Planck scale, hence paving the way for the use of the theory on more generic backgrounds. Our first objective will then be to explore the cosmological implications of (E)PN theory \cite{deRham:2021efp} in Chapters \ref{chap:CovEPN} and \ref{chap:Cosmo}. Although not consistent in full generality, we will exhibit two alternative, partial covariantization schemes that successfully describe a massive spin-1 field coupled to Einstein gravity, with no additional degrees of freedom, for cosmological solutions at the levels of both the homogeneous and isotropic background and of general linear perturbations. Our main result is that, in each setup, there exists a window of parameter values for which cosmological perturbations are free of ghost- and gradient-like instabilities and of superluminal propagation speeds. In particular, each scenario accommodates exactly luminal gravitational waves.

The first covariantization is particularly neat in that the coupling with gravity is minimal, unlike what occurs in GP theory. On the other hand, this model requires a technically-natural tuning of coefficients which has the advantage of providing a simple and tractable model with relatively few arbitrary functions. A particularly interesting property of this setup is that, without any further tuning or special choices of coefficients, tensor fluctuations propagate exactly as in GR. As a consequence, observational bounds on the production and propagation of gravitational waves do not impose any extra constraints on the theory. After deriving the stability conditions for all types of perturbations---tensor, vector and scalar---for the model coupled to perfect fluid matter, we then analyze the resulting cosmological solutions. We will see that the model exhibits hot Big Bang solutions with epochs of radiation, matter and dark energy domination, with the latter corresponding to a ``self-accelerating'' phase, being driven by the vector field condensate and not a cosmological constant. We further show that perturbations within this model are fully under control, stable and causal.

The second covariantization is more general but requires non-minimal couplings between the vector field and the curvature. These non-minimal terms are precisely those of GP, so this model has the virtue of accommodating the covariant GP theory as a particular case, which is known to be free of pathologies for various choices of parameters. We will show however that this general setup extends the cosmology of GP in interesting ways. For instance, we will prove that the dispersion relation of the Proca vector mode is non-linear, both in vacuum and when coupled to a perfect fluid. Similarly, the mixing of the perfect fluid with the extended PN sector results in a modification of the speed of propagation of the longitudinal fluctuation of the fluid, i.e.\ the phonon. As this effect is absent both in GR and in GP, it gives in principle a clean signature to test the theory and distinguish it from other vector-tensor models.

\section{Causality bounds}
\label{sec:introCausBounds}

From a bottom-up perspective, the construction of effective field theories (EFTs) based on symmetry principles allows us to compute observables in the infrared (IR) without the full knowledge of the ultraviolet (UV) completion of the theory. This has proven to be a useful approach not only in particle physics and cosmology but also when studying gravitational systems. While the EFT contains an infinite number of higher derivative interactions, at low energies only a finite number is relevant at a given order in the EFT expansion. Nevertheless, symmetry principles on their own are not sufficient to ensure that the EFT is unitary and causal. Imposing these physical principles leads to constraints on the possible values of the coefficients in the Wilsonian effective action of the low energy EFT \cite{deRham:2022hpx}. A well-known approach for bounding these Wilson coefficients consists of looking at dispersion relations for $2 \rightarrow 2$ scattering amplitudes and engineering positive bounded functions of the scattering amplitude \cite{Adams:2006sv,deRham:2017avq,deRham:2017zjm,Bellazzini:2015cra,Bellazzini:2016xrt,Vecchi:2007na,Nicolis:2009qm}\footnote{Earlier approaches in the chiral perturbation theory context are found in \cite{yndurain1972rigorous,PhysRevD.31.3027,Pennington:1994kc,Ananthanarayan:1994hf,Comellas:1995hq,Manohar:2008tc}.}. The associated positivity bounds require assumptions about the UV completion such as unitarity, locality, causality, Poincar\'e symmetry, and crossing symmetry. Additionally, one can obtain stronger bounds when considering weakly coupled tree-level UV completions. In recent years this has proven to be a fruitful approach (see for example \cite{Cheung:2016yqr,Bonifacio:2016wcb,deRham:2018qqo, Alberte:2019xfh, Alberte:2019zhd, Wang:2020xlt,deRham:2017xox,deRham:2017imi,Davighi:2021osh,Bellazzini:2019bzh,Sinha:2020win,Chandrasekaran:2018qmx,Wang:2020jxr,Du:2021byy,Li:2021lpe,Bellazzini:2020cot,Arkani-Hamed:2020blm,Guerrieri:2020bto,Tolley:2020gtv, Caron-Huot:2020cmc,Hebbar:2020ukp,Bellazzini:2021oaj,Chiang:2021ziz,Chiang:2022ltp}). Crucially in \cite{Tolley:2020gtv, Caron-Huot:2020cmc} it was shown that incorporating the constraints of full crossing symmetry, now referred to as null constraints, imposes two-sided positivity bounds generically on the space of all Wilson coefficients\footnote{This phenomenon was already noted in \cite{Cheung:2016yqr,Bonifacio:2016wcb,deRham:2018qqo,Alberte:2019xfh,Alberte:2019zhd,Wang:2020xlt} in the context of massive spin-1 and -2 theories where the two-sidedness comes from consideration of different external polarizations.}. The purpose of this thesis is to show that these two-sided bounds can be largely anticipated from low-energy causality considerations alone. 

The extension of positivity bounds to (massless) gravitational theories and arbitrarily curved spacetimes, and more specifically to time-dependent gravitational backgrounds is not straightforward \cite{Alberte:2020jsk,Alberte:2020bdz}. Gravitational amplitudes in Minkowski spacetime have recently been incorporated by using dispersive arguments that evade the $t$-channel pole inevitable in gravitational amplitudes \cite{Caron-Huot:2022ugt,Bern:2021ppb,Chowdhury:2021ynh,Caron-Huot:2022jli,Chiang:2022jep} or account for it by its implied Regge behaviour \cite{Tokuda:2020mlf,Aoki:2021ckh,Noumi:2021uuv,Alberte:2021dnj,Herrero-Valea:2020wxz,Herrero-Valea:2022lfd}. While perturbative unitarity rules can be generalized on curved spacetime \cite{deRham:2017aoj}, analyticity has proven more challenging and some initial explorations of positivity bounds to curved spacetimes were proposed in \cite{Baumann:2015nta,Baumann:2019ghk,Afkhami-Jeddi:2018own}. Further analyses considering that the bounds arising from positivity constraints around a Minkowski vacuum can be translated into bounds for Wilson coefficients around a curved vacuum are examined in \cite{Melville:2019wyy,deRham:2021fpu,Traykova:2021hbr,Kim:2019wjo,Herrero-Valea:2019hde,Ye:2019oxx}. The main difficulties in constructing dispersion relations in curved backgrounds arise due to the broken Lorentz symmetries and the lack of an S-matrix. Some progress has been made recently for broken Lorentz boost theories \cite{Grall:2021xxm} and in de Sitter spacetimes where there is an equivalent notion of positivity of spectral densities \cite{Bros:1995js,Sleight:2019hfp,Sleight:2020obc,Sleight:2021plv,Hogervorst:2021uvp,DiPietro:2021sjt}. By contrast the causality approach discussed here is easily generalizable to curved spacetimes.

We will focus on applying this method to shift-symmetric scalars \cite{CarrilloGonzalez:2022fwg} in Chapter \ref{chap:causalScalar} before exploring EFTs of massless photons \cite{CarrilloGonzalez:2023cbf} in Chapter \ref{chap:causalVector}. In both cases, we will see that causality bounds are an efficient tool to constrain low-energy EFT coefficients and that apart from isolated special configurations, two-sided bounds can be obtained and are competitive with positivity bounds, if not better. In the scalar case, we will compare our results with positivity bounds derived in \cite{Tolley:2020gtv, Caron-Huot:2020cmc}, whereas we will compare them with \cite{Adams:2006sv,Arkani-Hamed:2020blm,Henriksson:2021ymi,Henriksson:2022oeu,Haring:2022sdp} (and the ones of \cite{CarrilloGonzalez:2023cbf} worked out by my collaborators) in the vector case.

\section{Outline and summary of new results of the thesis}
\label{sec:IntroOutline}

Before diving into the heart of the thesis, we would still like to dedicate Chapter \ref{chap:tools} to the introduction of several useful theoretical tools. In particular, we will start by reviewing Effective Field Theories in a very broad way before specializing in important building blocks (Galileons, Generalized Proca and Massive Gravity) for the construction of our new model. We finish this introductory part with a few words on positivity bounds to give the reader some context regarding theoretical bounds on EFT coefficients.

This thesis will then be divided into two parts. Part \ref{part:PN} will focus on introducing a new class of ghost-free massive vector fields: (Extended) Proca-Nuevo. After having reviewed the standard quadratic Proca and Generalized Proca in the Introduction, we will go beyond and relax one of the underlying hypotheses of such theories, i.e. having second-order equations of motion, to explore a new and highly non-linear way of realizing the sought-after Hessian constraint. The (E)PN interactions are constructed in a similar fashion as the ones of the vector sector in the decoupling limit of dRGT massive gravity, hence why we took care to introduce them in Section \ref{sec:IntroMG}. The construction of (E)PN will be followed by proofs of its inequivalent nature to GP by comparing specific scattering amplitudes in Chapter \ref{chap:PN}, and by an in-depth analysis of its constraint structure, both in the Lagrangian and Hamiltonian pictures in Chapter \ref{chap:Constraint}. This will show that (E)PN is a genuinely new and ghost-free theory. We will use Chapter \ref{chap:CovEPN} to propose several covariantization schemes for EPN and prove that the constraint can be maintained on curved backgrounds. This will then lead us to use the theory in a cosmological setup to model dark energy in Chapter \ref{chap:Cosmo} and we will prove that the background is compatible with a self-accelerating hot big-bang scenario, while the linear perturbations can respect all stability conditions at once. To summarize the main results of Part \ref{part:PN}:

\begin{itemize}
	\item Chapter \ref{chap:PN}: based on publication \cite{deRham:2020yet}.
	\begin{itemize}
		\item We propose a new set of higher-derivative self-interactions of a massive spin-1 field where equations of motion are \textit{not} second-order but exhibiting a (second-class) constraint at the level of the Hessian matrix. 
		\item We find the exact analytic form of the (E)PN null eigenvector of the Hessian matrix in arbitrary spacetime dimension $D$.
		\item We prove that (E)PN and GP are inequivalent by comparing scattering amplitudes in both theories.
	\end{itemize}
	\item Chapter \ref{chap:Constraint}: based on publication \cite{deRham:2023brw}.
	\begin{itemize}
		\item We show that there exists a secondary second-class constraint in (E)PN both at the level of the Lagrangian and the Hamilton in spacetime dimension $D=2$.
		\item We also provide a full proof of its existence and a way to construct it in arbitrary higher dimensions.
		\item We rigorously extend the $2d$ proofs to standard quadratic Proca and Generalized Proca, where the constraints are well known but their full analysis seems to lack in the literature. 
	\end{itemize}
	\item Chapter \ref{chap:CovEPN}: based on publications \cite{deRham:2020yet,deRham:2021efp}
	\begin{itemize}
		\item We propose two different covariantizations of (E)PN.
		\item We prove that the (E)PN Hessian constraint is maintained on arbitrarily-curved backgrounds, and that it gets broken in a Planck-suppressed way when coupling to dynamical gravitational degrees of freedom.
		\item We show that the theory can be fully covariantized in models of cosmological importance such as FLRW background.
	\end{itemize}
	\item Chapter \ref{chap:Cosmo}: based on publication \cite{deRham:2021efp}.
	\begin{itemize}
		\item We show that the theory propagates the correct number of degrees of freedom on FLRW and that its tensor, vector and scalar perturbation sectors are subluminal and stable when coupled to a perfect fluid matter. (E)PN is used to model dark energy.
		\item We prove that gravitational waves in (E)PN propagate in the exact same way as predicted by GR.
		\item We show that we can reproduce a hot big-bang scenario with a late-time self-accelerating branch, going successively through radiation, matter and dark energy domination eras.
	\end{itemize}
\end{itemize}

Having shown the consistency and stability of (E)PN, we will then proceed to explore different ways to impose consistency in Part \ref{part:CausalityBounds}. To be more specific, we will dedicate this part to causality bounds, which is a way of bounding low-energy EFT coefficients by the sole requirement of infrared (IR) causality. We start by showing the potential of this method in the very simple shift-invariant scalar field in Chapter \ref{chap:causalScalar}. There, we will compare our results with the well-studied positivity bounds (which will be soon introduced in Section \ref{sec:IntroPosBounds}) and conclude that both methods are in very close agreement. This represents a promising result as it confirms the validity of the approach and hence, opens new possibilities since causality bounds can be applied in some range that is out of reach of the positivity bounds. Indeed, arbitrarily curved spacetime and potential interactions (non-derivative) among other examples can be constrained using causality but not with positivity. In Chapter \ref{chap:causalVector}, we extend the analysis to massless spin-1 particles, making a connection with Part \ref{part:PN} and extending the consistency analysis of such models. The main results of Part \ref{part:CausalityBounds} are gathered below.

\begin{itemize}
	\item Chapter \ref{chap:causalScalar}: based on publication \cite{CarrilloGonzalez:2022fwg}.
	\begin{itemize}
		\item We compute the time delay of shift-symmetric scalars both on a homogeneous and a spherically symmetric background and impose our low-energy causality criteria that no resolvable time advance should be observed.
		\item We derive numerically optimized bounds that are consistent with the requirement of IR causality. We produce two-sided bounds that are in close agreement with previously derived compact positivity bounds.
		\item We rule out most of the configuration space of Galileon-symmetric scalar fields, in agreement with positivity bounds where no such theories satisfy positivity.
	\end{itemize}
	\item Chapter \ref{chap:causalVector}: based on publication \cite{CarrilloGonzalez:2023cbf}
	\begin{itemize}
	\item We perform a similar analysis to what was done in Chapter \ref{chap:causalScalar} \cite{CarrilloGonzalez:2022fwg} applied to an EFT of massless photons. The main difference is that they propagate two degrees of freedom. We adapt our formalism to extract the dynamics of the two uncoupled modes.
	\item We optimize our algorithm to derive tighter bounds in a more automatized way.
	\item We perform a consistency check of the validity of our results by showing that all (partial) UV completion lie within our bounds, often exactly on their boundary.
	\item We produce compact causality bounds (or at least double-sided at worst in special cases). We compare our results to positivity bounds and show that our bounds are often more competitive than the ones derived using positivity methods. More, their union is a powerful tool that can reduce the space of parameters to a single point, exactly corresponding to a viable partial UV completion.
	\end{itemize}
\end{itemize}

\chapter{Useful tools}
\label{chap:tools}

The focus of this thesis will be scalar and vector Effective Field Theories (EFTs) and ways to build them consistently. To this end, it is necessary to start with a short introduction to the very concept of an EFT in Section \ref{sec:IntroEFT}, where we will discuss the basic ideas and tools needed to navigate this thesis. In Part \ref{part:PN}, we will introduce Extended Proca-Nuevo (EPN), a new EFT of a massive spin-1 field with higher-order self-interactions and no ghostly propagating degrees of freedom. It was first realized in Galileon (and even Horndeski) theories that one could use a Levi-Civita structure to construct such higher-order operators while preserving at most second-order equations of motion. These interactions were then extended to massive spin-1 in the context of Generalized Proca (GP). Because Galileon and Generalized Proca were essential in the process of discovering EPN, we will introduce them in Sections \ref{sec:IntroGal} and \ref{sec:IntroGP} respectively. However, what differentiates EPN from Galileons and GP is the highly non-linear way in which the constraint is realized in EPN, and this specific aspect is inspired by the de Rham-Gabadadze-Tolley (dRGT) theory of massive gravity. This model has received a lot of interest in the recent past, especially in cosmology, but we will focus on its consistency and briefly discuss it in Section \ref{sec:IntroMG}.

In Part \ref{part:CausalityBounds}, we construct bounds on low-energy EFT coefficients based on the sole requirement that there should be no resolvable time advance in the theory. This relies on infrared (IR) causality only, as opposed to positivity bounds, which make use of causality in the ultraviolet (UV) too. Even though this thesis does not aim to derive any positivity bounds, we make sure our bounds are sensible by comparing them to the latter in the well-tested context of shift-symmetric Galileon theories and find very good agreement. We find it useful to briefly introduce the method behind these positivity bounds in Section \ref{sec:IntroPosBounds}. We then use our causality bounds on gauge-symmetric vector fields and obtain competitive results in some sectors, while even surpassing positivity bounds in others. It is worth noting that positivity methods have limitations due to the very definition of the S-matrix. Our method does not rely on such a definition, and hence causality bounds can be used to constrain theories beyond the reach of positivity. To be more concrete, EFTs on arbitrarily curved backgrounds can be constrained by causality bounds for instance. We will conclude this Chapter by providing an outline of the thesis in Section \ref{sec:IntroOutline}.

\section{Effective Field Theories}
\label{sec:IntroEFT}

\begin{figure}[h]
  \centering
  \includegraphics[width=0.2\textwidth]{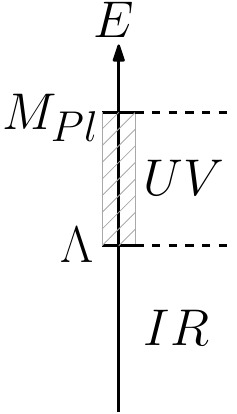}
  \caption{Ultraviolet and infrared regimes in Effective Field Theories.}
  \label{fig:EnergyScales}
\end{figure}

An Effective Field Theory (EFT) is a theory that does not pretend to model the physics of processes at all energies but only below a given energy cut-off $\Lambda$, which constitutes the regime of validity of the EFT. The description of physical phenomena at energies above the cut-off $\Lambda$ is known as a UV completion of the EFT. Here UV refers to ultraviolet and is an analogy to electromagnetic waves with higher energies than the visible spectrum. Describing a physical process with an EFT could be used in one of the following two situations:

\begin{enumerate}
	\item There exists a (partial) UV completion but the energy at which the experiment is performed is well within the regime of validity of the EFT. In this case, the EFT prediction is very close to the prediction of the UV complete theory while being considerably simpler. The high-energy (or equivalently microscopic) details are not necessary for a low-energy (macroscopic) experiment and the description becomes effective. This is the basis of the top-down approach.
	\item The full description at higher energies (the UV completion) is not known. The explicit form of the low-energy EFT has some free coefficients (known as Wilson coefficients) that need to be fixed by experiment. Pushing these experiments to higher and higher energies allows us to better approximate the unknown UV complete theory. This constitutes the basis of the bottom-up approach.
\end{enumerate}
In full generality, any EFT in $D$ spacetime dimensions can be written in the following way
\begin{equation}
	\L_{\rm EFT} = \Lambda^D \sum_n c_n \mathcal{O}_n[\psi] \,,
	\label{eq:LEFTintro}
\end{equation}
where $c_n$ are the so-called low-energy Wilson coefficients and $\mathcal{O}_n$ are derivative operators acting on the field content of the model, collectively denoted as $\psi$. In this picture, the Wilson coefficients are dimensionless and the typical strength of each operator $\mathcal{O}_n$ is roughly given by $(E/\Lambda)^n$ where $E$ is the energy at which the experiment is performed (see Figure \ref{fig:EnergyScales}). In the same fashion as any Taylor expansion, the smaller the control parameter is (here the fraction of the energy cut-off $\epsilon=E/\Lambda$) the better the approximation gets. This also means that one can safely operate a truncation at a lower-order in the dimension of the operators $\mathcal{O}_n$ and maintain a high level of accuracy in their prediction.

\paragraph{Top-down approach.} 

This aspect of the EFT is not essential to this thesis and hence will be quickly reviewed before going to the main appeal of using an EFT: the bottom-up approach. Indeed, the top-down approach supposes that the UV completion is known and hence the EFT is nothing else but an approximation, which nevertheless can be useful, but is simply a calculation trick. In the case where $E/\Lambda \ll 1$, the EFT can be thought of as a Taylor series of the full theory with Taylor coefficients corresponding to the Wilson ones and the expansion parameter being $\epsilon = E/\Lambda \ll 1$. One can get arbitrary precision by including the correct number of terms in the series, whose coefficients are computed with the help of the full UV complete function known to arbitrarily high energies.

To give a simple physical example, let's focus on the fact that an increasing energy scale corresponds to probing matter with higher precision. Consider the example of protons and neutrons, which were thought to be elementary but are now known to possess an inner structure made of quarks and gluons. This microscopic information only becomes accessible when probing these particles (in a particle collider for instance) with an energy at least comparable to the inverse of their spatial extension. Macroscopically, the effect of the massive particles is `integrated out' in a sense that will be made clearer below and one gets an effective description of low-energy quantities, such as the quantum numbers. It is worth noting that this effective description, though not fully satisfying, is enough to understand any physical phenomena ever measured before CERN's construction.

In more precise terms, the quarks can be `integrated out' from the partial UV complete theory, leaving an EFT of the proton where the Wilson coefficients find their roots in the properties of the massive quarks that were integrated out. If we consider a theory with a collection of light fields $\psi_L$ whose masses are lower than the cut-off $\Lambda$ and heavy fields $\psi_H$ with masses beyond the regime of validity of the EFT, then the path integral reads
\begin{equation}
	\mathcal{Z} = \int \mathcal{D}\psi_H \mathcal{D} \psi_L e^{S[\psi_H,\psi_L]} \,.
\end{equation}
One can proceed to the integration of all the heavy fields $\psi_H$, leaving an effective action for the light fields $\psi_L$ with cut-off scale $\Lambda$
\begin{equation}
	\mathcal{Z} = \int \mathcal{D}\psi_L e^{S^{\rm EFT}_{\Lambda}[\psi_L]} \,.
\end{equation}
The effective action now only depends on the light degrees of freedom $\psi_L$ and is only valid up to the energy scale $\Lambda$, which will parametrize the strength of the interactions of the light degrees of freedom. Generically, such an EFT might suffer from issues such as being non-renormalizable, non-unitary, having ghostly degrees of freedom, etc. However, all such violations happen at energy scales parametrically above the cut-off $\Lambda$, where the EFT breaks down anyway.

\paragraph{Bottom-up approach.}

We have seen that an Effective Field Theory can be thought of as an approximation of the physics below a given energy scale by integrating out the effects of micro-physics above the cut-off. This is only possible when one knows the full (or at least partial) UV completion of the studied theory. More often than not, physicists find themselves wanting to describe phenomena for which they don't have a complete model. The natural solution is then to resort to EFTs. The strategy is to write down every possible operator compatible with the symmetry of the problem and to sort them in an expansion in the dimension of these operators, such as written in Eq.~\eqref{eq:LEFTintro}. To make the dimensions more explicit, it is often useful to express the Wilson coefficients in a dimensionless way,
\begin{equation}
	\L_{\rm EFT} = \Lambda^D \sum_n c_n \frac{\mathcal{O}_n[\psi]}{\Lambda^n} = \sum_n c_n \, \Lambda^{D-n} \mathcal{O}_n[\psi] \,,
	\label{eq:LEFTdimlessintro}
\end{equation}
where the mass dimension of the operators $\mathcal{O}_n$ is equal to $n$. Note that to make this identification, we work in natural units where $\hbar = c = 1$, such that energies have dimensions of mass and lengths of inverse mass. Hence, we can write $\left[ \mathcal{O}_n \right]=n$. In this context, operators with dimension $n<D$ are called relevant as their contribution grows when the energy decreases, whereas the ones with $n>D$ are called irrelevant. While deep inside the EFT regime of validity, the contribution of the irrelevant operators becomes vanishingly small and can then safely be truncated to a potentially low order, to be determined by the expected precision of the result. 

Once the complete set of linearly independent operators \footnote{The operators should be linearly independent at the level of the equations of motion as this avoids any redundancy given by integration by parts of potential total derivatives in the action.} is found and the order of truncation is agreed upon, some physical observables can then be computed. Of notable importance, especially in relation to particle physics, one can compute scattering amplitudes within an EFT. Note that at this step, the results are not predictive since they depend on a potentially very large number of Wilson coefficients. The way to circumvent this issue is to match these low-energy EFT coefficients by comparing the scattering amplitudes calculated within the EFT framework with experiments, e.g. data from particle colliders. By comparing enough scattering amplitudes (say at different energies, between different particle species or different helicities), all Wilson coefficients can be fixed \textit{once and for all}. Once all freedom is completely eliminated from the low-energy EFT, the latter becomes a predictive tool whose robustness can be tested against different sets of experiments.

In the remainder of this thesis, we will be working with a bottom-up approach as we will try to model degrees of freedom whose partial or full UV completions have not been found yet.

\paragraph{Independence.}

The crucial starting point of any EFT building is to write down the complete set of independent operators respecting a given symmetry. This task can be cumbersome but represents nothing else than a rigorous classification, and as such, should be straightforward. However, how can one be sure that the chosen basis isn't redundant? 

The first subtlety to be aware of is the inclusion of total derivatives in the Lagrangian. It is a well-known fact that for any theory on flat spacetime with appropriate boundary conditions (such that the value of the fields and their derivatives vanish at the boundary), a total derivative term in the (unphysical) Lagrangian doesn't contribute at the level of the (physical) action. It follows that two Lagrangians that only differ by total derivatives are physically equivalent. This can be an obstruction to the clarity of the linear independence of the basis of operators chosen.

To evade this problem, it is more physical to work at the level of the equations of motion rather than the Lagrangian. First, start by noting that the equations of motion are defined to be the variation of the action $S$ with respect to the fields $\psi$,
\begin{equation}
	\mathcal{E} = \frac{\delta S}{\delta \psi} \,.
\end{equation}
Now, if we have two Lagrangians $\L_1$ and $\L_2$ such that
\begin{equation}
	\Delta \L = \L_2 - \L_1 = \p_{\mu} X^{\mu} \,,
\end{equation}
for a given Lorentz tensor $X^{\mu}$, then their respective equations of motion $\mathcal{E}_1$ and $\mathcal{E}_2$ exactly coincide. With this in mind, one should construct a set of independent operators by 
\begin{itemize}
	\item Listing \textit{all} operators compatible with the symmetry of the action.
	\item Computing the equations of motion.
	\item Selecting the linearly independent operators.
\end{itemize}
However, in practice, some redundancies might remain. Indeed, the only physical quantities that one is interested in are the various scattering amplitudes that can be derived for the set of degrees of freedom and helicities described by the EFT. Hence, it is at the level of the complete set of amplitudes that the number of physically independent operators of the EFT can be found.

\paragraph{Ghosts.}

There exist some `good' ghosts (like the Fadeev-Popov ghosts, that bear no instabilities and are used in regularization procedures) but we will not be talking about these in this thesis. The focus is on the `bad' ghosts, whose main physical signature is that they destabilize the theory and make it unhealthy. There are several ways to think about this type of ghost, the first of which is the appearance of the so-called \Ostro instability. Such instability appears when a given mode of the theory features a negative sign for its kinetic energy (or at the level of its propagator). This means that the total energy of the system is unbounded from below and hence it is physically favourable to infinitely excite such ghostly modes and destabilize the vacuum of the theory.

Conversely, the \Ostro instability is linked with the fact that one would in principle need more than $2$ initial conditions (for the field value and its first derivative) to solve the differential equation and hence there is propagation of extra non-physical and unwanted degrees of freedom. Such instability manifests itself with higher-order ($>2$) time derivatives acting on the field in the equations of motion. It was proven by \Ostro that theories with higher-order derivatives acting on a field (up to the caveat of field-redefinition and other subtleties explained below) inevitably lead to the propagation of more than one degree of freedom, with at least one entering with the wrong sign for its kinetic term, hence linking high-order time derivatives to the unboundedness of the Hamiltonian. However, one needs to take great care because the inverse is not true. Indeed, having higher-order time derivatives at the level of the equation of motion is not necessarily a diagnosis of an instability. For example, the Lagrangian
\begin{equation}
	\L = - \frac12 m^2 \left( \phi + \frac{1}{\Lambda^2} \Box \phi \right)^2 - \frac12 \left( \p_{\mu} \phi + \frac{1}{\Lambda^2} \Box \p_{\mu} \phi \right)^2 \,,
\end{equation}
is nothing else than the free massive theory for a single scalar degree of freedom after the field-redefinition
\begin{equation}
	\phi \rightarrow \phi + \frac{1}{\Lambda^2} \Box \phi \,,
\end{equation}
and as such, is perfectly stable since physical observables are invariant under field-redefinition. However, it leads to manifestly higher equations of motion ($6^{\rm th}$ order!)
\begin{equation}
	\mathcal{E} = \left( 1 + \frac{\Box}{\Lambda^2} \right)^2 (\Box - m^2) \phi := 0 \,.
\end{equation}
A cleaner way to diagnose the presence of ghosts is to study the theory in the Hamiltonian picture, i.e. after Legendre-transforming the Lagrangian. Note that we will not enter into much detail about this procedure in the case of higher-order Lagrangians but this will be explained in Chapter \ref{chap:Constraint}. For such Lagrangians, the Legendre transform is not enough to reach the physically equivalent Hamiltonian picture, one needs to include possible auxiliary fields and (second-class) constraints and make sure that the equations of motion are preserved.

Let's now go back to the simplest case and illustrate our point schematically. Usually, the different degrees of freedom are mixed and one needs to diagonalize them in order to get a sum of decoupled kinetic terms for each propagating degree of freedom,
\begin{equation}
	\mathcal{H} \supset \sum_{i} \frac12 \lambda_i m^2 \dot{\psi}_i^2 \,,
\end{equation}
where $\psi_i$ are the degrees of freedom and $\lambda_i$ their associated Hessian eigenvalue. The sign of these eigenvalues is crucial in the understanding of the nature of the polarizations. A positive eigenvalue is associated with a physical propagating degree of freedom whereas a vanishing one signals that the degree of freedom is frozen or does not propagate. The instability arises when $\lambda_i < 0$. Indeed, in this case, the total energy balance of the system decreases as the mode $\psi_i$ gets excited. Since it is physically favourable to lower the total energy, one ends up in a situation where an infinity of excitations of the mode $\psi_i$ arise out of vacuum and destabilize the theory. Such degrees of freedom are what we will refer to as ghosts in this thesis.

In the following, we will not diagnose ghosts at the level of the equations of motion but really at the level of the kernel of the Hamiltonian kinetic term. This kernel can easily be extracted and corresponds to the Hessian matrix which we introduce in Eq.~\eqref{eq:HessianFirstTime} and discuss at length in the bulk of the thesis.

\section{Galileons}
\label{sec:IntroGal}

Modified theories of gravity or additional degrees of freedom can be motivated by the will to solve the cosmological constant problem. The easiest and most natural addition to General Relativity is the inclusion of a scalar field, which trivially preserves homogeneity and isotropy of the matter distribution while explicitly providing an accelerating universe. Hence, scalar fields coupled to GR are interesting from a cosmological point of view. In this context, the Galileon was first introduced in \cite{Nicolis:2008in} before being generalized to arbitrarily curved backgrounds in \cite{Deffayet:2009mn}.

The original idea behind the Galileon was inspired by the DGP model, a $5d$ modified gravity theory named after its discoverers Dvali, Gabadadze and Porrati \cite{Dvali:2000hr}. This model makes the hypothesis that our $4d$ spacetime is a localized brane embedded in a $5$-dimensional space. The action is composed of a $4d$ Einstein-Hilbert term on the brane and an equivalent $5d$ term in the bulk. Furthermore, there exists a hierarchy of scales between the $5d$ and $4d$ scales of interaction of gravity, namely the Planck mass, such that $M_5 \ll M_{\rm Pl}$. This means that the usual $4d$ Planck mass is much larger than its $5d$ equivalent. Such a hierarchy leads to gravity behaving as GR on short distances before transitioning to a regime where the $5d$ Einstein-Hilbert term dominates at larger distances.

The appeal of such a model is that it possesses a so-called self-accelerating branch of solution, meaning that it doesn't require any cosmological constant to provide cosmic acceleration of the dark energy. The number of degrees of freedom propagated by the particle carrying the gravitation interaction, the graviton, is now equal to $5$ in this intrinsically $5$-dimensional theory. On the $4d$ brane, it is possible to show that one of the degrees of freedom behaves like a spin-0 particle or a scalar. The DGP model admits a limit where effectively the higher-dimensional gravity is turned off while preserving the number of degrees of freedom in the theory. This is achieved by taking both interaction scales to infinity, with $M_5 / M_{\rm Pl} \rightarrow 0$, while preserving a certain composite scale constant. This scale will act as the strong coupling scale for the helicity-$0$ mode in the so-called Decoupling Limit (DL). As its name suggests, the scalar degree of freedom is maintained while its dynamics are decoupled from the rest of the content of the theory and thus acts as a single scalar field on top of a $4$-dimensional theory of gravity.

In this DL, all notions of an initial embedding in a $4+1$ Minkowski spacetime are captured in the self-interactions of the helicity-$0$ mode, denoted by $\pi$. Note that $\pi$ will correspond to the cubic Galileon. The name reflects the fact that the scalar theory is invariant under the transformations $\p_{\mu} \pi \rightarrow \p_{\mu} \pi + b_{\mu}$ (where $b_{\mu}$ is a constant vector), resembling the non-relativistic Galilean transformation $\dot{x} \rightarrow \dot{x} + v$. It appears that the Galileon interactions are the most general scalar interactions that do not lead to any Ostrogradsky instabilities. Furthermore, the equations of motion remain second-order even though the interactions include higher-order derivatives of the field. Note that Galileons were actually first introduced decades earlier in the context of Horndeski scalar-tensor theories \cite{Horndeski:1974wa} and are now present in a large class of models where gravity is modified at large distances \cite{Dvali:2000hr,Luty:2003vm,Nicolis:2004qq,Nicolis:2008in,deRham:2009rm,deRham:2010gu,deRham:2010eu,deRham:2010ik}. It is also worth noting that Galileon interactions are technically natural in the sense that they are stable under quantum corrections \cite{Luty:2003vm,Nicolis:2004qq,Nicolis:2008in,deRham:2010eu,Burrage:2010cu,Burrage:2011bt,deRham:2012ew,deRham:2012az}.

Let us now review the Galileon interactions. The DL of the DGP model produced the term $(\p \pi)^2 \Box \pi$ which is cubic in the field and hence corresponds to the cubic Galileon. However, the class of Galileon interactions in the sense of the most general scalar theory free of Ostrogradsky instabilities is wider and includes terms having up to $(D+1)$ copies of the field $\pi$ in $D$ dimensions. There are a variety of ways to write down its Lagrangian but the most compact one is to make use of the Levi-Civita symbol. In arbitrary spacetime dimension $D$, there exist $D-1$ independent interacting terms reading
\begin{equation}
	\L^{\rm (Gal,D)} = - \frac12 (\p \pi)^2 + \Lambda^D \sum_{n=3}^{D+1} c_n \frac{\L_n^{\rm (Gal)}}{\Lambda^{n(D/2+1)-2}} \,,
	\label{eq:IntroGalLagr}
\end{equation}
where each individual term $\L_n^{\rm (Gal)}$ is built out of symmetric polynomials, and where $\Lambda$ corresponds to the scale of the Galileon interactions. The total Lagrangian doesn't rely on specific tunings of the individual Lagrangians since they all respect the Galileon symmetry on their own. Hence, they can all be added with an arbitrary low-energy coefficient $c_n$. Before writing explicitly each individual term, we note that they schematically follow $\L_n^{\rm (Gal)} \sim (\p \pi)^2 (\p^2 \pi)^{n-2}$. Now, in arbitrary dimension $D$, the mass dimension of the scalar field $\pi$ is $D/2-1$ and hence the mass dimension of $\L_n^{\rm (Gal)}$ is $n(D/2+1)-2$ which motivates the effective field theory (EFT) expansion provided in Eq.~\eqref{eq:IntroGalLagr}. Note that we omitted $\L_1^{\rm (Gal)} = - \pi$ which corresponds to a tadpole, and that we explicitly wrote down the kinetic term $\L_2^{\rm (Gal)} = - \frac12 (\p \pi)^2$.

Before writing down the full Galileon Lagrangian, let's introduce the symmetric polynomials of a given symmetric tensor $\K\ud{\mu}{\nu}$ in arbitrary dimension $D$,
\begin{equation}
	\L_n[\K] = -\frac{1}{(D-n)!} \varepsilon^{\mu_1 \cdots \mu_n \mu_{n+1} \cdots \mu_D} \varepsilon_{\nu_1 \cdots \nu_n \mu_{n+1} \cdots \mu_D} \K^{\nu_1}_{\phantom{\nu_1}\mu_1} \cdots \K^{\nu_n}_{\phantom{\nu_n}\mu_n}\,.
	\label{eq:refLnK}
\end{equation}
Note that $n$ can take values between $0$ and $D$. These interactions can be written down explicitly in terms of various traces of the tensor $\K$ and when specializing to $4d$ we get
\begin{equation}
\begin{aligned}
	\L_0[\K] &= 1 \\
	\L_1[\K] &= [\K]  \\
	\L_2[\K] &= ([\K]^2 - [\K^2])  \\
	\L_3[\K] &= [\K]^3 - 3[\K][\K^2] + 2[\K^3]\\
	\L_4[\K] &= [\K]^4 - 6[\K]^2[\K^2] + 3[\K^2]^2 + 8[\K][\K^3] - 6[\K^4] \,,
\end{aligned}
\label{eq:refLnK0to4}
\end{equation}
where we use square brackets to denote the trace of a tensor. The remaining Galileon interactions are then given by
\begin{equation}
	\L_n^{\rm (Gal)} = (\p \pi)^2 \L_{n-2}[\Pi] \,,
	\label{eq:IntroLGal}
\end{equation}
where we introduced $\Pi\mn = \p_{\mu} \p_{\nu} \pi$. For a generic term of order $n$ in dimension $D$, there are $2(n-1)$ derivatives acting on $n$ field $\pi$. If one were to write an arbitrary term of the form $\p^{2(n-1)} \pi^n$, it would inevitably produce higher-order equations of motion, which would generically result in an Ostrogradsky instability. However, the specific antisymmetric structure of the Levi-Civita symbols prevents this from happening and instead produces second-order equations of motion. Indeed, if we write $\mathcal{E}_n \equiv \delta \L_n^{\rm (Gal)}/ \delta \pi$, the equations of motion are surprisingly simple and imply reduce to $\mathcal{E}_n \propto \L_{n-1}[\Pi]$, which only depend on $\Pi\mn = \p_{\mu} \p_{\nu} \pi$ and hence are at most second order in time derivatives. This ensures the absence of any Ostrogradsky instability and hence the theory is consistent. From a pure EFT point of view, the Galileon theory is an interesting realization of ghost-free higher-order derivative self-interactions.

\section{Generalized Proca}
\label{sec:IntroGP}

\subsection{A primer on vector fields}
\label{ssec:PrimerVector}

After having introduced the Galileon class of scalar fields, let's now turn to their vector counterpart, namely Generalized Proca (GP). But before this, let's go back to the basics of spin-1 fields. Spin-1 particles are bosons found everywhere around us and  mediating different elementary forces. They can be either massive or massless, the latter enjoying an additional $U(1)$ gauge symmetry and describing a photon, the particle that mediates electromagnetism. If one defines $A_{\mu}$ as the vector field, then the field-strength tensor 
\begin{equation}
	F\mn = \p_{\mu} A_{\nu} - \p_{\nu} A_{\mu} \,,
\end{equation}
is the only linear gauge (and parity) invariant quantity, i.e. is invariant under $A_{\mu} \rightarrow A_{\mu} + \p_{\mu} \chi$ for a generic scalar $\chi$ (if one forgets about its dual tensor $\tilde{F}\mn$ which is parity-breaking). Any (parity-preserving) massless vector field theory needs to be written as a function of this tensor only, and in particular, the kinetic term needs to be canonically normalized in the following way
\begin{align}
	\L_{\rm (massless)}[A] &= - \frac14 F\mnup F\mn + \L^{\rm (higher)}(F, \p F, ...).
\end{align}
By antisymmetry, it is clear that the temporal component of the vector field $A_0$ does not propagate since $F_{00}=0$ and hence the Lagrangian is independent of $\dot{A}_0$. It is easy to show that the Hamiltonian density is linear in the constant $A_0$ and includes the term $- A_0 \vec{\nabla} \cdot \vec{E}$ where $E_j = - F_{0j}$ is the electric field. The temporal component of the vector field hence acts as a Lagrange multiplier for the Gauss Law $\vec{\nabla} \cdot \vec{E}=0$ that has to be imposed as a constraint. The $U(1)$ gauge symmetry removes two extra degrees of freedom in phase space. Indeed, one can use the Coulomb gauge $\vec{\nabla} \cdot \vec{E}=0$ to fix the gauge and use the residual gauge freedom to set $A_0=0$, leaving two propagating degrees of freedom, or polarizations, for the photon \cite{Henneaux:1992ig}. These correspond to the electric and magnetic fields and are normal to the direction of propagation of the photon.

Adding a mass term of the form of $-\frac12 m^2 A^2$ explicitly breaks the gauge invariance and downgrades the first-class constraint to a second-class constraint, meaning that the field now propagates $3$ degrees of freedom in $4$ dimensions. Now that the theory is no longer gauge-invariant, the form of the higher-order derivative operators is less restrictive and, in principle, one can have any contraction of the field and its derivatives (even though we will see that there exist some constraints)
\begin{equation}
	\L_{\rm (massive)}[A] = - \frac14 F\mnup F\mn - \frac12 m^2 A^2 + \L^{\rm (higher)}(A, \p A, ...).
	\label{eq:LmassiveA}
\end{equation}

\paragraph{Hessian matrix.}

Let's come back to the form of Eq.~\eqref{eq:LmassiveA} and clarify what types of derivative self-interactions are allowed. Indeed, we will soon see that not \textit{any} operator can lead to a healthy theory and some will excite the last degree of freedom, effectively giving rise to the propagation of a ghost and an Ostrogradsky instability. A simple and efficient way to diagnose the existence of a ghost is to compute the eigenvalues of the Hessian matrix. The latter is nothing else than the kernel of the kinetic term, i.e. isolates the part of the Lagrangian that is quadratic in the velocities, and takes the form
\begin{equation}
	\mathcal{H}\mnup[A] = \frac{\p^2 \L[A]}{\p \dot{A}_{\mu} \p \dot{A}_{\nu}} \,.
	\label{eq:HessianFirstTime}
\end{equation} 
The eigenvalues of this matrix give information on the propagating degrees of freedom and the theory is healthy if and only if the would-be ghostly degree of freedom is absent, which translates into the fact that at least one of the eigenvalues vanishes. There are at least $3$ degrees of freedom for any massive theory whose vacuum state is perturbatively related to the Proca one. Hence, a ghost-free massive vector theory possesses one and exactly one vanishing eigenvalue which implies
\begin{equation}
	{\rm det}(\mathcal{H}) = 0 \,.
\end{equation}
This second-class constraint can equivalently be written down as the existence of a null eigenvector (NEV) $V_{\mu}[A]$ such that
\begin{equation}
	\mathcal{H}\mnup V_{\mu} = 0 \,.
\end{equation}
In the standard quadratic Proca theory, it remains true that $A_0$ does not propagate and the NEV takes the simple form $V_{\mu} = \delta^0_{\mu}$. Hence, any massive vector model needs to have its NEV perturbatively related to $(1,\vec{0})$.

\paragraph{\stu fields.}

Before moving on to the formulation of the Generalized Proca theory, let's introduce an important EFT tool, the \stu field. The idea is to restore a given symmetry, gauge symmetry in this case, by introducing an extra field to the theory with a carefully-chosen transformation law. This way, the identification of the degrees of freedom becomes more automatic and straightforward. For the simple example of the quadratic Proca, the mass term breaks $U(1)$ symmetry and the vector field $A_{\mu}$ acquires a third polarization, which is longitudinal, i.e. aligned with the momentum of the particle.

The gauge symmetry one would like to restore is given by $A_{\mu} \rightarrow A_{\mu} + \p_{\mu} \chi/m$ (where the mass normalization is not strictly necessary but will prove convenient for the normalization of the kinetic term of the \stu field) and the idea is to write a theory that doesn't depend solely on $A_{\mu}$ anymore but on a combination of the vector field and a new scalar \stu field $\phi$, such that the field-strength tensor $F\mn$ remains invariant. It is clear that the only possibility is to work with a combination of the form $(A_{\mu} - \p_{\mu} \phi / m )$. The new Lagrangian now reads
\begin{equation}
	\L_{\rm \text{(\stu)}} = - \frac14 F\mnup F\mn - \frac12 m^2 \left( A_{\mu} - \frac{\p_{\mu} \phi}{m} \right)^2 \,,
\end{equation}
which is invariant under the following simultaneous transformation of the vector and \stu fields
\begin{equation}
\begin{aligned}
	A_{\mu} \rightarrow& \; A_{\mu} + \p_{\mu} \chi/m \\
	\phi \rightarrow& \; \phi - \chi \,.
\end{aligned}
\end{equation}
The new Lagrangian now enjoys the sought-after $U(1)$ gauge symmetry and hence the vector field $A_{\mu}$ propagates $2$ physical degrees of freedom as in electrodynamics. The third degree of freedom that arose due to the inclusion of a mass term for the vector field can then be absorbed by a \stu scalar field, hence its common denomination as the $0$-mode of the massive vector field. The total number of degrees of freedom remains unchanged but the longitudinal mode of the massive vector field has been absorbed into the \stu scalar field.

\subsection{Formulation of Generalized Proca}
\label{ssec:GP}

Generalized Proca is the most general theory of a massive vector field $A_{\mu}$ including an arbitrary number of derivative self--interactions such that its equations of motion remain second order and is free of \Ostro instabilities when including the helicity--0 part $\phi$ of the \stu field $A_\mu \to A_\mu +\p_\mu \phi/m$. This property  ensures that the theory has three propagating degrees of freedom in four dimensions\footnote{As we shall see the requirement that the equation of motion for $\phi$ remains second order in derivatives is a sufficient condition for the absence of \Ostro instabilities but not always a necessary one.}. In this language,  the helicity--0 mode $\phi$ is then nothing other than a Galileon.

Requiring the equations of motion to be at most second order in derivatives implies that GP interactions include at most one derivative per field at the level of the action and are hence solely expressed in terms of $A_\mu$ and $\p_\mu A_\nu$. In deriving the full action, it is useful to separate out the gauge--invariant building blocks \ie the Maxwell strength field $F\mn$ and its dual $\tilde F\mn$ and the gauge--breaking contributions that involve the \stu field $\phi$. One can then parameterize the GP Lagrangians in terms of the powers of the gauge--breaking contribution $\p A$. The advantage of this ordering is that it is finite in the sense that \textit{all} the interactions are listed, and the remaining infinite freedom is captured by arbitrary functions. In this language, we have \cite{Tasinato:2014eka, Heisenberg:2014rta}
\begin{equation}
	\L_{\text{GP}} = \sum_{n=2}^6 \L_n\,,
\label{eq:LGP}
\end{equation}
where,
\begin{align}
	\L_2 &= f_2(A_{\mu}, F\mn, \tilde{F}\mn)
		\label{eq:LGP2} \\
	\L_3 &= f_3(A^2) (\partial \cdot A)
		\label{eq:LGP3} \\
	\L_4 &= f_4(A^2) [(\partial \cdot A)^2 - \partial_{\mu} A_{\nu} \partial^{\nu} A^{\mu}]
		\label{eq:LGP4} \\
	\L_5 &= f_5(A^2) [(\partial \cdot A)^3 -3 (\partial \cdot A) \partial_{\mu} A_{\nu} \partial^{\nu} A^{\mu} + 2 \partial_{\mu} A_{\nu} \partial^{\nu} A^{\rho} \partial_{\rho} A^{\mu} ] + \tilde{f}_5(A^2) \tilde{F}^{\mu \alpha} \tilde{F}^{\nu}_{\phantom{\nu} \alpha} \partial_{\mu} A_{\nu}
		\label{eq:LGP5} \\
	\L_ 6 &= \tilde{f}_6(A^2) \tilde{F}^{\mu \nu} \tilde{F}^{\alpha \beta} \partial_{\alpha} A_{\mu} \partial_{\beta} A_{\nu}\,.
		\label{eq:LGP6}
\end{align}
All the functions $f_n$'s and $\tilde{f}_n$'s are arbitrary polynomial functions so these Lagrangians span an infinite family of operators depending on the form of these functions\footnote{Notice that this formulation differs ever so slightly with that originally introduced in \cite{Heisenberg:2014rta}. For instance, the contribution to $\L_4$ proportional to $c_2$ in Eq.~(2.2) of \cite{Heisenberg:2014rta} is here absorbed into the function $f_2$, however, both formulations are entirely equivalent.}. For comparison with other theories, and to compute scattering amplitudes, it is convenient to expand all the functions $f_n$ and $\tilde{f}_n$ in the most generic possible way and repackage the Lagrangian \eqref{eq:LGP} perturbatively in a field expansion. In this case, the theory is expressed perturbatively as
\begin{equation}
	\L_{\text{GP}} = \sum_{n=2}^\infty \frac{1}{\Lambda_2^{2(n-2)}}\L_{\text{GP}}^{(n)} \,,
\label{eq:LGPbis}
\end{equation}
where $\Lambda_2$ is introduced as the dimensionful scale  for the interactions and
where up to quartic order
\begin{align}
	\L_{\text{GP}}^{(2)} &= -\frac{1}{4}F\mnup F\mn - \frac{1}{2}m^2 A^2
		\label{eq:LGP2bis} \\
	\L_{\text{GP}}^{(3)} &=  a_1 m^2 A^2 \partial_{\mu} A^{\mu} + a_2 \tilde{F}^{\mu \alpha} \tilde{F}^{\nu}_{\phantom{\nu} \alpha} \partial_{\mu} A_{\nu}
		\label{eq:LGP3bis} \\
	\L_{\text{GP}}^{(4)} &=  b_1 m^4 A^4 + b_2 m^2 A^2 F\mnup F\mn + b_3 m^2 A^2 \left[ (\partial \cdot A)^2 - \partial_{\alpha} A_{\beta} \partial^{\beta} A^{\alpha} \right] + b_4 m^2 F^{\mu \alpha}F^{\nu}_{\phantom{\nu} \alpha} A_{\mu} A_{\nu}  \nonumber \\
													 & + b_5 F\mnup F\abup F_{\mu \alpha} F_{\nu \beta} + b_6 F\mnup F\mn F\abup F\ab + b_7 \tilde{F}\abup \tilde{F}\mnup \p_{\alpha} A_{\mu} \p_{\beta} A_{\nu}\,,
		\label{eq:LGP4bis}
\end{align}
with the coefficients $a_{i}$ and $b_j$ being dimensionless constants. The scaling is introduced so as to `penalize' the breaking of gauge--invariance with the scale $m$ (see \cite{deRham:2018qqo} for the appropriate scaling of operators in gauge--breaking effective field theories).
Note that there exist various different but equivalent ways to express the Lagrangian perturbatively depending on how total derivatives are included, nevertheless, irrespectively on the precise formulation, there exist $2$ linearly independent terms at cubic order and $7$ at quartic order (ignoring total derivatives).

\subsection{Generalized Proca in the Decoupling Limit}
\label{ssec:GPDL}

For any theory, its decoupling limit (DL) is determined by scaling parameters of the theory so as to be able to focus on the irrelevant operators that arise at the lowest possible energy scale while maintaining all the degrees of freedom alive in that limit. Hence by definition, the number of degrees of freedom remains the same in the DL. Taking a DL is different from taking a low--energy effective field theory and also differs from switching off interactions or degrees of freedom. See for instance Refs.~\cite{deRham:2014wfa,deRham:2014zqa,deRham:2016wji} for more details on the meaning of a DL.

In the particular case of GP, the DL is taken by first introducing the \stu field explicitly in a canonically normalized way,
\begin{equation}
	A_{\mu} \rightarrow A_{\mu} + \frac{1}{m}\partial_{\mu} \phi\,,
\label{eq:Stuck}
\end{equation}
so that the kinetic term for the helicity--0 mode is explicitly manifest in \eqref{eq:LGP2bis}, indeed $\L^{(2)}_{\rm GP}\supset -\frac 12 (\p \phi)^2$. We then take the DL by sending the mass $m$ to zero and $\Lambda_2\to \infty$ in such a way as to keep the lowest interaction scale finite in that limit. Denoting generic interactions scales $\Lambda_p$ by $\Lambda_p=(m^{p-2}\Lambda_2^2)^{1/p}$  (with $\Lambda_3\equiv (m \Lambda_2^2)^{1/3}$), one can check that the lowest scale at which interactions appear is $\Lambda_3$. The $\Lambda_3$-DL of GP is then taken by sending
\begin{equation}
	m \rightarrow 0, \quad \Lambda_2 \rightarrow \infty \quad \text{keeping} \quad \Lambda_3 \equiv  (m \Lambda_2^2)^{1/3} = \text{const.}\,,
\label{eq:Lambda3DL}
\end{equation}
once all the fields are properly normalized.

Upon taking this DL, one notices that out of all the interactions that entered the quartic GP Lagrangian $\L_{\text{GP}}^{(4)} $ in  \eqref{eq:LGP4bis} only terms proportional to $b_3$ and $b_7$ survive and one ends up with
\begin{equation}
	\L_{\text{DL GP}} = \L_{\text{DL GP}}^{(2)} + \frac{1}{\Lambda_3^3} \L_{\text{DL GP}}^{(3)} + \frac{1}{\Lambda_3^6} \L_{\text{DL GP}}^{(4)} + \frac{1}{\Lambda_3^9} \L_{\text{DL GP}}^{(5)}\,,
\label{eq:LDLGP}
\end{equation}
where the first four Lagrangians are given by
\ba
	\L_{\text{DL GP}}^{(2)} &=& -\frac{1}{4}F\mnup F\mn - \frac{1}{2} (\partial \phi)^2
		\label{eq:LDLGP2} \\
	\L_{\text{DL GP}}^{(3)} &=& a_1 (\p \phi)^2 [\Phi] + a_2 \tilde{F}^{\mu \alpha} \tilde{F}^{\nu}_{\phantom{\nu} \alpha} \Phi\mn
		\label{eq:LDLGP3}
= a_1 (\p \phi)^2 [\Phi] + a_2 F^{\mu \alpha} F^{\nu}_{\phantom{\nu} \alpha} \(\Phi\mn - \frac{1}{2} [\Phi]\eta\mn \)
		\label{eq:LDLGP3b} \\
	\L_{\text{DL GP}}^{(4)} &=& b_3 (\p \phi)^2 \([\Phi]^2-[\Phi^2]\) + b_7 \tilde{F}^{\alpha \beta} \tilde{F}\mnup \Phi_{\alpha\mu}\Phi_{\beta\nu}
		\label{eq:LDLGP4} \\
&=& b_3 (\p \phi)^2 \([\Phi]^2-[\Phi^2]\) \nn \\
&&+ b_7 F^{\alpha \beta} F\mnup \Bigg[ \Phi_{\mu\alpha}\Phi_{\nu \beta}
+2 \eta\ab \left( \Phi^2\mn - \Phi\mn [\Phi] \right)
+\frac{1}{2} \eta_{\mu \alpha}\eta_{\nu \beta} \left( [\Phi]^2 - [\Phi^2] \right) \label{eq:LDLGP4b} \Bigg]\,,\nn\hspace{-0.5cm} 										
\ea
where we used the notation $\Phi\mn=\p_\mu \p_\nu \phi$.
In contrast with the $9$ parameters family of interactions up to quartic order for GP, its DL up to quartic order only includes the cubic and quartic Galileon interactions as well as two genuine mixings between the helicity--0 and --1 modes, parametrized by $a_2$ and $b_7$. The quintic Lagrangian $\L_{\text{DL GP}}^{(5)}$ involves the quintic Galileon and can include interactions between the helicity--0 and --1 modes although the precise form of these interactions is not relevant for this study.

\section{Massive gravity}
\label{sec:IntroMG}

\subsection{A primer on tensor fields}
\label{ssec:PrimerTensor}

Having explored the scalar and vector sectors through interesting examples of non-ghostly models exhibiting higher-order operators, it is natural to finally turn to the spin-2 or tensor sector. In parallel with the vector case where the massless theory reduced to Maxwell's electromagnetism, the kinetic term for tensor models is unique, as long as the theory is local, Lorentz-invariant, and ghost-free. For a Lorentz tensor $h\mn$, the unique and canonically normalized quadratic kinetic term is given by
\begin{equation}
	\L[h] \supset - \frac{1}{4} h\mnup \hat{\mathcal{E}}\mn^{\alpha \beta} h_{\alpha \beta} \,,
\end{equation}
where the Lichnerowicz operator reads
\begin{equation}
	\hat{\mathcal{E}}\mn^{\alpha \beta} h_{\alpha \beta} = - \frac12 \left( \Box h\mn - 2 \p_{(\mu} \p_{\alpha} h^{\alpha}_{\nu)} + \p_{\mu} \p_{\nu} h - \eta\mn (\Box h - \p_{\alpha} \p_{\beta} h^{\alpha \beta}) \right) \,,
\end{equation}
and where we used $h=h^{\alpha}_{\alpha}$ as a short-hand notation for the trace of the spin-2 field. Recall that the Maxwell kinetic term was gauge-invariant by construction, a symmetry that any inclusion of a mass term explicitly broke. Here, it is easy to show that the spin-2 kinetic term is invariant under
\begin{equation}
	h\mn \rightarrow h\mn + \p_{(\mu} \xi_{\nu)} \,,
	\label{eq:lindiff}
\end{equation}
which is a gauge transformation too, also known as linear diffeomorphism. Note that any mass term would explicitly break this symmetry, in a similar fashion to the vector case previously introduced. This gauge symmetry represents $4$ constraints ($1$ per component of the arbitrary gauge vector $\xi_{\mu}$), each removing $2$ degrees of freedom in the $10$-dimension field-space of symmetric $4$-dimensional tensors. Hence, gravitational waves propagate $2$ degrees of freedom in $4$ spacetime dimensions.

Giving a mass to the spin-2 field breaks gauge invariance and it is possible to show that the only possible candidate that does not generate any Ostrogradsky instability is the so-called Fierz-Pauli mass term and reads
\begin{equation}
	\L_{\rm \text{(FP mass term)}} = - \frac18 m^2 (h\mn^2 - h^2) \,.
\end{equation}
The gauge symmetry breaking has consequences at the level of the constraint structure and hence the number of propagating degrees of freedom rises. It is possible to show that the massive spin-2 field can be split into,
\begin{equation}
	h\mn = h\mn^T + \p_{(\mu} \chi_{\nu)} \,, \qquad {\rm where} \qquad \chi_{\mu} = \frac{1}{m} A_{\mu} + \frac{1}{m^2} \p_{\mu} \pi \,,
\end{equation}
where
\begin{itemize}
	\item $h\mn^T$ is a symmetric transverse tensor and hence has $6$ independent entries
	\item $A_{\mu}$ is a transverse vector carrying $3$ independent components
	\item $\pi$ is a longitudinal single scalar mode.	
\end{itemize}
Hence, the $10$ components of the symmetric tensor $h\mn$ are preserved. One can also check that after suitable field-redefinition and diagonalization into mass eigenstates, the matter content of the theory reduces to 
\begin{itemize}
	\item the GR helicity-2 mode contributing to $2$ degrees of freedom,
	\item a helicity-1 mode, also contributing to $2$ degrees of freedom,
	\item a helicity-0 mode counting for $1$ degree of freedom.
\end{itemize}
It is then clear that the massive gravitational waves propagate $5$ degrees of freedom in $4$ spacetime dimensions. This is all very well but one needs to go beyond the linear level when coupling it to external matter or simply including higher-order operators. In this case, the exact linear diffeomorphism of Eq.~\eqref{eq:lindiff} needs to be generalized fully non-linearly to give a covariant theory. By including such higher-order terms, then the kinetic term also needs to include non-linearities and it is well known that there exists a unique fully non-linear kinetic term for massless and diffeomorphism-invariant spin-2 particles which is given by the Einstein-Hilbert term,
\begin{equation}
	\L_{\rm EH} = \frac{M_{\rm Pl}^2}{2} \sqrt{-g} R[g] \,,
\end{equation}
where $g$ is the determinant of the metric $g\mn = \eta\mn + h\mn/M_{\rm Pl}$ and $R$ is the Ricci scalar. A very non-trivial question now arises: how do we write down a covariant theory of a massive spin-2 field?

\subsection{A brief introduction to de Rham-Gabadadze-Tolley massive gravity}
\label{ssec:dRGT}

In this Section, we would like to introduce the de Rham-Gabadadze-Tolley (dRGT) theory of massive gravity \cite{deRham:2010ik, deRham:2010kj, Hassan:2011vm, Hassan:2011hr, Hassan:2011tf, Hassan:2011ea, Creminelli:2005qk, Hassan:2012qv} which provides a satisfying and elegant framework of a ghost-free theory of a massive spin-2 field. We will not enter into the details of the cosmological challenges it is facing and how the constraint structure is realized at the level of the Hamiltonian but will simply introduce its Lagrangian and will establish the relevance of this theory in relation to this thesis.

Interestingly, dRGT massive gravity relies on a physical metric $g\mn$ together with a reference metric $f\mn$ which doesn't necessarily need to reduce to the flat space metric, and breaks the full non-linear diffeomorphism in the same way as the mass term broke $U(1)$ gauge symmetry for vector fields. We start by defining the following quantity
\begin{equation}
	\K\ud{\mu}{\nu} = \delta^{\mu}_{\nu} - \X\ud{\mu}{\nu} \,,
	\label{eq:refKtensor}
\end{equation}
where the tensor $\X$ is the square root $\sqrt{g^{-1}f}$ in the sense that
\begin{equation}
	\X\ud{\mu}{\alpha} \X\ud{\alpha}{\nu} = (g^{-1}f)\ud{\mu}{\nu} \,.
	\label{eq:refXtensor}
\end{equation}
The dRGT theory of massive gravity can be written as a sum of the usual Einstein-Hilbert term plus a new potential term $\mathcal{U}$ that encompasses highly non-trivial and non-linear interactions for the metric $g\mn$ and the reference metric $f\mn$,
\begin{equation}
	\L_{\rm dRGT} = \frac{M_{\rm Pl}^2}{2} \sqrt{-g} R[g] - m^2 \mathcal{U} \left[ g,f \right] \,,
\end{equation}
where in $4$ spacetime dimensions the potential reads
\begin{equation}
	\mathcal{U}\left[ g,f \right] = - \frac{M_{\rm Pl}^2}{4} \sqrt{-g} \sum_{n=0}^4 \alpha_n \L_n \left[ \K \left[ g,f \right] \right] \,.
\end{equation}
The individual Lagrangians $\L_n$ are symmetric polynomials of their argument $\K[g,f]$ and are introduced in Eq.~\eqref{eq:refLnK}.  

\subsection{\stu trick in dRGT massive gravity}
\label{ssec:DLdRGT}

Now that we have introduced the dRGT massive gravity, we are interested in its vector sector. Indeed, the interactions used for the Proca-Nuevo model developed in Part \ref{part:PN} are inspired by that of dRGT vector's sector. The fact that the latter is ghost-free was the original motivation to study dRGT-like interactions for a massive vector theory. 

In the same fashion as what was done for standard or Generalized Proca in order to reintroduce gauge symmetry, one can make use of \stu fields in order to restore full covariance of the Lagrangian. The breaking of covariance is materialized in the choice of a reference metric $f\mn$, which means that $\K\mn$ does \textit{not} transform as a tensor. Let us introduce a set of $4$ \stu field $\phi_{\mu}$ such that the reference metric is promoted to the tensor $\tilde{f}\mn$, defined by
\begin{equation}
	\tilde{f}\mn = \p_{\mu} \phi^{\alpha} \p_{\nu} \phi^{\beta} f_{\alpha\beta} \,.
\end{equation}
For the following, we will focus on a flat reference metric, i.e. $f\mn=\eta\mn$. This allows one to identify the tensor as a spin-2 particle, as the notion of spin only makes sense in a maximally symmetric spacetime.

Notice now that we are in the presence of a theory which has been proven to be ghost-free and where Lorentz-invariance has been restored through the means of introducing $4$ \stu fields $\phi^{\alpha}$. It is quite natural to ask what happens if one thinks about them as a single vector field with $4$ spacetime components instead, and this is precisely the reasoning that went behind the introduction of Proca-Nuevo, a theory which will be extensively studied in Part \ref{part:PN}.

\section{Positivity bounds in a nutshell}
\label{sec:IntroPosBounds}

In this section, we would like to briefly introduce positivity bounds. The latter are not the focus of this thesis and will not be used to constrain any EFT coefficients. However, we will be working with causality bounds in Part \ref{part:CausalityBounds} and will compare our results to the ones obtained by positivity, when known or possible, hence we take this opportunity to give a short overview of this well-known method. To this end, we start by looking at the K\"{a}ll\'{e}n-Lehmann \cite{Kallen:1952zz, Lehmann:1954xi} spectral representation for the 2-particle amplitude. 

\paragraph{K\"{a}ll\'{e}n-Lehmann spectral representation.} Let's begin with a scalar 2-point function for simplicity's sake. Thanks to the work of K\"{a}ll\'{e}n and Lehmann, one can re-write the 2-point function in momentum space using a spectral representation in the following form
\begin{equation}
	\mathcal{A}_2(-p^2) = \int_0^{\infty} {\rm d}\mu \, \frac{\rho(\mu)}{p^2 + \mu} \,,
	\label{eq:2ptA}
\end{equation}
where the integral in the full interacting theory is nothing else than the free propagator modulated by the spectral density function $\rho(\mu)$. This function should have a pole at $\mu=m^2$, where $m$ is the mass of the scalar field, and should also only have support above the  2-particle production threshold $\mu>4m^2$. This means the spectral representation can be recast into
\begin{equation}
	\rho(\mu) = Z\, \delta(\mu-m^2) + \sigma(\mu) \theta(\mu-4m^2) \,.
\end{equation}
Note that the free theory only possesses the first term and hence $Z$ simply is a normalization factor of the wavefunction, whereas $\sigma(\mu)$ encodes information on the interactions and, as such, represents the $1 \rightarrow 1$ cross-section. Plugging this definition back in the one for the 2-point amplitude in Eq.~\eqref{eq:2ptA}, and identifying the Mandelstam variable $s=-p^2$ standing for the center of mass energy square, we get
\begin{equation}
	\mathcal{A}_2(s) = \frac{Z}{m^2-s} + \int_{4m^2}^{\infty} {\rm d}\mu \, \frac{\sigma(\mu)}{\mu-s} \,.
\end{equation}

\paragraph{Unitary and optical theorem.} We would like now to make use of the unitarity of the S-matrix. In a similar fashion as in the K\"{a}ll\'{e}n-Lehmann spectral representation, one can separate the S-matrix in a free contribution and one coming from interactions, namely the transition matrix, denoted by $iT$,
\begin{equation}
	S = \mathbb{I} - i T \,.
\end{equation}
Unitarity is the statement that the sum of probabilities should be equal to unity, and translates into $S^{\dag} S = \mathbb{I}$, which can equivalently be written as an equation for the transition matrix
\begin{equation}
	i (T^{\dag} - T) = T^{\dag} T \,.
\end{equation}
This statement can be used for scattering amplitudes. To be more concrete, if we specialize to the case of $1 \rightarrow 1$ scattering with the same initial and final state $|f \rangle=|i \rangle$, we find
\begin{equation}
	2 {\rm Im} \langle i|T|i \rangle = \sum_{|j\rangle } \langle i|T|j \rangle \langle j|T|i \rangle = \sum_{|j \rangle} \left| \langle i|T|j \rangle \right|^2 > 0 \,.
\end{equation}
Note that the quantity computed above is exactly the imaginary part of the 2-point amplitude, which is also nothing else than the $1 \rightarrow 1$ scattering amplitude and hence the optical theorem establishes the positivity of $\sigma(\mu)$ thanks to the unitarity of the S-matrix.

\paragraph{Integration contour.} One could also derive the previous results by using analyticity of the scattering amplitude in the complex $\mu$ plane. Let $C$ be a counter-clockwise contour around the unique pole $s$, then the Cauchy theorem states
\begin{equation}
	\mathcal{A}_2(s) = \frac{1}{2\pi i} \oint_C {\rm d}\mu \frac{\mathcal{A}_2(\mu)}{\mu-s} \,.
\end{equation}
Let's recall that the analytic structure of the 2-point amplitude $\mathcal{A}_2(\mu)$ is such that it has a pole at $m^2$ and a branch cut along the real axis for $\mu>4m^2$. Let us now deform the contour $C$ such that it encircles the pole and runs along the branch cut from below and above. We get
\begin{equation}
\begin{aligned}
	\mathcal{A}_2(s) =& \frac{{\rm Res}(\mathcal{A}_2,s=m^2)}{m^2-s} + \frac{1}{2\pi i} \int_{4m^2}^{\infty} {\rm d}\mu \frac{\mathcal{A}_2(\mu+i 0) - \mathcal{A}_2(\mu-i0)}{\mu-s} \\
	=& \frac{{\rm Res}(\mathcal{A}_2,s=m^2)}{m^2-s} + \frac{1}{\pi} \int_{4m^2}^{\infty} {\rm d}\mu \frac{{\rm Im} \mathcal{A}_2(\mu)}{\mu-s} \,,
\end{aligned}
\end{equation}
from which we easily identify
\begin{equation}
	Z = {\rm Res}(\mathcal{A}_2,s=m^2) \,, \qquad \sigma(\mu) = \frac{1}{\pi} {\rm Im} \mathcal{A}_2(\mu) > 0 \,.
\end{equation}
Using positivity, we end up with
\begin{equation}
	\mathcal{A}_2(s) - \frac{{\rm Res}(\mathcal{A}_2,s=m^2)}{m^2-s} > 0 \,.
\end{equation}
However, this has been derived using the underlying hypothesis that the contribution to the integral from the infinite radius circle vanished, which is only true if $|\sigma(s)|$ is strictly bounded by unity as the modulus of $s$ goes to infinity. Nevertheless, one can relax this hypothesis and still derive bounds if $|\sigma(s)| < s^N$ when $|s| \rightarrow \infty$ instead, where $N$ is a positive integer. In this case, the integrand is bounded by $s^{N-1}$ and does not converge. One can always take $N$ derivatives (also called `subtractions' in this context) of the integrand to make it converge. We then find
\begin{equation}
	\frac{\p^M}{\p s^M} \left[ \mathcal{A}_2(s) - \frac{{\rm Res}(\mathcal{A}_2,s=m^2)}{m^2-s} \right] = \frac{M!}{\pi} \int_{4m^2}^{\infty} {\rm d}\mu \frac{{\rm Im} \mathcal{A}_2(\mu)}{(\mu-s)^{M+1}} > 0 \,, \qquad \forall M \geq N \,.
\end{equation}

\paragraph{4-point amplitude.} We will not derive any new results but will simply draw an analogy to understand the more interesting 4-point function. Indeed, our computations in Part \ref{part:CausalityBounds} will be based on $2 \rightarrow 2$ scattering processes and we will then compare our results to such positivity bounds. Note that the kinematical space is enlarged and $\mathcal{A}_4$ now depends on $2$ independent Mandelstam variables (in $4d$), e.g. $s$ and $t$. Let's simply quote a result known as the Froissart-Martin bound \cite{Froissart:1961ux, Martin:1962rt} which states that the amplitude is asymptotically quadratically bounded in $s$, for any $0 \leq t < 4m^2$, meaning that positivity bounds need to be at least twice subtracted. If we call $\tilde{\mathcal{A}}_4$ the twice subtracted amplitude, then we have
\begin{equation}
	\frac{\p^N}{\p s^N} \tilde{\mathcal{A}}_4(s) > 0 \,, \qquad \forall N \geq 2 \,.
	\label{eq:TwiceSubtracted}
\end{equation}

\paragraph{Limits of positivity bounds.} It is important to note that positivity bounds for spins $\bm{s} \leq 3/2$ and for massive spin-2 particles are well understood but issues arise in the case of massless spin-2 and other higher-spins. This is due to the fact that the exchange of a particle of mass $m$ and spin $\bm{s}$ will have a contribution to the 4-point amplitude in the form $s^{\bm{s}}/(t-m^2)$, which vanishes when differentiated at least twice as in Eq.~\eqref{eq:TwiceSubtracted}, if $\bm{s}<2$. Hence, such terms only become concerning when $\bm{s} \geq 2$ and furthermore, when $m \rightarrow 0$ we get a pathological t-channel pole. It is important to note that no such divergences arise in deriving causality bounds. 

Then, positivity bounds rely on the ability to even define the S-matrix, which in turn relies on asymptotic flatness of the spacetime. This means that positivity bounds cannot be extended to arbitrarily curved backgrounds, which is not a problem for causality bounds as they do not rely on scattering processes but simply on the notion of time delay. These provide some basic motivations to constrain low-energy EFT coefficients using a different approach, namely IR causality.

\part{Proca-Nuevo: a new ghost-free massive vector theory.}
\label{part:PN}

The first part of this thesis is dedicated to the presentation and the theoretical study of the (Extended) Proca-Nuevo model, before applying it to cosmology. Chapter \ref{chap:PN} is based on \cite{deRham:2020yet} and introduces the Proca-Nuevo model by building on both Generalized Proca and dRGT Massive Gravity. Before this analysis, GP was considered to be the widest and most general class of massive vector interactions that doesn't propagate any ghostly degree of freedom. However, as we shall see, this was made on the hypothesis that such a theory should exhibit second-order equations of motion, trivially avoiding any Ostrogradsky instability. The fact is that this is too strong a requirement and that it is perfectly possible to have degenerate higher-order equations of motion while describing a healthy theory. Such a case arises naturally in the simplest of cases when field-redefining a free theory for instance. The approach taken here is to constrain the determinant of the Hessian matrix rather than the order of the equations of motion, which gives more freedom and allows for the discovery of wider classes of healthy operators. This is what we will refer to as the Null Eigenvector or Hessian constraint. We prove that PN is genuinely inequivalent to GP by computing scattering amplitudes in both theories. This confirms that the class of PN operators discovered, inspired by the vector sector arising in the decoupling limit of dRGT massive gravity, includes some GP operators but more importantly, explores a new part of the space of massive vector models. 

Chapter \ref{chap:Constraint}, which is based on \cite{deRham:2023brw}, is a pure mathematical constraint analysis of the proposed EPN model. There, we prove at the level of the Lagrangian and the Hamiltonian that there exists another second-class constraint arising from the consistency of the second-class NEV constraint. This pair of constraints formally removes a pair of conjugate coordinates in the Hamiltonian phase-space, hence fully removing the would-be Boulware-Deser ghost in field space. This provides yet another example of a parity and Lorentz-invariant theory where the existence of a second-class constraint automatically ensures the existence of its companion, so that no half degrees of freedom can propagate. The analysis is performed in $2d$ first before being extended to arbitrary dimensions.

After having shown the viability of the theory on flat spacetime, we turn to covariantization of EPN in Chapter \ref{chap:CovEPN}. We show that Generalized Massive Gravity is the most natural way to couple EPN to gravity given the fact that EPN interactions arise from the $\Lambda_3$ decoupling limit of the latter. However, there exists a variety of other covariantization schemes and we explore other options. We then specialize in cosmological backgrounds and prove that we can maintain the constraint with or without the addition of non-minimal couplings. This result will be the main foundation of the final Chapter of Part \ref{part:PN}.

Indeed, we then consider the application to cosmology in Chapter \ref{chap:Cosmo} (based on \cite{deRham:2021efp}), where (E)PN is used to model a dark energy fluid coupled to matter described by the Schutz-Sorkin action. We discuss the couplings to gravity before specialising to a FLRW background metric which is well-suited for cosmology purposes. The background equations of motion are derived and we prove that we can qualitatively recover a hot big-bang scenario with a self-accelerating branch. We get a universe which is successively dominated by radiation, matter and finally dark energy, as expected from observational data. Linear perturbations are then studied in each of the tensor, vector and scalar sectors. The tensor sector behaves in the same way as GR predicts and hence we get no modification to the propagation of the very tightly constrained gravitational waves. We find theoretical constraints on the parameters of the model such that the vector and scalar perturbations remain free of any ghosts and instability throughout the universe's history and exhibit results for a given set of data consistent with our constraints.

\chapter{Proca-Nuevo}
\label{chap:PN}

In this Chapter, we propose a new class of Proca interactions that enjoys a non--trivial constraint and hence propagates the correct number of degrees of freedom for a healthy massive spin--1 field. We will show that the scattering amplitudes always differ from those of the Generalized Proca. This implies that the new class of interactions proposed here are genuinely different from the Generalized Proca and there can be no local field redefinitions between the two. 


\section{\Pro}
\label{sec:NewProca}

\subsection{Full non--linear theory}
\label{sssec:NonLinTh}

We shall now build our intuition from massive gravity to derive a new type of fully non--linear Proca interactions. The DL of massive gravity includes an infinite number of scalar--vector interactions whose exact form was provided in \cite{Ondo:2013wka}. Interestingly, the scalar--vector sector of the DL of massive gravity can in principle be thought of as the DL of a Proca theory, similarly to what was considered in Section \ref{ssec:GPDL} for GP. On another hand, the scalar--vector interactions included in the DL of massive gravity involve higher derivatives acting on the fields and thus violate the original assumption in deriving the most general GP operators. Yet massive gravity has been proven to be ghost--free in many different languages \cite{deRham:2010kj,Hassan:2011hr,deRham:2011rn,deRham:2011qq,Hassan:2012qv} and hence so is its DL. Indeed, as emphasized in \cite{deRham:2011rn,deRham:2011qq,deRham:2016wji} the constraint can manifest slightly differently in theories with multiple fields and the existence of higher derivatives in the equations of motion does not necessarily imply an \Ostro ghost instability. For instance, there can be a linear combination of the equations of motion which is free from higher derivatives so that no higher--order \Ostro ghost instability occurs \cite{deRham:2011qq,deRham:2016wji}. This phenomenon is similar to what is observed in Beyond--Horndeski theories and other extensions \cite{Gleyzes:2014dya,Zumalacarregui:2013pma,
Langlois:2015cwa,Langlois:2015skt,BenAchour:2016cay,Crisostomi:2016tcp,
Crisostomi:2016czh,Ezquiaga:2016nqo,Motohashi:2016ftl,deRham:2016wji}.

Massive gravity is the theory of an interacting massive spin-2 field $h\mn$. In terms of a gravitational dynamical  metric $g\mn$, the spin-2 field $h\mn$ is expressed as $\mpl^{-1}h\mn=g\mn-\eta\mn$ in unitary gauge. The fact that the Minkowski metric $\eta\mn$ is not diffeomorphism invariant implies that expressed in this way $h\mn$ is not a tensor. However gauge invariance can be easily restored through the introduction of four \stu fields $\phi^a$ which transform as scalars under coordinate transformations. Indeed, expressed in terms of the tensor $f\mn$
\ba
\mpl^{-1}h\mn&=&g\mn-f\mn\\
{\rm with}\quad f\mn&=&\eta_{ab}\p_\mu \phi^a \p_\nu \phi^b\,,
\label{eq:fmn1}
\ea
the quantity $h\mn$ is now a tensor under diffeomorphisms.  In the limit where $\mpl \to \infty$ we may identify the index $a$ as a Lorentz index.  Splitting the fields $\phi^a$ as $=x^a+A^a$, the field $A^a$ can then be associated with a Lorentz vector which is anchored in the very formulation of massive gravity.

However, at this stage, the link between massive gravity and Proca interactions is not necessarily immediately manifest as massive gravity always includes the tensor modes.
 In fact there is no limit of {\it pure} massive gravity that would lead to a massive vector theory on Minkowski. Indeed, for such a  limit to occur, the helicity-0 mode of the massive spin-2 field of massive gravity should play the role of the helicity-0 mode of the massive vector field. However in {\it pure} massive gravity on Minkowski, the helicity-0 mode only acquired its kinetic term from mixing with the tensor mode \cite{deRham:2010ik}. 

 Instead one can consider the DL of massive gravity on AdS \cite{deRham:2018svs} where the helicity--0 mode acquires its own kinetic term without the need for a coupling with the tensor modes. Alternatively one can consider generalized massive gravity \cite{deRham:2014lqa,deRham:2014gla} where the scalar mode also acquires its own kinetic term. In both cases a new type of $\Lambda_2$--decoupling limit that only involves couplings between the scalar and vector modes can be considered \cite{deRham:2015ijs,deRham:2016plk,Gabadadze:2017jom,deRham:2018svs,Gabadadze:2019lld}. In \cite{deRham:2018svs} it was shown that on AdS, the resulting scalar--vector interactions could never be expressed as a local and Lorentz invariant field redefinition of the scalar--vector interactions that arise in the DL of GP, suggesting that these classes of interactions were indeed distinct from GP. In what follows we shall build from these results to provide a new class of non--linear ``\Pro" massive Proca interactions that rely on the same structure as the decoupling limit of massive gravity.
%
%
We start with a Lorentz vector field $A_\mu$ and work on flat spacetime with the Minkowski metric $\eta\mn$ (coupling to gravity is considered in section~\ref{sec:CouplGrav}).

 These considerations are mainly motivational for this context and following our intuition from massive gravity, we may consider the tensor $f\mn$ defined in \eqref{eq:fmn1} where the $\phi^a$'s are expressed in terms of the vector field as follows
 \ba
\label{eq:phia}
\phi^a=x^a+\frac{1}{\Lambda_2^2}A^a\,,
\ea
so that in terms of the vector field, the quantity $f\mn$ is expressed as \footnote{The object $f\mn$ is simply a Lorentz tensor constructed out of the first derivative of the Lorentz vector $A_\mu$ and at this level has no connection with any type of auxiliary metric. Note that in this context of a massive vector field, introducing the quantity $\phi^a$ in terms of the coordinate $x^a$ may be misleading as it suggests a breaking of Poincar\'e invariance, however, the quantity we shall be interested in, $f\mn$, is manifestly a Poincar\'e tensor if $A_\mu$ is itself a Poincar\'e vector as is clear from the expression \eqref{eq:deff2}.}
\ba
	f\mn[A] = \eta\mn + 2 \frac{\p_{(\mu} A_{\nu)}}{\Lambda_2^2} + \frac{\p_{\mu} A_\alpha \p_{\nu} A_\beta \eta^{\alpha \beta}}{\Lambda_2^4}\,.
		\label{eq:deff2}
\ea
Next we recall the definitions of the (Poincar\'e) tensor  $\K\mupn$ and $\X\mupn$ presented in Eqs.~\eqref{eq:refKtensor} and \eqref{eq:refXtensor} where, in the gravitational context,  $\K\mupn$ would be playing the role of the extrinsic curvature \cite{deRham:2013awa,deRham:2014zqa} and $\X$ that of the vielbein \cite{Hinterbichler:2012cn}.

In four dimensions, the theory of the vector field $A_\mu$ we propose is then expressed as
\begin{equation}
	\L_{\K}[A] = \Lambda_2^4 \sum_{n=0}^4 (4-n)! \alpha_n(A^2) \L_n[\K[A]]\,,
\label{eq:defLK}
\end{equation}
where the order by order Lagrangians are defined in Eqs.~\eqref{eq:refLnK} and \eqref{eq:refLnK0to4} and the factor $(4-n)!$ is introduced for convenience. As mentioned before, the theory \eqref{eq:defLK} has no gravitational degrees of freedom, rather it is a pure vector theory with an infinite tower of self--interactions. We shall prove in section~\ref{ssec:Hessian} that this vector--field theory corresponds to a Proca theory with at most three propagating degrees of freedom.

 Note that $\L_0$ is just a potential for the vector field, $\alpha_0(A^2)\L_0=V(A^2)$, which is where the vector field will carry its mass from and so it is essential for the consistency of this theory that $\alpha_0$ includes at the very least a contribution going as $\alpha_0\supseteq -\frac 12 (m^2/\Lambda_2^4) A^2$.

\subsection{Perturbative Action}
\label{sssec:PertAction}

The exact non--perturbative Lagrangian is expressed in \eqref{eq:defLK} but it is instructive to consider its  perturbative expression and we shall provide it up to quartic order in the field (as needed for the $2 \rightarrow 2$ tree--level scattering amplitudes).
To provide such a perturbative expression, we first Taylor expand the functions $\alpha_n(A^2)$ as follows
\begin{equation}
	\alpha_n(A^2) = \bar{\alpha}_n + \frac{m^2}{\Lambda_2^4} \bar{\gamma}_n A^2 + \frac{m^4}{\Lambda_2^8} \bar{\lambda}_n A^4 + \cdots\,.
\label{eq:betans}
\end{equation}
Plugging it into \eqref{eq:defLK} and requiring the canonical normalization for the quadratic Lagrangian (Maxwell with a mass term) requires the following normalization:
\ba
\bar \alpha_1=-\frac 13 \( 1 - 2 \bar \alpha_2 \) \qquad {\rm and}\qquad \bar \gamma_0=-\frac 1{48}\,.
\ea
The perturbative expansion up to quadratic order then takes the form
\begin{equation}
	\L_{\K} = \L_{\K}^{(2)} + \frac{1}{\Lambda_2^2} \L_{\K}^{(3)} + \frac{1}{\Lambda_2^4} \L_{\K}^{(4)} + \cdots\,,
\label{eq:LK}
\end{equation}
with
\ba
	\L_{\K}^{(2)} &=& -\frac{1}{4}F\mnup F\mn - \frac{1}{2}m^2 A^2
		\label{eq:LK2} \\
	\L_{\K}^{(3)} &=& \frac 14 \(2\bar \alpha_2-3 \bar \alpha_3\) [F^2] [\p A]
+ \frac14\(1-4 \bar \alpha_2+6\bar \alpha_3\) F^2\mn \p^{\mu} A^{\nu} +6 \bar \gamma_1 m^2 A^2 [\p A]\label{eq:LK3} \\
	\L_{\K}^{(4)} &=& \frac 1{32}\(\bar \alpha_2-3 \bar \alpha_3+6 \bar \alpha_4\) [F^2]^2
+ \frac{1}{64}\(5-20 \bar \alpha_2-12 \bar \alpha_3+168 \bar \alpha_4\) F^2\mn F^2{}^{\mu\nu} \label{eq:LK4} \\
&+& \frac 38 \(\bar \alpha_3-4 \bar \alpha_4\) [F^2] \left( [\p A]^2 - \p_{\alpha}A_{\beta} \p^{\beta} A^{\alpha} \right)
-\frac 18 F^2\mn \p^{\beta} A^{\mu} \p_{\beta} A^{\nu} \nonumber \\
&+&\(\frac 12 \bar \alpha_2+\frac 34 \bar \alpha_3-6 \bar \alpha_4\)F^2{}^{\mu\nu}\(\p^{\beta} A_{\mu} \p_{\beta} A_{\nu}-[\p A]\p_\mu A_\nu\)\nn\\
&+&\(-\frac 18+\frac 12 \bar \alpha_2-3 \bar \alpha_4\)F^{\mu\nu}F^{\alpha \beta}\p_\mu A_\alpha \p_\nu A_\beta\nn\\
&+&m^2 A^2 \left[2 \bar \gamma_2 [\p A]^2-\(\frac 32 \bar \gamma_1+\bar \gamma_2\)\p_\mu A_\nu \p^\nu A^\mu
+\(\frac 32 \bar \gamma_1-\bar \gamma_2\)\p_\mu A_\nu \p^\mu A^\nu\right]\nn\\
&+&24 \bar \lambda_0 m^4 A^4\nn\,,
\ea
where we use the notation $F^2\mn=F_{\mu}{}^{\alpha}F_{\nu \alpha}$ and $[F^2]=F^{\mu\nu}F\mn$\,.

\subsection{Decoupling Limit}
\label{sssec:NewDL}
It will also be instructive to consider the DL of this \Pros theory. In the same fashion as what has been done for GP in Section \ref{ssec:GPDL} and inspired by the $\Lambda_3$-DL of massive gravity (where the scales taken to infinity and to $0$ are respectively the Planck mass $M_{\rm Pl}$ and the graviton mass $m_g$.), we will be taking the $\Lambda_3$ decoupling limit of Proca-Nuevo, where $\Lambda_3 = (m \Lambda_2^2)^{1/3}$ is kept finite and is the lowest interaction scale of the theory. Introducing the helicity--0 \stu field $\phi$ as in \eqref{eq:Stuck}
using the same scaling as in \eqref{eq:Lambda3DL}, we get
\begin{equation}
	\L_{\K {\text{DL}}} = \L_{\K {\text{DL}}}^{(2)} + \frac{1}{\Lambda_3^3} \L_{\K {\text{DL}}}^{(3)} + \frac{1}{\Lambda_3^6} \L_{\K {\text{DL}}}^{(4)} + \cdots\,,
\label{eq:LKDL}
\end{equation}
with
\ba
	\L_{\K {\text{DL}}}^{(2)} &=& -\frac{1}{4}F\mnup F\mn - \frac{1}{2} (\p \phi)^2
		\label{eq:LKDL2} \\
	\L_{\K {\text{DL}}}^{(3)} &=& \frac 14 \(2\bar \alpha_2-3 \bar \alpha_3\) [F^2] \Box \phi
+ \frac14\(1-4 \bar \alpha_2+6\bar \alpha_3\) F^2\mn \Phi^{\mu\nu} +6 \bar \gamma_1 (\p \phi)^2 \Box \phi
		\label{eq:LKDL3} \\
	\L_{\K {\text{DL}}}^{(4)} &=&
\frac 38 \(\bar \alpha_3-4 \bar \alpha_4\) [F^2] \left( [\Phi]^2-[\Phi^2] \right)
-\frac 18 F^2\mn \Phi^2{}^{\mu\nu} \label{eq:LKDL4}
 \\&+&
\(\frac 12 \bar \alpha_2+\frac 34 \bar \alpha_3-6 \bar \alpha_4\)F^2{}^{\mu\nu}\(\Phi^2\mn-[\Phi]\Phi\mn\)\nn\\
&+&\(-\frac 18+\frac 12 \bar \alpha_2-3 \bar \alpha_4\)F^{\mu\nu}F^{\alpha \beta}\Phi_{\mu \alpha}\Phi_{\nu \beta}\nn\\
&+&2 \bar \gamma_2 (\p\phi)^2\([\Phi]^2-[\Phi^2]\)\nn\,.
\ea
In this DL, we see that the coefficients $\bar \gamma_{1,2}$ govern the pure cubic and quartic Galileon interactions while the other $\bar \alpha_{2,3,4}$ coefficients govern the interactions between the vector and the scalar sector. This scalar--vector mixing matches precisely those that arise in the DL of massive gravity \cite{Ondo:2013wka} up to a trivial redefinition of the coefficients (see Appendix \ref{sec:appDL}). While the DL of GP truncates at quintic order (see Eq.~\eqref{eq:LDLGP}), we note that the DL of \Pros does not truncate and involves an infinite number of interactions in the scalar--vector sector. Moreover one can check that these interactions are never exactly of the GP form even after local and Lorentz invariant field redefinitions \cite{deRham:2018svs}.

While GP was constructed so as to ensure that its DL leads to second order equations of motion one can check explicitly that the \Pro's DL involves higher derivatives in its equations of motion. At first sight, one may worry that those higher derivatives are related to \Ostro ghost--like instabilities however we shall see below that the constraint remains in the \Pros theory and in four dimensions, only three degrees of freedom are excited. Since the theory enjoys the same vacuum as a free Proca theory with no ghost, this ensures that there can be no ghost excitations when working about configurations that are connected to the standard Proca vacuum when remaining within the regime of validity of the theory. In what follows we start by proving that the Hessian in two dimensions has a vanishing eigenvalue. We then prove the existence of a null eigenvector for the Hessian in arbitrary dimensions, hence signaling the existence of a constraint. We note that since we are dealing with a parity preserving Lorentz--invariance theory, there can be no half number of propagating degrees of freedom and hence the existence of a primary second class constraint automatically ensures the existence of a secondary constraint (see Ref.~\cite{deRham:2014zqa} for more details on that point).

\subsection{Hessian}
\label{ssec:Hessian}

We shall now show that the Hessian of \Pros always includes a vanishing eigenvalue hence implying the existence of a constraint that removes the would--be \Ostro ghost.

\subsubsection{Example}
\label{sssec:2d}

To start with, we may consider the theory in two dimensions and focus on the Lagrangian given by
\ba
	\L^{(\text{2d})} = -2[\K] -\frac 12 m^2 A^2\,.\label{eq:L2d}
\ea
 In two dimensions, an interactive massive vector field could in principle excite two degrees of freedom, but a healthy Proca theory should only excite one. We shall thus determine the Hessian of \Pros in two dimensions and prove that it only involves one non--vanishing eigenvalue. For simplicity we define
\ba
	x =\frac{1}{\Lambda_2^2} \p_\mu A^\mu \qquad {\rm and }\qquad
    y =\frac{1}{\Lambda_2^2} F_{01}=\frac{1}{\sqrt{2}\Lambda_2^2} \sqrt{-[F^2]}\,. \label{eq:xdef}\label{eq:ydef}
\ea
Then the Lagrangian takes the very simple form
\begin{align}
	\L^{(\text{2d})} = -2[\K] -\frac 12 m^2 A^2&= - 4 \sqrt{1 + x + \frac{x^2-y^2}{4}} +4 -\frac 12 m^2 A^2 \\
														&= - 4 \sqrt{1 + \frac{\p_{\mu} A^{\mu}}{\Lambda_2^2} + \frac{2(\p_{\mu} A^{\mu})^2 + [F^2]}{8\Lambda_2^4}}+4 -\frac 12 m^2 A^2\,,\label{eq:L2db}
\end{align}
and the Hessian matrix is given by
\ba
 \mathcal{H}^{ab} =  \frac{\p^2 \L}{\p \dot A_a \p \dot A_b}
= \frac{2}{[\X]^3 \Lambda_2^4}
		\begin{pmatrix}
			y^2 &  y (2+x) \\
			 y(2+x) & (2+x)^2
		\end{pmatrix}\,.
\label{eq:Hessmat}
\ea

It is straightforward to check that the determinant of the Hessian does indeed vanish, signaling that one of the vector components  is non--dynamical and leaving only one propagating degree of freedom in two dimensions. The null eigenvector simply reads
\begin{equation}
v_a =
		\begin{pmatrix}
			1 \\
			0
		\end{pmatrix}+\frac{1}{2}
		\begin{pmatrix}
			x \\
			-y
		\end{pmatrix}\,.
\label{eq:eigvec}
\end{equation}
We see that this null eigenvector is perturbatively connected with the vector $(1,0)$ and still ensures that $A_0$ is not dynamical. Next we shall prove the existence of a similar type of null eigenvector for any \Pros theory in any number of dimensions.

\subsubsection{Null Eigenvector in arbitrary dimensions}
\label{sssec:Null4d}

We shall now give a non--perturbative proof of the absence of ghost in four or any other dimensions, for the full theory, by deriving analytically the Hessian matrix and giving an expression for a null eigenvector. The proof for the absence of ghost follows from the arguments provided in \cite{deRham:2011rn,deRham:2014lqa,deRham:2014gla,deRham:2016plk} and generalizes the proof given in \cite{Hassan:2012qv} beyond the minimal model. We recall that $\K = \X-1$ with $\X=\sqrt{\eta^{-1}f}$ and we introduce the matrix $\Z$ defined as
\begin{equation}
\Z = \X^{-1} \eta^{-1}\,.
\label{eq:defZ}
\end{equation}
One can check that $\Z$ is symmetric, using the same similarity transformation as introduced in \cite{deRham:2014naa},
\ba
\Z^{-1}=\eta \X= \(\eta \sqrt{\eta^{-1}f}\eta^{-1}\)\eta
=\sqrt{f \eta^{-1}}\eta=\X^T \eta=\(\Z^{-1}\)^{T}\,.
\label{eq:Zinv}
\ea
It follows that $\Z=\Z^T$ and
\begin{align}
	\Z\abup f_{\beta \gamma} &= \X^{\alpha}_{\phantom{\alpha} \gamma} \label{eq:Zf} \\
	\Z\mnup f_{\nu \alpha} \Z\abup &= \eta^{\mu \beta} \label{eq:ZfZ}
\end{align}
Now if we evaluate the $00$-component of \eqref{eq:ZfZ} and differentiate it with respect to the time--derivative of the vector field $\dot A^a$, we find
\begin{equation}
\frac{\p }{\p \dot A^a }\(\Z^{0 \mu} f_{\mu \nu} \Z^{\nu 0} \)= 2\frac{\p \Z^{0 \mu}}{\p \dot A^a }f_{\mu \nu} \Z^{\nu 0}+\frac{2}{\Lambda_2^{2}} \Z^{00}\Z^{0\mu}\p_\mu \phi_a=0
\quad \Rightarrow \quad
\frac{\p \Z^{0 \mu}}{\p \dot A^a } \X^0_{\phantom{0} \mu} = - \Lambda_2^{-2} \Z^{00} V_a\,,
\label{eq:dZ}
\end{equation}
where $\phi_a=\eta_{ab}\phi^b$ is introduced in \eqref{eq:phia} and where we have introduce the normalized time--like vector $V_a$  defined as
\begin{equation}
	V_a = \Z^{0 \mu} \p_{\mu} \phi_a\,,
\label{eq:Va}
\end{equation}
so that $V^a V_a = -1$.
It is then straightforward to show that
\begin{equation}
	\p_{\mu} \phi_a V^a = \X^0_{\phantom{0} \mu}\,.
\label{eq:dphiV}
\end{equation}
Using these relations, we find the following expressions for the generic derivatives,
\ba
\label{eq:dK1}
	\frac{\partial}{\partial \dot{A}^a} [\X^n] = n \Lambda_2^{-2} \(\X^{n-2}\)^0{}_\mu \, \partial^{\mu} \phi_a\,,
\ea
for any $n\ge 1$. In particular for $n=1$, this implies $\frac{\partial}{\partial \dot{A}^a} [\X] = \Lambda_2^{-2}V_a$.
Now that every element has been introduced, we can compute the momenta first and then the Hessian matrices for each order in $\K$ or $\X$. Since $\K$ and $\X$ are linearly related to one another, the $\L_n[\K]$ can be expressed as linear combinations of the $\L_n[\X]$ as summarized in \cite{deRham:2014zqa} and we may use either choice for the following argument without loss of generality. We will then show that $V_a$ is actually the null eigenvector for the Hessian derived for any linear combination of  $\L_n[\K]$ or equivalently any linear combination of $\L_n[\X]$ hence proving the existence of a constraint. 

Let us start with the easiest case by considering $\L_1[\X]$. The conjugate momentum associated to $\phi_a$ is already given in \eqref{eq:dK1} and we have
\begin{equation}
	p_a^{(1)} =\Lambda_2^{4}\frac{\p \L_1[\X]}{\p \dot A^a}=\Lambda_2^{2} V_a\,.
\label{eq:pa1}
\end{equation}
The Hessian associated with this Lagrangian is then
\ba
\mathcal{H}_{ab}^{(1)}=\Lambda_2^{4}\frac{\p^2 \L_1[\X]}{\p \dot A^a \p \dot A^b}=\Lambda_2^{2}\frac{\p V_a}{\p \dot A^b}\,.
\ea
Rather than computing this Hessian explicitly, it is actually easier to simply make use of the property of $V_a$ (and the fact that it has constant norm),
\ba
\Lambda_2^{-2}	\mathcal{H}_{ab}^{(1)}V^a = \frac{\partial V_a}{\partial \dot{A}^b}V^a
	                          = \frac{1}{2} \frac{\partial (V_a V^a)}{\partial \dot{A}^b}
											      = \frac{1}{2} \frac{\partial (-1)}{\partial \dot{A}^b}
											      = 0\,,
\label{eq:eig1}
\ea
hence proving that $V_a$ is indeed a null eigenvector of $\H^{(1)}_{ab}$.\\

Generalizing this result for any \Pros Lagrangian is straightforward and the details are provided in appendix~\ref{app:Null}, where we show that for any Lagrangian of the form \eqref{eq:defLK}, the associated Hessian carries the {\it same} null eigenvector $V_a$
for all linear combinations of Lagrangians  $\L_n[\X]$.
It follows that any linear combination of $\L_n[\X]$ or $\L_n[\K]$ carries a constraint and only excites three degrees of freedom in four dimensions. Interestingly, the way the constraint manifests itself differs from the way it does in GP (their respective null eigenvectors differ). This implies that considering a hybrid theory composed of GP {\it and} \Pros interactions would not enjoy a constraint.

Remarkably, the existence of a constraint is now manifest irrespectively of the choices of $\alpha_n$. The argument provided here, therefore, extends prior proofs for the absence of ghost in massive gravity in the \stu language beyond what was proposed in \cite{deRham:2011rn} and \cite{Hassan:2012qv}. Such a general proof was previously missing in the literature. Interestingly with the exact form of the null eigenvector at hand, one should now be able to determine the full non--linear version of the \stu field in terms of which massive gravity and \Pros can be manifestly expressed in first order form.

\section{Inequivalence with Generalized Proca}
\label{sec:Ineq}

The aim of this Section is to show that the \Pros theory provided in \eqref{eq:defLK} does not enter the scope of GP. It is clear that \Pros includes an infinite number of operators with arbitrarily high order in $(\p A)$ while GP only includes a finite number of those (putting aside the gauge--invariant interactions). However by itself, this does not imply that both theories may not still be the same in disguise for instance through a sophisticated field redefinition or even an analogue to the Galileon duality proposed in \cite{deRham:2013hsa,deRham:2014lqa}. In \cite{deRham:2018svs} it was shown that on AdS, there were no local and Poincar\'e invariant field redefinitions between GP and the DL of massive gravity. In what follows we shall show that this result is generic, and even account for more subtle types of space--dependent field redefinitions like generalized Galileon dualities, there can be no local field redefinition that maps GP with \Pros theories. This will be done in full generality by computing and comparing the S matrix of both theories in section~\ref{ssec:scattering} but to start with we shall start by recalling that the very way the constraint gets satisfied differs in GP and \Pros theories as can be seen very easily in two dimensions.

\subsection{Appetizer}
\label{ssec:naive2d}

By definition, a GP is a theory carrying a constraint and thus propagating only $d-1$ degrees of freedom in $d$ spacetime dimensions. However the existence of a constraint can take various different forms and the non--dynamical variable does not necessarily need to be  $A_0$ itself, it may be a linear combination of $A_0$ and other components of the vector field. In GP, the Hessian is always of the form
\begin{equation}
 \tilde{\mathcal{H}}^{\rm (GP)}_{ab} =
		\begin{pmatrix}
			0 & 0 \\
			0 & \#
		\end{pmatrix}\,.
\label{eq:Hessmat2}
\end{equation}
In \Pro, on the other hand, while the Hessian still carries a null eigenvalue, its form differs from \eqref{eq:Hessmat2} at least when expressed in terms of the components of the field $A_\mu$,
indeed, the Hessian for the two--dimensional Lagrangian $\L^{\rm(2d)}$ \eqref{eq:L2d} is expressed in \eqref{eq:Hessmat} and is not of the form \eqref{eq:Hessmat2} even though both Hessians have null determinant. 

 Let us now suppose there could exist a field redefinition $A_\mu \to \tilde{A}_\mu(A)$ such that the Hessian for $\tilde A$ is of the form \eqref{eq:Hessmat2}.
After the field--redefinition, the Hessian matrix takes the form
\begin{equation}
	\tilde{\mathcal{H}}_{ab} =  \frac{\delta \tilde{A}_a}{\delta A_c} \mathcal{H}_{cd}  \frac{\delta \tilde{A}_b}{\delta A_d}\,.
\label{eq:Hessmat3}
\end{equation}
Asking for $\tilde{\mathcal{H}}$ to be of the form \eqref{eq:Hessmat2} would require the field redefinition to be such that
\begin{equation}
	(\dot{A}_1-A_0') \frac{\delta \tilde{A}_0}{\delta A_0} = (2\Lambda_2^2+A_1'-\dot{A}_0) \frac{\delta \tilde{A}_0}{\delta A_1}\,.
\label{eq:nonlocal}
\end{equation}
which cannot be satisfied without imposing a non--local expression for $\tilde A_0$ in terms of $A_0$ and $A_1$. At this stage, one can already expect there to be no local field redefinition that brings \Pros back to a GP form. The same conclusion was highlighted in AdS in Ref.~\cite{deRham:2018svs}. We shall make this statement more rigorous in what follows.

\subsection{Scattering amplitudes}
\label{ssec:scattering}
To consolidate the previous argument on the absence of local field redefinition that would bring \Pros into a GP form, we shall compare here the tree--level $2 \rightarrow 2$ scattering amplitudes for both theories.

First we emphasize that at the linear level, GP and \Pros are identical, indeed $\L_{\text{GP}}^{(2)}$  in \eqref{eq:LGP2bis} is identical to $\L_{\K}^{(2)}$ in \eqref{eq:LK2}. This implies that the free asymptotic states defined in both theories are the same and one can meaningfully compare the amplitudes computed for each model.
Computing the indefinite $2 \rightarrow 2$ tree--level amplitudes in both theories is straightforward but for conciseness, we only present here the results for scatterings of some specific definite helicity states. As we shall see, these definite amplitudes are by themselves sufficient to show that the new Proca interactions we introduced in section~\ref{sec:NewProca} differ from those of GP theories.

For simplicity, we choose to describe the kinematic space with the Mandelstam variable $s$ (center of mass energy$^2$) and the scattering angle $\theta$, see Appendix~\ref{sec:appKin}.

Starting with $++\rightarrow--$, the respective scattering amplitudes in \Pros and GP  are given by
\begin{align}
\label{eq:ampKppmm}
	\mathcal{A}^{++\rightarrow--}_{\K}(s,\theta)
		&= -\frac{i}{64\Lambda_2^4} \left(\frac{s^3}{m^2} (1 + 4\bar{\alpha}_2 - 6 \bar{\alpha}_2)^2 \right. \\
		& \quad \quad \quad \quad - 2 s^2 \left(4(\bar{\alpha}_2 - 3 \bar{\alpha}_3 + 6\bar{\alpha}_4) + (1+ 8\bar{\alpha}_2 - 12\bar{\alpha}_3)^2 - 96(1 + 4\bar{\alpha}_2 - 6\bar{\alpha}_3) \bar{\gamma}_1 \right) \nonumber \\
		& \quad \quad \quad \quad + 8 m^2 s \left(1 + 4(\bar{\alpha}_2 - 3\bar{\alpha}_3 + 6\bar{\alpha}_4 - 12\bar{\gamma}_1 + 8\bar{\gamma}_2) + 2(4\bar{\alpha}_2 - 6\bar{\alpha}_3 - 24\bar{\gamma}_1)^2 \right) \nonumber \\
		& \quad \quad \quad \quad - 16 m^4 \left(1 - 48\bar{\gamma}_1 + 32\bar{\gamma}_2 + 768\bar{\lambda}_0 \right) \nonumber \\
		& \quad \quad \quad \quad + (s-4m^2)(-8(1 - 4\bar{\alpha}_2 + 6\bar{\alpha}_3)m^2 + 3(1 - 4\bar{\alpha}_2 + 4\bar{\alpha}_3 + 8\bar{\alpha}_4)) \sin(\theta)^2 \nonumber \\
		& \quad \quad \quad \quad - \left. \frac{4(s-m^2)^2 (s-2m^2) (s-4m^2) (1 - 4\bar{\alpha}_2 + 6\bar{\alpha}_3)^2}{4m^2(s-3m^2) + (s-4m^2)^2 \sin(\theta)^2} \sin(\theta)^2 \right)\,,\nonumber
\end{align}
and
\begin{align}
\label{eq:ampGPppmm}
	\mathcal{A}^{++\rightarrow--}_{\text{GP}}(s,\theta)
		&= -\frac{i}{4\Lambda_2^4} \left(\frac{s^3}{m^2} a_2^2 \right.  \\
		& \quad \quad \quad \quad + 8 s^2 \left(a_2 (a_1 - a_2) - b_5 - 4b_6 \right) \nonumber \\
		& \quad \quad \quad \quad + 8 m^2 s \left(-4b_2 + b_4 + 4b_5 + 16b_6 + 2(a_1-a_2)^2 \right) \nonumber \\
		& \quad \quad \quad \quad - 32 m^4 \left(b_1 - 2b_2 + 2b_5 + 4b_6 \right) \nonumber \\
		& \quad \quad \quad \quad - 4 (s-4m^2)(b_5 s + b_4 m^2) \sin(\theta)^2 \nonumber \\
		& \quad \quad \quad \quad - \left. \frac{4(s-m^2)^2 (s-2m^2) (s-4m^2) a_2^2}{4m^2(s-3m^2) + (s-4m^2)^2 \sin(\theta)^2} \sin(\theta)^2 \right)\,.\nonumber
\end{align}
Remarkably we see that perturbative unitarity gets broken when $s^3\sim \Lambda_2^4 m^2 \sim \Lambda_3^6$, hence confirming the existence of non-trivial operators at the scale $\Lambda_3$.
If both theories were equivalent they would predict the same scattering amplitudes for any incoming and outgoing polarization states. We will note any amplitude difference for a given set of polarizations $\Delta \mathcal{A}$ and ask them to vanish for all $(s,\theta)$, in particular
\begin{align}
	\Delta \mathcal{A}^{++\rightarrow--}(s,\theta) &= \mathcal{A}^{++\rightarrow--}_{\K}(s,\theta) - \mathcal{A}^{++\rightarrow--}_{\text{GP}}(s,\theta) \nonumber \\
		&= \sum_{n=0}^3 C_n s^n m^{4-2n} + (s-4m^2) \sin(\theta)^2 (C_4 m^2 + C_5 s) \label{eq:deltaA} \\
		& \quad \quad + C_6 \frac{(s-m^2)^2 (s-2m^2) (s-4m^2)}{4m^2(s-3m^2) + (s-4m^2)^2 \sin(\theta)^2} \sin(\theta)^2 \,, \nonumber
\end{align}
where the constants $C_{n}$ are expressed in terms of the coupling constants of the GP and \Pros only.
For the scatterings \eqref{eq:ampKppmm} and \eqref{eq:ampGPppmm} to be equivalent, one should have $C_n=0$ for all $n = 0, \dots, 6$.
Imposing these relations in terms of the coupling constants then sets
\begin{equation}
	\begin{cases}
		a_2 &= \pm \frac{1}{4} \\
		b_1 &=\frac 18 ( 1 - 2a_1 + 8a_1^2- 12 \bar{\gamma}_1 -288 \bar \gamma_1^2+192 \bar{\lambda}_0)\\
		b_2 &= \frac14\(2a_1^2 -3 \bar{\gamma}_1- 72\bar{\gamma}_1^2-2 \bar{\gamma}_2 \)\\
		b_4 &= \frac{1}{8} \\
		b_5 &= - \frac{1}{64} (3 - 4\bar{\alpha}_2 + 24 \bar{\alpha}_4 ) \\
		b_6 &= \frac{1}{32} (2a_1 - \bar{\alpha}_2 + 6 \bar{\alpha}_4 - 12\bar{\gamma}_1) \\
		\bar{\alpha}_3 &= \frac{2}{3} \bar{\alpha}_2
	\end{cases}
\label{eq:Cnsol}
\end{equation}
From these relations, it is clear that the most generic \Pros theory cannot be put in the form of GP since one already needs to impose $\bar{\alpha}_3 = \frac{2}{3} \bar{\alpha}_2$ but looking at other polarizations makes it clear that even within this choice of coefficients the theories are never equivalent. Indeed, turning now to $+- \rightarrow +-$ scatterings then upon imposing the solution \eqref{eq:Cnsol}, we find
\begin{equation}
	\Delta \mathcal{A}^{+-\rightarrow+-}(s,\theta=0) = \frac{i}{4\Lambda_2^4}(4m^2-s)s\,,
\label{eq:deltaAbis}
\end{equation}
at this stage there are no further couplings one can dial to ensure the equivalence and so irrespectively of the choice of coefficients  $\left\{\bar{\beta}_i,\bar{\gamma}_i,\bar{\lambda}_i,a_i,b_i\right\}$ the \emph{full} tree--level $2 \rightarrow 2$ scattering amplitude of our new Proca interactions never matches that predicted by GP. This concludes the proof that both theories are fundamentally different and are \emph{not} equivalent.

\section{Extended Proca-Nuevo theory}
\label{sec:ExtPN}

\begin{figure}
	\center{\includegraphics[width=6cm]{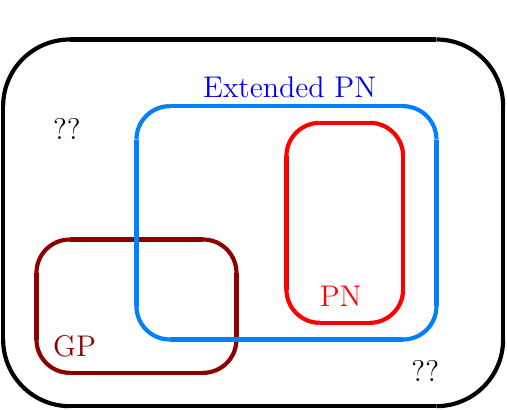}}
	\caption[Space of ghost-free massive vector field theories]{Charting the space of massive spin-1 self-interacting theories that exhibit a constraint.}
	\label{fig:proca theory space}
\end{figure}

Given the inequivalence between GP and PN, it is natural to ask whether a still more general theory exists from which {\it both} GP and PN could arise as particular ``corners'' in the space of models, i.e.\ through particular choices (possibly in a limiting sense) of coupling constants. Exploring this question is the first aim of this Chapter. We shall show that such extension does exist, in a model that we imaginatively call ``Extended Proca-Nuevo''. While this proposal succeeds in furnishing a link between GP and PN, we should warn the reader of two caveats. First, in the ``GP limit'' of extended PN not all of the operators belonging to the GP class are obtained, although the whole PN class is included; this is represented artistically in Fig.\ \ref{fig:proca theory space}.

The PN model defined previously is special due to its link to massive gravity, but it is actually straightforward to include additional interactions within the same class. The main observation is that any operator that leaves the Hessian fully invariant can be added to the theory without affecting the form of the NEV.
In 4D there exist precisely five operators  built out of the tensor $\partial^{\mu}A_{\nu}$ that respect that condition, namely the operators $d_n(X) \L_n[\p A]$, defined according to the rule in Eqs.~\eqref{eq:refLnK}--\eqref{eq:refLnK0to4}.

Although the $\L_n[\p A]$ operators are total derivatives and thus trivial in isolation, when added to the Lagrangian with a field-dependent coefficient, they produce non-trivial phenomenological effects, while being trivial at the level of the Hessian. These operators are precisely those that define the novel derivative interactions of GP theory (with the exception of $\L_4$, on which we will comment later). What we have uncovered here is that they may be added to the complete PN Lagrangian without thwarting the constraint structure. It is worth pointing out that the other members of the GP class, namely those that are not constructed solely in terms of elementary symmetric polynomials of $\partial^{\mu}A_{\nu}$, do in general contribute to the Hessian matrix and can therefore not trivially be added within this setup.

We note that some redundancies are introduced through the construction we have just outlined: (i) $\L_0[\p A]$ is a constant, therefore its coefficient will contribute to the non-derivative potential and hence can be absorbed into $\alpha_0$; (ii) because of the identity
\begin{equation}
	\sum_{n=1}^4 \frac{\L_n[\K]}{n!} = \sum_{n=1}^4 \frac{\L_n[\p A]}{\Lambda^{2n} n!} \,,
\end{equation}
it follows that only three among the four remaining terms are linearly independent from the PN operators; (iii) moreover, it has been proved \cite{BeltranJimenez:2019wrd} that $f(X) \L_4[\p A]$ is a total derivative for \textit{any} function $f$, therefore this term is always redundant. However, properties (ii) and (iii) hold only in flat spacetime, and are no longer true in a generic curved background upon replacing $\p A\to\nabla A$. Since our aim is to use this setup as a starting point for building a covariant theory, we are thus led to consider all four GP terms $\L_1[\p A]$ through $\L_4[\p A]$.

With these considerations in mind, we now introduce the following Lagrangian,
\beq
	\L_{\text{EPN}} = \tilde{\Lambda}^4 \sum_{n=0}^4 \alpha_n(\tilde{X}) \L_n[\tilde{\K}[A]] + \Lambda^4 \sum_{n=1}^4 d_n(X) \frac{\L_n[\p A]}{\Lambda^{2n}} \,,
	\label{eq:LextPNflat}
\eeq
which we refer to as ``Extended Proca-Nuevo'' (EPN) theory. In four dimensions, it includes four additional arbitrary functions, $d_n(X)$, besides the original $\alpha_n$. Note that we have allowed for the possibility for the two families of operators to enter at different scales, namely at the scale $\Lambda$ and $\tilde \Lambda$, and denote as $\tilde \K$ and $\tilde X$ the same quantities as defined previously but suppressed with the scale $\tilde \Lambda$.

Obviously, the difference in scaling could be absorbed into the functions $\alpha_n$ (and we will do so later), but for now, we keep treating both scales independently as the relation between them can be used as a ``dialing'' parameter to interpolate between the respective PN and GP Lagrangians. In this sense, PN provides a perhaps unexpected link between the two previously known models.

This interpolation between PN and GP is seen most explicitly at the level of the NEV. Given that the additional  $d_n \L_n[\p A]$ operators do not affect the Hessian matrix, the NEV of EPN coincides with that of PN defined above in \eqref{eq:Va}. Denoting as $V_a^{\rm EPN}(\tilde \Lambda, \Lambda)$ the NEV of EPN, it is straightforward to check that
\ba
V_a^{\rm EPN}(\tilde \Lambda, \Lambda)= V_a^{\rm PN}(\tilde \Lambda)\,,
\ea
where the NEV for PN is defined in \eqref{eq:Va}. In the limit where $\tilde \Lambda \to \infty$, keeping the vector mass and the scale $\Lambda$ fixed, we have $\chi^\mu{}_\nu (\tilde \Lambda)\to \delta^\mu{}_\nu$, so it is clear that we recover the respective GP and PN null eigenvectors by taking the following limits for $\tilde \Lambda$
\begin{equation}
	\begin{cases}
		V_a^{\rm EPN}(\tilde \Lambda,\Lambda) \xrightarrow[\tilde \Lambda \rightarrow \infty]{} (1,\vec{0}) \,, \qquad &\text{(GP case)} \\
		V_a^{\rm EPN}(\tilde \Lambda,\Lambda) \xrightarrow[\tilde \Lambda \rightarrow \Lambda]{} V^{\rm PN}_a (\Lambda) \,, \qquad &\text{(PN case)}
	\end{cases}
\end{equation}
that is, the NEVs of GP and PN are obtained from the NEV of EPN in particular limits.

Let us emphasize that the GP limit is however non-trivial. Even though the limit $\tilde \Lambda\to \infty$ is well-defined and unambiguous at the level of the NEV, taking that limit at the level of the Lagrangian is on the other less trivial. Nevertheless, it is straightforward to see that one can indeed isolate all the $\L_n(\p A)$ GP operators via this procedure so long as the scale $\Lambda$ and the vector mass are both kept fixed in that limit. See Appendix \ref{sec:GPfromEPN} for details.

Having understood the relation with GP, we will focus on the full EPN theory for the remainder of the Chapter and without further loss of generality, we may set $\tilde{\Lambda}=\Lambda$.

\section{Conclusions}
\label{sec:Outlook}

In this Chapter, we proposed a new interactive theory for a single massive vector field with derivative self--interactions and free of \Ostro ghost instability. The \Pros Lagrangian is heavily inspired by massive gravity and is genuinely different from the GP classes of interactions. We started by proving that \Pros exhibits a constraint in two dimensions before providing the exact non--perturbative form of the null eigenvector of the Hessian matrix in any dimensions. \Pros provides an insightful example of an \Ostro ghost--free theory with a non--trivial null eigenvector. Indeed, whereas GP imposes $A_0$ to be non--dynamical, \Pro's constraint arises as a combination of $A_0$ and the spatial field components. This is already a strong hint indicating that both theories are fundamentally different. To complete the proof more rigorously, we computed the $2 \rightarrow 2$ scattering amplitudes in GP and \Pros theories and showed that they could never be matched irrespectively of the choice of coefficients. This proves that their respective $S$--matrices are different and thus \Pros cannot be related to GP by any local field redefinition.

Throughout this Chapter, we have focused our analysis on the existence of a constraint and on the counting of the number of propagating degrees of freedom, which will be studied in much more depth in Chapter \ref{chap:Constraint}. In itself this question is distinct from whether or not the theory provided here can ever enjoy a standard analytic, unitary, local, Lorentz--invariant and causal high energy completion\footnote{We emphasis that the absence of such high energy completion does not necessarily rule out the existence of other consistent completions, see Refs.~\cite{Keltner:2015xda,deRham:2017xox} for relevant discussions.} although some connections were previously established for massive spin--2 interactions \cite{deRham:2018qqo} using the so--called beyond--forward positivity bounds \cite{deRham:2017zjm}. Applying the forward bounds to a specific class of spin--1 effective field theory was considered in \cite{Bonifacio:2016wcb} and implications to GP and other types of massive spin--1 effective field theories in and beyond the forward limit was considered in \cite{deRham:2018qqo}. Interestingly the positivity bounds on GP requires the introduction of very specific operators and it would be interesting to understand whether the same type of arguments applies to the theory at hand. 

Finally, we note that in proving the existence of a constraint for \Pro, we have generalized the proof for the absence of ghost in massive gravity in the \stu language beyond what had previously been proposed in the literature. Remarkably, we now have the full non--linear expression for the null eigenvector of the Hessian. With this knowledge at hand, one should now be able to determine the full non--linear expressions for the \stu fields in terms of which massive gravity can be express in a manifestly first order form. 

With this proposal of a new massive vector theory enjoying a Hessian constraint, we would now like to rigorously study the full constraint structure and prove that there exists a secondary second class constraint that fully removes the would-be ghostly degree of freedom. This analysis is the subject of Chapter \ref{chap:Constraint} to follow.

\chapter{Constraint Analysis for Proca-Nuevo}
\label{chap:Constraint}

Chapter \ref{chap:PN} showed how one can introduce a set of higher-order self-interactions for a massive vector field that do \textit{not} lead to second-order equations of motion and yet realize a Hessian constraint in a highly-non linear way. This constraint is second class and formally only removes half a degree of freedom in field space. However, it has recently been argued that half degrees of freedom could emerge in Lorentz and parity invariant field theories, using Proca-Nuevo as a specific example. In this Chapter, we will provide two proofs, using the Lagrangian and Hamiltonian pictures, that the theory possesses a pair of second class constraints, leaving $D-1$ degrees of freedom in $D$ spacetime dimensions, as befits a consistent Proca model. Our proofs are explicit and straightforward in two dimensions and we discuss how they generalize to an arbitrary number of dimensions. We also clarify why local Lorentz and parity invariant field theories cannot hold half degrees of freedom.

\section{Half--Being}
\label{sec:HalfDegree}

Before diving into the subtleties related to interacting Proca theories and how the secondary constraint is realized, it is worth asking ourselves if and when half a number of dynamical field space degrees of freedom can propagate in a generic local field theory. We also refer to \cite{Golovnev:2022rui} for a useful review of the role played by constraints and degrees of freedom.

As explained in \cite{deRham:2014zqa}\footnote{See specifically the explanations below Eq.~(7.30) of
\href{https://link.springer.com/article/10.12942/lrr-2014-7}{Living Rev. Rel. 17 (2014) 7}.}, half a field space degree of freedom or an odd number of degrees of freedom in phase space corresponds to a system of first order differential equations that cannot be recombined into a fully second order system. This implies the existence of a {\it field space} degree of freedom (say $\chi$) whose dynamics (upon appropriate diagonalization) would be governed by a first order time derivative equation, $\p_t \chi=\hat{O}(\p_i) \chi+\cdots$, where ellipses involve other fields with no time derivatives acting on them and $\hat O$ is an operator that only involves functions of spacetime and spatial derivatives of any order but no additional time derivatives.

If the theory is fundamentally {\it Lorentz invariant} (even if considering a solution that spontaneously breaks Lorentz invariance), then this differential equation can necessarily be recast as $L^\mu \p_\mu \chi=f(x^\mu) \chi+\cdots$, for a Lorentz vector $L^\mu$ meaning that the evolution equations should contain terms that are linear in space derivatives, $L^i\p_i \chi$.  Under a parity transformation $x^i \to -x^i$, these terms are odd and the theory as a whole would therefore break parity. Alternatively, an odd number of dynamical phase variables can be propagating in theories that preserve parity at the price of breaking Lorentz invariance. Phrased in terms of the last odd phase space variable, locality joined together with Lorentz and parity invariance impose $L^\mu=0$ meaning that the equation for $\chi$ is none other than the additional constraint responsible for ensuring an even-dimensional phase space.

As a result, in complete generality, we can infer that a local, Lorentz and parity invariant field theory can never propagate an odd number of phase space degrees of freedom or equivalently half a number of field space degrees of freedom\footnote{A constraint-based proof of this fact was given in \cite{Crisostomi:2017aim} in the context of degenerate scalar field theories.}. An exception to this rule was suggested recently in \cite{ErrastiDiez:2022qvd}, which if correct would hint towards a potential undiagnosed loophole to the previous argument. However, as we shall see below, those conclusions were premature and upon appropriately identifying the constraint algebra the example suggested in \cite{ErrastiDiez:2022qvd} follows precisely the logic highlighted above. Rather than being a dynamical equation for half a field space degrees of freedom, the remaining equation is nothing other than a secondary second class constraint that projects out the other half field space variable, as expected from Lorentz and parity invariance. We shall see this more precisely in what follows.

\section{(Extended) Proca-Nuevo in arbitrary dimensions}
\label{sec:EPNConstraintArbitraryD}

Note that mathematically there exists multiple branches of solutions satisfying the relation $\X \X =\eta^{-1}f$ \cite{Comelli:2015ksa}, but only solutions which are continuously connected to the trivial one for which $\X\mupn[0]=\delta\mupn$ should be considered. With this choice in mind, the EPN theory for a massive vector field $A_{\mu}$ in arbitrary spacetime dimension $D$ is defined by the Lagrangian \cite{deRham:2020yet, deRham:2021efp}\footnote{Note that on flat spacetime, the $d_D$ term is a total derivative independently of the form of the arbitrary function of $X$ \cite{BeltranJimenez:2019wrd}.}
\begin{equation}
	\L_{\rm EPN}[A] = \Lambda^D \sum_{n=0}^D \alpha_n(X) \L_n[\K] + \Lambda^D \sum_{n=1}^{D-1} d_n(X) \frac{\L_n[\p A]}{\Lambda^{n D/2}} \,,
\label{eq:defLKconstraints}
\end{equation}
where the first sum contains the pure PN terms and the second involves the GP interactions introduced in \cite{Tasinato:2014eka,Heisenberg:2014rta}, with the individual Lagrangians following the notation of Eqs.~\eqref{eq:refLnK}. 
The dimensionless generic functions $\alpha_n$ and $d_n$ depend on the Lorentz scalar $X$ defined as
\begin{equation}
	X = \frac{1}{\Lambda^{D-2}}\,A^{\mu}A_{\mu} \,.
	\label{eq:defXargument}
\end{equation}

\subsection{Trivial vacuum}

We are only interested in theories for which we recover the standard free Proca theory perturbatively about the vacuum $\langle A_\mu \rangle=0$, so to quadratic order in the vector field, it is understood that the Lagrangian \eqref{eq:defLKconstraints} must reduce to
\ba
\L_{2}=-\frac 14 F\mn^2-\frac 12 m^2 A^2 + \mathcal{O}\(\frac{\p A^3, A^4}{\Lambda^{(D-4)/2}}\)\,.
\ea
In particular, as emphasized in \cite{deRham:2020yet}, the theory only makes sense if the helicity-0 mode of the vector field carries a kinetic term on the trivial  standard Lorentz invariant vacuum $\langle A_\mu \rangle=0$, which implies that the potential should always include a mass term. The functions $\alpha_n(X)$ and $d_n(X)$ should therefore be analytic functions of their argument about $X=0$, so we can express them in terms of their Taylor expansion
\ba
\label{eq:expansion_alpha_d}
\alpha_n(X)=\sum_{n\ge 0} \frac{1}{k!}\bar \alpha_{n,k}X^k\,, \qquad {\rm and}\qquad
d_n(X)=\sum_{n\ge 0} \frac{1}{k!}\bar d_{n,k}X^k\,,
\ea
with the convention $\bar \alpha_{0,0}=0$, and
\ba\label{eq:alpha0}
\bar \alpha_{0,1}=-\frac 12 \frac{m^2}{\Lambda^2}\,, \quad {\rm and} \quad
\bar\alpha_{2,0}=1+\frac 12 \bar \alpha_{1,0}\,.
\ea
If instead one had for instance a theory where $\alpha_0$ is constant and does not carry the mass term linear in $X$, \ie setting $\bar \alpha_{0,1}=0$ unlike what is indicated in \eqref{eq:alpha0}, would result in an infinitely strongly coupled vacuum solution  $\langle A_\mu \rangle =0$ which would be against the logic of the model presented here.

\subsection{Null Eigenvector}

There exist multiple ways to show that PN and GP are genuinely different theories (not even related by Vector dualities \cite{deRham:2014lqa}). It was first proven in \cite{deRham:2020yet} that their scattering amplitudes differ, hence establishing their inequivalent nature. It was then shown in \cite{deRham:2021efp} that the class of cosmological predictions of PN could differ from the  GP ones. However, the most immediate and natural hint at the fact that these theories are truly different is to envisage how the constraint is realized. This can be identified by considering the null eigenvector (NEV) of their respective Hessian matrices of field velocities. One of the underlying hypotheses of GP is that all modes in the decoupling limit should have at most second-order equations of motion, which is related to the fact that the component $A_0$ of the massive vector field remains non-dynamical when adding the GP interactions to the standard Proca theory. This corresponds to a NEV in the field-space direction $(1,\vec{0}\,)$. On the other hand, the very construction of the pure PN interactions inhibits such a clean identification of the non-propagating degree of freedom. Yet it was proven in \cite{deRham:2020yet} that their Hessian was still degenerate as it enjoys a null eigenvector. Since we shall be interested in the presence of a secondary constraint, it is beneficial to first identify the degenerate direction of (E)PN and hence review how the NEV can be identified.

To this end we first recall the tensor $\mathcal{Z}=\X^{-1} \eta^{-1}$ which was shown to be symmetric in Eq.~\eqref{eq:Zinv}, $\Z=\Z^T$, and to satisfy the properties Eqs.~\eqref{eq:Zf}--\eqref{eq:ZfZ}. We further define a tensor $W^\mu{}_\nu$ via
\begin{equation}
	W\ud{\mu}{\nu} = \mathcal{Z}^{\mu \alpha} \p_{\alpha} \phi_{\nu} \,,
	\label{eq:defWtensor}
\end{equation}
which can be seen to belong to the Lorentz group,
\begin{equation}
	W^{\mu}{}_{\alpha}\eta^{\alpha\beta} W^{\nu}{}_{\beta} = \eta\mnup \,.
\end{equation}
The generalized Hessian matrix is given by
\begin{equation}
	\mathcal{H}^{\mu \nu, \alpha \beta} = \frac{\p^2 \L_{\rm EPN}}{\p \p_\mu {A}_{\alpha} \p \p_\nu {A}_{\beta}} = \mathcal{H}^{\nu \mu, \beta \alpha} \,,
\end{equation}
and $\mathcal{H}^{\mu \nu}\equiv \mathcal{H}^{00,\mu \nu}$ plays the role of the kernel of the kinetic term of the Lagrangian, hence encoding information about the propagating degrees of freedom of the theory. One can verify that the vector
\begin{equation}
	V_{\mu} = W\ud{0}{\mu} \,,
	 \label{eq:PN NEV constraints}
\end{equation}
is the normalized time-like NEV of the Hessian matrix of PN \cite{deRham:2020yet},
\begin{equation}
	\mathcal{H}\mnup V_{\mu} = 0 \,, \qquad V^{\mu} V_{\mu} = - 1 \,.
\end{equation}
In addition, as proven in \cite{deRham:2021efp}, the inclusion of the GP operators $\L_n[\p A]$ leaves the Hessian matrix $\mathcal{H}^{\mu \nu, \alpha \beta}$ invariant and can therefore be safely added to the PN Lagrangian without affecting the constraint structure, resulting in the EPN model.

\subsection{Pair of second class constraints}

The existence of a NEV in (E)PN theory implies the existence of a constraint, which must be second class as the theory does not have any gauge symmetries. The constraint, therefore, removes one phase-space variable corresponding to half a Lagrangian degree of freedom. The removal of the other half then necessitates the existence of another second class constraint. While the analyses of \cite{deRham:2020yet,deRham:2021efp} did not derive the latter, it was however argued that the Hessian constraint was enough to prove the absence of  the full Ostrogradsky ghost in (E)PN. This follows for multiple physically motivated reasons:
\begin{enumerate}
\item  First of all, since PN is a DL of generalized massive gravity (see \cite{deRham:2014gla,deRham:2016plk}) for which the secondary constraint was already derived in the literature (proven fully non-linearly in \cite{Hassan:2011hr,Golovnev:2011aa,Alexandrov:2013rxa,Golovnev:2017iix}), it directly follows that the secondary constraint has to be realized in PN (see \cite{deRham:2014zqa} on what it means physically to take a DL).
\item About the trivial vacuum $\langle A_\mu \rangle =0$, we recover a standard Proca theory at the linear level, which as is well known propagates $D-1$ degrees of freedom, implying that PN has to propagate {\it at least} $D-1$ degrees of freedom. Since it is expressed in terms of $D$ vector field components with only first derivatives acting on them at the level of the action, it can {\it at most} propagate $D$ modes. However, the presence of a NEV for the Hessian matrix implies that the system is degenerate and must in fact propagate strictly fewer than $D$ field space degrees of freedom. Moreover, as explained in \cite{deRham:2014zqa} and reviewed already in Section~\ref{sec:HalfDegree}, a local, Lorentz and parity preserving field theory cannot propagate an odd number of physical field space degrees of freedom. Together, these arguments imply that about any solution analytically connected to the standard Lorentz invariant vacuum $\langle A_\mu \rangle =0$, there should be precisely $D-1$ field space degrees of freedom. This result is consistent with previous arguments reminiscent of massive gravity \cite{Hassan:2011hr,Alexandrov:2013rxa}.
\end{enumerate}
The above points are also consistent with further analyses:
\begin{enumerate}\setcounter{enumi}{2}
\item Perturbations of (E)PN on cosmological backgrounds were analyzed in \cite{deRham:2021efp}, where they were shown to exhibit the expected number of physical degrees of freedom, all of which were identified as being stable (even though following the logic of the analysis presented in \cite{ErrastiDiez:2022qvd} a mismatch would already have been identified at that level).
\item Positivity bounds, which rely on unitarity (which would be broken if half a ghost degree of freedom was propagating) together with Lorentz invariance and crossing symmetry were shown to be satisfied for generic (E)PN parameters in \cite{deRham:2022sdl}.
\end{enumerate}
These points suggest (if not prove) that the Hessian constraint in (E)PN is actually part of a pair of second class constraints fully removing the unwanted ghostly degree of freedom. Nevertheless, it remains an interesting open question to establish how the constraint structure manifests itself more precisely. In the rest of this work, we turn our attention to this question.

\subsection{Extended Proca-Nuevo in two dimensions}
\label{sec:EPN2dConstraint}

In this Section, we will focus on the example analyzed in \cite{ErrastiDiez:2022qvd} corresponding to the EPN model in $D=2$ dimensions. This model is interesting because the structure of the Lagrangian can be made very explicit and allows for a tractable analytical treatment.

The most general two-dimensional EPN Lagrangian is given by \cite{deRham:2020yet}
\begin{equation}
\label{eq:LEPN2d}
	\L^{\rm (2d)}_{\rm EPN} = \Lambda^2 \( \alpha_0(X) +\alpha_1(X)  [\K] + \alpha_2(X) \left([\K]^2-[\K^2] \right) + \frac{d_1(X)}{\Lambda}[\p A] \) \,,
\end{equation}
where the analytic functions $\alpha_{0,1,2}(X)$ and $d_{1}(X)$ satisfy the expansion given in \eqref{eq:expansion_alpha_d} with the convention \eqref{eq:alpha0}. Note however that even though $\L_2[\p A]$ is multiplied by an arbitrary function of $A^2$, namely $d_2$, the whole term $d_2(X) ([\p A]^2-[(\p A)^2])$ is a total derivative and hence can be discarded \cite{BeltranJimenez:2019wrd}.

As discussed in \cite{deRham:2020yet}, in $D=2$ dimensions, the Lagrangian can be expressed in closed form in terms of the  following variables:
\begin{equation}
	x_{\pm} = 1 \pm 1 + \frac{A_1' \mp \dot{A}_0}{\Lambda} \,, \qquad y_{\pm} = \frac{\dot{A}_1 \mp A_0'}{\Lambda} \,, \qquad N_{\pm} = \sqrt{x_{\pm}^2 - y_{\pm}^2} \,,
\end{equation}
where a dot (prime) stands for time (space) derivative. We also define their reduced versions
\begin{equation}
	\bar{x}_{\pm} = \frac{x_{\pm}}{N_{\pm}} \,, \qquad \bar{y}_{\pm} = \frac{y_{\pm}}{N_{\pm}} \,,
\end{equation}
allowing us to rewrite the Lagrangian in the form
\begin{equation}
\label{eq:LEPN_2d}
	\L^{\rm (2d)}_{\rm EPN} = \Lambda^2 \left[ \tilde{\alpha}_0 + \tilde\alpha_1  N_{+}+ d_1  x_{+} + \frac{\alpha_2}{2} \(N_{+}^2 - N_{-}^2\) \right] \,,
\end{equation}
with $\tilde{\alpha}_0 = \alpha_0-2(\alpha_1+d_1)+2\alpha_2$ and $\tilde \alpha_1 = \alpha_1 - 2\alpha_2$.

In principle, there is another branch of solutions to the defining matrix square root equation, although as already discussed we only commit to theories which are continuously related to the standard Proca one at linear order and so implicit in the expression for $\X\mupn=(\sqrt{\eta^{-1}f})\mupn$ is the choice of solution satisfying $\X\mupn[0]=\delta\mupn$. On the other hand, the theory corresponding to the other branch does not reduce to the standard Proca one on the trivial vacuum, and in fact, is not even properly formulated about the $\langle A_\mu\rangle =0$ vacuum and we do not consider it any further. Explicitly, the physical branch has the following perturbative expansion:
\begin{equation}
\begin{aligned}
	\L^{\rm (2d)}_{\rm EPN} =&- \frac14 F^2- \frac12 m^2 A^2 +\Lambda \p A\left( \left( \bar \alpha_{1,1}+\bar d_{1,1} \right)A^2+\frac{1}{8\Lambda^2} F^2 \right) \\
	&+\frac {\Lambda^2}2 \bar \alpha_{0,2}A^4+\frac 18 \left( \bar \alpha_{1,1} -2\bar \alpha_{2,1} \right)F^2 A^2+\frac{1}{128\Lambda^2}(F^2)^2-\frac{1}{16 \Lambda^2}F^2 (\p A)^2+\Lambda^2\mathcal{O}\(A^6,  \frac{\p A^5}{\Lambda}\)\,,
\end{aligned}
\end{equation}
where we use the notation $F^2 = F\mnup F\mn$ and $A^2=A^{\mu}A_{\mu}$.

The Hessian matrix is given fully non-linearly by
\begin{equation}
\label{eq:EPN2dHessian}
	\mathcal{H}\mnup \equiv \frac{\p \L^{\rm (2d)}}{\p \dot{A}_{\mu} \p \dot{A}_{\nu}} = - \frac{\tilde\alpha_1}{N_+} \begin{pmatrix}
	\bar{y}_+^2 &\bar{x}_{+} \bar{y}_{+} \\
	\bar{x}_{+} \bar{y}_{+} & \bar{x}_{+}^2
	\end{pmatrix} \,,
\end{equation}
whose determinant vanishes and thus admits a NEV. The latter can easily be inferred by inspection but can also be derived from the tensor $W\ud{\mu}{\nu}$, which here takes the form
\begin{equation}
	{W}\ud{\mu}{\nu} = \begin{pmatrix}
		 \bar{x}_{+} & - \bar{y}_{+} \\
		- \bar{y}_{+} & \bar{x}_{+}
	\end{pmatrix} \,.
\end{equation}
The NEV is then
\begin{equation}
	V_{\mu} =W^0{}_\mu = \left(\bar{x}_{+} , - \bar{y}_{+} \right) \,, \qquad \text{so that} \qquad V_{\mu} \eta\mnup V_{\nu} = - 1 \,,
\label{eq:VPN2d}
\end{equation}
and it is easy to check that it indeed annihilates the Hessian matrix,
\begin{equation}
	\mathcal{H}\mnup V_{\mu} = 0 \,.
\end{equation}
For later use we also introduce the vector $V^\perp$ normal to the NEV,
\begin{equation}
\label{eq:EPN2dNEV}
V^\perp_\mu =  W^1{}_\mu= \left( - \bar{y}_{+} , \bar{x}_{+} \right) \,, \qquad \text{so that} \qquad
V^\perp_\mu \eta\mnup V^\perp_\nu = 1 \qquad \text{and} \qquad V_{\mu} \eta\mnup V^\perp_\nu = 0 \,.
\end{equation}
The Euler-Lagrange equations for EPN in $D=2$ have the form
\begin{equation}
	\mathcal{E}^{\mu} \equiv \mathcal{H}\mnup \ddot{A}_{\nu} + u^{\mu} = 0 \,,
	\label{eq:EL}
\end{equation}
where the explicit expression for the acceleration-free part $u^{\mu}$ is provided in Appendix \ref{app:EPNLagrangian}. One can now contract the equation $\mathcal{E}^{\mu}$ with the vector $V_{\mu}$ and make use of the fact that the latter is the NEV of the Hessian matrix to find a constraint
\begin{equation}
	\mathcal{C}_{V1} \equiv V_{\mu} \mathcal{E}^{\mu} = V_{\mu} u^{\mu} \approx 0 \,.
	\label{eq:secondclassEVa}
\end{equation}
Here and in what follows, the symbol ``$\approx$" is used to designate ``on the constraint surface''.  It is also possible to show that the constraint takes the following compact form
\begin{equation}
	\mathcal{C}^{\rm (EPN)}_{V1} = \Lambda^2 \left[ \Phi_1 \tilde\alpha_1 + \phi_0 ( \tilde\alpha_{0,X} + d_{1,X} ) + \phi_1 \tilde\alpha_{1,X} + \phi_2 \left( \alpha_{2,X} + \frac12 d_{1,X} \right) \right] \approx 0 \,,
	\label{eq:secondclassEVb}
\end{equation}
and we refer the reader to Appendix \ref{app:EPNLagrangian} for the definitions of the functions $\phi_i$ and $\Phi_1$.

This establishes the existence of a primary constraint, as was already proven to be true in two and four dimensions in \cite{deRham:2020yet, deRham:2021efp}. The extension of the proof to arbitrary dimensions is straightforward. The following sections are devoted to the proof that the model also has a secondary constraint, leaving precisely one dynamical degree of freedom in $D=2$ dimensions (or $D-1$ degrees of freedom in $D$ dimensions), thus avoiding the existence of any half degree of freedom in this local, Lorentz and parity invariant theory.

\section{Constraint analysis in the Lagrangian picture}
\label{sec:2dConstraint}
\label{sec:LagrConstraint}

In this Section, we demonstrate the existence of a secondary second class constraint using the Lagrangian formalism, before turning to the formal computation of the constraint algebra of the system using the Hamilton-Dirac formalism in Section \ref{sec:HamConstraint}. In each case, we begin with the analysis of linear Proca theory for the sake of pedagogy and to fix notation, followed by a warm-up treatment of GP theory, before turning our attention onto the EPN model of interest.

\subsection{Linear Proca}
\label{ssec:ProcaLagrConstraint}

The linear Proca theory in two dimensions has the Lagrangian,
\ba
	\L^{\rm (2d)}_{\rm Proca} = - \frac{1}{4} F^2 - \frac12 m^2 A^2 = \frac12 (\dot{A}_1-A_0')^2 - \frac12 m^2 (-A_0^2 + A_1^2) \,.
\ea
The Lagrangian is independent of $\dot{A}_0$, hence $A_0$ is non-propagating and we immediately obtain the following expressions for the Hessian matrix and NEV:
\begin{equation}
	\mathcal{H}\mnup = \begin{pmatrix}
	0 & 0 \\
	0 & 1
	\end{pmatrix} \,, \qquad V_{\mu} = (1,0) \,, \qquad V^\perp_{\mu}=(0,1) \,.
\end{equation}
The Euler-Lagrange equations read
\begin{equation}
\label{eq:EmuProca}
	\mathcal{E}^{\mu} = \mathcal{H}\mnup \ddot{A}_{\nu} + u^{\mu} = 0 \,,
\end{equation}
with
\begin{equation}
	u^{\mu} = \begin{pmatrix}
	A_0'' - m^2 A_0 - \dot{A}_1' \\
	m^2 A_1 - \dot{A}_0'
	\end{pmatrix} \,,
\end{equation}
and the primary constraint is given by
\begin{equation}
	\mathcal{C}_{V1} \equiv V_{\mu} u^{\mu} = A_0'' - m^2 A_0 - \dot{A}_1' \,.
\end{equation}
Consistency of the primary constraint under time evolution yields a secondary constraint\footnote{The constraints presented here agree with the well-known textbook results; see for instance \cite{Henneaux:1992ig} for electromagnetism (the massive case is given as an exercise) and \cite{Banerjee:1994pp}. These however differ from those given in \cite{ErrastiDiez:2020dux}.}
\begin{equation}
	\dot{\mathcal{C}}_{V1} = - \left( V^{\perp}_{\mu} \mathcal{E}^{\mu} \right)' + m^2 (A_1' - \dot{A}_0) \approx 0 \,,
\end{equation}
which, on-shell and for $m^2\neq0$, gives
\begin{equation}
	\mathcal{C}_{V2} \equiv A_1' - \dot{A}_0 \approx 0 \,,
\end{equation}
the familiar Lorenz condition which is now not a gauge choice but a constraint. We see that, once initial conditions are specified for $A_1$ and its time derivative, the variable $A_0$ is then completely determined (modulo spatial boundary conditions) and is therefore non-dynamical. This leaves us with a single Lagrangian degree of freedom in two dimensions.

We can now only make use of integrations by part to show that our Lagrangian can be recast in a way that explicitly shows that the associated phase space will simply be $\{ A_1 , p^1 \}$ where $p^1$ is the conjugate momentum to $A_1$.
\begin{align}
	\L^{\rm (2d)}_{\rm Proca} 	=& \frac12 ( A_0' - \dot{A}_1 )^2 + \frac12 m^2 (A_0^2 - A_1^2) \nonumber \\
	=& \frac12 \dot{A}_1^2 - \frac12  ( A_1')^2 - \frac12 m^2 A_1^2 - \frac12 A_0 \mathcal{C}_{V1}^{\rm (Proca)} + \frac12 A_1' \mathcal{C}_{V2}^{\rm (Proca)} + (\text{total derivatives}) \,.
\end{align}
Note that the Lagrangian includes two linear Lagrange multipliers $A_0$ and $A_1'$ and hence simply reduces to $\frac12 \dot{A}_1^2 - \frac12  ( A_1')^2 - \frac12 m^2 A_1^2$ on the constraint surface. It is now independent of both $A_0$ and $\dot{A}_0$ and hence the dynamics will be fully fixed by only specifying $2$ initial conditions for $A_1$ and $\dot{A}_1$ (or its conjugate momentum $p^1$ equivalently in the Hamiltonian formalism). This shows that there is only one propagating physical degree of freedom in the standard Proca model in $D=2$ dimensions.

\subsection{Generalized Proca}
\label{ssec:GenProcaLagrConstraint}

Next, we review the constraint analysis of GP \cite{Tasinato:2014eka, Heisenberg:2014rta}. In $D=2$ the Lagrangian reads
\ba
	\L^{\rm (2d)}_{\rm GP}  = - \frac{1}{4} F^2 - \frac12 m^2 A^2 + \Lambda^2 d_0(X) + \Lambda d_1(X) \p_\alpha A^\alpha  \,.
\ea
For a generic functions $d_{1}$, the Lagrangian is no longer independent of $\dot{A}_0$, yet it is still true that the equations of motion are independent of $\ddot{A}_0$, hence $A_0$ is non-propagating. The resulting Hessian matrix, associated NEV and normal vector take again the exact same form as for the free Proca theory
\begin{equation}
	\mathcal{H}\mnup = \begin{pmatrix}
	0 & 0 \\
	0 & 1
	\end{pmatrix} \,, \qquad V_{\mu} = (1,0) \,, \qquad V^{\perp}_{\mu} = (0,1) \,.
\end{equation}
The Euler-Lagrange equations are given as in \eqref{eq:EmuProca}
\begin{equation}
\label{eq:EmuGP}
	\mathcal{E}^{\mu} = \mathcal{H}\mnup \ddot{A}_{\nu} + u^{\mu} = 0 \,,
\end{equation}
with now
\begin{equation}
	u^{\mu} = \begin{pmatrix}
	A_0'' - m^2 A_0 -\dot{A}_1' + 2 \Lambda^2 d_{0,X} A_0 + 2 \Lambda d_{1,X} \left( A_0 A_1' - A_1 \dot{A}_1 \right) \\
	m^2 A_1 - \dot{A}_0' - 2 \Lambda^2 d_{0,X} A_1 - 2 \Lambda d_{1,X} \left( A_0 A_0' - A_1 \dot{A}_0 \right)
	\end{pmatrix} \,,
\end{equation}
and the constraint spells
\begin{equation}
	\mathcal{C}_{V1} \equiv V_{\mu} u^{\mu} = A_0'' - m^2 A_0 - \dot{A}_1' + 2 \Lambda^2 d_{0,X} A_0 + 2 \Lambda d_{1,X} \left( A_0 A_1' - A_1 \dot{A}_1 \right) \approx 0 \,.
\end{equation}
Taking the time derivative of this constraint, it is straightforward to show that all second time derivatives of the fields can be eliminated  using combinations of the equations of motion, so that
\begin{equation}
	\dot{\mathcal{C}}_{V1} + V^{\perp}_{\mu} (\mathcal{E}^{\mu})' + 2 \Lambda A_1 d_{1,X} V^{\perp}_{\mu} \mathcal{E}^{\mu} \equiv \mathcal{C}_{V2} \approx 0 \,,
\end{equation}
where $\mathcal{C}_{V2}$ is free of any higher-order time derivatives (or accelerations) and does \textit{not} vanish on the primary constraint surface. Hence it is a genuinely new second class constraint. Its explicit expression reads
\begin{equation}
\begin{aligned}
	\mathcal{C}_{V2} =&\; (m^2 - 2\Lambda^2 d_{0,X}) (\p_{\mu} A^{\mu} + 2 \Lambda d_{1,X} A_1^2) - 4 \Lambda^2 A_1 d_{1,X}^2 (A_0 A_0' - A_1 \dot{A}_0)  \\
	&- 2 \Lambda d_{1,X} \left[ (\p_{\mu} A^{\mu})^2 - \p_{\mu} A_{\nu}\p^{\mu} A^{\nu} + A_0 \left( A_0' - \dot{A}_1 \right)' \right] - 4\Lambda^2 d_{0,XX} A_{\mu} A_{\nu} \p^{\mu} A^{\nu} \\
	&- 4 \Lambda d_{1,XX} A_{\mu} A_{\nu} \left[ \p^{\mu} A^{\nu} \p_{\sigma} A^{\sigma} - \p_{\sigma} A^{\mu} \p^{\sigma} A^{\nu} \right] \,.
\end{aligned}
\end{equation}
This is sufficient to establish the consistency of GP theory from the point of view of the constraint structure.

\subsection{Extended Proca-Nuevo}
\label{ssec:EPNLagrConstraint}
We finally turn to the analysis of EPN in $D=2$ dimensions. The primary constraint, Hessian matrix, NEV and normal vector are given above in Section \ref{sec:EPN2dConstraint}, see Eqns.~(\ref{eq:EPN2dHessian}--\ref{eq:EPN2dNEV}). Following the previous warm-up examples, the strategy is now clear: Take the time derivative of the primary constraint and add combinations of the equations of motion so as to remove all field accelerations (and derivatives thereof). The result is the secondary constraint.

\paragraph{Secondary Constraint for General EPN:}
Carrying out this procedure we obtain
\begin{equation}
\label{eq:C22dEPN}
	\dot{\mathcal{C}}_{V1} + V^{\perp}_{\mu}(\mathcal{E}^{\mu})' + 2 \frac{\Lambda}{\tilde\alpha_1} \beta \left( V^{\perp}_{\mu} \mathcal{E}^{\mu} \right) \equiv \mathcal{C}_{V2} \approx 0 \,,
\end{equation}
where
\begin{equation}
\begin{aligned}
	\beta =&\; (\bar{y}_+ A_0 + \bar{x}_+ A_1) \left( \tilde\alpha_{0,X} + \tilde{d}_{1,X} \right) -(\Delta A_0 - (1-\Sigma)A_1) \tilde\alpha_{1,X} \\
	& - \left[ (\bar{x}_+ \Delta - \bar{y}_+ \Sigma) A_0 + (\bar{x}_+ \Sigma - \bar{y}_+ \Delta) A_1 \right] \left( 2 \tilde\alpha_{2,X} + \tilde{d}_{1,X} \right) \,.
\end{aligned}
\end{equation}
The exact expression for the secondary constraint $\mathcal{C}_{V2}$  is given in the minimal model below. For the generic EPN, while straightforward to derive, its exact expression is rather formidable and not particularly illuminating, we thus refer the reader to Appendix~\ref{app:EPNSecConstraint} for its full expression. We emphasize however that it is a true independent constraint: it does not involve any accelerations and does not vanish on the primary constraint surface.

We conclude that EPN theory possesses a pair of constraints that together remove a full Lagrangian degree of freedom, since there are no gauge symmetries, thus defining a consistent massive spin-1 model. While this proof applies only in $D=2$ dimensions, a partial proof in generic dimension will be given in Section \ref{sec:HigherDConstraint}.

\paragraph{Minimal Model:} As a special example, we can consider the minimal model which will also be studied in generic dimensions in Section~\ref{sec:HigherDConstraint}. Focusing for now in $D=2$ dimensions, the minimal model corresponds to setting $d_1(X)=\alpha_2(X)\equiv0$ and $\alpha_1(X)\equiv 2$  in the EPN Lagrangian \eqref{eq:LEPN2d}, while keeping the potential arbitrary $\alpha_0(X)=-\frac 12 m^2 X + V(X)$. Note that it would not make sense to set $\alpha_0$ to a constant as it would set $m=0$ and the field would lose its mass term on the trivial vacuum, leading to an infinitely strong coupling. The minimal model in two dimensions is then given by
\ba
\L^{\rm (2d)}_{\rm Minimal}=\Lambda^2\(\alpha_0(X)-2[\mathcal{K}]\)=\Lambda^2\(\alpha_0(X)-2[\mathcal{X}]+4\)\,.
\ea
As in the previous cases, the Euler-Lagrange equations are given by $\mathcal{E}^{\mu} = \mathcal{H}\mnup \ddot{A}_{\nu} + u^{\mu} = 0 $ with the vector $u^\mu$ now given by (see Appendix~\ref{app:EPNLagrangian})
\ba
u^\mu_{\rm Minimal}=-\frac{2}{N_+^3} \begin{pmatrix}
 x_{+} y_{+} \left( 2  \dot{A}_0' -  A_1'' \right) - x_{+}^2  A_0'' + \left( x_{+}^2 + y_{+}^2 \right)  \dot{A}_1' \\
 x_{+} y_{+} \left( 2  \dot{A}_1' -  A_0'' \right) - y_{+}^2  A_1'' + \left( x_{+}^2 + y_{+}^2 \right)  \dot{A}_0'
\end{pmatrix}-2\Lambda^2 \alpha_{0,X}A^\mu\,.
\ea
In this case, the primary constraint is given by
\begin{equation}
	\mathcal{C}^{\rm (Minimal)}_{V1} = -2 \Lambda \left( \bar{x}_{+} \p_1 \bar{y}_{+} - \bar{y}_{+} \p_1 \bar{x}_{+}  \right) + 2 \alpha_{0,X} \Lambda^2 \left( \bar{x}_{+} A_0 + \bar{y}_{+} A_1 \right)   \approx 0 \,,
	\label{eq:minimalCV1}
\end{equation}
and  the secondary constraint $\mathcal{C}_{V2}$  takes the form (see \eqref{eq:C22dEPN}),
\begin{equation}
\begin{aligned}
\Lambda^{-3}\mathcal{C}^{\rm (Minimal)}_{V2}=&- \frac{2}{\Lambda^2} \left( \bar{x}_{+} \p_1 \bar{y}_{+} - \bar{y}_{+} \p_1 \bar{x}_{+}  \right)^2  \\
&+ 2 \alpha_{0,X} \left(2\bar{x}_+-N_+\right) +2 \alpha_{0,X}^2 (\bar{y}_+ A_0 + \bar{x}_+ A_1) ^2 \\
&+ \frac{2}{\Lambda}\alpha_{0,XX} \left(\dot{X}\left( \bar{x}_{+} A_0 + \bar{y}_{+} A_1 \right) - 2\(\bar{y}_+ A_0 + \bar{x}_+ A_1\)  \left(-A_0 A_0'+A_1 A_1'\right)\right) \,.
\end{aligned}
\end{equation}
Even in the case where  the potential reduces to a quadratic mass term, $\alpha_0=-\frac 12 m^2 A_\mu^2$ and $\alpha_{0,XX}=0$, we see that both constraints are independent. Once again,  the minimal massless case $\alpha_{0,X}\equiv 0$ is infinitely strongly coupled and meaningless.

\section{Constraint analysis in the Hamiltonian picture}
\label{sec:HamConstraint}

In this Section, we carry out the Hamilton-Dirac analysis of the two-dimensional EPN theory. We will demonstrate that the model enjoys a pair of second class constraints, leaving a two-dimensional reduced phase space, or equivalently a single Lagrangian degree of freedom. We warm up again with the examples of linear Proca and GP.

\subsection{Linear Proca}
\label{ssec:ProcaHamConstraint}

We perform a 1+1 decomposition of the linear Proca Lagrangian density,
\beq
\L^{\rm (2d)}_{\rm Proca}=\frac{1}{2}\left(\dot{A}_1-A_0'\right)^2+\frac{m^2}{2}\left(A_0^2-A_1^2\right) \,,
\eeq
so the canonical momenta read
\beq
p^0=0 \,,\qquad p^1=\dot{A}_1-A_0' \,,
\eeq
and we infer the primary constraint
\beq
\mathcal{C}_1\equiv p^0\approx 0 \,.
\eeq

The canonical or base Hamiltonian density reads
\begin{equation}
\begin{aligned}
	\mathcal{H}_{\rm base}&= p^{\mu}\dot{A}_{\mu}-\L \\
&=\frac{1}{2}(p^1)^2+p^1A_0'-\frac{m^2}{2}\left(A_0^2-A_1^2\right) \,,
\end{aligned}
\end{equation}
and the ``augmented'' or ``primary'' Hamiltonian is obtained by adding the primary constraints with arbitrary Lagrange multipliers. In this case, $\mathcal{H}_{\rm aug}=\mathcal{H}_{\rm base}+\lambda_1\mathcal{C}_1$.

Consistency of the constraint $\mathcal{C}_1$ under time evolution, $\dot{\mathcal{C}}_1\approx0$,  may either fix the Lagrange multiplier $\lambda_1$ or else produce a secondary constraint\footnote{
There is a possibility that the secondary constraint is simply inconsistent with the primary one, signaling a fundamentally pathological theory. To our knowledge, examples of this kind are all contrived or trivial and the theory can be seen to be inconsistent without performing a constraint analysis. In any case, this outcome will not occur for the models we investigate here.}.
The latter happens when the Poisson bracket $\{\mathcal{C}_1(x),\mathcal{C}_1(y)\}$ vanishes weakly, as is obviously the case for linear Proca\footnote{It is worth emphasizing that the vanishing of the Poisson bracket of a constraint with itself is not automatic. An example of a ``self-second class'' constraint arises in Lorentz-breaking Ho\v{r}ava-Lifshitz gravity \cite{Li:2009bg}.}. Thus we obtain a secondary constraint:
\beq\bal
\dot{\mathcal{C}}_1&= \{\mathcal{C}_1,H_{\rm aug}\} \\
&=(p^1)'+m^2A_0 \qquad \Rightarrow\qquad \mathcal{C}_2\equiv (p^1)'+m^2A_0\approx0\,.
\eal\eeq
We observe that $\mathcal{C}_1$ and $\mathcal{C}_2$ are second class as they do not commute with each other,
\beq
\{\mathcal{C}_1(x),\mathcal{C}_2(y)\}=-m^2\delta(x-y) \,.
\eeq
Preservation in time of $\mathcal{C}_2\approx0$ thus fixes the multiplier $\lambda_1$,
\beq
\dot{\mathcal{C}}_2=-m^2A_1'+m^2\lambda_1\approx0 \qquad \Rightarrow\qquad \lambda_1=A_1' \,.
\eeq
Finally, the total Hamiltonian from which the equations of motion are derived (which must be supplemented with initial conditions consistent with \textit{all} the constraints) is obtained by substituting the solutions for the Lagrange multipliers into the augmented Hamiltonian density,
\beq
\mathcal{H}_{\rm tot}=\mathcal{H}_{\rm base}+p^0A_1' \,.
\eeq
Since each second class constraint reduces the dimensionality of the physical phase by unity, we are left in the end with a two-dimensional phase space or a single degree of freedom in field space.

\subsection{Generalized Proca}
\label{ssec:GenProcaHamConstraint}

We refer the interested reader to \cite{Heisenberg:2014rta,BeltranJimenez:2019wrd} for a more complete constraint analysis of GP, while here we content ourselves with the derivation of the primary and secondary constraints, in addition to the proof that they are second class. The 1+1-decomposed GP Lagrangian density is
\beq
\L^{\rm (2d)}_{\rm GP}=\frac{1}{2}\left(\dot{A}_1-A_0'\right)^2+\frac{m^2}{2}\left(A_0^2-A_1^2\right)+\Lambda^2d_0(X)+\Lambda d_1(X)\left(-\dot{A}_0+A_1'\right) \,.
\eeq
From the canonical momenta,
\beq
p^0=-\Lambda d_1(X) \,,\qquad p^1=\dot{A}_1-A_0' \,,
\eeq
we infer the primary constraint $\mathcal{C}_1\equiv p^0+\Lambda d_1(X)\approx 0$ as well as the base and augmented Hamiltonian densities,
\beq\bal
\mathcal{H}_{\rm base}&= \frac{1}{2}(p^1)^2+p^1A_0'-\frac{m^2}{2}\left(A_0^2-A_1^2\right)-\Lambda^2d_0(X)-\Lambda d_1(X)A_1' \,,\\
\mathcal{H}_{\rm aug}&= \mathcal{H}_{\rm base}+\lambda_1\mathcal{C}_1 \,.
\eal\eeq
Although less obvious in this case, it can again be easily checked that $\mathcal{C}_1$ commutes with itself, so its consistency under time evolution generates a secondary constraint:
\beq\bal
\mathcal{C}_2\equiv \dot{\mathcal{C}}_1 &= \{\mathcal{C}_1,H_{\rm aug}\} \\
&=(p^1)'+m^2A_0-2\Lambda^2d_{0,X}A_0+2\Lambda d_{1,X}\left(-A_0A_1'+A_1A_0'+p^1A_1\right)\approx0 \,.
\eal\eeq

It is clear that $\mathcal{C}_1$ and $\mathcal{C}_2$ do not Poisson-commute (since they do not in the case of linear Proca, and the GP functions $d_{0,1}$ are generic), although we will not derive the explicit result. It follows that they are independent, second class constraints, implying the absence of further constraints and the determination of the Lagrange multiplier $\lambda_1$ in terms of the phase space variables, and hence of the total Hamiltonian.

\subsection{Extended Proca-Nuevo}
\label{ssec:EPNHamConstraint}

The 1+1-decomposed EPN Lagrangian was already given above in \eqref{eq:LEPN_2d}. Working out the canonical momenta, we find
\beq\bal
p^0&= -\Lambda\left[\tilde{\alpha}_1\bar{x}_{+}+2\alpha_2\Sigma+d_1\right] \,,\\
p^1&= -\Lambda\left[\tilde{\alpha}_1\bar{y}_{+}+2\alpha_2\Delta\right] \,,
\eal\eeq
with $\Sigma=1+A_1'/\Lambda$ and $\Delta=-A_0'/\Lambda$.
We infer the following primary constraint:
\beq
\mathcal{C}_1\equiv \left[\frac{p^0}{\Lambda}+2\alpha_2\Sigma+d_1\right]^2-\left[\frac{p^1}{\Lambda}+2\alpha_2\Delta\right]^2-\tilde{\alpha}_1^2 \approx0 \,.
\eeq
This is an interesting novelty of (E)PN theory relative to GP: the primary constraint is non-linear in the momenta\footnote{Of course any function of a constraint is also a constraint, defining the same hypersurface in phase space. However, some caution is needed when dealing with a constraint which is non-linear in all variables. Namely, one must check the so-called regularity condition, i.e.\ the requirement that the Jacobian of the constraints must have constant rank throughout phase space \cite{Henneaux:1992ig}. It can be easily checked that the regularity condition is satisfied by the constraint $\mathcal{C}_1$.}.
The resulting base and augmented Hamiltonians are therefore
\beq\bal
\mathcal{H}_{\rm base}&= \Lambda^2\left[\frac{p^0}{\Lambda}(1+\Sigma)-\frac{p^1}{\Lambda}\Delta-\tilde{\alpha}_0+2\alpha_2(\Sigma^2-\Delta^2)\right] \,,\\
\mathcal{H}_{\rm aug}&= \mathcal{H}_{\rm base}+\lambda_1\mathcal{C}_1 \,.
\eal\eeq
The next question is whether $\mathcal{C}_1$ commutes with itself. Since the consistency of the theory hinges on this question, we shall provide explicit details. Observe first that terms arising from derivatives of $\tilde{\alpha}_{1}$, $\alpha_2$ and $d_1$ yield zero. Indeed, such terms do not involve spatial derivatives of the field and hence give a contribution of the form
\beq
\{\mathcal{C}_1(x),\mathcal{C}_1(y)\}\supset \int \d z\left[F(x,y)\delta(x-z)\delta(y-z)-(x\leftrightarrow y)\right]=\left[F(x,y)-F(y,x)\right]\delta(x-y) \,,
\eeq
which vanishes, as can be seen more explicitly by integrating with an arbitrary test function. Contributions from the remaining terms give
\ba
\{\mathcal{C}_1(x),\mathcal{C}_1(y)\}
&=&\int \d z\bigg\{ 8\left[\alpha_2\left(\frac{p^1}{\Lambda}+2\alpha_2\Delta\right)\right]_x\left(\frac{p^0}{\Lambda}
+2\alpha_2\Sigma+d_1\right)_y\delta'(x-z)\delta(y-z) \nn \\
&&\phantom{\int \ \ }-8\left[\alpha_2\left(\frac{p^0}{\Lambda}+2\alpha_2\Sigma+d_1\right)\right]_x
\left(\frac{p^1}{\Lambda}+2\alpha_2\Delta\right)_y\delta'(x-z)\delta(y-z)\nn\\
&&\phantom{\int  \ \ }-(x\leftrightarrow y) \bigg\} \nn \\
&=&8\delta'(x-y)\bigg\{ \left[\alpha_2\left(\frac{p^1}{\Lambda}+2\alpha_2\Delta\right)\right]_x\left(\frac{p^0}{\Lambda}+2\alpha_2\Sigma+d_1\right)_y \nn \\
&&\quad -\left[\alpha_2\left(\frac{p^0}{\Lambda}+2\alpha_2\Sigma+d_1\right)\right]_x\left(\frac{p^1}{\Lambda}+2\alpha_2\Delta\right)_y \bigg\} -(x\leftrightarrow y) \nn \\
&=&8\delta'(x-y)\left[F(x,y)+F(y,x)\right] \,,
\ea
where now
\beq
F(x,y)\equiv \alpha_2(x)P^1(x)P^0(y)-\alpha_2(x)P^0(x)P^1(y) \,,
\eeq
\beq
P^0\equiv \frac{p^0}{\Lambda}+2\alpha_2\Sigma+d_1 \,,\qquad P^1\equiv \frac{p^1}{\Lambda}+2\alpha_2\Delta \,.
\eeq
Integrating with a test function, we get
\beq
\int \d yf(y)\{\mathcal{C}_1(x),\mathcal{C}_1(y)\}=8f(x)\left[F^{(1)}(x,x)+F^{(2)}(x,x)\right] \,,
\eeq
where $F^{(n)}$ denotes differentiation w.r.t.\ to the $n$-th argument of the function $F$. It is now easy to see that $F^{(2)}(x,x)=-F^{(1)}(x,x)$, establishing the consistency of the constraint $\mathcal{C}_1\approx 0$ under time evolution. Thus we obtain the secondary constraint
\begin{align}
\mathcal{C}_2 \equiv& \; \frac{\dot{\mathcal{C}}_1}{\Lambda}=\frac{1}{\Lambda}\{\mathcal{C}_1,H_{\rm aug}\}=\frac{1}{\Lambda}\{\mathcal{C}_1,H_{\rm base}\} \\
=& \; 2\left(\frac{p^0}{\Lambda}+2\alpha_2\Sigma+d_1\right) \left[\frac{\{p^0,H_{\rm base}\}}{\Lambda^2}-2\left(2\alpha_{2,X}\Sigma+d_{1,X}\right)\left(A_0(1+\Sigma)+A_1\Delta\right)-2\alpha_2\frac{\Delta'}{\Lambda}\right] \nn  \\
&- 2\left(\frac{p^1}{\Lambda}+2\alpha_2\Delta\right)\left[\frac{\{p^1,H_{\rm base}\}}{\Lambda^2}-4\alpha_{2,X}\Delta\left(A_0(1+\Sigma)+A_1\Delta\right)-2\alpha_2\frac{\Sigma'}{\Lambda}\right]  \\
&+ 4\tilde{\alpha}_1\tilde{\alpha}_{1,X}\left(A_0(1+\Sigma)+A_1\Delta\right) \vphantom{\frac{p^0}{\Lambda}} \,,\nn
\end{align}
and
\beq\bal
\{p^0,H_{\rm base}\}&= \Lambda^2\left[-2\tilde{\alpha}_{0,X}A_0+4 \alpha_{2,X}A_0\left(\Sigma^2-\Delta^2\right)\right]+\left(p^1+4\Lambda \alpha_2\Delta\right)' \,,\\
\{p^1,H_{\rm base}\}&=\Lambda^2\left[2\tilde{\alpha}_{0,X}A_1-4 \alpha_{2,X}A_1\left(\Sigma^2-\Delta^2\right)\right]+\left(p^0+4\Lambda \alpha_2\Sigma\right)' \,.
\eal\eeq
To complete the analysis of the constraint algebra it remains to verify the absence of tertiary constraint and the second class nature of $\mathcal{C}_1$ and $\mathcal{C}_2$. This is a straightforward\footnote{Let us remind the reader of the argument given in Section \ref{sec:EPNConstraintArbitraryD} that the theory must propagate \textit{at least} $D-1$ degrees of freedom in view of the fact that (E)PN reduces to linear Proca theory upon linearization about the trivial vacuum. We therefore should not expect the presence of tertiary constraints, and given the absence of gauge symmetries, $\mathcal{C}_1$ and $\mathcal{C}_2$ must be second class and have non-zero Poisson bracket among each other.} but cumbersome task, so for brevity let us consider a minimal PN model with $\tilde{\alpha}_0,\alpha_2,d_1=0$ and $\tilde{\alpha}_1\propto X$. In this case, we find
\begin{align}
	\{\mathcal{C}_1(x),\mathcal{C}_2(y)\} =& \; \frac{8\tilde{\alpha}_{1,X}}{\Lambda}\left[2\tilde{\alpha}_{1,X}\left( A_0(1+\Sigma)+A_1\Delta\right)\left(p^0A_0+p^1A_1\right) -\tilde{\alpha}_1\left(p^0(1+\Sigma)+p^1\Delta\right)\right]\delta(x-y) \nn \\
&+ \frac{8\tilde{\alpha}_1\tilde{\alpha}_{1,X}}{\Lambda^2}\left[\left((p^0)'A_1+(p^1)'A_0\right)\delta(x-y)+2\left(p^0A_1+p^1A_0\right)\delta'(x-y)\right] \,,
\end{align}
which is not weakly zero. The final tally of degrees of freedom is now the familiar one: in two dimensions we have four phase space variables, reduced by two due to the presence of two second class constraints, hence a single Lagrangian degree of freedom.

\section{Minimal PN model in arbitrary dimensions}
\label{sec:HigherDConstraint}

In arbitrary spacetime dimension, it is unfortunately not possible to avoid the matrix square root structures that define (E)PN theory, making the problem of its constraint analysis technically more challenging.
The derivation of the primary and secondary constraint in massive gravity was derived in generality in arbitrary dimensions in \cite{Hassan:2011hr,Hassan:2011ea} and since PN follows the same structure as massive gravity, the same logic will apply. In what follows we restrict our attention to the minimal $D$-dimensional PN model given by the Lagrangian
\beq \label{eq:LEPND}
\L=\Lambda^D \left( \alpha_0(X) - 2 [\K] \right)=\Lambda^D \left( \alpha_0(X) - 2 [\X]+2D \right) \,,
\eeq
where we chose $\alpha_1 = -2$ to recover the canonical kinetic term $- \frac14 F\mnup F\mn$ at quadratic order in perturbation and the pure potential term $\alpha_0(X)$ should at the very least include the mass term, $\Lambda^D \alpha_0(X)=-\frac 12 m^2 A^2+\cdots$ and hence $\alpha_{0,X}\not \equiv 0$, (in fact if the theory admits a solution where $\alpha_{0,X}=0$ at some point in spacetime, then the theory is infinitely strongly coupled at that point and the classical solution cannot be trusted in the vicinity of that point). In the following we present the constraint analysis of this minimal model both in the Lagrangian and Hamiltonian pictures, finding in both cases the presence of a pair of constraints.

\subsection{Lagrangian picture}

Let us first collect some preliminary results. Given the definition of $W\ud{\mu}{\nu}$ in Eq.~\eqref{eq:defWtensor}, it is possible to show that
\begin{equation} \label{eq:trace derivative identity}
	\frac{\p \left[ \X^n \right]}{\p (\p_{\alpha} A_{\beta})} = \frac{n}{\Lambda^{D/2}} (\X^{n-1})^{\alpha}_{\phantom{\alpha}\mu} W^{\mu \beta} \,,
\end{equation}
which will be useful for $n=1$ here. From the definition of the generalized Hessian, applied to the Lagrangian \eqref{eq:LEPND}, we infer the relations
\begin{equation}
\label{eq:HW}
	\p_{\mu} W^{\alpha\beta} = -\frac{1}{2 \Lambda^{D/2}} \mathcal{H}^{\nu \alpha, \rho \beta} \p_{\mu} \p_{\nu} A_{\rho} \,,\qquad \mathcal{H}^{\hat{\mu}\nu, \alpha \beta} W\ud{\hat{\mu}}{\alpha} =\mathcal{H}^{\mu  \hat{\nu}, \alpha \beta} W\ud{\hat{\nu}}{\beta} =0 \,,
\end{equation}
where in the last expression the indices $\hat{\mu}$ and $\hat{\nu}$ have a fixed value and are not summed over. In particular this implies that the vector $V_\beta=W^0{}_\beta$ is a NEV for all $\mathcal{H}^{\mu 0, \alpha\beta}$ (\ie for any $\mu$, $\alpha$),
\ba
\label{eq:HW2}
\mathcal{H}^{\mu 0, \alpha\beta} V_\beta=0\,\quad\forall \ \mu, \alpha\,.
\ea

To proceed further, it will prove useful to write the equation of motion in two ways as
\beq
\mathcal{E}^{\mu} = \mathcal{H}\mnup \ddot{A}_{\nu} + u^{\mu} =  -2 \Lambda^{D/2} \dot{V}^{\mu} + \tilde{u}^{\mu} \,,
\eeq
where
\begin{align}
	u^{\mu} &= \p_j \dot{A}_{\alpha} \mathcal{H}^{0j,\mu \alpha} -2 \Lambda^{D/2} \p_i W^{i\mu} - 2\Lambda^2 \alpha_{0,X} A^{\mu} \,, \\
	\tilde{u}^{\mu} &= -2 \Lambda^{D/2} \p_i W^{i\mu} - 2\Lambda^2 \alpha_{0,X} A^{\mu} \,.
\end{align}
Given the NEV $V_{\mu}$, we infer the constraint
\begin{equation}
	\mathcal{C}_{V1} = V_{\mu}\mathcal{E}^{\mu} = V_{\mu} u^{\mu} = V_{\mu} \tilde{u}^{\mu} \,.
\end{equation}
The next step is to calculate the time derivative of $\mathcal{C}_{V1}$ and attempt to eliminate all instances of the field acceleration using combinations of the equations of motion. Inspired by the two-dimensional case, we first consider
\begin{align}
	\dot{\mathcal{C}}_{V1}+W\ud{i}{\mu} \p_i \mathcal{E}^{\mu} =& \p_i \ddot{A}_{\alpha} \left[ V_{\mu} \mathcal{H}^{i0,\mu\alpha} + W\ud{i}{\mu} \mathcal{H}^{\mu\alpha} \right] + \ddot{A}_{\alpha} \left[ V_{\mu} \p_i \mathcal{H}^{i0,\mu\alpha} + W\ud{i}{\mu} \p_i \mathcal{H}^{\mu\alpha} \right] \nn \\
	&+ V_{\mu} \p_i \left( \mathcal{H}^{ij,\mu \alpha} \p_j \dot{A}_{\alpha} \right) + W\ud{i}{\mu} \p_i \left( \mathcal{H}^{0j,\mu \alpha} \p_j \dot{A}_{\alpha} + \tilde{u}^{\mu} \right) \\
	&- 2 \Lambda^2 \p_t \left(\alpha_{0,X} A^{\mu} \right) V_{\mu} + \frac{1}{2\Lambda^{D/2}} \tilde{u}_{\mu} \tilde{u}^{\mu} \,,\nn
\end{align}
and observe that the last two lines do not involve second time derivatives, while the coefficient of $\p_i \ddot{A}_{\alpha}$ is in fact zero since
\ba
	0 = \frac{\p \eta^{i0}}{\p \dot{A}_{\alpha}} = \frac{\p (W^{i\mu} V_{\mu})}{\p \dot{A}_{\alpha}} = \left( \frac{\p W^{i\mu}}{\p \dot{A}_{\alpha}} V_{\mu} + W\ud{i}{\mu} \frac{\p V^{\mu}}{\p \dot{A}_{\alpha}}  \right) 
	\quad \Rightarrow \quad V_{\mu} \mathcal{H}^{i0,\mu\alpha} + W\ud{i}{\mu} \mathcal{H}^{\mu\alpha} = 0 \label{eq:HVplusHW} \,,
\ea
using the above identities.

It remains to eliminate the terms proportional to $\ddot{A}_{\alpha}$, which can in principle be achieved by adding a linear combination $\Upsilon_{\mu} \mathcal{E}^{\mu}$ of the undifferentiated equation of motion. It simplifies the calculation to separate the term $\p_i W\ud{i}{\mu} \mathcal{E}^{\mu}$ from the unknown vector $\Upsilon_{\mu}$, i.e.\
\begin{align}
	\dot{\mathcal{C}}_{V1} +\p_i \left(W\ud{i}{\mu} \mathcal{E}^{\mu} \right) + \Upsilon_{\mu} \mathcal{E}^{\mu} &=  \ddot{A}_{\alpha} \left[ \Upsilon_{\mu} \mathcal{H}^{\mu\alpha} -(\p_i V_{\mu}) \mathcal{H}^{i0,\mu\alpha} \right] \nn \\
	&\quad + V_{\mu} \p_i \left( \mathcal{H}^{ij,\mu \alpha} \p_j \dot{A}_{\alpha} \right) + \p_i \left( W\ud{i}{\mu} u^{\mu} \right) \\
	&\quad - 2 \Lambda^2 \p_t \left(\alpha_{0,X} A^{\mu} \right) V_{\mu} + \frac{1}{2\Lambda^{D/2}} \tilde{u}_{\mu} \tilde{u}^{\mu} + \Upsilon_{\mu} u^{\mu} \,.\nn
\end{align}
From this last result, we see that a second constraint exists provided the equation
\begin{equation}
	\Upsilon_{\mu} \mathcal{H}^{\mu\alpha} = (\p_i V_{\mu}) \mathcal{H}^{i0,\mu\alpha} \,,
	\label{eq:defUpsilon}
\end{equation}
admits a solution for $\Upsilon_{\mu}$.

\paragraph{Two-dimensions:}
    Note that in $D=2$ dimensions, we have the simple relation $\p_1 V_{\mu} = - \Lambda \Phi_1 W^1_{\mu}$. Since from \eqref{eq:HW}, we have $\mathcal{H}^{10,\mu \alpha}W^1{}_\mu=0$,  the RHS of \eqref{eq:defUpsilon} cancels exactly and there is  no need for a vector $\Upsilon$ in two dimensions. This agrees with the explicit two-dimensional derivation performed previously.

\paragraph{Higher Dimensions:} More generically, i.e. in arbitrary dimensions, since the Hessian $\mathcal{H}^{\mu\alpha}$ is non-invertible, it is clear that the solution of the previous equation is degenerate and indeed, $\Upsilon_{\mu}$ is not unique. Naturally, for any $\Upsilon_{\mu}$ that solves \eqref{eq:defUpsilon}, the vector $\Upsilon_{\mu}+f V_{\mu}$ is also a solution for any function $f$. This is perfectly consistent as it simply encodes the fact that the constraint $\mathcal{C}_{V2}$ can be shifted by  $fV_{\mu}u^{\mu}=f\mathcal{C}_{V1}\approx0$.

With this in mind, it will be useful to separate our Hilbert space into the NEV direction and its $(D-1)$-normal plane to derive the generic solution of \eqref{eq:defUpsilon}. We use the eigenvectors of the Hessian $\mathcal{H}^{\mu\nu}$ to span over our full $D$-dimensional Hilbert space and choose the set of $D$ eigenvectors $\{\tilde{V}_\mu^{(\sigma)}\}_{\sigma=0,\cdots,D-1} =  \{V_\mu, V^{\perp\, (a)}_\mu\}_{a=1,\cdots,D-1}$ which forms a complete orthonormal basis, satisfying the following properties
\beq\bal
\label{eq:norm}
& \mathcal{H}^{\mu \alpha}  \tilde V^{(0)}_{\alpha}=  \mathcal{H}^{i0,\mu \alpha}  \tilde V^{(0)}_{\alpha} = 0\,,  \\
& \mathcal{H}^{\mu \alpha} \tilde V^{(a)}_\alpha =\lambda^{(a)}\tilde V^{(a)\mu }\ne 0 \,, \ \   \forall \ a=1,\cdots,D-1,\\
 \text{and normalized as} &\  \tilde V^{(\sigma)}_\mu \tilde V^{(\sigma')\mu}=\eta^{\sigma \sigma'}\,, \ \  \forall \ \ \sigma, \sigma'=0\cdots,D-1\,.
\eal\eeq
Expanded in this basis, we have $\Upsilon_\mu=\upsilon_0 V_\mu +\upsilon_a V^{\perp (a)}_\mu$, with the component $\upsilon_0$ being arbitrary as discussed earlier, and the coefficients $\upsilon_a$ given by
\ba
\label{eq:upsilona}
\upsilon_a=\frac{1}{\lambda^{(a)}} (\p_i V_{\mu}) \mathcal{H}^{i0,\mu\alpha} V^{\perp (a)}_\alpha\ \forall \ a=1,\cdots, D-1\,.
\ea
Since $\lambda^{(a)} \ne 0$ for any situation  continuously related to the trivial vacuum, there is no ambiguity in uniquely identifying each coefficient $\upsilon_a$. See Appendix~\ref{app:Upsilon} for the explicit proof that the vector  $\Upsilon_\mu = \upsilon_a V_\mu^{\perp (a)}$ satisfies the relation \eqref{eq:defUpsilon}.
  This proves the existence of the required vector $\Upsilon_\mu$ and hence the existence of a secondary constraint in arbitrary dimensions.

Perturbatively, $\Upsilon_\mu$ is given by (up to  $\upsilon_0 V_\mu$),
\begin{align}
	\Upsilon_{\mu} =& \frac{1}{8\Lambda^D} \delta_{\mu}^i (\p^j A_0 + \dot{A}^j) (\p_i F_{j0} - \p_j F_{i0} ) \nonumber \\
	&+ \frac{1}{32 \Lambda^{3D/2}} \delta_{\mu}^i \left[ 2 \left( 2 \dot{A}_0 \dot{A}^j + \dot{A}^k F\ud{j}{k} - \p^k A_0 B\ud{j}{k} \right) (\p_i F_{j0} -\p_j F_{i0} ) \right. \nonumber \\
	& + 2 \left( \p_i A^k B\ud{j}{0} - \p_i A^j B\ud{k}{0} \right) (\p_j F_{k0} -\p_k F_{j0} )  \\
	&\left. + B\ud{j}{0} \left( 2\p_i \dot{A}_0 F_{j0} - 2\p_j \dot{A}_0 F_{i0} + B\du{j}{k} \p_i F_{0k} + 2 F\ud{k}{j} \p_i \dot{A}_k + B\ud{k}{0} \p_j F_{ik} + B\du{i}{k} \p_k F_{j0} + 2 F\du{i}{k} \p_j \dot{A}_k \right) \right] \nonumber \\
	&+ \mathcal{O}\left( \frac{(\p \p A) (\p A)^3}{\Lambda^{2D}} \right) \nn \,.
\end{align}
Note that this expression trivially cancels in $D=2$ dimensions, as previously highlighted.

To summarize this analysis, we have therefore proven the presence of a second Lagrangian constraint in arbitrary dimensions,
\begin{equation}
	\mathcal{C}_{V2} \equiv V_{\mu} \p_i \left( \mathcal{H}^{ij,\mu \alpha} \p_j \dot{A}_{\alpha} \right) + \p_i \left( W\ud{i}{\mu} u^{\mu} \right) - 2 \Lambda^2 \p_t \left(\alpha_{0,X} A^{\mu} \right) V_{\mu} + \frac{1}{2\Lambda^{D/2}} \tilde{u}_{\mu} \tilde{u}^{\mu} + \Upsilon_{\mu} u^{\mu} \approx0 \,,
\end{equation}
with $\Upsilon_{\mu} u^{\mu}$  given in terms of the non-null eigenvectors $V^{\perp (a)}_\mu$ and related eigenvalues $\lambda^{(a)}\ne 0$ of the Hessian by
\ba
\Upsilon_{\mu} u^{\mu}= \sum_{a=1}^D \frac{1}{\lambda^{(a)}} (\p_i V_{\nu}) \mathcal{H}^{i0,\nu\alpha} V^{\perp (a)}_\alpha  V^{\perp (a)}_\mu u^{\mu} \,.
\ea

\subsection{Hamiltonian picture}

Establishing the existence of a pair of constraints in the Hamilton-Dirac analysis is very simple for the minimal model \eqref{eq:LEPND} and in fact, proceeds very analogously to the two-dimensional case studied above. Using the identity \eqref{eq:trace derivative identity} with $n=1$ we derive the canonical momenta,
\beq
p^{\mu}=-2\Lambda^{D/2} V^{\mu} \,,
\eeq
in terms of the NEV $V^{\mu}$. The normalization of the latter then immediately allows us to infer the primary constraint
\beq
\mathcal{C}_1\equiv p^{\mu}p_{\mu} + 4 \Lambda^D \approx0 \,.
\eeq
This constraint trivially commutes with itself (recall that we take $\alpha_1$ constant for simplicity, but the conclusion also easily follows if $\alpha_1$ is a generic function), thus proving the existence of a secondary constraint, since $\mathcal{C}_1$ cannot be first class (since it is already not first class at the linear level).
Deriving the full expression of the canonical Hamiltonian in closed form is technically more challenging but not required to ascertain the existence of a secondary constraint. To close the algebra, one should then in principle check whether a tertiary or further constraints could exist (which would occur if $\mathcal{C}_1$ and $\mathcal{C}_2$ commute), however the arguments given in Section \ref{sec:EPNConstraintArbitraryD} are proof enough that, at least for generic parameters, the theory cannot have \textit{fewer} than $D-1$ degrees of freedom (and if did at some point for some solutions and choices of parameters, these would not be trustable as the theory would then be infinitely strongly coupled at those points). We therefore conclude that there must be precisely $D-1$ degrees of freedom given the two constraints we have inferred.

\section{Discussion and conclusions}
\label{sec:ConclConstraint}

We have performed a constraint analysis of the recently proposed (Extended) Proca-Nuevo theory, with the particular aim of establishing the existence of a pair of constraints responsible for removing the Ostrogradsky ghost mode and thus rendering the model consistent from the point of view of the degree of freedom count, i.e.\ that a massive spin-1 system must describe $D-1$ dynamical modes in $D$ spacetime dimensions.

We devoted the first part of the Chapter to explaining why this outcome had to be expected. We showed through several formal and physical arguments why local, Lorentz and parity invariant field theories cannot hold a half number of Lagrangian degrees of freedom, equivalently an odd-dimensional physical phase space. Although these arguments are not new, they are certainly worth being recollected and emphasized in view of the fact that opposing claims have been made in the literature.

Summarizing our results concerning the analysis of EPN theory, we proved in full detail the existence of two, and only two, constraints in the general two-dimensional model, using both Lagrangian and Hamiltonian approaches. In the latter case, by deriving the full constraint algebra we moreover established the second class nature of the constraints and the absence of tertiary and further constraints. The generalization to arbitrary dimensions proved to be technically challenging, but we successfully analyzed a minimal version of the theory, concluding again the existence of a pair of constraints.

In addition to the main result regarding the counting of degrees of freedom, we find it worthwhile to remark on the interesting structure of the constraints we found in the canonical formalism. In particular, for the models we studied, the constraints are non-linear in all phase space variables and thus do not appear to smoothly deform the structures found in GP theory or linear Proca. We think this motivates a revisiting of the Hamiltonian analysis of EPN in arbitrary dimensions, where it may shed light on the issues related to the coupling to gravity of the theory.

The analysis performed in Chapters \ref{chap:PN} and \ref{chap:Constraint} prove that Proca-Nuevo is a genuinely new and ghost-free theory of a massive spin-1 field. It remains to understand how to covariantize such a theory and this will be the focus of Chapter \ref{chap:CovEPN}.

\chapter{Covariantization of Extended Proca-Nuevo}
\label{chap:CovEPN}

In this Chapter, we are interested in ways to covariantize Extended Proca-Nuevo. In curved spacetime,  massive gravity is the natural covariantization but we show how other classes of covariantizations can be considered. We further prove that the EPN flat spacetime constraint is maintained on any fixed curved background. Upon mixing extended EPN dynamically with gravity, we show that the constraint gets broken in a Planck scale suppressed way.

\section{(Re)coupling to gravity}
\label{sec:CouplGrav}

The covariantization of \Pros is very similar to that of the Galileon \cite{Nicolis:2008in}. Originally derived from the DL of the gravitational Dvali--Gabadadze--Porrati model \cite{Dvali:2000hr,Luty:2003vm}, the natural covariantization of the Galileon is hence the DGP model itself, or generalized massive gravity. Remarkably, it was indeed shown in Ref.~\cite{Garcia-Saenz:2019yok} that massive gravity is the natural way the Galileon symmetry can be gauged.

However taken as a scalar field in its own right, one may envisage a covariantization of the Galileon where the fields transform as a diffeomorphism (diff) scalar in the embedding gravitational theory. Such types of covariantizations lead to the `Covariant Galileon',  \cite{Deffayet:2009wt}, proxy theories of massive gravity \cite{deRham:2011by} or more generically to Horndeski \cite{Horndeski:1974wa} and were then further extended to Beyond--Horndeski and more generic classes of degenerate higher order theories \cite{Gleyzes:2014dya,Zumalacarregui:2013pma,
Langlois:2015cwa,Langlois:2015skt,BenAchour:2016cay,Crisostomi:2016tcp,
Crisostomi:2016czh,Ezquiaga:2016nqo,Motohashi:2016ftl}.

Viewed as Effective Field Theories, the Galileon just like GP or \Pros have a very low cutoff at the scale $\Lambda_3$ (or lower \cite{deRham:2017xox}) and there can be a continuum of interactions between the scale $\Lambda_3$ and the Planck scale so that the question of what the natural covariantization of these theories is may not be particularly meaningful. However for many of these classes of theories, one may postulate the existence of a Vainshtein--type of mechanism that may allow us to push their regime of applicability beyond the scale $\Lambda_3$.

\subsection{Generalized Massive Gravity as the Natural Covariantization}
\label{ssec:MGCov}

As introduced in Section~\ref{sec:NewProca}, \Pros is heavily inspired by massive gravity. When considering the coupling of \Pros to gravity (or when considering \Pros in curved spacetime) a natural covariantization is therefore simply the theory of massive gravity introduced in \cite{deRham:2010kj} (or rather its generalized form introduced in \cite{deRham:2014lqa,deRham:2014gla}) where the Lorentz vector $A_\mu$ is not promoted to a diff vector (ie to a vector under general coordinate transformations) but rather is considered as being part of a diff scalar $\phi^a$ as introduced in \eqref{eq:phia}.

In this covariantization of  \Pro, the quantity $f\mn$ remains identical as that defined in  \eqref{eq:deff2}, still expressed in terms of the Minkowski metric,
\ba
\label{eq:fmn2}
f\mn =\p_\mu \phi^a \p_\nu \phi^b \eta_{ab}  =\eta\mn + 2 \frac{\p_{(\mu} A_{\nu)}}{\Lambda_2^2} + \frac{\p_{\mu} A_\alpha \p_{\nu} A_\beta \eta^{\alpha \beta}}{\Lambda_2^4}\,,
\ea
 even though the field is living on an arbitrary spacetime with dynamical metric $g\mn$. The metric $g\mn$ enters the definition of $\K$ which is now defined as \cite{deRham:2010kj}
\ba
\K\mupn=\(\sqrt{g^{-1}f}\)\mupn-\delta\mupn\,,
\ea
leading to the lagrangian for massive gravity including the dynamics of the metric,
\ba
\label{eq:cov}
\L_{\rm Cov}=\frac{\mpl^2}{2}\sqrt{-g}R[g]+\Lambda_2^4 \sqrt{-g}\sum_{n=0}^4 \alpha_n(\phi) \L_n[\K]\,.
\ea
This generalized theory of massive gravity reduces to \Pros in the limit where gravity is `switched off' or decoupled, $\mpl\to \infty$ so long as $\alpha_0$ includes a quadratic term in the vector field. The absence of ghost in this covariantization follows from the absence of ghost in massive gravity \cite{deRham:2010kj,Hassan:2011hr,deRham:2014lqa,deRham:2014gla}.

\begin{figure}[h]
  \centering
  \includegraphics[width=0.8\textwidth]{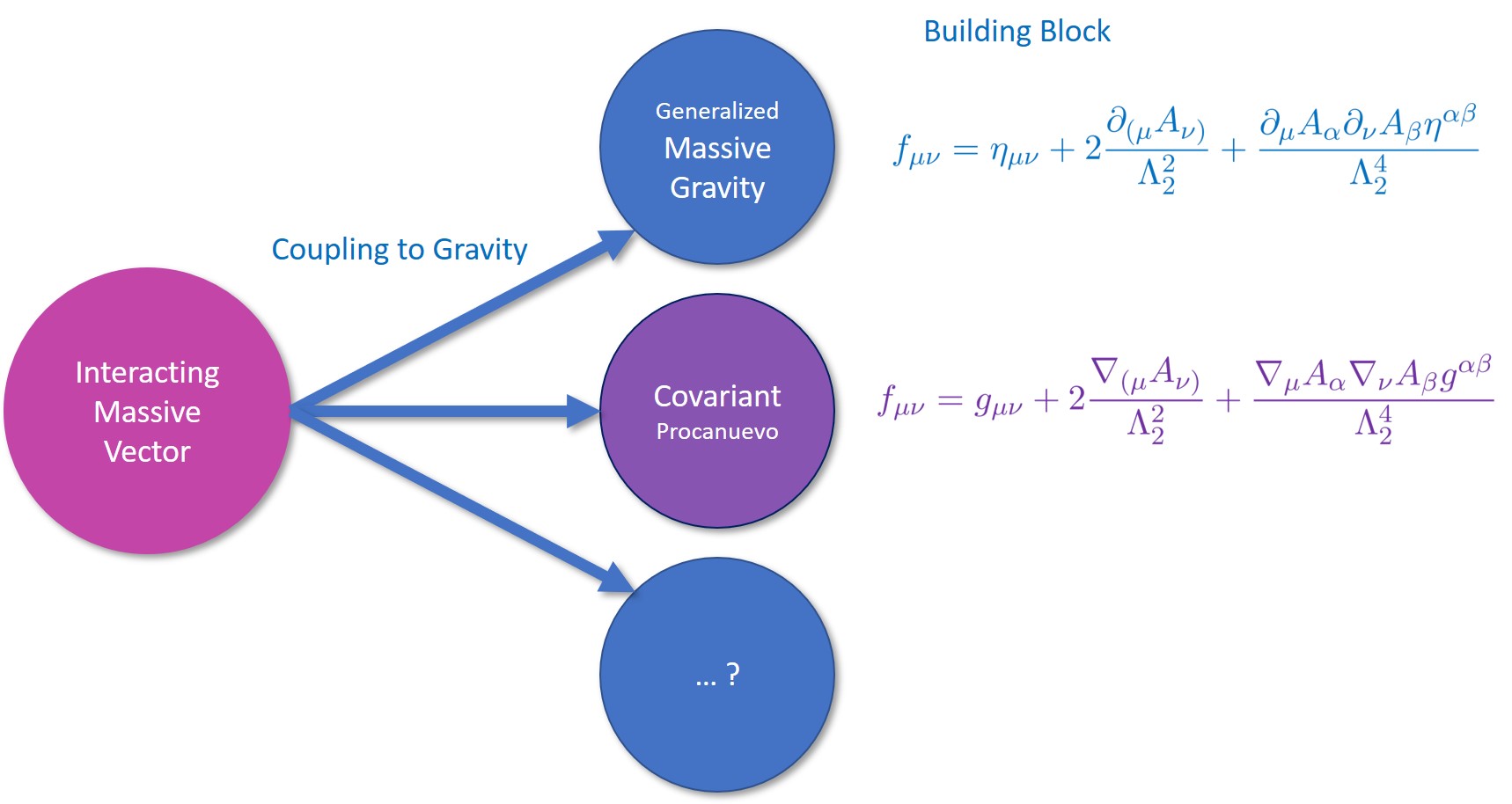}
  \caption[Potential covariantization classes of Proca-Nuevo.]{Any theory proposed in flat space can admit various potential classes on different covariantization. Generalized massive gravity is a natural one to consider for \Pros since this is where it was originally inspired from, but other non--equivalent covariantizations can be considered. See Ref.~\cite{deRham:2019wjj} for related arguments. }\label{Fig:Covs}
\end{figure}

\subsection{Alternative Covariantization}
\label{ssec:AltCov}

When coupling to gravity, an alternative approach is to treat $A_\mu$ as a diff vector. In doing so, instead of using the quantity $f\mn$ defined in \eqref{eq:fmn2}, the building block of the covariant theory would then be the diff tensor $f^{(g)}\mn$ defined as
\ba
\label{eq:Fmn}
f^{(g)}\mn = g\mn + 2 \frac{\nabla_{(\mu} A_{\nu)}}{\Lambda_2^2} + \frac{\nabla_{\mu} A_\alpha \nabla_{\nu} A_\beta g^{\alpha \beta}}{\Lambda_2^4}\,.
\ea
In this covariantization, the gravitational--vector theory would be given by an expression similar to \eqref{eq:cov} but with $\K$ now being a diff tensor defined as
\ba
\label{eq:KF}
\K\mupn=\(\sqrt{g^{-1}f^{(g)}}\)\mupn-\delta\mupn\,.
\ea
The absence of ghost in this covariantization is non--trivial and indeed non--minimal couplings to gravity, for instance of the form $G^{\mu\nu}A_\mu A_\nu$ may in principle need to be included to ensure the absence of \Ostro ghost. Proving the existence of such a class of covariantization which is entirely free of the \Ostro ghost is beyond the scope of this work however it can easily be done in two dimensions.

For concreteness, consider the covariant version of the two--dimensional Lagrangian $\L^{(\text{2d})}$ introduced in \eqref{eq:L2d},
\ba
	\L^{(\text{2d})}_{\rm cov} = \sqrt{-g}\(-2[\K] -\frac 12 m^2 A_\mu A_\nu g^{\mu \nu}\)\,,\label{eq:L2dc}
\ea
with $\K$ now as defined in \eqref{eq:KF}. This theory includes five variables that may be split into the lapse $N$, shift $n_1$ and 1--dimensional spatial metric $\gamma_{11}=\gamma$, and the two components of the vector field $A_0$ and $A_1$. For the theory to avoid any type of \Ostro ghost, out of these five variables, only one of them ought to be dynamical (in practise the helicity--0 mode of the massive vector).
To check that the theory \eqref{eq:L2dc} does indeed satisfy this property, we may compute the five--dimensional field space Hessian given by
\ba
\label{eq:Hab}
\H_{AB}=\frac{\p^2 \L^{(\text{2d})}_{\rm cov}}{\p \dot \Psi ^A \dot \Psi^B}\,,
\ea
with $\Psi^A=\{N, n_1, \gamma, A_0, A_1\}$ and check that it is of rank--1.

Upon defining the following two quantities,
\begin{align}
	B &= A_0 - A_1 n_1 \label{eq:functionB} \\
	C &= 4 N^3 \gamma^2 + 4 A_0' n_1 N \gamma^2 + 2 A_1' N \gamma (N^2 - n_1^2 \gamma) + 2 n_1' N \gamma^2 (A_0 - 2 A_1 n_1)  \\
		& \quad + (2 N' \gamma - N \gamma') (A_1(N^2 + n_1^2 \gamma) - A_0 n_1 \gamma) - 2 \dot{A}_0 N \gamma^2 + 2 A_1 \dot{n}_1 N \gamma^2 + (A_0 - A_1 n_1) (2 \dot{N} \gamma - N \dot{\gamma}) \gamma\,,\nn
\end{align}
one can check explicitly that the Hessian defined in \eqref{eq:Hab} can actually  be written in the form
\begin{equation}
	\H_{ab} =-\frac{1}{2 \gamma^5 N^8} P_a P_b\,,
\label{eq:hess2dgrav}
\end{equation}
with the field space vector $P$ defined as
\ba
P_a=\(-2 B \gamma^2 F_{01}, -2 N \gamma^2 A_1 F_{01}, N \gamma B F_{01}, C, 2 N \gamma^2 F_{01}\)\,.
\ea
This directly implies that the Hessian is of rank--1 and hence the theory \eqref{eq:L2dc} only propagates one degree of freedom in two dimensions.
This shows that the direct covariantization of the quantity $f\mn$ as in \eqref{eq:Fmn} is a `consistent choice'  in two dimensions in the sense that it maintains all the constraints required both for gravity and for the Proca field. Extending the covariantization more generically to four dimensions is beyond the scope of this work as the argument provided was merely to illustrate the presence of different types of alternative covariantizations as illustrated in Fig.~\ref{Fig:Covs}.

\section{Coupling Extended Proca-Nuevo with gravity}
\label{sec:couplextPN}

As the theory of a Lorentz massive spin-1 field, the previous analysis naturally constructed the EPN theory on Minkowski, where the symmetries of the Lorentz and Poincar\'e groups make sense. However, in order to make contact with astrophysics and cosmology, we can also attempt to first extend the theory on arbitrarily curved spacetime-dependent backgrounds and then further include the coupling with the gravitational dynamical degrees of freedom in the constraint analysis. We will address this by studying the existence of an NEV associated with the Hessian matrix of the covariantized version of the theory.

We will first prove that the suggested covariantized version of  EPN does possess an NEV on any arbitrarily curved background metric no matter its spacetime dependence. However, this vector fails to be an eigenvector as soon as the metric is taken as a dynamical variable. By itself this simply indicates that the vector ought to be modified to include the non-trivial mixing with gravity, however, any modification of this vector would necessarily result in a non-vanishing eigenvalue and hence a breaking of the constraint. The presence of an additional degree of freedom is then inexorably linked to this loss of constraint, and standard analysis shows that when such additional degrees of freedom enter, they are always of ghostly nature. However since this loss of constraint is related to the mixing between the gravitational degrees of freedom and the Proca field, the resulting effects are Planck scale-suppressed. Moreover, we will see that the case of FLRW (Friedmann-Lema\^itre-Robertson-Walker) spacetime is special in that the additional degree of freedom is absent due to the isometries of the background so that the issue can be evaded at that level. This statement is strengthened by the fact that linear perturbations on cosmological backgrounds are free from any additional ghost degrees of freedom as will be shown explicitly.

\subsection{Covariant EPN theory}
\label{ssec:CovExtPN}

We define the covariant EPN theory by the action
\begin{equation}
	S = \int \d^4 x \sqrt{-g} \left(\frac{M_{\rm Pl}^2}{2}\,R + \L_{\rm EPN} + \L_M \right) \,,
	\label{eq:SextPN}
\end{equation}
where $R$ is the curvature scalar, $\L_M$ is the matter Lagrangian and
\begin{equation}
\label{eq:covEPN}
	\L_{\rm EPN} = - \frac14 \,F\mnup F\mn + {\Lambda}^4 \left( \L_0 + \L_1 + \L_2 + \L_3 \right) \,,
\end{equation}
with the definitions
\begin{align}
	 \L_0 &= \alpha_{0}(X) \label{eq:L0} \,,\\
	 \L_1 &=  \alpha_{1}(X) \L_1[\K] +  d_{1}(X) \frac{\L_1[\nabla A]}{\Lambda^2} \label{eq:L1} \,,\\
	 \L_2 &= \left( \alpha_2(X) + d_2(X) \right) \frac{R}{\Lambda^2} + \alpha_{2,X}(X) \L_2[\K] + d_{2,X}(X) \frac{\L_2[\nabla A]}{\Lambda^4} \label{eq:L2} \,,\\
	 \L_3 &= \left( \alpha_3(X) \K\mnup +d_3(X) \frac{\nabla^{\mu} A^{\nu}}{\Lambda^2} \right) \frac{G\mn}{\Lambda^2} - \frac16 \alpha_{3,X}(X) \L_3[\K] - \frac{1}{6} d_{3,X}(X) \frac{\L_3[\nabla A]}{\Lambda^6} \label{eq:L3} \,.
\end{align}
Note that for later convenience we are considering arbitrary functions of a  slightly modified argument than the one introduced in Eq.~\eqref{eq:defXargument},
\beq
X=-\frac{1}{2\Lambda^2}\,A^{\mu}A_{\mu}\,,
\eeq
Here the subscript $X$ on the coefficient functions denotes differentiation w.r.t.\ $X$. While the Einstein-Hilbert term could be absorbed into the definition of $\alpha_2$ or $d_2$, we have chosen to write it independently in order to distinguish the Planck scale from the scale controlling the EPN interactions.

Some comments are in order regarding our definition of $\L_{\rm EPN}$. First, the Lagrangian includes non-minimal coupling terms, proportional to $R$ in $\L_2$ and to the Einstein tensor $G_{\mu\nu}$ in $\L_3$. These are motivated by the known non-minimal couplings of GP theory \cite{Tasinato:2014eka, Heisenberg:2014rta}. Second, and also related to the question regarding non-minimal couplings, our Lagrangian omits the $\L_4$ term that was present in flat spacetime. As remarked before, this term does not belong to the GP class and neglecting this term  has the advantage of simplifying the analysis of cosmological perturbations, which is our main scope.

\subsection{EPN on an arbitrary background}
\label{ssec:ExtPNFix}

Before proceeding with the constraint analysis of the covariant EPN action defined in \eqref{eq:SextPN}, we make a brief digression here to point out that EPN admits a simpler covariantization in the case when the metric is non-dynamical (in the sense that we do not include the gravitational degrees of freedom in the counting of degrees of freedom), yet with a background spacetime that is arbitrarily curved. Indeed, in this situation, the minimal coupling prescription applied to the full flat-space Lagrangian, Eq.\ \eqref{eq:LextPNflat}, is already enough to furnish a fully consistent theory. The proof is similar to that used in \cite{deRham:2020yet} to establish the consistency of PN theory in flat spacetime, i.e.\ through the explicit construction of the NEV of the Hessian matrix associated to the Lagrangian.

Unsurprisingly, this NEV is nothing but the minimal covariantization of the flat-space NEV. Starting with the vector field we define the auxiliary metric
\beq
f_{\mu\nu}=g\mn + 2 \frac{\nabla_{(\mu} A_{\nu)}}{\Lambda^2} + \frac{g^{\alpha \beta}\nabla_{\mu} A_\alpha \nabla_{\nu} A_\beta}{\Lambda^4} \,,
\eeq
where the covariant derivatives are taken with respect to the arbitrary metric $g\mn$ and the tensor
\beq
\X\mupn = \left( \sqrt{g^{-1}f}  \right)\mupn \,.
\eeq
The claim is that the vector
\beq
V_\mu = (\mathcal{X}^{-1})^{0}{}_{\alpha}\left(\delta^\alpha{}_\mu+\frac{g^{\alpha \beta}}{\Lambda^2}\,\nabla_{\beta}A_{\mu}\right) \,,
\label{eq:NEVcovariant}
\eeq
is the desired NEV. We will prove this here for the EPN term $\L_1$; the proof for the other terms can be found in Appendix \ref{ssec:NEVFix}.

As explained previously, the ``extended'' terms $\L_n[\nabla A]$ do not contribute to the Hessian matrix (again, in the absence of dynamical gravity), so it suffices to focus on $\L_1[\K]$. Actually, it is more convenient for the proof to consider instead $\L_1[\X]$, which entails no loss of generality given that the set $\L_n[\K]$ is linearly related to the set $\L_n[\X]$. Note that this statement is only true for the complete sets of operators $\L_n$, with $n$ going from $1$ to $4$. However, we prove in Appendix \ref{ssec:NEVFix} that the vector $V^{\mu}$ is the common NEV to each $\L_n[\X]$, including $\L_4[\X]$, and hence it is also the desired NEV for each $\L_n[\K]$.

We then define, for each $\L_n[\X]$, the associated canonical momentum conjugate to the vector field as
\begin{equation}
	p^{(n)}_{\alpha} = \Lambda^4 \frac{\p \L_n[\X]}{\p \dot{A}^{\alpha}} \,,
\end{equation}
and the corresponding Hessian matrix
\begin{equation}
	\mathcal{H}^{(n)}_{\alpha \beta} = \frac{\p p^{(n)}_{\alpha}}{\p \dot{A}^{\beta}} \,.
	\label{eq:HessianOrderN}
\end{equation}
For $\L_1[\X]$ we find $p^{(1)}_{\alpha} = {\Lambda}^{2} V_{\alpha}$, and therefore
\begin{equation}
	\mathcal{H}^{(1)}_{\alpha \beta} V^{\alpha} = {\Lambda}^2 \frac{\p V_{\alpha}}{\p \dot{A}^{\beta}} V^{\alpha} = \frac{{\Lambda}^2}{2} \frac{\p (V_{\alpha} V^{\alpha})}{\p \dot{A}^{\beta}} = 0 \,,
	\label{eq:HV1equals0}
\end{equation}
which follows because $V_{\alpha} V^{\alpha}=g^{00}$ is independent of the vector field velocity. A similar proof applies to the other $\L_n[\X]$ terms, see Appendix \ref{ssec:NEVFix} for details.

\subsection{Coupling with gravitational degrees of freedom}
\label{ssec:ExtPNDyn}

We now extend the previous analysis to accommodate a dynamical metric, in the sense where the dynamical degrees of freedom of the metric are included in the constraint analysis. While the degeneracy of the full Hessian matrix is preserved by the $\L_1$ EPN term upon minimal coupling to gravity, this property will be shown to fail for the other covariant EPN terms, $\L_2$ and $\L_3$, with or without the GP-inspired non-minimal couplings. Once again, because the GP-like contributions $\L_n[\nabla A]$ (with the appropriate non-minimal coupling terms) are known to be ghost-free, it suffices to focus on the PN terms, and without loss of generality we consider the set $\L_n[\X]$ instead of the set $\L_n[\K]$.

To proceed, we start by decomposing the metric in ADM variables,
\begin{equation}
	g_{00} = - N^2 + N^{i} N_i \,, \qquad g_{i0} = N_i \,, \qquad g_{ij} = \gamma_{ij} \,,
\end{equation}
where $N$ is the lapse, $N^i$ is the shift (defined with an upper index) and $\gamma_{ij}$ is the three-dimensional spatial metric, used to raise and lower indices on any spatial tensor. The vector field is parametrized by the time and spatial components of $A_{\mu}^{*}$, related to the original field via
\begin{equation}
	A_{\mu} = M_{\mu}^{\phantom{\mu} \nu} A_{\nu}^{\ast} \,,
\end{equation}
with
\begin{equation}
	M_{\mu}^{\phantom{\mu} \nu} \equiv
	\begin{pmatrix}
		N & N^i \\
		0 & \delta^i_j
	\end{pmatrix} \,.
\end{equation}
Even though $\dot{A}_{\mu}$ and $\dot{A}^*_{\mu}$ are not linearly related (because $M_{\mu}^{\phantom{\mu} \nu}$ is time-dependent), the corresponding conjugate momenta do satisfy a linear relation,
\beq
p^{\ast(n)}_{\mu} \equiv \Lambda^4 \frac{\p \L_n[\X]}{\p \dot{A}^{\ast\mu}} = M^{\nu}_{\phantom{\nu} \mu} p^{(n)}_{\nu} \,,
\eeq
and similarly for the Hessian submatrix
\beq
\mathcal{H}^{\ast(n)}_{\mu\nu} \equiv \frac{\p p^{\ast(n)}_{\mu}}{\p \dot{A}^{\ast\nu}} = M^{\rho}_{\phantom{\rho} \mu} M^{\sigma}_{\phantom{\sigma} \nu} \mathcal{H}^{(n)}_{\rho\sigma} \,.
\eeq

The full Hessian matrix of field velocities is now a $10\times 10$ matrix spanning the four components of the vector field $A^*_\mu$  and the six components of the spatial metric. In this analysis, we ignore the lapse and shift since it can be shown that no instance of $\dot{N}$ and $\dot{N}^i$ appears in the Lagrangian after performing the redefinition $A_\mu \mapsto A^*_\mu$ \cite{Langlois:2015skt}.

We claim that a natural candidate for the Hessian NEV of the covariant EPN theory is
\beq \label{eq:full EPN NEV}
\bm{\mathcal{V}} \equiv \left( V^*_\mu \,,\, 0\right) \,,
\eeq
where the null entry runs over the metric components and
\begin{equation}
	V^{\ast}_{\mu} \equiv \left( M^{-1} \right)_{\mu}^{\phantom{\mu} \nu} V_{\nu} \,.
	\label{eq:Vast}
\end{equation}
The vector $\bm{\mathcal{V}}$ indeed annihilates both the pure vector and pure metric subsectors. The latter property is trivial, while the former holds because
\beq
\mathcal{H}^{*(n)}_{\mu\nu}V^{*\nu} = M^{\rho}_{\phantom{\rho} \mu}\mathcal{H}^{(n)}_{\rho\sigma}V^{\sigma}=0 \,,
\eeq
where the last equality follows from the results of the previous subsection. Thus the outstanding question is whether $\bm{\mathcal{V}}$ annihilates the mixed vector-metric components of the Hessian.

It is easy to verify this for the $\L_1[\X]$ term. Defining
\beq
\mathcal{H}^{*(n)}_{\mu,ij} \equiv \frac{\p p^{*(n)}_{\mu}}{\p \dot{\gamma}^{ij}} \,,
\eeq
we have
\beq
\mathcal{H}^{*(n)}_{\mu,ij}V^{*\mu} = \frac{\p p^{(n)}_{\mu}}{\p \dot{\gamma}^{ij}}V^{\mu} \,,
\eeq
and in particular for $n=1$
\beq
\mathcal{H}^{*(1)}_{\mu,ij}V^{*\mu} = {\Lambda}^2 \frac{\p V_{\alpha}}{\p \dot{\gamma}^{ij}} V^{\alpha} = \frac{{\Lambda}^2}{2} \frac{\p (V_{\alpha} V^{\alpha})}{\p \dot{\gamma}^{ij}} = 0 \,.
\eeq
Therefore $\L_1[\X]$ defines a consistent ghost-free theory when coupled to dynamical gravity. Since $\bm{\mathcal{V}}$ is proven to be the null eigenvector for $\L_1[\X]$, we can directly infer that if $\bm{\mathcal{V}}$ fails to also be a null eigenvector for any  other $\L_n[\X]$, then irrespectively of what the appropriate eigenvectors would then be, it cannot be a null eigenvector for all $\L_n[\K]$ and thus the constraint will always be necessarily lost for generic theories given by \eqref{eq:covEPN}.
And indeed, as it turns out, when considering  $\L_2[\X]$ and $\L_3[\X]$, we can show that in the absence of any non-minimal couplings then $\mathcal{H}^{*(n)}_{\mu,ij}V^{*\mu}\neq0$ for $n=2,3$ (see Appendix \ref{ssec:NEVDyn} for the explicit expressions).

At this stage, this means that the covariant theory must necessarily include non-minimal couplings between the vector field and the curvature, unsurprisingly since we know this to be the case in the simpler GP theory. Our proposed covariant version of EPN theory, Eq.\ \eqref{eq:SextPN}, contains the non-minimally coupled term
\beq \label{eq:L2 nonmin}
\L_2^{\rm(non-min)} = \alpha_{2,X}\,\L_2[\X]+\frac{\alpha_2}{\Lambda^2}\,R \,.
\eeq
As is shown in Appendix~\ref{ssec:NEVDyn}, while the vector $\bm{\mathcal{V}}$ fails to be a precise null eigenvector for the resulting Hessian matrix, our claim is that the addition of the curvature scalar operator allows us to consistently apply our model to cosmology.

The first virtue of the above choice \eqref{eq:L2 nonmin} of non-minimal coupling is that $\L_2^{\rm(non-min)}$ is indeed degenerate whenever the tensor $\nabla_{\mu}A_{\nu}$ is symmetric. For instance, this is the case for the cosmological backgrounds that we are interested in, namely the FLRW metric
\begin{equation}
	g_{\mu\nu}\d x^{\mu}\d x^{\nu} = -\d t^2 + a^2(t) \delta_{ij} \d x^i \d x^j \,,
	\label{eq:lineFLRW}
\end{equation}
and the vector field profile
\begin{equation} \label{eq:vector bkgd}
	A_{\mu}\d x^{\mu} = -\phi(t)\d t \,.
\end{equation}
In fact, for this background, the absence of additional degrees of freedom can be seen very directly by noting that\footnote{With some abuse of terminology, we will refer to the background defined by Eqs.\ \eqref{eq:lineFLRW} and \eqref{eq:vector bkgd} as ``FLRW''.}
\begin{equation}
	\K^{\mu}_{\phantom{\mu}\nu} = \frac{1}{{\Lambda}^2}\,\nabla^{\mu}A_{\nu} \,,\quad  \text{on FLRW}\,.
\end{equation}
It follows that $\L_n[\K]=\L_n[\nabla A]/\Lambda^{2n}$ for this background, implying that the EPN theory actually reduces to a subclass of GP theory when restricted to FLRW, however the perturbations themselves differ quite significantly. Yet, we will confirm in Section \ref{sec:fulltheory} that the Lagrangian \eqref{eq:L2 nonmin} propagates the correct number of degrees of freedom also at the level of general linear perturbations about this background, where the equivalence between EPN and GP no longer holds. Although reassuring as an explicit check, let us emphasize that the presence of a constraint and absence of additional Ostrogradsky ghost was indeed expected given our proof that the NEV $\bm{\mathcal{V}}$ indeed annihilates the Hessian on the FLRW background.

We can also extend the derivations beyond cosmological backgrounds, and another virtue of the above choice of non-minimal coupling is that when expanded perturbatively in higher-dimensional operators, the matrix product $\bm{\mathcal{H}}^{(2)}_{\rm(non-min)}\bm{\mathcal{V}}$ correctly vanishes at leading order, but does not vanish when the two operators in \eqref{eq:L2 nonmin} are taken separately (see Appendix~\ref{ssec:NEVDyn}).
Since the constraint is present at leading order in an operator expansion and only gets broken at higher-order, it is in principle possible that the addition of new higher-order curvature terms could cancel the left-over, and so on in a perturbative fashion. Such precise constructions are however beyond the scope of this work and are kept to a future work.
Note however that the scale at which the vector $\bm{\mathcal{V}}$ ceases to be a null eigenvector is crucial.  The loss of constraint is related to the presence of operators that mix between the gravitational and vector degrees of freedom. Following the result presented in Eq.~\eqref{eq:sumHcoupl}, and using the fact that at leading order the momentum is given by $p^*_\mu=\dot A^*_0 \delta^0_\mu$, we can infer that at the level of the Lagrangian, the loss of constraint is associated with an operator which behaves symbolically as
\ba
\L_{\rm ghost}\sim \frac{1}{\Lambda^2}\dot \gamma^{ij} \dot A_0^* F_{0i} A_j+\text{higher-dimensional operators}\,.
\ea
This term would be irrelevant if the gravitational degrees of freedom were not considered as dynamical, so including the gravitational tensor modes $h_T$ and the vector fluctuation $\delta A_0^*$,  this corresponds to a ghostly operator of the form
\ba
\L_{\rm ghost}\sim \frac{1}{\Lambda^2 \mpl}\dot h_{T} \delta\dot A_0^* \bar F_{0i} \bar A_j+\text{higher-dimensional operators}\,.
\ea
Remaining on the conservative side, this implies that a background configuration with vector vev $\bar A$ and field strength vev $\bar F$ would excite an additional ghost degree of freedom $\chi$, entering as $\dot A_0^*\sim \ddot \chi/\bar m$ at the symbolic cutoff scale
\ba
m_{\rm ghost}\sim \frac{\Lambda^2 \mpl \bar m}{\bar A_\perp \bar F}\,,
\ea
where $\bar m$ is the mass of the vector field on the background in question, and $\bar A_\perp$ is the dynamical part of the vector field. In particular on any background where the field strength tensor vanishes (i.e.\ where $\p_\mu \bar A_\nu$ is symmetric), we recover an absence of ghost, as is the case on the cosmological background we shall have in mind. Note that these considerations are meaningless on backgrounds where vector field happens to vanish $\bar m=0$ since the helicity-0 is then infinitely strongly coupled. On a background where $\bar A_\perp \sim \Lambda$, and $\bar F \sim \p \bar A_\perp \sim \bar m \Lambda$, the mass of the would-be ghost would be of order $\mpl$.

All the previous considerations also apply to the EPN term $\L_3$ and its associated non-minimal curvature coupling as given in \eqref{eq:SextPN}; details can be found in Appendix \ref{ssec:NEVDyn}. The upshot of this analysis is that our proposal for a covariantization of EPN theory, while not successful in complete generality, does indeed define a consistent cosmological model insofar as the number of degrees of freedom is concerned and as long as one is interested in linear perturbations about cosmological solutions defined by Eqs.\ \eqref{eq:lineFLRW} and \eqref{eq:vector bkgd}.


\section{Special model without non-minimal couplings}
\label{sec:SpecExCov}

We introduced in Eq.~\eqref{eq:SextPN} what we have argued to be a natural first step in the covariantization of the flat-space EPN theory derived in Section \ref{sec:ExtPN}. We will refer to this Lagrangian as the ``general'' model, because it reproduces all the operators in \eqref{eq:LextPNflat} in the flat space limit (with the exception of $\L_4[\K]$, which we omit as previously explained). We will study this general model in the next section.

In the present section we focus instead on an alternative covariantization in which all non-minimal coupling terms are omitted. We recall that in our analysis of the Hessian matrix we found that the non-minimal couplings were in fact necessary for the NEV ansatz to succeed at leading order in a standard EFT operator expansion. While this statement is true generically, there remains the possibility that other covariantization schemes may exist when the theory is restricted by a specific choice of the coefficient functions $\alpha_n$ and $d_n$. In this section we show that this is indeed the case.

The ``special'' model we consider is defined by the action
\begin{equation}
	\hat{S} = \int \d^4 x \sqrt{-g} \left(\frac{M_{\rm Pl}^2}{2}\,R + \hat{\L}_{\rm EPN} + \L_M \right) \,,
	\label{eq:Shat}
\end{equation}
where we use a hat to distinguish it from the general model in Eq.\ \eqref{eq:SextPN}. Here again $R$ is the curvature scalar, $\L_M$ is the matter Lagrangian, and the special EPN Lagrangian reads
\begin{equation}
\hat{\L} = - \frac14 \,F\mnup F\mn + \Lambda^4 \left( \hat{\L}_0 + \hat{\L}_1 + \hat{\L}_2 + \hat{\L}_3 \right) \,,
\end{equation}
where
\begin{align}
	 \hat{\L}_0 &= \alpha_{0}(X) \label{eq:L0example} \,,\\
	 \hat{\L}_1 &= \alpha_{1}(X) \L_1[\K] + d_{1}(X) \frac{\L_1[\nabla A]}{\Lambda^2} \label{eq:L1example} \,,\\
	 \hat{\L}_2 &= \alpha_{2,X}(X) \left(\L_2[\K] - \frac{\L_2[\nabla A]}{\Lambda^4}\right) \label{eq:L2example} \,,\\
	 \hat{\L}_3 &= - \frac16 \alpha_{3,X}(X) \left(\L_3[\K] - \frac{\L_3[\nabla A]}{\Lambda^6} \right)\label{eq:L3example} \,.
\end{align}

We remark that this Lagrangian can be formally obtained from that of the general model by setting $d_2=-\alpha_2$, $d_{2,X}=-\alpha_{2,X}$, $d_{3,X}=-\alpha_{3,X}$ and $\alpha_3=0=d_3$. But we emphasize that this is only a formal procedure, since the latter two conditions are obviously inconsistent as functional relations (except in the trivial case $\alpha_3(X)=0=d_3(X)$). With this choice of coefficients, the model has the advantage of being particularly simple, having no non-minimal couplings between the vector field and the metric and with comparatively few free coefficient functions. The precise constraint analysis for this special model is performed in details in Appendix~\ref{ssec:NEVspec}. It follows the precise same pattern as that discussed previously in the more general case in Section~\ref{sec:couplextPN}, and in particular the exact same conclusions as those of subsection~\ref{ssec:ExtPNDyn} hold here upon accounting for the dynamical mixing between the gravitational and vector degrees of freedom in this special example. Note in particular that this special model is free of ghost on cosmological backgrounds.

\section{Discussion}

Aside from generalized massive gravity \cite{deRham:2014lqa,deRham:2014gla}, the existence of another complete covariantization of EPN (and of PN) remains an open question. Here we have taken a first step toward its solution by proposing a covariant model (Eq.~\eqref{eq:SextPN}). With this covariantization in mind, one can find a null eigenvector (given in Eq.~\eqref{eq:full EPN NEV}) for the full Hessian of the first family of operators (namely $\L_1(\X)$) proving that it enjoys a constraint at all orders. For the other family of operators (namely $\L_{n\ge 2}(\X)$), we showed that the same ansatz for the null eigenvector correctly annihilates the Hessian matrix at leading order in an expansion in the strong coupling scale $\Lambda$ but the process fails when pushing it to higher order. The result is nevertheless non-trivial and provides a hint that a full covariantization is in principle feasible. We note also that the failure of the constraint only occurs from mixing with the gravitational degrees of freedom and is thus Planck scale-suppressed. Moreover, we show that our proposed eigenvector remains a null one for the Hessian of the full theory (including the gravitational degrees of freedom), on any background where the tensor $\nabla_\mu A_\nu$ is symmetric. This directly implies the presence of a constraint that would remove the unwanted ghostly additional degree of freedom at linear order in perturbations about any such backgrounds, including on FLRW.

These results for the covariant EPN theory are by themselves sufficient to motivate the study of the predictions of the model in the context of cosmology.

Having now shown that EPN is a viable theory that admits consistent covariantization schemes, we will now apply it to cosmology in Chapter \ref{chap:Cosmo} to follow, where we will use EPN to model dark energy. We will use the stability of the background and the perturbations to get more insight into the consistency of EPN as a suitable phenomenological model of dark energy.

\chapter{Cosmology of Extended Proca-Nuevo}
\label{chap:Cosmo}

In this Chapter, we show there exist (Extended) Proca-Nuevo models that allow for consistent and ghost-free cosmological solutions. We study these models in the presence of perfect fluid matter and show that they describe the correct number of dynamical variables and derive their dispersion relations and stability criteria. We also exhibit, in a specific setup, explicit hot Big Bang solutions featuring a late-time self-accelerating epoch, which are such that all the stability and subluminality conditions are satisfied and where gravitational waves behave precisely as in General Relativity.

\section{Cosmology of the special model without non-minimal couplings}
\label{sec:SpecEx}

In this Section, we will consider the ``special'' model introduced in Eq.~\eqref{eq:Shat}. This model is an alternative covariantization that is free of any non-minimal couplings. In the following, we derive the equations governing the dynamics of the FLRW background (defined by Eqs.\ \eqref{eq:lineFLRW} and \eqref{eq:vector bkgd}) and of its linear perturbations. The matter sector is assumed to be a perfect fluid, although at this stage we do not specify its equation of state. As expected, our analysis recovers the correct number of propagating degrees of freedom on this cosmological background. In the last subsection, we consider an admixture of pressureless matter and radiation and then solve the background equations for a particular choice of the EPN coefficients. We further show that, for this particular choice, all the stability and subluminality conditions for the perturbations are satisfied. The proposed example thus provides a proof of principle that a healthy candidate for the Big Bang history of our Universe can be accommodated within EPN. This model does not rely on the presence of any cosmological constant, but rather on the presence of non-trivial Proca field self-interactions that enter at a technically natural scale \cite{deRham:2021yhr}.

\subsection{Background}
\label{ssec:SpecExbkgeom}

We proceed  by deriving the background cosmological equations of motion.
We focus on the FLRW metric
\begin{equation}
	\d s^2 = -N^2(t)\d t^2 + a^2(t) \delta_{ij} \d x^i \d x^j \,,
	\label{eq:lineFLRW2}
\end{equation}
with the vector field profile
\begin{equation} \label{eq:vector bkgd2}
	A_{\mu}\d x^{\mu} = -\phi(t)\d t \,.
\end{equation}
The equation obtained from varying the action with respect to the lapse yields the modified Friedmann equation
\begin{equation}
	H^2 = \frac{1}{3M_{\text{Pl}}^2} \left( \rho_M + \hat{\rho}_{\text{EPN}} \right) \,,
	\label{eq:Friedmannhat}
\end{equation}
where from now on we may set $N=1$ and where
\beq
	\hat{\rho}_{\text{EPN}} \equiv \Lambda^4 \left[ - \alpha_{0} + \alpha_{0,X} \frac{\phi^2}{\Lambda^2} + 3 \left( \alpha_{1,X} + d_{1,X} \right) \frac{H \phi^3}{\Lambda^4} \right] \label{eq:rhoExtPNhat} \,,
\eeq
is the effective energy density of the vector field. Note that $\rho_M$ is the energy density of the usual content of the universe (including matter and radiation). It corresponds to the standard energy density appearing in the Friedmann equation and coincides with the definition of the $T^{00}$ entry of the stress-energy tensor of a perfect fluid in equilibrium. Furthermore, it will feature in the action for the for the perfect fluid matter in Eq.~\eqref{eq:SMSS}. The Friedmann equation may be combined with the equation that follows from varying the action with respect to the scale factor $a(t)$ to produce the modified Raychaudhuri equation
\begin{equation}
	\frac{\ddot{a}}{a} = \dot{H} + H^2 = - \frac{1}{6 M_{\text{Pl}}^2} \left(\rho_M + \hat{\rho}_{\text{EPN}} + 3 P_M + 3 \hat{P}_{\text{EPN}}\right) \,,
\end{equation}
where
\begin{equation}
	\hat{P}_{\text{EPN}} \equiv \Lambda^4 \left[ \alpha_{0} - \left( \alpha_{1,X} + d_{1,X} \right)  \frac{\phi^2 \dot{\phi}}{\Lambda^4} \right] \,,
	\label{eq:PExtPNhat}
\end{equation}
is interpreted as the effective pressure of the vector condensate. Finally, variation with respect to $\phi(t)$ gives
\beq
	\alpha_{0,X} + 3 \left( \alpha_{1,X} + d_{1,X} \right) \frac{H \phi}{\Lambda^2} = 0 \,,
	\label{eq:eom3hat}
\eeq
which is however not independent of the other two as a consequence of the Bianchi identity. The fact that Eq.\ \eqref{eq:eom3hat} is a constraint, enforcing an algebraic relation between $H$ and $\phi$, is no accident but follows from the precise form of the Lagrangian of the special model.

\subsection{Perturbations}
\label{ssec:SpecExPert}

\subsubsection{Definitions}
\label{sssec:SpecExIntroPert}

Next, we introduce perturbations about the FLRW background, following \cite{DeFelice:2016yws,DeFelice:2016uil}. Metric perturbations in the flat gauge are composed of two scalar modes $\alpha$ and $\chi$, one vector mode $V_i$ and one tensor mode $h_{ij}$. The line element reads
\begin{equation}
	g_{\mu\nu}\d x^{\mu}\d x^{\nu} = - \left(1+2 \frac{\alpha}{M_{\text{Pl}}} \right)\d t^2 + \frac{2}{M_{\text{Pl}}} \left(\frac{\p_i \chi}{M_{\text{Pl}}} + a V_i \right) \d t \d x^i + a^2(t) \(\delta_{ij} + \frac{h_{ij}}{\mpl}\) \d x^i \d x^j \,.
\end{equation}
The vector mode is transverse and the tensor mode is traceless and transverse, so that they each have two degrees of freedom. In here and what follows, spatial indices $i,j,\cdots$ are raised and lowered with respect to the spatial Euclidean metric $\delta_{ij}$.

The vector field $A_{\mu}$ is parametrized by two scalar perturbations $\delta \phi$ and $\chi_V$, together with a (transverse) vector mode $Z_i$. The perturbed vector field is then defined as
\begin{align}
	A^0 &= \phi(t) + \delta \phi \,,\\
	A^i &= \frac{1}{a^2} \delta^{ij}\left( aZ_j-\frac{a}{M_{\rm Pl}}\,\phi V_j +  \frac{\p_j \chi_V}{\Lambda} \right) \,.
\end{align}
The appearance of $V_i$ in the vector field perturbation may appear as unnecessary at this stage but will prove convenient later and prevent the need of additional field redefinitions.

For the perfect fluid matter we use the Schutz-Sorkin action \cite{SCHUTZ19771},
\begin{equation}
	S_M = - \int \d^4x \left[ \sqrt{-g} \, \rho_M(n) + J^{\mu} \left( \p_{\mu} l + \mathcal{A}_i \p_{\mu} \mathcal{B}^i \right) \right] \,.
	\label{eq:SMSS}
\end{equation}
Here
\begin{equation}
	n= \sqrt{\frac{J^{\mu} J_{\mu}}{g}} \,,
\end{equation}
is the fluid number density, whose background value is given by $\overline{n} = \mathcal{N}_0/a^3$, with $\mathcal{N}_0$ a constant. The current $J^{\mu}$ is decomposed as
\begin{align}
	J^0 &= \mathcal{N}_0 + M_{\text{Pl}}^2 \delta J \nonumber \,, \\
	J^i &= \frac{M_{\text{Pl}}}{a^2} \delta^{ik} \left( \p_k \delta j + M_{\text{Pl}} W_k \right) \,,
\end{align}
where $\delta J$ and $\delta j$ are scalars and $W_k$ is a transverse vector.

The scalar $l$ in \eqref{eq:SMSS} is such that on the background,
\begin{equation}
	\p_0 \overline{l} = - \rho_{M,n} \equiv - \frac{\p \rho_M}{\p n} \,,
\end{equation}
and we define its scalar perturbation $v$ by
\begin{equation}
	l = - \int^t \rho_{M,n} \d t' - \frac{\rho_{M,n} v}{M_{\text{Pl}}^2} \,,
\end{equation}
and note that on FLRW we have
\begin{equation}
	\rho_{M,n} = \frac{\rho_M + P_M}{\overline{n}} = a^3 \frac{\rho_M + P_M}{\mathcal{N}_0} \,.
	\label{eq:defrhoMn}
\end{equation}
Finally, the vectors $\mathcal{A}_i$ and $\mathcal{B}_i$ are also transverse. The canonical choice for their associated perturbations $\delta \mathcal{A}_i$ and $\delta \mathcal{B}_i$ reads
\begin{equation}
	\mathcal{A}_i = \frac{\delta \mathcal{A}_i}{M_{\text{Pl}}} \,, \qquad \mathcal{B}_i = M_{\text{Pl}} x_i + \frac{\delta \mathcal{B}_i}{M_{\text{Pl}}}  \,.
	\label{eq:dAidBi}
\end{equation}

For later use we note that the normalized $4$-velocity of the fluid can be found by varying the action \eqref{eq:SMSS} with respect to the current $J^{\mu}$, with the result
\begin{equation}
	u_{\mu} = \frac{J_{\mu}}{n \sqrt{-g}} = \frac{1}{\rho_{M,n}} \left( \p_{\mu} l + \mathcal{A}_i \p_{\mu} \mathcal{B}^i \right) \,,
	\label{eq:umu}
\end{equation}
and $u_i$ can be further split as
\begin{equation}
	u_i = - \frac{\p_i v}{M_{\text{Pl}}^2} + \frac{v_i}{M_{\text{Pl}}} \,,
	\label{eq:ui}
\end{equation}
where $v_i$ is transverse.

In the rest of this subsection we compute the quadratic part of the action for all perturbations, respectively for tensor, vector and scalar modes, and determine the conditions for every dynamical mode to be stable.

\subsubsection{Tensor perturbations}
\label{sssec:SpecExTens}

The quadratic action for the tensor perturbations is given by
\begin{equation}
	\hat{S}_T^{(2)} = \int \d^4 x \, a^3 \frac{1}{8} \left[ \dot{h}_{ij}^2 - \frac{1}{a^2} \(\p_k h_{ij}\)^2 \right] \,,
	\label{eq:SThat}
\end{equation}
We see that, in the special model, the EPN dynamics does not affect the quadratic action for the tensors, which are therefore entirely determined by the Einstein-Hilbert term. Thus not only is the tensor sector of \eqref{eq:Shat} free of instabilities, the speed of propagation of gravitational waves in this setup is also exactly luminal. We find this to be a remarkable property considering the fact that the vector field, even though minimally coupled, still interacts non-trivially with the metric. Having the same dispersion relation as that of GR is of course also a phenomenological virtue given the recent experimental measurements on the speed of gravitational waves \cite{LIGOScientific:2017zic}.

\subsubsection{Vector perturbations}
\label{sssec:SpecExVec}

For the vector sector it is convenient to consider first the matter action \eqref{eq:SMSS}. Expanding to quadratic order in perturbations we find
\begin{equation}
	S_{M,V}^{(2)} = \int \d^4x \frac{1}{M_{\text{Pl}}^2} \left[ \frac{\rho_{M,n}}{2a^2 \mathcal{N}_0} \left( M_{\text{Pl}}^3 W_i + a \mathcal{N}_0 V_i \right)^2 - \frac12 a^3 \rho_M V^2 - \left( \mathcal{N}_0 \delta \dot{\mathcal{B}}_i + \frac{M_{\text{Pl}}^4}{a^2} W_i \right) \delta \mathcal{A}_i  \right] \,,
	\label{eq:SMV2}
\end{equation}
in agreement with \cite{DeFelice:2016uil}. We now proceed to eliminate the non-dynamical variables so as to identify the dynamical degrees of freedom. Varying \eqref{eq:SMV2} with respect to $W_i$ we have
\begin{equation}
	W_i = \frac{\mathcal{N}_0 \left( \delta \mathcal{A}_i - \rho_{M,n} a \frac{V_i}{M_{\text{Pl}}} \right)}{\rho_{M,n} M_{\text{Pl}}^2} \,.
\end{equation}
Plugging the definitions of $\delta \mathcal{A}_i$ and $\delta \mathcal{B}_i$ in Eq.\ \eqref{eq:dAidBi} into Eq.\ \eqref{eq:ui} we find
\begin{equation}
	\delta \mathcal{A}_i = \frac{\rho_{M,n} v_i}{M_{\text{Pl}}} \,,
	\label{eq:dAi1}
\end{equation}
and hence
\begin{equation}
	W_i = \frac{\mathcal{N}_0}{M_{\text{Pl}}^3} \left( v_i - a V_i \right) \,.
\end{equation}
Varying \eqref{eq:SMV2} with respect to $\delta \mathcal{A}^i$ we get
\begin{equation}
	v_i = a \left( V_i - a \frac{\delta \dot{\mathcal{B}}_i}{M_{\text{Pl}}} \right) \,.
	\label{eq:vi1}
\end{equation}
Combining these results we may integrate out $W_i$ and $\delta \mathcal{A}_i$ to obtain
\begin{equation}
	S_{M,V}^{(2)} = \int \d^4x \frac{a^3}{2} \frac{1}{M_{\text{Pl}}^2} \left[ (\rho_M + P_M) \left( V_i - a \frac{\delta \dot{\mathcal{B}}_i}{M_{\text{Pl}}} \right)^2 - \rho_M V_i^2 \right] \,.
	\label{eq:SMV2b}
\end{equation}
Collecting Eq.\ \eqref{eq:SMV2b} with the expansion of the vector part of the EPN Lagrangian we arrive at
\begin{align}
	\hat{S}_V^{(2)} &= \int \d^4 x \frac{a^3}{2} \left[ \hat{q}_V \dot{Z}_i^2 - \frac{1}{a^2} \hat{\mathcal{C}}_1 (\p_i Z_j)^2 - H^2 \hat{\mathcal{C}}_2 Z_i^2 +\frac{1}{2a^2}(\p_i V_j)^2 + \frac{(\rho_M + P_M)}{M_{\text{Pl}}^2} \left( V_i - a \frac{\delta \dot{\mathcal{B}}_i}{M_{\text{Pl}}} \right)^2 \right] \,.
	\label{eq:SVhat2}
\end{align}
Note that to obtain this expression we made use of the background equations of motion. The coefficients appearing in \eqref{eq:SVhat2} are given by
\begin{align}
	\hat{q}_V &= 1 - \frac{1}{2 \left( 1 + \frac{\dot{\phi} + H \phi}{2 \Lambda^2} \right)} \left[ \alpha_1 - 2 \left( 1 - 2 \frac{H \phi}{\Lambda^2} \right) \alpha_{2,X} + \frac{H \phi}{\Lambda^2} \left( 2 - \frac{H \phi}{\Lambda^2} \right) \alpha_{3,X} \right] \label{eq:qVhat} \,,\\
	\hat{\mathcal{C}}_1 &= 1 - \frac{1}{2\left( 1 + \frac{H \phi}{\Lambda^2} \right)} \left[ \alpha_1 - 2 \left( 1 - \frac{H \phi + \dot{\phi}}{\Lambda^2} \right) \alpha_{2,X} + \left( \frac{H \phi}{\Lambda^2} + \left( 1 - \frac{H \phi}{\Lambda^2} \right) \frac{\dot{\phi}}{\Lambda^2} \right) \alpha_{3,X} \right] \label{eq:C1hat} \,, \\
	\hat{\mathcal{C}}_2 &= 2 \hat{q}_V + \frac{\p_t(\hat{q}_V H)}{H^2} + \frac{\dot{\phi}}{H^2} \left( \alpha_{1,X} + d_{1,X} \right) \label{eq:C2hat} \,.
\end{align}
The action \eqref{eq:SVhat2} describes two dynamical vector modes, since it is clear that $V_i$ is non-dynamical and may be integrated out (although the solution of its equation of motion involve non-linear instances of the 3-momentum). This integration could be performed formally but for what interests us here, namely the Proca vector mode $Z_i$, this degree of freedom is fully decoupled from $V_i$ and $\delta\mathcal{B}_i$, which are moreover independent of the parameters of the EPN model and thus evolve exactly as in GR.

Focusing then on the $Z_i$ mode, from \eqref{eq:SVhat2} we immediately infer the dispersion relation, with sound speed and effective mass being given by
\beq
\hat{c}_V^2 = \frac{\hat{\mathcal{C}}_1}{\hat{q}_V} \,,\qquad \hat{m}_V^2 = H^2 \frac{\hat{\mathcal{C}}_2}{\hat{q}_V} \,.
\eeq
Stability under ghosts and gradient modes then imposes the conditions
\beq
\hat{q}_V>0\,,\qquad \hat{\mathcal{C}}_1>0 \,.
\eeq
One may in addition ask for tachyon modes to be absent, which would then also require $\hat{\mathcal{C}}_2>0$.

\subsubsection{Scalar perturbations}
\label{sssec:SpecExScal}

Turning next to the scalar sector, we start again by expanding the matter action \eqref{eq:SMSS} to quadratic order. It proves useful to introduce
\begin{equation}
	\delta \rho_M \equiv \frac{\rho_{M,n}}{a^3} \delta J = \frac{\rho_M + P_M}{n_0 a^3} \delta J \,,
\end{equation}
in terms of which the scalar part of the action takes the form
\begin{align}
	S_{M,S}^{(2)} &= \int \d^4x \left[ M_{\text{Pl}}^2 \frac{\rho_{M,n}}{2a^5 \overline{n}} \left( \p_i \delta j + a^3 \frac{\overline{n}}{M_{\text{Pl}}^3} \p_i \chi \right)^2 + \frac{\rho_{M,n}}{a^2 M_{\text{Pl}}} \p_i \delta j \p_i v + a^3 \dot{v} \delta \rho_M - 3 \frac{a^3 \overline{n} \rho_{M,nn}}{\rho_{M,n}^2} H v \delta \rho_M \right. \nonumber \\
	& \quad \left. - \frac{a^3 M_{\text{Pl}}^4 \rho_{M,nn}}{2 \rho_{M,n}^2} \delta \rho_M^2 - a^3 M_{\text{Pl}} \alpha \delta \rho_M + \frac{a^3 \rho_M}{2 M_{\text{Pl}}^2} \left( \alpha^2 - \frac{(\p_i \chi)^2}{a^2 M_{\text{Pl}}^2} \right) \right] \,,
	\label{eq:SMS2}
\end{align}
where
\begin{equation}
	c_M^2 =  \frac{\overline{n} \rho_{M,nn}}{\rho_{M,n}} \,,
\end{equation}
is the squared sound speed of the fluid in pure GR. It also corresponds to the sound speed in GP theory and, as we will see, in the EPN special model, but not in the general model.

We may already integrate out at this stage the scalar mode $\delta j$. From its equation of motion we get
\begin{equation}
	\delta j = -a^3 \frac{\overline{n}}{M_{\text{Pl}}^3} \left( v + \chi \right) \,,
	\label{eq:eomdeltaj}
\end{equation}
and substituting into \eqref{eq:SMS2} furnishes
\begin{align}
	S_{M,S}^{(2)} &= \int \d^4x \, a^3 \left[ - \frac{\overline{n} \rho_{M,n}}{2M_{\text{Pl}}^4} \frac{(\p_i v)^2}{a^2} + \left( \frac{\overline{n} \rho_{M,n}}{M_{\text{Pl}}^4} \frac{\p^2 \chi}{a^2} - \delta \dot{\rho}_M - 3 H (1+ c_M^2) \delta \rho_M \right) v \right. \nonumber \\
	& \quad \left. - \frac{c_M^2 M_{\text{Pl}}^4}{2 \overline{n} \rho_{M,n}} (\delta \rho_M)^2 - M_{\text{Pl}} \alpha \delta \rho_M + \frac{\rho_M}{2 M_{\text{Pl}}^2} \left( \alpha^2 - \frac{(\p_i \chi)^2}{a^2 M_{\text{Pl}}^2} \right) \right] \,.
	\label{eq:SMS2b}
\end{align}
This result is to be combined with the expansion of the EPN Lagrangian. We eventually obtain (using again the background equations of motion)
\begin{align}
	\hat{S}_S^{(2)} &= \int \d^4 x \, a^3 \left\lbrace  - \frac{\overline{n} \rho_{M,n}}{2M_{\text{Pl}}^4} \frac{(\p_i v)^2}{a^2} + \left[ \frac{\overline{n} \rho_{M,n}}{M_{\text{Pl}}^4} \frac{\p^2 \chi}{a^2} - \delta \dot{\rho}_M - 3 H (1+ c_M^2) \delta \rho_M \right] v   \right.  \\
	& \quad  - \frac{c_M^2 M_{\text{Pl}}^4}{2 \overline{n} \rho_{M,n}} (\delta \rho_M)^2 - M_{\text{Pl}}\, \alpha\, \delta \rho_M - \hat{\omega}_3 \frac{(\p_i \alpha)^2}{a^2 M_{\text{Pl}}^2} + \hat{\omega}_4 \frac{\alpha^2}{M_{\text{Pl}}^2}   \\
	& \quad - \left[ \left(3 H \hat{\omega}_1 - 2 \hat{\omega}_4 \right) \frac{\delta \phi}{\phi} - \hat{\omega}_3 \frac{\p^2 (\delta \phi)}{a^2 \phi} - \hat{\omega}_3 \frac{\p^2 \dot{\psi}}{a^2 \phi \Lambda} + \hat{\omega}_6 \frac{\p^2 \psi}{a^2 \Lambda} \right] \frac{\alpha}{M_{\text{Pl}}}   \\
	& \quad - \frac{\hat{\omega}_3}{4} \frac{(\p_i \delta \phi)^2}{a^2 \phi^2} + \hat{\omega}_5 \frac{(\delta \phi)^2}{\phi^2} - \frac12 \left[ \left( \hat{\omega}_2 + \hat{\omega}_6 \phi \right) \psi - \hat{\omega}_3 \dot{\psi} \right] \frac{\p^2 (\delta \phi)}{a^2 \phi^2 \Lambda}  \\
	& \quad \left. - \frac{\hat{\omega}_3}{4 \phi^2} \frac{(\p_i \dot{\psi})^2}{a^2 \Lambda^2} + \frac{\hat{\omega}_7}{2} \frac{(\p_i \psi)^2}{a^2 \Lambda^2} + \left( \hat{\omega}_1 \frac{\alpha}{M_{\text{Pl}}} + \hat{\omega}_2 \frac{\delta \phi}{\phi} \right) \frac{\p^2 \chi}{a^2 M_{\text{Pl}}^2} \right\rbrace   \,,\\ \label{eq:SSnocoupl2}
\end{align}
for the complete quadratic action of scalar perturbations in the special model. Here we introduced
\begin{equation}
	\psi \equiv \chi_V + \frac{\Lambda}{M_{\text{Pl}}^2} \phi \chi \,,
\end{equation}
and the (time-dependent) coefficients $\hat{\omega}_I$ are given in Appendix \ref{ssec:DefCoefsExScal}.

We observe that the action \eqref{eq:SSnocoupl2} has the same structure as the quadratic scalar action derived in GP theory \cite{DeFelice:2016uil}, only with different $\hat{\omega}_I$ coefficients. We emphasize that this is a non-trivial result since the special model \eqref{eq:Shat} is manifestly not of the GP class. Indeed, if one were to ``detune'' the operators in $\hat{\L}_2$ and $\hat{\L}_3$ in \eqref{eq:Shat} then additional operators would appear in \eqref{eq:SSnocoupl2}. These extra operators modify the degree of degeneracy of the equations of motion and, as a consequence, additional degrees of freedom become active.

To see that the action \eqref{eq:SSnocoupl2} propagates two dynamical modes one can simply analyze the resulting equations of motion,
\begin{align}
	&(3 H \hat{\omega}_1 -2\hat{\omega}_4) \frac{\delta \phi}{\phi} -2\hat{\omega}_4 \frac{\alpha}{M_{\text{Pl}}} + M_{\text{Pl}}^2 \delta \rho_M + \frac{k^2}{a^2 \Lambda^2} \left[ \hat{\mathcal{Y}} + \hat{\omega}_1 \frac{\Lambda^2}{M_{\text{Pl}}^2} \chi - \hat{\omega}_6 \Lambda \psi \right] = 0 \label{eq:SpecExeomalpha} \,,\\
	&\frac{(\rho_M + P_M)}{M_{\text{Pl}}} v + \hat{\omega}_1 \alpha + M_{\text{Pl}} \hat{\omega}_2 \frac{\delta \phi}{\phi} = 0 \label{eq:SpecExeomchi} \,,\\
	&(3 H \hat{\omega}_1 - 2 \hat{\omega}_4 )\frac{\alpha}{M_{\text{Pl}}} - 2 \hat{\omega}_5 \frac{\delta \phi}{\phi} + \frac{k^2}{a^2 \Lambda^2} \left[ \frac{1}{2}\hat{\mathcal{Y}} + \hat{\omega}_2 \frac{\Lambda^2}{M_{\text{Pl}}^2} \chi - \frac{\Lambda}{2} ( \hat{\omega}_2 + \hat{\omega}_6 \phi ) \frac{\psi}{\phi} \right] = 0 \label{eq:SpecExeomdphi} \,,\\
	&\frac{\dot{\hat{\mathcal{Y}}}}{H} + \left( 1 - \frac{\dot{\phi}}{H \phi} \right) \hat{\mathcal{Y}} + \frac{\Lambda^2}{H} \left\lbrace \hat{\omega}_2 \frac{\delta \phi}{\phi} + 2 \hat{\omega}_7 \frac{\phi \psi}{\Lambda} + \hat{\omega}_6 \left( 2 \frac{\alpha \phi}{M_{\text{Pl}}} + \delta \phi \right) \right\rbrace = 0 \label{eq:SpecExeomppsi} \,,\\
	&\dot{\delta \rho}_M + 3 H (1 + c_M^2) \delta \rho_M + \frac{k^2}{a^2} \frac{(\rho_M + P_M)}{M_{\text{Pl}}^4} ( v+ \chi) = 0 \label{eq:SpecExeomv} \,,\\
	&\alpha M_{\text{Pl}} + c_M^2 \left( 3 H v + \frac{M_{\text{Pl}}^4}{(\rho_M + P_M)} \delta \rho_M \right) - \dot{v} = 0 \label{eq:SpecExeomdrhoM} \,,
\end{align}
respectively for $\alpha$, $\chi$, $\delta \phi$, $\psi$, $v$ and $\delta \rho_M$, and we defined
\begin{equation}
	\hat{\mathcal{Y}} \equiv \frac{\Lambda^2}{\phi} \hat{\omega}_3 \left(  \delta \phi + 2 \frac{\alpha \phi}{M_{\text{Pl}}} + \frac{\dot{\psi}}{\Lambda} \right) \label{eq:SpecExY} \,.
\end{equation}
It is straightforward to show that these equations can be solved algebraically for $\alpha$, $\delta \phi$, $\chi$ and $v$ in order to be left with a system of two second-order differential equations for $\psi$ and $\delta \rho_M$. This completes the proof that the special EPN model exhibits the correct number of degrees of freedom in the tensor, vector and scalar sectors.

To study the stability of the propagating scalar modes we integrate out all the non-dynamical variables. The resulting action is formidably lengthy, but for the purpose of deciding whether the fields exhibit ghost- or gradient-type instabilities it suffices to focus on the short wavelength limit $k \rightarrow \infty$ which focuses on perturbations deep inside the sound horizon. This regime is interesting for the study of the growth of large-scale structure and we use it to make contact with the analysis performed in \cite{DeFelice:2016uil}, and especially the explanation given in Appendix A there. In this regime, the action can be recast in the form
\begin{equation}
	\hat{S}_S^{(2)} = \int \d^4x \, a^3 \left[ \dot{\vec{\Omega}}^t \hat{\bm{K}} \dot{\vec{\Omega}} - \vec{\Omega}^t \left( \hat{\bm{M}} - \frac{k^2}{a^2} \hat{\bm{G}} \right) \vec{\Omega} - \vec{\Omega}^t \hat{\bm{B}} \dot{\vec{\Omega}} \right] \,,
	\label{eq:SpecExSMmatrix}
\end{equation}
where $\vec{\Omega}^t \equiv \left( \psi, \delta \rho_M / k \right)$ (note that $\delta \rho_M$ has mass dimension 2, hence the rescaling by $k$). For brevity we omit the explicit expressions for the matrices $\hat{\bm{K}}, \hat{\bm{M}}, \hat{\bm{G}}$ and $\hat{\bm{B}}$, but let us remark that they are independent of $k$ (again in the limit $k \rightarrow \infty$ as explained above).

Absence of ghosts requires the kinetic matrix $\hat{\bm{K}}$ to be positive definite. Thanks to the special form of the Schutz-Sorkin action it turns out that $\hat{\bm{K}}$ is diagonal \cite{Kase:2014cwa, Kase:2014yya}, and we find
\begin{equation}
\hat{Q}_{S,\psi} = \frac{M_{\text{Pl}}^2 H^2}{\Lambda^2 \phi^2} \frac{3 \hat{\omega}_1^2 + 4 M_{\text{Pl}}^2 \hat{\omega}_4}{(\omega_1 - 2 \omega_2)^2} \,,\qquad \hat{Q}_{S,M} = \frac{a^2}{2} \frac{M_{\text{Pl}}^4}{(\rho_M + P_M)} \,,
\end{equation}
for the eigenvalues associated to $\psi$ and $\delta\rho_M$, respectively. Positivity of $\hat{Q}_{S,M}$ requires $\rho_M + P_M > 0$, i.e.\ the (strict) null energy condition, while the condition $\hat{Q}_{S,\psi}>0$ is equivalent to
\begin{equation}
	3 \hat{\omega}_1^2 + 4 M_{\text{Pl}}^2 \hat{\omega}_4 > 0 \,.
\end{equation}

Absence of gradient-unstable modes requires the sound speeds square to be positive. From the dispersion relation,
\begin{equation}
	\text{det} \left[ \hat{\omega}^2 \hat{\bm{K}} - \left( \hat{\bm{M}} - \frac{k^2}{a^2} \hat{\bm{G}} \right) \right] = 0 \,,
	\label{eq:SpecExdispSM}
\end{equation}
we obtain that the fluid propagates with speed $c_M^2$, as previously claimed, while the Proca scalar mode $\psi$ has
\begin{equation}
	\hat{c}_{S,\psi}^2 = \frac{1}{M_{\text{Pl}}^2 H^2 \phi^2} \frac{\hat{\Gamma}}{8 \hat{q}_V (3 \hat{\omega}_1^2 + 4 M_{\text{Pl}}^2 \hat{\omega}_4)} \,,
	\label{eq:cSpsihat}
\end{equation}
where
\begin{align}
	\hat{\Gamma} &\equiv 2 \hat{\omega}_2^2 \hat{\omega}_3 ( \rho_M + P_M ) - \hat{\omega}_3 ( \hat{\omega}_1 - 2 \hat{\omega}_2) ( \hat{\omega}_1 \hat{\omega}_2 + \phi ( \hat{\omega}_1 - 2 \hat{\omega}_2 ) \hat{\omega}_6 ) \left( \frac{\dot{\phi}}{\phi}- H \right) - \hat{\omega}_3 ( 2 \hat{\omega}_2^2 \dot{\hat{\omega}}_1 - \hat{\omega}_1^2 \dot{\hat{\omega}}_2 ) \nonumber \\
	&\quad + \phi ( \hat{\omega}_1 - 2 \hat{\omega}_2)^2 ( \hat{\omega}_3 \dot{\hat{\omega}}_6 + \phi (2 \hat{\omega}_3 \hat{\omega}_7 + \hat{\omega}_6^2)) + \hat{\omega}_1 \hat{\omega}_2 \left( \hat{\omega}_1 \hat{\omega}_2 + (\hat{\omega}_1 - 2 \hat{\omega}_2) \left( 2 \phi\hat{\omega}_6 - \hat{\omega}_3 \frac{\dot{\phi}}{\phi} \right) \right) \,.
\end{align}
Note that $\hat{q}_V>0$ is already required by the stability of vector perturbations (cf.\ Eq.\ \eqref{eq:qVhat}), while $3 \hat{\omega}_1^2 + 4 M_{\text{Pl}}^2 \hat{\omega}_4 > 0$ from the above no-ghost condition. It therefore suffices to impose $\hat{\Gamma} > 0$ for Laplacian instabilities to be absent.

While tachyonic instabilities are less concerning---on the contrary, they are potentially interesting---later we will also examine the effective masses of the scalar modes. The expressions are somewhat lengthy and so we leave them for Section \ref{ssec:MassesExScal} in the Appendix.

Having derived the stability conditions for all the dynamical modes, the outstanding question is whether there exist choices of parameters of the special model Lagrangian such that all the criteria are satisfied while providing a consistent cosmological history. In the next subsection we show that this is the case.

\subsection{Cosmology of the special model}
\label{ssec:SpecExCosmo}

\subsubsection{Background}
\label{sssec:BkgSpec}

Having established that the simple model we analyzed could be stable on FLRW, we can push analysis to whether it could be relevant for the cosmological evolution of our Universe. For this, we specify the matter perfect fluid to be a mixture of pressureless matter and radiation, respectively denoted by subscripts ``$m$'' and ``$r$'', i.e.\ $\rho_M = \rho_m + \rho_r$ and $P_M = P_m + P_r$ with
\begin{align}
	P_m = 0 \qquad &\Rightarrow\qquad  \dot{\rho}_m + 3H \rho_m = 0 \nonumber \,, \\
	P_r = \frac13 \rho_r \qquad &\Rightarrow \qquad \dot{\rho}_r + 4H \rho_r = 0 \,.
\end{align}
The effective energy density and pressure of the EPN field were defined previosuly in \eqref{eq:rhoExtPNhat} and \eqref{eq:PExtPNhat}.

We wish to recast the background equations as a dynamical system, again following the analysis of \cite{DeFelice:2016yws,DeFelice:2016uil}. As a first step solve for $\phi$ in terms of $H$ by using the constraint equation \eqref{eq:eom3hat}. Next it is convenient to introduce the density parameters
\begin{equation}
	\Omega_r \equiv \frac{\rho_r}{3 M_{\text{Pl}}^2 H^2}\,, \qquad \Omega_m \equiv \frac{\rho_m}{3 M_{\text{Pl}}^2 H^2}\,, \qquad \hat{\Omega}_{\text{EPN}} \equiv \frac{\hat{\rho}_{\text{EPN}}}{3 M_{\text{Pl}}^2 H^2} \,,
\end{equation}
so that the Friedmann equation reads
\begin{equation}
	\Omega_r + \Omega_m + \hat{\Omega}_{\text{EPN}} = 1 \,,
\end{equation}
which we use to solve for $H$ as a function of $\hat{\Omega}_{\text{EPN}}$ (or equivalently $\Omega_r + \Omega_m$). At this stage the scalar field $\phi$ and the Hubble parameter $H$ are now fully determined by the Lagrangian parameters, the Planck mass $M_{\text{Pl}}$, the mass scale $\Lambda$ and the density parameters.

We are interested in the time evolution of the density parameters $\hat{\Omega}_{\text{EPN}}$ and $\Omega_r$ ($\Omega_m$ being trivially determined from these). We employ the e-folding number $N=\log(a)$ as the time variable, with derivatives with respect to $N$ being denoted by a prime. In order to express $\hat{\Omega}_{\text{EPN}}'$ and $\Omega_r'$ solely in terms of $\hat{\Omega}_{\text{EPN}}$ and $\Omega_r$ we first use the Raychaudhuri equation to write the EPN pressure as
\beq
	\hat{P}_{\text{EPN}} = 3M_{\text{Pl}}^2 H^2 \left( \hat{w}_{\rm eff} - \frac{\Omega_r}{3} \right) \,,
\eeq
where
\begin{equation}
	\hat{w}_{\rm eff} \equiv - 1 - \frac{2\dot{H}}{3H^2} \,,
\end{equation}
is the effective equation of state parameter of the universe. For later use let us also introduce the effective equation of state parameter for the vector condensate,
\begin{equation}
	\hat{w}_{\text{EPN}} \equiv \frac{\hat{P}_{\text{EPN}}}{\hat{\rho}_{\text{EPN}}} = \frac{\hat{w}_{\rm eff} - \frac{\Omega_r}{3}}{\hat{\Omega}_{\text{EPN}}} \,.
\end{equation}
The only other ingredients we need are the time derivatives $\dot{\hat{\rho}}_{\text{EPN}}$ and $\dot{\rho}_{r}$ which are easily obtained from the respective continuity equations.

Collecting these preliminary results we get
\begin{align}
	\hat{\Omega}_{\text{EPN}}' &= 3 \hat{w}_{\rm eff} ( \hat{\Omega}_{\text{EPN}} - 1 ) + \Omega_r = \mathcal{F}_1 ( \hat{\Omega}_{\text{EPN}}, \Omega_r)  \,,\\
	\Omega_r' &= (3\hat{w}_{\rm eff}-1) \Omega_r = \mathcal{F}_2 ( \hat{\Omega}_{\text{EPN}}, \Omega_r) \,.
\end{align}
The precise form of the functions $\mathcal{F}_1$ and $\mathcal{F}_2$ can only be determined once $\hat{w}_{\rm eff}$ is known in terms of the density parameters, and for this we need to specify the model.

Before doing so let us comment on the size of the scales involved in the problem. For consistency we require the cutoff scale $\Lambda$ of the EPN sector to be parametrically smaller than the Planck scale, i.e.\ $\Lambda\ll M_{\text{Pl}}$. In line with our aim of using the Proca field as the dark energy fluid responsible for the late-time cosmic acceleration, we take
\begin{equation}
	\Lambda^4 \sim M_{\text{Pl}}^2 H_{\rm dS}^2 \,,
\end{equation}
where $H_{\rm dS}$ is the Hubble parameter of the late-time de Sitter fixed point, roughly of the order of the present-day Hubble constant. Note that this implies $H_{\rm dS}/\Lambda\sim \Lambda/M_{\text{Pl}} \ll 1$. Finally, we assume the bare mass of the vector field to be of order $m^2\sim H_{\rm dS}^2$, and for convenience we introduce
\begin{equation}
	c_m \equiv \frac{m^2 M_{\text{Pl}}^2}{\Lambda^4} \sim 1 \,.
\end{equation}
Although not necessary, we will eventually set $c_m=1$ for the sake of simplicity.

We now specify the parameters of the model by making the following choice:
\ba \label{eq:special model choice of coeffs}
&	\alpha_0 = - \frac{m^2}{\Lambda^2} X \,, \qquad
	\alpha_1 = - \frac{\Lambda^4}{M_{\text{Pl}}^4} b_1 X^2 - \frac{\Lambda^2}{M_{\text{Pl}}^2} c_1 X \,,\qquad   d_1 =- \frac{\Lambda^4}{M_{\text{Pl}}^4} e_1 X^2 + \frac{\Lambda^2}{M_{\text{Pl}}^2} c_1 X\\
&	\alpha_{2,X} = \frac{\Lambda^4}{M_{\text{Pl}}^4} b_2 X^2 + \frac{\Lambda^2}{M_{\text{Pl}}^2} c_2 X \,,\quad {\rm and}\quad
	\alpha_{3,X} = \frac{\Lambda^4}{M_{\text{Pl}}^4} b_3 X^2 + \frac{\Lambda^2}{M_{\text{Pl}}^2} c_3 X \,.
\ea
We leave the constants $b_I$, $c_I$ and $e_1$ unspecified for the time being. It is convenient to introduce
\begin{equation}
	y \equiv \frac{c_m}{3(b_1 + e_1)} \,,
\end{equation}
where just like $c_m$, $y$ is a dimensionless parameter which sets relations between the various scales in our system. With the benefit of hindsight we set
\beq \label{eq:choice y}
y = 4 \sqrt{\frac{6}{c_m}} \,.
\eeq
The reason behind this choice is that the effective squared mass of the Proca scalar mode, in this particular model, generically goes as $\propto -H_{\rm dS}^2 (y - 4 \sqrt{6/c_m})^2/(\hat{\Omega}_{\text{EPN}}-1)^2$ near the de Sitter attractor $\hat{\Omega}_{\text{EPN}}\to 1$. The tuning in \eqref{eq:choice y} then has the purpose of eliminating this pathological behavior.

From the background equation \eqref{eq:eom3hat} we obtain
\begin{equation}
\label{eq:H2y}
H^2 = \frac{y^2 \Lambda^4 M_{\text{Pl}}^4}{\phi^6} \,.
\end{equation}
From \eqref{eq:rhoExtPNhat} we can also evaluate the dark energy density in the Proca field and its associated density parameter,
\begin{equation}
\hat{\rho}_{\text{EPN}} = \Lambda^4 \frac{c_m y^{2/3}}{2} \left( \frac{\Lambda^4}{M_{\text{Pl}}^2 H^2} \right)^{1/3} \,,\qquad \hat{\Omega}_{\text{EPN}} =  \frac{c_m y^{2/3}}{6} \left(\frac{\Lambda^4}{M_{\text{Pl}}^2 H^2} \right)^{4/3} \,.
\end{equation}
Observe that $\hat{\rho}_{\text{EPN}} \sim \Lambda^4$ when approaching the de Sitter point $H \to H_{\rm dS}$, justifying the choice of scales made in \eqref{eq:special model choice of coeffs}.

From these results we obtain the following expressions for the effective equation of state parameters:
\begin{align}
	\hat{w}_{\text{eff}} &= \frac{- 4 \hat{\Omega}_{\text{EPN}}+ \Omega_r}{3 + \hat{\Omega}_{\text{EPN}}} \,,\\
	\hat{w}_{\text{EPN}} &= - \frac{12 + \Omega_r}{9 + 3 \hat{\Omega}_{\text{EPN}}} \,,
\end{align}
so that the autonomous system determining the evolution of $\hat{\Omega}_{\text{EPN}}$ and $\Omega_r$ reads
\begin{align}
	\hat{\Omega}_{\text{EPN}}' &= \frac{4 \hat{\Omega}_{\text{EPN}} ( 3(1-\hat{\Omega}_{\text{EPN}}) + \Omega_r) }{3 + \hat{\Omega}_{\text{EPN}}} \,,\\
		\Omega_r' &= - \frac{\Omega_r \left( 3 ( 1 - \Omega_r ) + 13 \hat{\Omega}_{\text{EPN}} \right)}{3 + \hat{\Omega}_{\text{EPN}}} \,.
	\label{eq:dynsys}
\end{align}
A straightforward analysis shows that this system admits three fixed points corresponding to radiation domination, matter domination and dark energy domination (de Sitter fixed point). The results are summarized in Table \ref{tab:fixedpoints}. In the last column we show the eigenvalues of the Jacobian matrix of the system evaluated at the respective fixed points, from which we can infer their stability. We conclude in particular that the de Sitter fixed point is an attractor.
\begin{table}[h!]
\begin{center}
\begin{tabular}{ c | c  c  c | c  c | c }
  & $\Omega_r$ & $\Omega_m$ & $\hat{\Omega}_{\text{EPN}}$ & $\hat{w}_{\rm eff}$ & $\hat{w}_{\text{EPN}}$ & eigenvalues \\ \hline
  radiation & $1$ & $0$ & $0$ & $\frac{1}{3}$ & $- \frac{13}{9}$ & $\{\frac{16}{3},1\}$: unstable \\
  matter & $0$ & $1$ & $0$ & $0$ & $-\frac43$ & $\{4,-1\}$: saddle point \\
  de Sitter & $0$ & $0$ & $1$ & $-1$ & $-1$ & $\{ -4, -3 \}$: stable
\end{tabular}
\caption[Fixed points of the autonomous system for the Extended Proca-Nuevo dark energy.]{Fixed points of the autonomous system \eqref{eq:dynsys}, describing the cosmic density parameters carried respectively in radiation, matter and in the Proca field.}
\label{tab:fixedpoints}
\end{center}
\end{table}

\begin{figure}[!htb]
	\center{\includegraphics[width=8cm]{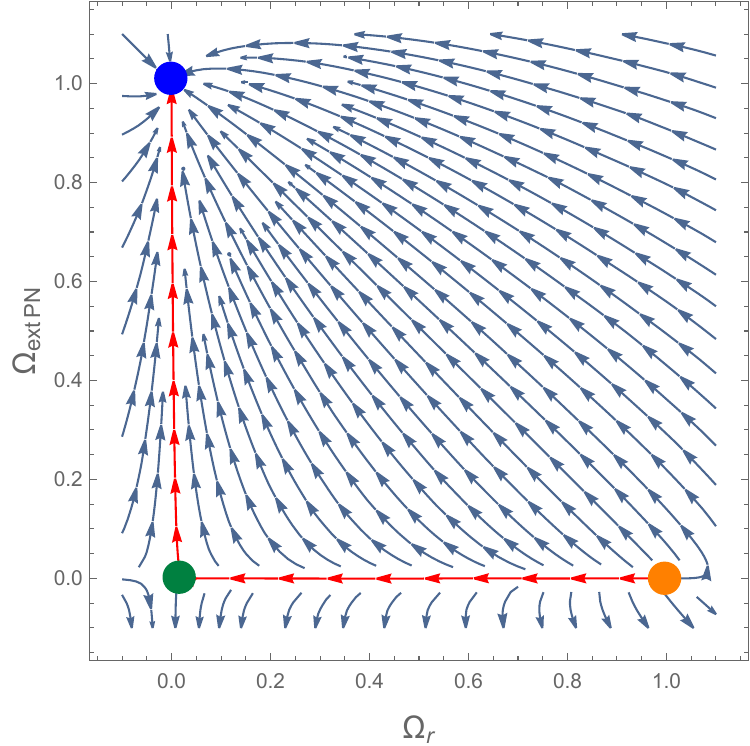}}
	\caption[Phase portrait for radiation and dark energy modelled by Extended Proca-Nuevo.]{\label{fig:plot_phase_portrait} Phase portrait associated to the autonomous system system \eqref{eq:dynsys}. The radiation, matter and dark energy fixed points are respectively indicated by the orange, green and blue dots. The red trajectory is a particular solution resembling the hot Big Bang phase of our universe with epochs of radiation, matter and dark energy domination.}
\end{figure}

In Fig.\ \ref{fig:plot_phase_portrait} we show the phase portrait of the dynamical system \eqref{eq:dynsys}, with the radiation, matter and dark energy fixed points shown as colored dots. The red trajectory is a particular solution that qualitatively mimics our universe's hot Big Bang phase, starting very close to the radiation point, flowing toward the matter point, and then asymptotically approaching the de Sitter point.

\begin{figure}[!htb]
	\center{\includegraphics[width=12cm]{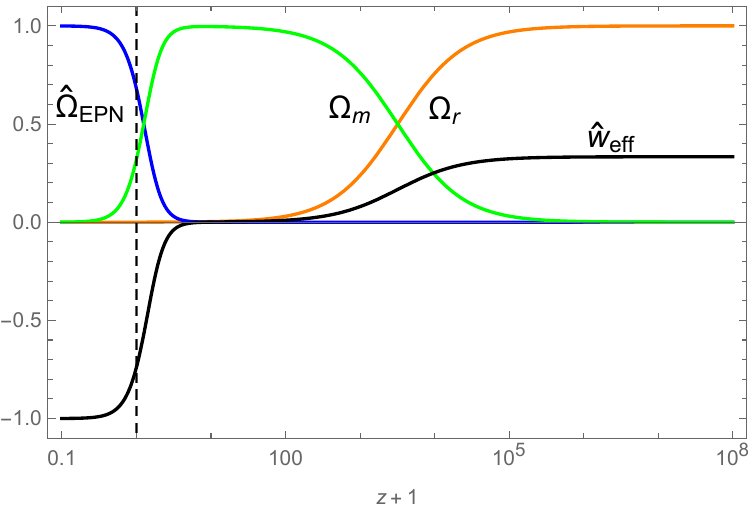}}
	\caption[Temporal evolution of the density parameters and effective equation of state]{\label{fig:plot_universe_content} Evolution of the density parameters and effective equation of state parameter $\hat{w}_{\rm eff}$ as functions of redshift, with initial conditions chosen such that $\hat{\Omega}_{\text{EPN}} = 0.68$ and $\Omega_r = 10^{-4}$ at $z=0$, indicated by the vertical dashed line.}
\end{figure}

The dynamical system \eqref{eq:dynsys} can also be solved numerically to obtain the time evolution of the density parameters. Rather than cosmic time we will show the results as functions of redshift $z$, setting $\hat{\Omega}_{\text{EPN}} = 0.68$ and $\Omega_r = 10^{-4}$ at $z=0$ (the present time), approximately the experimentally measured values. The solution for the three density parameters is shown in Fig.\ \ref{fig:plot_universe_content}, along with the effective equation of state parameter of the universe.

\subsubsection{Perturbations}
\label{sssec:PertSpec}

Finally we examine the stability  conditions for the perturbations as well as their speed. Even though all the stability conditions are time-dependent, we will evaluate them in the early- and late-time limits as way to derive a reduced set of algebraic constraints, and then verify numerically that there exists a choice of coefficients such that the constraints are satisfied at all times.

We already remarked that tensor perturbations propagate exactly as in GR. Starting then with the vector modes, we evaluate the kinetic term coefficient $\hat{Q}_V$ and squared sound speed $\hat{c}_V^2$ at both the radiation and dark energy fixed points:
\begin{equation}
	\hat{Q}_V = \begin{cases}
		1 - \frac{3 c_3 y}{10} + \mathcal{O}\left( 1 - \Omega_r, \hat{\Omega}_{\text{EPN}}^{1/4} \right) \qquad &\text{radiation} \\
		- \frac{y \left[ y (b_1 + 10b_2 + 8b_3) + 4 (c_1 + 10c_2 + 8c_3) \right]}{20 (1 - \hat{\Omega}_{\text{EPN}})}+ \mathcal{O}\left( \Omega_r, (1 - \hat{\Omega}_{\text{EPN}})^{0} \right) \qquad &\text{dS} \,,
	\end{cases}
\end{equation}
\begin{equation}
	\hat{c}_V^2 = \begin{cases}
		1 + \frac{4 c_3 y}{3 (10 - 3 c_3 y)} + \mathcal{O}\left( 1 - \Omega_r, \hat{\Omega}_{\text{EPN}}^{1/4} \right) \qquad &\text{radiation} \\
		0 + \frac{5y \left[ y (b_1 + 6b_2 + 2b_3) + 4 (c_1 + 6c_2 + 2c_3) \right] - 160}{8y \left[ y (b_1 + 10b_2 + 8b_3) + 4 (c_1 + 10c_2 + 8c_3) \right]} \left( 1 - \hat{\Omega}_{\text{EPN}} \right) + \mathcal{O}\left( \Omega_r, 1 - \hat{\Omega}_{\text{EPN}} \right) \qquad &\text{dS} \,.
	\end{cases}
\end{equation}
Notice that $\hat{Q}_V$ actually diverges while $\hat c_V$ asymptotes zero when one approaches the de Sitter point. Such a behaviour can be indicative of reaching strong coupling, however we analyze carefully the scale at which perturbative unitarity breaks down in Appendix \ref{sec:OperatordS} and show that the model becomes weakly coupled in the asymptotically de Sitter fixed point.

Recalling that we will later set $y$ as in \eqref{eq:choice y}, the conditions that $\hat{Q}_V>0$ and $0<\hat{c}_V^2\leq 1$ at early times\footnote{Note that since the tensor modes behave as in GR in this example and can be trivially decoupled, the relation between causality and subluminality of the other fields is more straightforward (see Ref.~\cite{deRham:2019ctd}).} imposes the constraint $c_3<0$, while positivity of $\hat{Q}_V$ at late times is clearly easy to achieve, for instance by taking all the coefficients $b_I$'s and $c_I$'s negative in \eqref{eq:special model choice of coeffs}. We will present a specific choice of values below. Note that this choice also ensures that $\hat{c}_V^2$ tends to $0$ from above about the dS point.

The Hubble-normalized effective mass of the vector mode is given, in our example, by
\begin{equation}
	\frac{\hat{m}_V^2}{H^2} = \begin{cases}
		0 + \left( \frac35 - \frac{1}{10 - 3 c_3 y} \right) (1 - \Omega_r ) + \mathcal{O}\left( 1 - \Omega_r, \hat{\Omega}_{\text{EPN}}^{1/4} \right) &\text{radiation} \\
		5 + \mathcal{O}\left( \Omega_r, 1 - \hat{\Omega}_{\text{EPN}} \right) &\text{dS}\,.
	\end{cases}
\end{equation}
Interestingly, the normalized mass of the vector mode approaches zero at early times and it acquires a Hubble-scale value at late times. Moreover, we will see below that it remains positive for all finite times, at least for this model and for a certain choice of coupling constants.

Continuing with the scalar sector, we focus on the Proca scalar mode $\psi$ as its stability has been shown to be independent of that of the matter fluid. The kinetic coefficient actually takes a very compact form with no need to evaluate at specific times,
\begin{equation}
	\hat{Q}_{S,\psi} = \frac{48 (3 + \hat{\Omega}_{\text{EPN}})}{y^2 (1 - \hat{\Omega}_{\text{EPN}})^2}\left(\frac{\Lambda}{M_{\rm Pl}}\right)^2 \,,
\end{equation}
while the expressions for the sound speed at early and late times read
\begin{equation}
	\hat{c}_{S,\psi}^2 = \begin{cases}
		\frac{11}{27} + \mathcal{O}\left( 1 - \Omega_r, \hat{\Omega}_{\text{EPN}}^{3/4} \right) \quad &\text{radiation} \\
		0 + \frac{5}{24y^2}\frac{y^2 \left[ y (b_1 + 10b_2 + 8b_3) + 4 (c_1 + 10c_2 + 8c_3) \right] - 32}{y (b_1 + 10b_2 + 8b_3) + 4 (c_1 + 10c_2 + 8c_3)} \left( 1 - \hat{\Omega}_{\text{EPN}} \right) + \mathcal{O}\left( \Omega_r, 1 - \hat{\Omega}_{\text{EPN}} \right) \quad &\text{dS}\,. \\
	\end{cases}
\end{equation}
We see that $\hat{Q}_{S,\psi}$ is always positive, although again we have a divergence about the de Sitter point. Similarly, $\hat{c}_{S,\psi}^2$ is manifestly positive and subluminal at early times, but tends to zero (while keeping positive values) at late times. As with the vector mode, it can be shown that this poses no problem, see Appendix \ref{sec:OperatordS}.

Finally, the effective mass of the scalar mode is given by
\begin{equation}
	\frac{\hat{m}_{S,\psi}^2}{H^2} = \begin{cases}
		0 - \frac{200}{y(10 - 3 c_3 y)} \hat{\Omega}_{\text{EPN}}^{3/4} + \mathcal{O}\left( 1 - \Omega_r, \hat{\Omega}_{\text{EPN}}^{3/4} \right) \quad &\text{radiation} \\
		\frac{10}{y^5}\, \frac{\left[ 3y^3 (b_1 + 10b_2 + 8b_3) + 12 y^2 (c_1 + 10c_2 + 8c_3) + 64 \right]^2 - 96^2}{\left[ y(b_1 + 10b_2 + 8b_3) + 4(c_1 + 10c_2 + 8c_3) \right]^2} \frac{\Lambda^4}{M_{\text{Pl}}^2 H_{dS}^2} + \mathcal{O}\left( \Omega_r, 1 - \hat{\Omega}_{\text{EPN}} \right) \quad &\text{dS} \,. \\
	\end{cases}
\end{equation}
Having $\hat{m}_{S,\psi}^2>0$ is clearly easy to achieve at late times, while at early times the Hubble-normalized effective mass approaches zero. It turns out that $\hat{m}_{S,\psi}^2$ actually tends to zero at early times from below, but being strongly suppressed relative to the time-dependent Hubble scale shows that the associated tachyonic instability is harmless.

\begin{figure}[!htb]
	\center{\includegraphics[width=7cm]{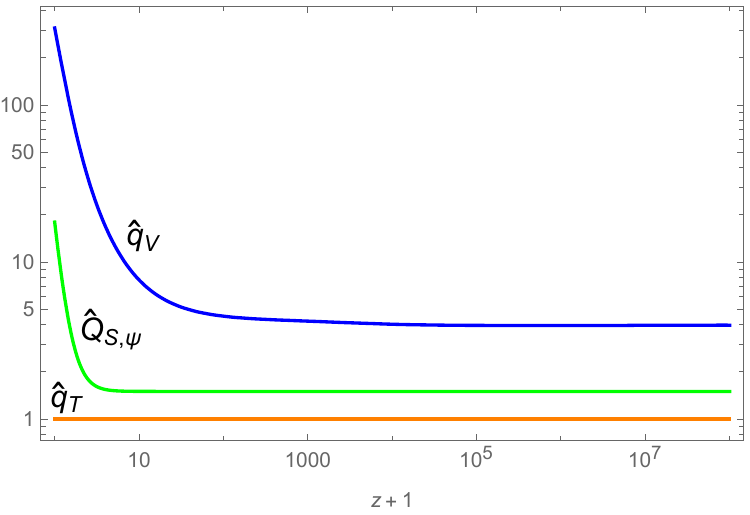}~~~~~~\includegraphics[width=7cm]{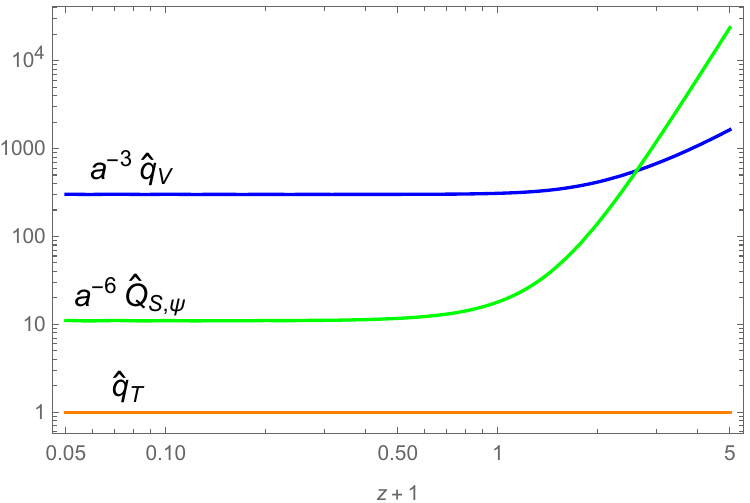}}
	\caption[Temporal evolution of the kinetic coefficients for various modes.]{\label{fig:plot_kinetic_terms} Left panel: Kinetic coefficients of the tensor, vector and scalar perturbations for $1 \leq z+1 \leq 10^8$. Right panel: The same kinetic coefficients rescaled by an appropriate power of the scale factor $a(t)$ in order to exhibit their scaling about the dS point. We observe they follow a power law scaling for small $z+1$.}
	\label{fig:kin}
\end{figure}

\begin{figure}[!htb]
	\center{\includegraphics[width=7cm]{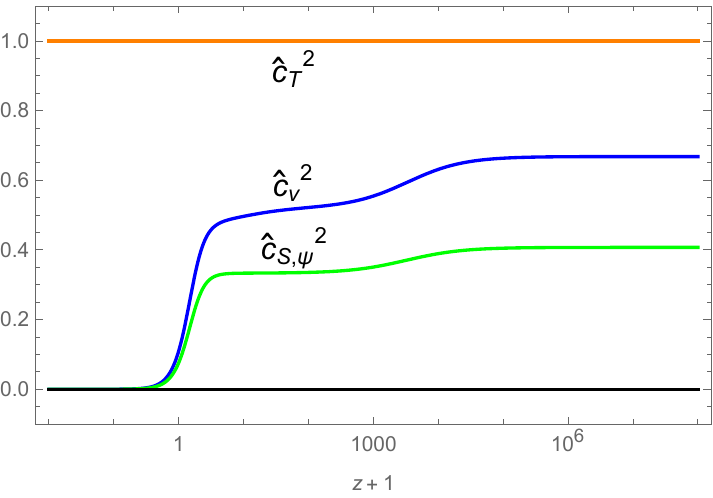}~~~~~~\includegraphics[width=7cm]{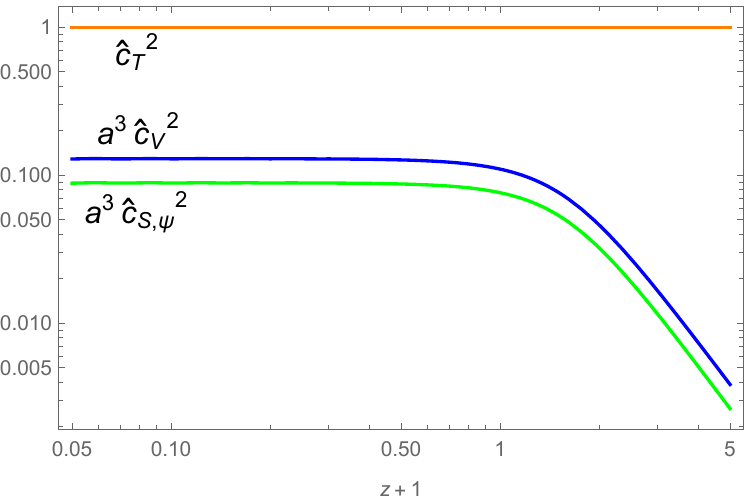}}
	\caption[Temporal evolution of the squared sound speed for various modes.]{\label{fig:plot_velocities} Left panel: Squared sound speeds of the tensor, vector and scalar perturbations. Right panel: The same speeds rescaled by an appropriate power of the scale factor $a(t)$ in order to exhibit their scaling about the dS point. We observe they follow a power law scaling for small $z+1$.}
	\label{fig:velocities}
\end{figure}

\begin{figure}[!htb]
	\center{\includegraphics[width=8cm]{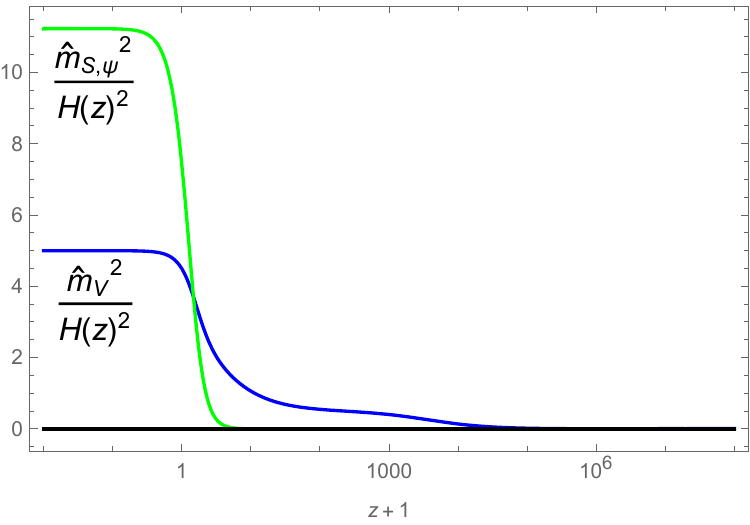}}
	\caption[Temporal evolution of the effective squared mass for various modes.]{\label{fig:plot_masses} Effective squared masses of the vector and scalar perturbations, normalized by the time-dependent Hubble parameter $H(z)$. Note that $\hat{m}_{S,\psi}^2/H^2$ is proportional to the ratio $M_{\rm{Pl}}^2 H_{\rm dS}^2 / \Lambda^4$, which we have kept generic in our analysis but have set to 1 in this particular plot.}
\end{figure}

We have argued that all the stability conditions for the dynamical modes can be met, at least in the epochs of radiation and dark energy domination and the fields are then all also subluminal. We can further show numerically that there exist coefficients such that no pathologies arise at any times. One simple explicit choice is
\begin{equation}
	c_m = 1\,, \qquad c_I = b_I = -1, \quad \text{for } I=1,2,3 \,.
\end{equation}
The results for the time evolution (plotted as functions of redshift) of the kinetic coefficients, squared sound speeds and effective squared masses are shown respectively in Figs.\ \ref{fig:plot_kinetic_terms}, \ref{fig:plot_velocities} and \ref{fig:plot_masses}. The cosmological history is the same as that of the previous subsection, with $\hat{\Omega}_{\text{EPN}} = 0.68$ and $\Omega_r = 10^{-4}$ at $z=0$.

To summarize, we have demonstrated that the EPN special model, with the particular choice of coefficient functions given in \eqref{eq:special model choice of coeffs}, admits a window of parameters such that all perturbations are free of ghost- and gradient-type instabilities and propagate subluminally. Although the velocities $\hat{c}_V^2$ and $\hat{c}_{S,\psi}^2$ approach zero in the late-time de Sitter limit, we have given an argument in Appendix \ref{sec:OperatordS} which shows that this is not a pathology. Furthermore, $\hat{c}_V^2$ and $\hat{c}_{S,\psi}^2$ are finite and positive for all $z\geq0$, so all the degrees of freedom behave in a smooth, stable and subluminal way throughout the cosmological history. While gravitational waves behave identically as in GR, the presence of the vector and scalar modes could have intriguing signatures for instance at the level of structure formation. The study of those is beyond the scope of this work, and saved for future considerations.


\section{General model}
\label{sec:fulltheory}

Having focused on a specific model for sake of concreteness, we now  return to the more generic covariant EPN theory given by Eq.\ \eqref{eq:SextPN}, what we refer to as the general model. Since the procedure follows a very similar pattern to what was given in the previous section, in what follows, we will  omit intermediate steps in most cases and highlight only the final results. We also refer the reader to the previous section for our parametrization of perturbations and other conventions.

\subsection{Background}
\label{ssec:bkgeom}

The Friedmann and Raychaudhuri equations take the same form as before,
\begin{equation}
	H^2 = \frac{1}{3M_{\text{Pl}}^2} \left( \rho_M + {\rho}_{\text{EPN}} \right) \,,\qquad \dot{H} + H^2 = - \frac{1}{6 M_{\text{Pl}}^2} \left(\rho_M + \rho_{\text{EPN}} + 3 P_M + 3 P_{\text{EPN}}\right) \,,
\end{equation}
with effective density and pressure for the dark energy fluid given by
\begin{align}
	\rho_{\text{EPN}} &= \Lambda^4 \left\{ - \alpha_{0} + \alpha_{0,X} \frac{\phi^2}{\Lambda^2} + 3 \left( \alpha_{1,X} + d_{1,X} \right) \frac{H \phi^3}{\Lambda^4} \right.  \\
	& \quad \left. + 6 \left[ - \left( \alpha_2 + d_2 \right) + 2 \left( \alpha_{2,X} + d_{2,X} \right) \frac{\phi^2}{\Lambda^2} + \left( \alpha_{2,XX} + d_{2,XX} \right) \frac{\phi^4}{\Lambda^4} \right] \frac{H^2}{\Lambda^2} \right.  \\
	& \quad \left. - \left[ 5 \left( \alpha_{3,X} + d_{3,X} \right) + \left( \alpha_{3,XX} + d_{3,XX} \right) \frac{\phi^2}{\Lambda^2} \right] \frac{H^3 \phi^3}{\Lambda^6} \right\} \,,\label{eq:rhoExtPN}
\end{align}
\begin{align}
	P_{\text{EPN}} &= \Lambda^4 \left\{ \alpha_{0} - \left( \alpha_{1,X} + d_{1,X} \right) \frac{\phi^2 \dot{\phi}}{\Lambda^4} + 2 \left(\alpha_2 + d_2 \right) \frac{3H^2 + 2\dot{H}}{\Lambda^2} \right.  \\
	&\quad - 2 \left( \alpha_{2,X} + d_{2,X} \right) \frac{\phi \left( 3 H^2 \phi + 2 H \dot{\phi} + 2 \dot{H} \phi \right)}{\Lambda^4} - 4 \left( \alpha_{2,XX} + d_{2,XX} \right) \frac{H \phi^3 \dot{\phi}}{\Lambda^6}  \\
	& \quad \left. + \left[ \left( \alpha_{3,X} + d_{3,X} \right) \frac{2 H^2 \phi + 3 H \dot{\phi} + 2 \dot{H} \phi}{\Lambda^3} + \left( \alpha_{3,XX} + d_{3,XX} \right) \frac{H \phi^2 \dot{\phi}}{\Lambda^5} \right] \frac{H \phi^2}{\Lambda^3} \right\} \,.
	\label{eq:PExtPN}
\end{align}

Finally, from the variation of the action with respect to $\phi$ one infers
\begin{align}
	&\alpha_{0,X} + 3 \left( \alpha_{1,X} + d_{1,X} \right) \frac{H \phi}{\Lambda^2} + 6 \left[ \left( \alpha_{2,X} + d_{2,X} \right) + \left( \alpha_{2,XX} + d_{2,XX} \right) \frac{\phi^2}{\Lambda^2} \right] \frac{H^2}{\Lambda^2} \nonumber \\
	& - \left[ 3 \left( \alpha_{3,X} + d_{3,X} \right) + \left( \alpha_{3,XX} + d_{3,XX} \right) \frac{\phi^2}{\Lambda^2} \right] \frac{H^3 \phi}{\Lambda^4} = 0 \,,
	\label{eq:eom3}
\end{align}
after discarding the trivial solution $\phi=0$. We observe that \eqref{eq:eom3} is again a ``constraint'' equation, relating $H$ and $\phi$ algebraically. While for the simple model this property was a consequence of the specific tuning of coefficients, in the general model this follows from the particular form of the non-minimal couplings and their coefficients.

We also note that this set of background equations is equivalent to those of GP theory \cite{DeFelice:2016uil}. As mentioned previously, this is simply because PN and GP coincide at the level of the FLRW background, and therefore so does the general EPN model.

\subsection{Perturbations}
\label{ssec:Perturbations}

\subsubsection{Tensor perturbations}
\label{sssec:TensPert}

Interestingly, in the presence of tensor perturbations the relation
\begin{equation}
	\K_{\mu\nu} = \frac{1}{\Lambda^2}\,\nabla_{\mu} A_{\nu} \qquad\qquad \mbox{(tensor modes, quadratic order)}\,,
\end{equation}
remains true up to quadratic order in the tensor modes. As for the background, it follows as an immediate result that the quadratic action for tensor perturbations in the general model will match that of GP:
\begin{equation}
	S_T^{(2)} = \int \d^4 x \, a^3 \frac{q_T}{8} \left[ \dot{h}_{ij}^2 - \frac{c_T^2}{a^2} \(\p_i h_{jk} \)^2 \right] \,,
	\label{eq:ST2}
\end{equation}
with
\begin{align}
	q_T &= 1 + 2 \frac{\Lambda^2}{M_{\text{Pl}}^2} \left[ (\alpha_2 + d_2) - \frac{\phi^2}{\Lambda^2}( \alpha_{2,X} + d_{2,X} ) \right] + \frac{H \phi^3}{M_{\rm{Pl}}^2 \Lambda^2} \left( \alpha_{3,X} + d_{3,X} \right) \label{eq:qT} \,,\\
	c_T^2 &= \frac{1 + 2 \frac{\Lambda^2}{M_{\rm{Pl}}^2} (\alpha_2 + d_2) + \frac{\phi^2 \dot{\phi}}{M_{\text{Pl}}^2 \Lambda^2} \left( \alpha_{3,X} + d_{3,X} \right)}{q_T} \,. \label{eq:cTsq}
\end{align}
These results  imply that the general model describes the expected two degrees of freedom in the tensor sector. Stability of tensor perturbations then dictates $q_T,c_T^2>0$. Subluminality of the tensor modes would also require $c_T^2<1$ but we refer to Refs.~\cite{deRham:2019ctd,deRham:2020zyh,deRham:2021fpu} for a word of caution on applying generic subluminality criteria to gravitational waves propagation without other further considerations. Imposing the speed of gravitational waves to be exactly luminal requires setting $\alpha_{2,X}+d_{2,X}=\alpha_{3,X}+d_{3,X}=0$ at all times, meaning for all values of the argument $X$ of those functions (unless $\dot \phi$ is constant). Such a choice would correspond to the example explored in details in the previous section.
We point out however that there may exist some subtleties related to the frequency at which the existing constraints on the speed of gravitational waves are satisfied \cite{deRham:2018red}, and in principle one would only require $\alpha_{2,X}+d_{2,X}=\alpha_{3,X}+d_{3,X}=0$ for a given range of arguments.

\subsubsection{Vector perturbations}
\label{sssec:VecPert}

Continuing with the vector perturbations, combining the expansion of \eqref{eq:SextPN} with the matter action derived before in \eqref{eq:SMV2b}, we find
\begin{align}
	S_V^{(2)} &= \int \d^4 x \, \frac{a^3}{2} \Bigg[ q_V \dot{Z}_i^2 - \frac{1}{a^2} \mathcal{C}_1 (\p_i Z_j)^2 - H^2 \mathcal{C}_2 Z_i^2 + \frac{1}{a^2} \mathcal{C}_3 \p_i V_j \p_i Z_j  + \frac{1}{a^2} \mathcal{C}_4 \p_i V_j \p_i \dot{Z}_j \nonumber \\
	& \quad + \frac{q_T}{2a^2}(\p_i V_j)^2 + \frac{(\rho_M + P_M)}{M_{\text{Pl}}^2} \bigg( V_i - a \frac{\delta \dot{\mathcal{B}}_i}{M_{\text{Pl}}} \bigg)^2 \Bigg] \,,
	\label{eq:SV2}
\end{align}
where the coefficients entering in this result are given in Appendix \ref{ssec:DefCoefsVec}. The structure of \eqref{eq:SV2} matches that of GP \cite{DeFelice:2016uil} except for the presence of the operator proportional to $\mathcal{C}_4$. Nevertheless, we see that this extra term does not spoil the counting of degrees of freedom, since the mode $V_i$ is still non-dynamical. This establishes that the general EPN model propagates the correct number of vector modes, namely one.

The extra operator proportional to $\mathcal{C}_4$ is interesting in that it modifies the dispersion relation of the dynamical field $Z_i$ in a way that is qualitatively different from GP. In order to highlight this effect we will ignore matter for the moment and return to the general case at the end.
Taking $\rho_M,P_M=0$ in \eqref{eq:SV2} and integrating out $V_i$ we obtain, after Fourier transforming and performing a partial integration,
\begin{align}
	S_V^{(2)}=\int \d t\,\frac{\d^3k}{(2\pi)^3}\,\frac{a^3}{2}\bigg\{ & q_V\bigg[1-\frac{k^2}{a^2}\,\frac{\mathcal{C}_4^2}{2q_Tq_V}\bigg]|\dot{Z}_i(k)|^2 \\
	&-\bigg[\mathcal{C}_2H^2+\frac{k^2}{a^2}\left(\mathcal{C}_1+\frac{\mathcal{C}_3^2}{2q_T}-a^{-1}\partial_t\left(a \frac{\mathcal{C}_3\mathcal{C}_4}{2q_T}\right)\right)\bigg]|Z_i(k)|^2\bigg\} \,. \nn
\end{align}
For a localized sub-Hubble perturbation we then infer the dispersion relation
\beq
\omega_V^2=\frac{\mathcal{C}_2H^2+\frac{k^2}{a^2}\left(\mathcal{C}_1+\frac{\mathcal{C}_3^2}{2q_T}-a^{-1}\partial_t\left(a \frac{\mathcal{C}_3\mathcal{C}_4}{2q_T}\right)\right)}{q_V\left(1-\frac{k^2}{a^2}\,\frac{\mathcal{C}_4^2}{2q_Tq_V}\right)} \qquad\qquad \mbox{(no matter)}\,.
\eeq
We see that the presence of the new coefficient $\mathcal{C}_4$ makes the dispersion relation non-linear. Expanding at small momenta (more precisely for $k^2/a^2 \ll |q_T q_V| / \mathcal{C}_4^2$) we have the linear approximation
\beq
\omega_V^2\simeq m_V^2+c_V^2\,\frac{k^2}{a^2} \,,
\eeq
with effective mass and speed of sound
\beq \label{eq:gen model mass speed vector without matter}
m_V^2\equiv \frac{\mathcal{C}_2}{q_V}\,H^2 \,,\qquad c_V^2\equiv \frac{1}{q_V} \left( \mathcal{C}_1+\frac{\mathcal{C}_3^2}{2q_T}- a^{-1}\partial_t\left(a \frac{\mathcal{C}_3\mathcal{C}_4}{2q_T}\right) + \frac{\mathcal{C}_2 \mathcal{C}_4^2}{2q_T q_V}\,H^2 \right) \,.
\eeq
In this approximation and remembering that we are neglecting matter, absence of gradient instabilities requires  $q_V, c_V^2>0$. Similarly one may also wish to demand the absence of tachyonic modes, which is achieved if $m_V^2>0$. Note that while the coefficient of the kinetic term is also modified by the $\mathcal{C}_4$ coupling, at low energies we still have the simple no-ghost condition $q_V>0$.

Returning to the general setup with matter present, we proceed again to integrate out $V_i$ from its equation of motion. The resulting action is non-diagonal in the fields $Z_i$ and $\delta\mathcal{B}_i$,
\begin{align}
	S_V^{(2)}&=\int \d t\,\frac{\d^3k}{(2\pi)^3}\,\frac{a^3}{2}\bigg\{q_V\bigg[1-\frac{\frac{k^4}{a^4}\,\mathcal{C}_4^2}{2q_V\left(\frac{k^2}{a^2}\,q_T+M^2\right)}\bigg]|\dot{Z}_i|^2+\frac{1}{2}\,\frac{\frac{k^2}{a^2}\,q_T}{\frac{k^2}{a^2}\,q_T+M^2} \frac{M^2}{M_{\text{Pl}}^2}\,|a\delta\dot{\mathcal{B}}_i|^2 \\
&\quad -\bigg[\mathcal{C}_2H^2+\frac{k^2}{a^2}\,\mathcal{C}_1+\frac{\frac{k^4}{a^4}\,\mathcal{C}_3^2}{2\left(\frac{k^2}{a^2}\,q_T+M^2\right)}-a^{-3}\partial_t\bigg(a^3\frac{\frac{k^4}{a^4}\,\mathcal{C}_3\mathcal{C}_4}{2\left(\frac{k^2}{a^2}\,q_T+M^2\right)}\bigg)\bigg]|Z_i|^2 \\
&\quad +\frac{1}{2}\,\frac{\frac{k^2}{a^2}\,\mathcal{C}_3}{\frac{k^2}{a^2}\,q_T+M^2}\frac{M^2}{M_{\text{Pl}}}\left(a\delta\dot{\mathcal{B}}_i^{*}Z_i+{\rm c.c.}\right)+\frac{1}{2}\,\frac{\frac{k^2}{a^2}\,\mathcal{C}_4}{\frac{k^2}{a^2}\,q_T+M^2}\frac{M^2}{M_{\text{Pl}}}\left(a\delta\dot{\mathcal{B}}_i^{*}\dot{Z}_i+{\rm c.c.}\right)\bigg\} \,,
	\label{eq:SV2 after int out}
\end{align}
and we introduced
\begin{equation}
	M^2\equiv 2 \frac{\rho_M+P_M}{M_{\rm Pl}^2} \,.
\end{equation}
Observe that the scale $M$ acts as a sort of infrared regulator modifying the long wavelength behavior of the coefficients in the action. Fourier transforming with respect to time in the sub-Hubble limit we find the following dispersion relation for the Proca vector mode:
\begin{equation}
	\omega_V^2=\frac{\mathcal{C}_2H^2+\frac{k^2}{a^2}\left(\mathcal{C}_1+\frac{\mathcal{C}_3^2}{2q_T}\right)-\frac{1}{2}a^{-3}\partial_t\left(a^3\frac{\frac{k^4}{a^4}\,\mathcal{C}_3\mathcal{C}_4}{\frac{k^2}{a^2}\,q_T+M^2}\right)}{q_V\left(1-\frac{k^2}{a^2}\,\frac{\mathcal{C}_4^2}{2q_Tq_V}\right)} \,.
\label{eq:gen model vec disp rel}
\end{equation}
Expanding at small momenta, assuming $k^2/a^2 \ll |q_T q_V| / \mathcal{C}_4^2$ and $k^2/a^2 \ll M^2/q_T$, we find the same effective mass as before (cf.\ Eq.\ \eqref{eq:gen model mass speed vector without matter}) and a speed of sound
\beq
c_V^2 = \frac{1}{q_V} \left( \mathcal{C}_1+\frac{\mathcal{C}_3^2}{2q_T} + \frac{\mathcal{C}_2 \mathcal{C}_4^2}{2q_T q_V}\,H^2 \right) \,,
\eeq
which curiously is a simpler expression than in the case without matter, as a consequence of the modified infrared behavior mentioned before.

The dispersion relation for the matter perturbation $\delta\mathcal{B}_i$ is $\omega^2=0$. We emphasize that this result only assumes that the fluctuation is localized on sub-Hubble scales but is otherwise exact. This degenerate dispersion relation may seem pathological but was in fact expected. The variable $\delta\mathcal{B}_i$ corresponds physically to the vorticity field of the fluid, which is indeed gapless and has no gradient energy (see \cite{Endlich:2010hf} for a discussion of this aspect in an EFT context).

The condition for the vector mode not to be ghostly is less immediate because of the non-trivial derivative couplings appearing in \eqref{eq:SV2 after int out}. To determine the norm of the propagating field we compute the residue matrix (see for instance \cite{Garcia-Saenz:2021uyv} for a review of this method),
\begin{equation}
	\lim_{\omega^2\to\omega_V^2}(\omega^2-\omega_V^2) \mathcal{P}(\omega,k) \,,
\end{equation}
where $\mathcal{P}$ is the matrix of propagators that we read off from \eqref{eq:SV2 after int out}. By construction the residue matrix has a single non-zero eigenvalue, which we find to be
\begin{align}
	\frac{1}{Q_V}&\equiv \frac{1}{q_V}\left[\frac{1+M_{\text{Pl}}^2\frac{\mathcal{C}_4^2}{q_T^2}}{1-\frac{k^2}{a^2}\,\frac{\mathcal{C}_4^2}{2q_Tq_V}}+\frac{M_{\text{Pl}}^2\frac{\mathcal{C}_3^2q_V}{q_T^2}}{\mathcal{C}_2H^2+\frac{k^2}{a^2}\left(\mathcal{C}_1+\frac{\mathcal{C}_3^2}{2q_T}\right)-\frac{1}{2}a^{-3}\partial_t\left(a^3\frac{\frac{k^4}{a^4}\,\mathcal{C}_3\mathcal{C}_4}{\frac{k^2}{a^2}\,q_T+M^2}\right)}\right] \\
&\simeq \frac{1}{q_V}\left[1+M_{\text{Pl}}^2 \left( \frac{\mathcal{C}_4^2}{q_T^2}+\frac{\mathcal{C}_3^2q_V}{\mathcal{C}_2q_T^2H^2} \right) \right] \,,
\end{align}
where in the second line we have neglected $k$-dependent corrections. Absence of ghosts in the vector sector then implies the condition $Q_V>0$. Note that this does not necessarily imply $q_V>0$ as one might have naively inferred from the action in the form \eqref{eq:SV2}.

\subsubsection{Scalar perturbations}
\label{sssec:ScalPert}

The analysis of scalar perturbations in the general model proceeds very analogously to that of the special model. Expanding the full action including matter we find
\begin{align}
	S_S^{(2)} &= \int \d^4 x \, a^3 \bigg[  - \frac{\overline{n} \rho_{M,n}}{2M_{\text{Pl}}^4} \frac{(\p_i v)^2}{a^2} + \left( \frac{\overline{n} \rho_{M,n}}{M_{\text{Pl}}^4} \frac{\p^2 \chi}{a^2} - \delta \dot{\rho}_M - 3 H (1+ c_M^2) \delta \rho_M \right) v \label{eq:SS2} \\
	& \quad  - \frac{c_M^2 M_{\text{Pl}}^4}{2 \overline{n} \rho_{M,n}} (\delta \rho_M)^2 - M_{\text{Pl}}\, \alpha\, \delta \rho_M  - \left( \omega_3 -2 \omega_8 +2 \omega_9 \right) \frac{(\p_i \alpha)^2}{a^2 M_{\text{Pl}}^2} + \omega_4 \frac{\alpha^2}{M_{\text{Pl}}^2}  \nn \\
	& \quad - \bigg( (3 H \omega_1 - 2 \omega_4) \frac{\delta \phi}{\phi} - \left( \omega_3 - 3\omega_8 + \omega_9 \right) \frac{\p^2 (\delta \phi)}{a^2 \phi} - \left( \omega_3 - \omega_8 + \omega_9 \right) \frac{\p^2 \dot{\psi}}{a^2 \phi \Lambda} + \omega_6 \frac{\p^2 \psi}{a^2 \Lambda} \bigg) \frac{\alpha}{M_{\text{Pl}}}  \nn \\
	& \quad  -  ( \omega_3 - 4 \omega_8 ) \frac{(\p_i \delta \phi)^2}{ 4a^2 \phi^2} + \omega_5 \frac{(\delta \phi)^2}{\phi^2} - \frac12 \left( \left( \omega_2 + \omega_6 \phi \right) \psi -  (\omega_3 - 2\omega_8) \dot{\psi} \right) \frac{\p^2 (\delta \phi)}{a^2 \phi^2 \Lambda} - \frac{\omega_3}{4} \frac{(\p_i \dot{\psi})^2}{a^2 \phi^2 \Lambda^2} \nn \\
	&  \quad + \frac{\omega_7}{2} \frac{(\p_i \psi)^2}{a^2 \Lambda^2} + \left( \omega_1 \frac{\alpha}{M_{\text{Pl}}} + \omega_2 \frac{\delta \phi}{\phi} \right) \frac{\p^2 \chi}{a^2 M_{\text{Pl}}^2} \bigg] \,. \nn
\end{align}
The coefficients $\omega_I$ are given in Appendix \ref{ssec:DefCoefsScal}. We have defined them in a way that highlights the differences with the result of GP theory \cite{DeFelice:2016uil}, in which case the coefficients $\omega_8$ and $\omega_9$ vanish. Although these parameters do not introduce any new operators (as it occurred in the vector sector), they do have the effect of ``detuning'' the relative coefficients among some of the terms. In \eqref{eq:SS2} we introduced
\begin{equation}
	c_M^2 \equiv \frac{\overline{n} \rho_{M,nn}}{\rho_{M,n}} \,,
\end{equation}
which we recall is the GR value of the matter fluid speed of sound. As anticipated previously, the actual speed of sound in EPN will turn out to be different.

The counting of degrees of freedom is again most easily performed by examining the equations of motion. Varying the action with respect to $\alpha$, $\chi$, $\delta \phi$, $\p \psi$, $v$ and $\delta \rho_M$, respectively, we derive
\begin{align}
	&(3 H \omega_1 -2\omega_4) \frac{\delta \phi}{\phi} -2\omega_4 \frac{\alpha}{M_{\text{Pl}}} + M_{\text{Pl}}^2 \delta \rho_M + \frac{k^2}{a^2 \Lambda^2} \left[ \mathcal{Y}_1 + \omega_1 \frac{\Lambda^2}{M_{\text{Pl}}^2} \chi - \omega_6 \Lambda \psi \right] = 0 \,, \label{eq:eomalpha} \\
	&\frac{(\rho_M + P_M)}{M_{\text{Pl}}} v + \omega_1 \alpha + M_{\text{Pl}} \omega_2 \frac{\delta \phi}{\phi} = 0 \,, \label{eq:eomchi} \\
	&(3 H \omega_1 - 2 \omega_4 )\frac{\alpha}{M_{\text{Pl}}} - 2 \omega_5 \frac{\delta \phi}{\phi} + \frac{k^2}{a^2 \Lambda^2} \left[ \frac{1}{2} \mathcal{Y}_2 + \omega_2 \frac{\Lambda^2}{M_{\text{Pl}}^2} \chi - \frac{\Lambda}{2} ( \omega_2 + \omega_6 \phi ) \frac{\psi}{\phi} \right] = 0 \,, \label{eq:eomdphi} \\
	&\frac{\dot{\mathcal{Y}}_3}{H} + \left( 1 - \frac{\dot{\phi}}{H \phi} \right) \mathcal{Y}_3 + \frac{\Lambda^2}{H} \left\lbrace \omega_2 \frac{\delta \phi}{\phi} + 2 \omega_7 \frac{\phi \psi}{\Lambda} + \omega_6 \left( 2 \frac{\alpha \phi}{M_{\text{Pl}}} + \delta \phi \right) \right\rbrace = 0 \,, \label{eq:eomppsi} \\
	&\dot{\delta \rho}_M + 3 H (1 + c_M^2) \delta \rho_M + \frac{k^2}{a^2} \frac{(\rho_M + P_M)}{M_{\text{Pl}}^4} ( v+ \chi) = 0 \,, \label{eq:eomv} \\
	&\alpha M_{\text{Pl}} + c_M^2 \left( 3 H v + \frac{M_{\text{Pl}}^4}{(\rho_M + P_M)} \delta \rho_M \right) - \dot{v} = 0 \,, \label{eq:eomdrhoM}
\end{align}
where
\begin{align}
	\mathcal{Y}_1 &\equiv \frac{\Lambda^2}{\phi} \left[  \left(\omega_3 - 3\omega_8 + \omega_9 \right) \delta \phi + 2 \left(\omega_3 - 2\omega_8 + 2 \omega_9 \right) \frac{\alpha \phi}{M_{\text{Pl}}} + \left(\omega_3 - \omega_8 + \omega_9 \right) \frac{\dot{\psi}}{\Lambda} \right] \,, \label{eq:Y1} \\
	\mathcal{Y}_2 &\equiv \frac{\Lambda^2}{\phi} \left[  \left(\omega_3 - 4\omega_8 \right) \delta \phi + 2 \left(\omega_3 - 3\omega_8 + \omega_9 \right) \frac{\alpha \phi}{M_{\text{Pl}}} + \left(\omega_3 - 2 \omega_8 \right) \frac{\dot{\psi}}{\Lambda} \right] \,, \label{eq:Y2} \\
	\mathcal{Y}_3 &\equiv \frac{\Lambda^2}{\phi} \left[  \left(\omega_3 - 2\omega_8 \right) \delta \phi + 2 \left(\omega_3 - \omega_8 + \omega_9 \right) \frac{\alpha \phi}{M_{\text{Pl}}} + \omega_3 \frac{\dot{\psi}}{\Lambda} \right] \,. \label{eq:Y3}
\end{align}
Note that $\mathcal{Y}_1=\mathcal{Y}_2=\mathcal{Y}_3$ when $\omega_8=\omega_9=0$. The equations for the variables $\alpha$, $\delta \phi$, $\chi$ and $v$ can be solved algebraically in terms of $\psi$ and $\delta \rho_M$. These expressions can be plugged back into \eqref{eq:eomppsi} and \eqref{eq:eomdrhoM} leading to a system of two second-order differential equations for $\psi$ and $\delta \rho_M$. This concludes the proof that the covariant EPN theory is completely free from unwanted degrees of freedom at the level of linear perturbations about the FLRW background.

To determine the dispersion relations and stability conditions we proceed as in Section \ref{sssec:SpecExScal}. After integrating out the non-dynamical modes, and focusing from the outset on the long wavelength approximation, we may recast the resulting action in the same form as in Eq.\ \eqref {eq:SpecExSMmatrix} for the propagating fields $\psi$ and $\delta \rho_M$. Recall that $\bm{K}$, $\bm{M}$, $\bm{G}$ and $\bm{B}$ are all independent of $k$ in this approximation. Moreover, we find that the kinetic matrix $\bm{K}$ is still diagonal. The no-ghost conditions are therefore immediately inferred from its entries, which we denote by $Q_{S,\psi}$ and $Q_{S,M}$. We find
\begin{align}
	Q_{S,\psi}&=\frac{1}{2 \Lambda^2 \phi^2 \left[ ( \omega_1 -2 \omega_2)^2 \omega_3 - 4 ( \omega_1 - 2\omega_2) ( ( \omega_1 - \omega_2 ) \omega_8 + \omega_2 \omega_9 ) - 2 ( \rho_M + P_M) (\omega_8 + \omega_9 )^2 \right]^2} \nn \\
&\quad\times \bigg\{ - (\omega_1 - \omega_2 ) \big(3H \omega_1^2 - 2(\omega_1-\omega_2) \omega_4\big) \big( \omega_1 (\omega_3 - 2 \omega_8) - 2 \omega_2 (\omega_3 - \omega_8 + \omega_9) \big)^2 \nn \\
&\quad + 4 (\rho_M + P_M) \big(3H \omega_1^2 - 2(\omega_1-\omega_2) \omega_4\big) \big( \omega_1 (\omega_3 - 2 \omega_8) - 2 \omega_2 (\omega_3 - \omega_8 + \omega_9) \big)^2 (\omega_8 + \omega_9) \omega_9 \nn \\
&\quad + 4 (\rho_M + P_M)^2 (\omega_8 + \omega_9)^2 \Big[ 2 \omega_4 \omega_9^2 + 3H (\omega_8 - \omega_9) \big( \omega_1 (\omega_8 + \omega_9) - \omega_2 (\omega_8 - \omega_9) \big) \Big] \bigg\} \,,
\end{align}
\begin{equation}
	Q_{S,M}=\frac{a^2}{2} \frac{M_{\text{Pl}}^4}{(\rho_M + P_M)}\,\frac{1}{1-\Delta} \,,
\end{equation}
where
\begin{equation}
	\Delta \equiv 2 (\rho_M + P_M) \frac{(\omega_8 + \omega_9)^2}{(\omega_1 - 2 \omega_2) \left[ \omega_1 (\omega_3 - 4 \omega_8) - 2 \omega_2 (\omega_3 - 2 \omega_8 + 2 \omega_9) \right]} \,.
\end{equation}
We observe that the ``new'' coefficients $\omega_8$ and $\omega_9$ have the interesting effect of inducing a modification of the kinetic term of the Proca scalar mode $\psi$ that depends on the matter density and pressure. These coefficients similarly affect the matter fluid's kinetic term through the parameter $\Delta$. In particular, we see that $Q_{S,M}$ now depends on the EPN Lagrangian parameters, whereas in GP the result would coincide with that of pure GR.

From the long-wavelength expansion of the dispersion relations we obtain the sound speeds $c_{S,\psi}^2$ and $c_{S,M}^2$ respectively for the Proca scalar and the fluid. The fluid speed of sound can be written as
\beq \label{eq:gen model phonon speed}
c_{S,M}^2 = (1 - \Delta) c_M^2 \,,
\eeq
showing that the parameter $\Delta$ has the interesting effect of modifying the GR (and also GP) value of the sound speed. On the other hand, the precise expression for $c_{S,\psi}^2$ is  not particularly illuminating, so we choose to omit it.
However, one can get insight on the difference between the GP and the EPN predictions by going to a minimal example where we set some of the coefficients to $0$ for simplicity's sake. A particularly simple example that is consistent with the GP constraints \cite{DeFelice:2016yws,DeFelice:2016uil} is reached when taking $\omega_2 = \omega_4 = \omega_6 = 0$. As a result, $\omega_1$ and $\omega_4$ are written solely in terms of $q_T$, and hence the problem is fully described by the variables $\{q_T, q_V, \omega_7,\omega_8,\omega_9\}$. Furthermore, we will redefine the variables $\omega_8$ and $\omega_9$ into the dimensionless $W_8 \equiv \omega_8/(q_V \phi^2)$ and $W_9 \equiv \omega_9/(q_V \phi^2)$. With these definitions, we can write
\begin{equation}
\label{eq:cpsi}
	c_{S,\psi}^2 = c_{S,\psi}^{\text{(GP)}2} \left[ 1 +2 W_8 + \frac{q_V}{4q_T^2} \frac{\phi^2}{M_{\text{Pl}}^2} \frac{\rho_M + P_M}{M_{\text{Pl}}^2 H^2} ( W_8 + W_9 )^2 \right]^2  \Upsilon^{-1} \,,
\end{equation}
where
\begin{align}
 c_{S,\psi}^{\text{(GP)}2}  =& - \frac{\omega_7 \phi^2}{6 M_{\text{Pl}}^2 H^2 q_T} \,, \\
	\Upsilon =& \left[ 1 + W_8 + \frac{q_V}{4q_T^2} \frac{\phi^2}{M_{\text{Pl}}^2} \frac{\rho_M + P_M}{M_{\text{Pl}}^2 H^2} ( W_8 + W_9 )^2 \right] \\
	& \times \left[ 1 + W_8 - \frac{q_V}{4q_T^2} \frac{\phi^2}{M_{\text{Pl}}^2} \frac{\rho_M + P_M}{M_{\text{Pl}}^2 H^2} ( W_8^2 - W_9^2 ) \right] \,. \nonumber
\end{align}
One can see that the positivity of $c_{S,\psi}^{\text{(GP)},\, 2} $ necessarily implies $\omega_7 < 0$, whereas this condition is relaxed to be $\omega_7 / \Upsilon < 0$ in the EPN case.

One can now turn to the masses and derive their expressions in all generality, however once again their expressions are not particularly illuminating.  However under the same limiting choice of coefficients as we did previously, one can check explicitly that the scale of the mass of the fluid is set by the Hubble parameter $H$. In principle, we would require the fluid's mass to be positive to avoid tachyonic instabilities but as we have already discussed in the cosmological context, a negative square mass of order $H^2$ is not worrisome. As for the mass of the $\psi$-mode (or vector helicity-0 mode), it happens to vanish for that particular choice of parameters, however relaxing this choice (for instance choosing a non-zero $\omega_6$), one can check that the mass of this mode is also of order $H$, and there is therefore no risk of a faster than $H^2$ tachyonic instability in the scalar sector.

The stability of the matter fluid is easy to analyze. The condition $c_{S,M}^2>0$ is equivalent to $\Delta<1$, which in turn implies the null energy condition, $\rho_M + P_M>0$, in order to have $Q_{S,M}>0$. For the Proca mode $\psi$ to be stable we similarly require $Q_{S,\psi}>0$ and $c_{S,\psi}^2>0$. While these conditions are difficult to dissect given the long expressions, it is worth remembering that they include the results of GP theory as a particular case, in which context it has been shown that stability can be achieved for a wide range of parameters \cite{DeFelice:2016uil}.


\section{Discussion}
\label{sec:discussion}

We started by showing that the theory describes the correct number of degrees of freedom at the level of cosmological backgrounds, defined by Eqs.~\eqref{eq:lineFLRW2} and \eqref{eq:vector bkgd2}, as well as at the level of linear perturbations about these solutions. In addition to establishing this result, in Section \ref{sec:fulltheory} we also derived the dispersion relations for the propagating variables in the presence of perfect fluid matter. Interestingly, EPN has some qualitative differences relative to GP in the dynamics of perturbations. Two particular results to highlight are that the Proca vector mode exhibits a non-linear dispersion relation (cf.\ Eq.~\eqref{eq:gen model vec disp rel}) and that the sound speed of the longitudinal matter perturbation (the phonon) is modified in the EPN setup relative to its GR value (cf.\ Eq.~\eqref{eq:gen model phonon speed}). We also found that the kinetic coefficient of the phonon differs in EPN from its GR and GP values, an effect which may in principle percolate to higher-point interactions and hence be potentially observable. While we did not explore explicit solutions in this general model, we remark again that EPN contains GP as a particular case, in which setup consistent cosmological solutions do exist. It would be interesting to perform a dedicated study of solutions and comparison with data within the complete theory.

In addition to investigating the possibility of covariantizing the full EPN theory, we have also considered the option that a subclass of the theory may admit a simpler covariantization, even if only a partial one in the sense we have described. Our so-called special model of Section \ref{sec:SpecEx} shows that this is the case, providing a particularly neat setup with few unspecified functions and which has the virtue that the Proca field interacts with gravity only through minimal coupling terms. To our knowledge, this is the first instance of a generalized Einstein-Proca theory (i.e.\ models with non-trivial derivative self-interactions beyond those given by contractions of the Maxwell field strength and/or the undifferentiated field) with this property. As with the general model, the caveat is that the covariantization scheme is only a partial one, but it is again sufficient for cosmological applications as long as one is interested in linear perturbations about homogeneous and isotropic backgrounds. Our results of Section \ref{sec:SpecEx} show that the special model indeed describes the expected dynamical degrees of freedom. Moreover, we have shown that explicit solutions exist such that all the dynamical variables are ghost-free, gradient-stable and subluminal. We believe that these results motivate further scrutiny of the setup.

These closing remarks conclude Part \ref{part:PN}. To briefly summarize, we have exhibited a new theory of massive vector fields, (Extended) Proca-Nuevo, and have explicitly shown that this theory is ghost-free and inequivalent to GP. Furthermore, we proposed some covariantization schemes in view of working with (E)PN on an FLRW background for cosmological purposes. There, we have shown that the theory provides us with a consistent and stable model of dark energy. We would like now to move to Part \ref{part:CausalityBounds} where we will introduce causality bounds and will apply them to scalar and vector fields as yet another tool to study the consistency of effective field theories.

\part{Causality bounds in scalar and vector theories.}
\label{part:CausalityBounds}

The second part of this thesis will focus on alternative methods to the ones explored earlier to constrain EFT coefficients. We have seen in Part \ref{part:PN} that requirements such as evading the would-be ghosts and imposing stability of the cosmological perturbations are tools that can be used to maintain the consistency of an EFT. We will now turn to a different but complementary approach consisting in imposing infrared (IR) causality, i.e. at low energy, where we will clarify our definition of causality. This method, which we will call ``causality bounds'' is similar in essence to the well-known positivity bounds that were briefly introduced in Section \ref{sec:IntroPosBounds}. The main difference lies in the fact that the latter also makes use of ultraviolet (UV) causality, i.e. above the regime of validity of the EFT. Furthermore, positivity bounds rely on the very definition of the S-matrix whereas our definition of causality is based on the sole classical time delay a wave acquires compared to a freely-propagating wave. 

We will start by formally introducing our definition of causality and its relation to the notion of time delay in Chapter \ref{chap:causalScalar}, which is based on \cite{CarrilloGonzalez:2022fwg}. This will lay the ground for the method we use to derive these so-called causality bounds for a low-energy shift-symmetric scalar theory, which we will first use to reproduce some very leading-order and known results on the sign of the lowest-order EFT coefficient. However, we will not stop there and will then be able to push the analysis to higher-order EFT operators and obtain strong constraints in the space of parameters of the EFT. Interestingly, and somewhat surprisingly, we will be able to prove that this method can be competitive with positivity bounds and give compact regions of parameters allowed by IR causality. The bounds are not as strong as the ones obtained by positivity but are consistent, giving a proof of principle of the power of this method.

In Chapter \ref{chap:causalVector}, we extend the range of applicability of the method to a theory of a massless photon, as seen in \cite{CarrilloGonzalez:2023cbf}. We show that the two propagating modes of the massless vector field decouple from one another at the level of the linear perturbations and each of them can be treated separately in a similar fashion as the scalar case. Once again, we derive some tight and compact regions of causality in the parameter space and we show that any known (partial) UV completion lies within these bounds, confirming their consistency. More, positivity bounds for the same theory are also derived in \cite{CarrilloGonzalez:2023cbf} and compared to the causality bounds with the result that the latter are much more constraining in some cases. This is a powerful result in the vector case as our bounds were less constraining than the ones obtained by positivity for scalars.

Part \ref{part:CausalityBounds} is used to demonstrating that EFT coefficients can be theoretically constrained by the sole requirement of causality in the IR. This method does \textit{not} rely on the definition of an S-matrix and as such, does not suffer from its pathologies. This means that causality bounds can be defined on any arbitrarily curved background, which in itself could be sufficient to consider them as an interesting tool. However, they don't only have the merit to exist but also to be competitive with positivity bounds in some cases, and even better in others. We are not claiming that our method will supplant positivity bounds but that both can be used in parallel. It is interesting to see that two ways of imposing causality by scanning physics in different energy regimes can lead to consistent results. By combining them, one can then take the union of the allowed causal regions and get even tighter results. Hence, rather than proving the power of one or the other method, we show that both are consistent, complementary, and provide an excellent theoretical constraining tool.

\chapter{Causal scalar effective field theory}
\label{chap:causalScalar}

Physical principles such as unitarity, causality, and locality can constrain the space of consistent effective field theories (EFTs) by imposing two-sided bounds on the allowed values of Wilson coefficients. In this Chapter, we consider the bounds that arise from the requirement of low-energy causality alone, without appealing to any assumptions about UV physics. We focus on shift-symmetric theories and consider bounds that arise from the propagation around both a homogeneous and a spherically-symmetric scalar field background. We find that low-energy causality, namely the requirement that there are no resolvable time advances within the regime of validity of the EFT, produces two-sided bounds in agreement with compact positivity constraints previously obtained  from $2 \rightarrow 2$ scattering amplitude dispersion relations using full crossing symmetry.

\section{Introduction}

In this Chapter, we will focus on constraints arising purely from causality in the low energy regime.  Our central tool is the scattering time delay well studied in non-relativistic scattering \cite{Eisenbud,Wigner:1955zz,Smith:1960zza,Martin:1976iw,DECARVALHO200283} and gravitational scattering \cite{Shapiro:1964uw,Gao:2000ga} which describes in the semi-classical (WKB) or eikonal approximation the delay of a scattered wave relative to a freely propagating wave. Causality violation is associated with the presence of a resolvable time advance, and this criterion has in recent years been utilized to impose similar bounds on Wilson coefficients \cite{Camanho:2014apa,Camanho:2016opx,Bai:2016hui,Goon:2016une,Hinterbichler:2017qcl,Hinterbichler:2017qyt,AccettulliHuber:2020oou,Bellazzini:2021shn,Serra:2022pzl,Chen:2021bvg,deRham:2021bll} which, importantly, do not require any assumption on the UV behaviour of the theory.
Indeed, since small superluminalities could lead to correlation functions having support outside of the light-cone when present at large distances, the violations of causality can be measured within the low energy EFT that describes the infrared physics. In a generic EFT, the higher-derivative interactions will modify the equation of motion for the propagation of a perturbation around an arbitrary background rendering a sound speed $c_s\neq 1$. Note however that a small superluminal low-energy speed is not necessarily in contradiction with causality since the would-be observation detecting violations of causality could turn out to be unmeasurable within the regime of validity of the EFT \cite{Hollowood:2015elj,deRham:2019ctd,deRham:2020zyh}. 

For a local field theory in Minkowski spacetime, causality tells us that the retarded Green's function evaluated in an arbitrary quantum state does not have support outside of the forward Minkowski light-cone. For a generic EFT, locally the propagation of information is encoded in effective metric arising in the hyperbolic equations of motion for small fluctuations around a given background which at leading order is determined by the sound speed $c_s$ and reads
\begin{equation}
	\mathrm{d} s_{\text{eff.}}^{2}=-c_{s}^{2}(x^\mu,\omega) \mathrm{d}t^{2}+\mathrm{d} \vec{x}^{2} \ . \label{eq:effg}
\end{equation}
Generically this speed is dependent on the momentum scale/frequency of propagating fluctuations $\omega$. Causality does not directly impose constraints on the phase velocity, but it requires that its high frequency (high $\omega$) limit, that is, the front velocity is luminal $c_{s}^{2}(x^\mu,\infty)=1$. This determines the support of the retarded propagator and implies that information propagates (sub)luminally. Furthermore, it can be shown that causality implies analyticity of the scattering amplitude and refractive index in the upper half complex $\omega$-plane \cite{Nussenzveig:1972tcd,HamiltonBook,PhysRev.104.1760}. 

Here, we will only focus on the causal properties of the EFT as encoded on light-cones defined by the effective metric in Eq.~\eqref{eq:effg} in the low frequency regime where the EFT is under control. In the EFT, the true front velocity is unknown, as is whether there is a Lorentz invariant UV completion. Furthermore demanding locally the strict bounds $c_{s}^{2}(x^\mu,\omega) \le 1$ is too strong since the associated apparent superluminality may be unresolvable within the EFT (furthermore in the gravitational context the local speed is sensitive to field redefinitions, although this last subtlety will not be relevant here\footnote{On curved backgrounds, the notion of asymptotic causality \cite{Gao:2000ga,Camanho:2014apa} (requiring the absence of superluminalities as compared to the asymptotic flat metric which imposes bounds on the net scattering time delay) is a physical requirement, but it does not always capture the full implications of causality. In fact, it leads to weaker bounds than the notion of infrared causality \cite{Hollowood:2015elj,deRham:2019ctd,deRham:2020zyh,deRham:2021bll,Chen:2021bvg}, (requiring the absence of superluminalities as compared to the local metric which imposes bounds on the net scattering time delay minus the Shapiro time delay).}). The presence of local low energy superluminality does not in itself imply the possibility of creating closed time-like curves. For that these superluminalities ought to be maintained for sufficiently large regions of spacetime. A cleaner diagnostic is the scattering time delay which is defined from the S-matrix and is hence independent of field redefinitions.
The scattering time delay for a given incident state containing a particle of energy $\omega$ may be defined in terms of the $S$-matrix by
\be
\Delta T = -i \left\langle {\rm in} \right| \hat S^{\dagger} \frac{\p }{\p \omega} \hat S \left|{\rm  in} \right\rangle \, .
\ee
The scattering phase shifts may be defined as the eigenvalues of the $S$-matrix, $\hat S|{\rm in}\rangle=e^{2i\delta}|{\rm in}\rangle$, so that in an incident eigenstate the time delay is simply
\be
\Delta T =2 \frac{\p  \delta }{\p \omega}\, .
\ee
For example, for one-particle scattering in a spherically symmetric background, the $S$-matrix diagonalizes in multipoles $\ell$ and we may define the associated multipole time delays
\be
\Delta T_{\ell} =2 \frac{\p  \delta_{\ell} }{\p \omega}\Big|_\ell \, .
\ee
In the large-$\ell$ limit, we may consider scattering at fixed impact parameter $b=(\ell+1/2)\omega^{-1}$, giving the time delay traditionally calculated in the Eikonal approximation \cite{Wallace:1973iu,Wallace:1973ni}
\be
\lim_{\ell \rightarrow \infty} \delta_{\ell=b \omega-1/2}(\omega) = \delta_{\rm Eikonal}(\omega,b) \, ,
\ee
for which the time delay is (see for example \cite{Camanho:2014apa})
\be
\Delta T_{b} =2 \frac{\p  \delta_{\ell} }{\p \omega} \Big|_b \, .
\ee

The signature of true causality violation would be the manifest existence of closed-time-like-curves within the regime of validity of the EFT, however it is understood that such phenomena are akin to experiencing a {\it resolvable}\footnote{Strict positivity of the scattering time delay is sometimes incorrectly imposed. This is not required since the time delay is only a meaningful indication of causality in the semi-classical region (WKB or eikonal).} scattering time advance, (within the regime of validity of the EFT). The resolvability requirement comes from the uncertainty principle which is reflected in the fact that a time advance no bigger than the uncertainty $\Delta t \sim\omega^{-1}$ is clearly not in conflict with causality.
Indeed in general, as is well understood, scattering time advances can be mildly negative without contradicting causality, but only in a bounded way. For example for s-wave (monopole) scattering in a spherically symmetric potential which vanishes for $r>a$, causality imposes the bound on the scattering time delay of the form \cite{Eisenbud,Wigner:1955zz,Smith:1960zza,Martin:1976iw,DECARVALHO200283}
\be
\Delta T_{\ell =0} \ge - \frac{2a }{v} + \frac{1}{k v}\sin(2 k a+\delta_0)  \ge - \frac{2a }{v} - \frac{1}{k v}\, ,
\ee
with $v$ the group velocity and $k$ the momentum with $\omega \sim \mathcal{O}(k v)$. The first term gives the allowed time advance associated with the spherical waves scattering of the boundary $r=a$, and the second term gives an allowed time advance due to the wave nature of propagation, i.e. the uncertainty principle. For the intermediate scale frequencies and smooth backgrounds considered in what follows the first term will be absent (see Appendix \ref{timeadvance} for a discussion) but we must still allow for the uncertainty principle. In other words, we will consider frequencies larger than the scale of variation of the background (within the WKB semi-classical region) and sufficiently high such that we do not encounter any potential barriers, but within the regime of validity of the EFT. All these conditions will be carefully monitored throughout the analysis performed below. Note that lower frequencies do not probe the support of the retarded Green's function and hence are not probing causality. Working in the regime of validity of the WKB approximation, our de facto relativistic causality requirement is that
\be
\label{defacto}
\Delta T  \gtrsim -\frac{1}{\omega} \, .
\ee
applied in the relativistic region where the background is sufficiently smooth and no potential barrier is encountered on scales set by the wavelength $\omega^{-1}$ such that the hard sphere type time advances $- 2a/v$ are absent.

The goal of this Chapter is then to determine constraints we obtain on a given EFT by imposing \eqref{defacto} around different backgrounds. Since our primary concern will be non-gravitational scalar field theories, we can choose to probe the EFT by adding an external source. This device allows us to consider backgrounds which are not solutions of the unsourced background equations of motion. By choosing different sources, we can adjust the background solution to probe different possible scattering phases, and by extremizing over the choices of backgrounds we will be able to obtain competitive constraints from the scattering time delay.

The rest of the Chapter is structured as follows. In Section \ref{sec:speed}, we introduce the shift-symmetric low energy scalar EFT we will be considering and discuss the positivity constraints that arise from consideration of their scattering amplitudes. We also provide generic arguments for the expected time delay within a WKB approach on generic backgrounds. For concreteness, we then focus  on specific profiles for the rest of the Chapter. In Section \ref{sec:hom}, we consider  the simple case of a homogeneous background and argue for the need of less symmetric configurations to make further contact with positivity bounds. We then proceed to consider the scattering of perturbations around a spherically-symmetric background in Section \ref{sec:Spherical}. We examine two limits: one where the waves have no angular dependence and the other where they have large angular momentum. For each of these cases, we spell out carefully the conditions for the validity of the EFT and the WKB approximation. After computing the time delay and requiring that we cannot obtain a resolvable violation of causality we obtain bounds on the Wilson coefficients of the EFT. The case of no angular momentum gives rise to a lower bound while the large angular momentum case draws an upper bound that approaches the non-linear positivity bounds obtained in \cite{Tolley:2020gtv,Caron-Huot:2020cmc}. Lastly, we discuss our results and conclude in Section \ref{sec:Concl}. In the Appendices, we show details of our calculations at higher orders in the EFT and for large angular momentum. We also explain our setup for obtaining bounds on the Wilson coefficients.

\section{Low energy effective field theory and propagation speed} \label{sec:speed}

In this Chapter, we consider the requirements for a scalar effective field theory to be causal. For pedagogical simplicity we focus on theories invariant under a shift symmetry $\phi\rightarrow\phi+c$. Since we are interested in comparing the constraints arising from $2 \rightarrow 2$ tree-level scattering, we will consider only operators up to quartic order in the field $\phi$, and we will ensure to work in a regime where operators that are higher order in the field remain irrelevant to our causality considerations.
In the following, we work with a minimal set of such independent operators up to dimension-12, so that our Lagrangian is given by \cite{Solomon:2017nlh}
\begin{align}
	\L = - \frac12 (\p \phi)^2 - \frac12 m^2 \phi^2+ \frac{g_8}{\Lambda^4} (\p \phi)^4
	  + \frac{g_{10}}{\Lambda^6} (\p \phi)^2 \Big[ (\phi_{, \mu \nu})^2  -  (\Box \phi)^2 \Big] + \frac{g_{12}}{\Lambda^8} (( \phi_{, \mu \nu} )^2 )^2 - g_{\text{matter}} \phi J\, ,
	\label{eq:L}
\end{align}
where $(\phi_{,\mu \nu})^2= \p_{\mu} \p_{\nu} \phi \p^{\mu} \p^{\nu} \phi$, $(\p \phi)^2=\p_\mu\phi\p^\mu\phi$, $g_{\text{matter}}$ is the coupling strength to external matter and $J$ is an arbitrary external source. Note that for convenience we choose to write down the dimension-10 operator as the quartic Galileon\footnote{The time delay remains manifestly invariant under field redefinitions as long as we can neglect boundary terms. This can be seen for instance explicitly in Section \ref{sec:monopole} for the zero angular momentum case up to the EFT order that we consider here.} \cite{Nicolis:2008in}. The scale $\Lambda$ has been introduced as the standard cutoff of this low energy EFT. Note that even though some EFTs may be reorganized so as to remain valid beyond $\Lambda$ (see for instance \cite{deRham:2014wfa} for a discussion), here we take the more conservative approach and consider the low energy EFT to break down at $\Lambda$. Except when we consider the case $g_8=0$, it proves convenient to redefine $\Lambda$ so that $g_8=1$.

\paragraph{Positivity Bounds:}
The aim of what follows is to establish to which extent positivity bounds constraining $\{ g_8 , g_{10}, g_{12} \}$ can be reproduced using low energy infrared causality arguments (i.e. the statement of causality as manifested directly at the level of the low energy EFT without any prior knowledge on the embedding of this EFT within a unitary high energy completion).
Technically, the derivation of the positivity bounds requires the presence of a mass gap and to this purpose, in principle, we can always introduce a shift-symmetry-breaking mass term in Eq.~\eqref{eq:L}. The mass term can indeed be treated as an irrelevant deformation of the shift-invariant Lagrangian which, at the quantum level, does not induce any further symmetry-breaking operators \cite{Burrage:2010cu,deRham:2017imi}.
In the following, we will be working in the limit  $m\ll \omega$ where the mass term can be neglected (hence effectively restoring shift symmetry). The positivity bounds from \cite{Tolley:2020gtv,Caron-Huot:2020cmc} can be translated into bounds on the Wilson coefficients appearing in the Lagrangian of Eq.~\eqref{eq:L} by using Table \ref{tab:dictionary} and read
\begin{equation}
	g_8 > 0, \qquad g_{12} > 0, \qquad g_{10} < 2 g_8, \qquad g_{12} < 4 g_8, \qquad - \frac{16 }{3} \sqrt{g_8 g_{12}} < g_{10} < \sqrt{g_8 g_{12}} \,. \label{eq:pos}
\end{equation}
From the above, only the left hand side of the last bound is derived by using full crossing symmetry whereas the other bounds follow from standard fixed $t$ dispersion relations. Note that every time we will be referring to positivity bounds in this Chapter, we will mean Eq.~\eqref{eq:pos} above. Note as well that these bounds were the work of \cite{Tolley:2020gtv,Caron-Huot:2020cmc} and not mine, and are only used as a reference to which we will compare our results.

\paragraph{Causality:}
Violations of causality can occur when superluminal speeds can be consistently maintained within a region of spacetime so as to lead to a physical support of the retarded propagator outside the standard Minkowski light-cone. In this section, we will compute the low frequency propagation speed of a perturbation $\psi=\phi-\bar \phi$ living on an arbitrary background $\bar{\phi}$  created by an external source $J$. For this, we work within the WKB approximation such that the background's scale of variation, $r_0$, is much larger than the scale on which the perturbation varies ($\omega^{-1}$).  In Section~\ref{sec:hom} we will perform a precise analysis of the possible violations of causality arising in a homogenous background and in a static and spherically-symmetric background in Section~\ref{sec:Spherical}, but for now it instructive to consider perturbations given by a plane wave $\partial_\mu\psi=i k_\mu \psi$.

The equation of motion for the scalar field $\phi$ is given by
\begin{align}
	\Box \phi =&  \frac{4g_8}{\Lambda^4} \left(\phi_{,\mu} (\p \phi)^2 \right)^{,\mu}
	-  \frac{2g_{10}}{\Lambda^6} \left[(\Box \phi)^3- 3 \Box \phi (\phi_{,\mu \nu})^2 + 2 (\phi_{,\mu \nu})^3 \right]
	 -  \frac{4g_{12}}{\Lambda^8} \left( \phi_{, \alpha \beta} (\phi_{,\mu \nu})^2 \right)^{, \alpha \beta} \nn \\
&- g_{\text{matter}} J  \,.
\label{eq:lag}
\end{align}
In the WKB approximation, we assume that perturbations can be characterized by plane waves with wave-vector $k_\mu=(\omega,\bf{k})$. In the regime of validity of the EFT, the $g_{8,10,12}$ operators considered in \eqref{eq:L} are treated perturbatively implying $-k_\mu k^\mu=\omega^2-|{\bf{k}}|^2=(c_s^2-1)|{\bf{k}}|^2\ll |{\bf{k}}|^2$. One should note that remaining within the regime of validity of the EFT requires
\begin{equation}
	\frac{\partial {\phi}}{\Lambda^2}\equiv\delta_1 \ll 1 \ ,\qquad  \frac{\partial^{p+1} {\phi}}{\Lambda^{p+2}}\equiv \delta_1 \, \delta_2^p \ll 1 \ , \label{eq:validity}
\end{equation}
where $p\in \mathbb{N}$ and the derivatives can hit the background or the perturbation. The most stringent bounds are obtained when $p\rightarrow \infty$. Since we are interested in contributions up to dimension-12 operators, we will consider the expansion up to order $\delta_1^2\delta_2^2,\delta_1^4$ and assume $\delta_1^2\ll\delta_2, \ \delta_2^2\ll\delta_1$. Thus, at order $\delta_1^2\delta_2^2,\delta_1^4$ and leading order in $\omega r_0$ we have:
\begin{align}
	c_s^2|{\bf{k}}|^2=&|{\bf{k}}|^2-g_8\frac{8}{\Lambda^4}(k^\mu\partial_\mu\bar{\phi})^2+g_8^2\frac{32}{\Lambda^8}(k^\mu\partial_\mu\bar{\phi})^2(\partial\bar{\phi})^2-g_{12}\frac{8}{\Lambda^8}(k^\mu k^\nu\partial_\mu\partial_\nu\bar{\phi})^2 \\
	& +g_{10}\frac{12 k^\mu k^\nu}{\Lambda^6}\left(\partial_{\mu}\partial_\rho\bar{\phi}\partial^\rho\partial_\nu \bar{\phi}-\square\bar{\phi}\partial_\mu\partial_\nu\bar{\phi}\right) \nn \ ,
\end{align}
where we can immediately see that the $g_8$ and $g_{12}$ contributions are sign definite which are directly equivalent to the first two positivity bounds included in \eqref{eq:pos}. This direct equivalence was pointed out for the $g_8$ operator in  \cite{Adams:2006sv}.
In what follows we shall attempt to make contact with the remaining bounds in \eqref{eq:pos} but note that the contributions of $g_8$ and $g_{12}$ to the speed imply that, within this framework  we consider here, it will be impossible to reproduce the bound $g_{12} < 4 g_8$  from pure infrared causality considerations since, in the absence of $g_{10}$, positivity of both $g_8$ and $g_{12}$ is sufficient to prevent causality violation in this limit.

To establish the basic setup, it will be useful to start by looking at the simple example of a homogeneous background first (even though no further bounds will be derived), before proceeding to a more instructive  spherically-symmetric situation which will allow us to make further contact with the remaining bounds of \eqref{eq:pos}.

\paragraph{Time Delay:}

To understand whether the perturbations are causal around an arbitrary background, we need to consider a hierarchy between the scales of variation of the background and the perturbations, namely, $\lambda_{\text{background}}\gg \lambda_{\text{perturbation}}$. Hence, we can use the WKB approximation to obtain the phase shift experienced by the scattered perturbation from which the time delay is easily computed. Considering the wave nature of the scattering and the uncertainty principle, we define a resolvable time advance as one that satisfies
\begin{equation}
	\omega	\Delta T  < - 1 \ , \label{eq:NoCausality}
\end{equation}
where $\omega$ is the asymptotic energy of the scattered state. This states that a resolvable time advance needs to be larger than the resolution scale of {\it{geometric optics}}\footnote{The geometric optics or eikonal limit assumes that the scattering problem can be described in terms of particle trajectories with large impact parameters and that the energies of the asymptotic states are large.}, \cite{Hollowood:2015elj,deRham:2020zyh}. If Eq.~\eqref{eq:NoCausality} is satisfied within the WKB approximation and the regime of validity of the EFT, then we have an observable violation of causality. At leading order in the EFT, the requirement in Eq.~\eqref{eq:NoCausality} can be equivalently written in terms of the scattering phase shift as $\delta^{\text{EFT}}<-1$. However, this does not hold when including higher EFT corrections that modify the speed of sound with $\omega$-dependent contributions. In such cases, one should simply consider the bound in Eq.~\eqref{eq:NoCausality}. A commonly taken approach to understanding causality bounds consists of working in the eikonal (geometric optics) limit. This amounts to considering the scattering of waves with large angular momentum $\ell$ so that the dynamics can be described in terms of the scattering of particle trajectories with fixed impact parameter $b=(\ell+1/2) \omega^{-1}$. We shall consider both this region and the small $\ell$ region which can also be described semi-classically. We will see that exploration of both limits is complementary when imposing bounds on Wilson coefficients from the requirement of causality in the EFT and will give rise to the two-sided bounds known from other considerations.

As already noted, demanding locally that $c_s(\omega) \le 1$ is too strict a requirement. The leading-order contributions to violations of causality, as encoded in the support of the retarded Green's function, are determined by the light-cones defined by the effective metric in the EFT equations of motion \cite{Caldwell:1993xw,deRham:2019ctd} which is given by Eq.~\eqref{eq:effg}. Thus, acausality can be measured by integrating the effects of the sound speed. More precisely, by measuring whether the scattered waves can propagate outside the Minkowski light-cone and if this effect is observable within the regime of validity of the EFT. It is clear that a subluminal speed will always lead to a causal theory. On the other hand, small (as dictated by the validity of the EFT) superluminalities do not violate causality if they do not have support in a large region of spacetime. 

In what follows we perform a careful treatment of causality for homogeneous backgrounds and static and spherically symmetric ones. Note that we always use the same definition of causality, but depending on the symmetries of the background it might be more natural to express the support outside the Minkowski lightcone as given by a timelike or a spatial observable. For example, when we work with a homogeneous background we have spatial momentum conservation and hence the natural observable is the spatial displacement of the lightcone defined by the effective metric. On the other hand, when we work with spherically symmetric backgrounds, we have energy conservation and it is more natural to describe the support outside the Minkowski lightcone with a timelike displacement. Both of these quantities encode the same physics and capture the support of the retarded Green's function; hence both encode the same causality criterion.

\section{Homogeneous background} \label{sec:hom}
In this section, we will derive the dispersion relation for a homogeneous background. To do so, we consider the equation of motion in Eq.~\eqref{eq:lag} and perturb the field around a homogeneous background $\bar{\phi}(t)$ which varies on a time scales of order $H^{-1}$, which we shall consider as being constant to a first approximation. To access the information encoded in the retarded Green's function, we consider a perturbative setup in which we derive a perturbative, second-order in time, hyperbolic, equation of motion for the perturbations. Any higher-order ($>2$) time derivative can be iteratively removed at each  order in the EFT expansion. Schematically, the equations of motion for the perturbation reads
\begin{equation}
	\ddot{\psi} + A \dot{\psi} + B \psi = 0 \,,
\end{equation}
where $A$ and $B$ are functions of the background $\bar{\phi}(t)$ and its derivatives, the wavenumber $k$, the coupling constants $g_I$, and the energy scale $\Lambda$. For convenience, the friction term can be removed by performing a field redefinition of the form $\psi(t) = f(\bar{\phi}(t)) \psi_0(t)$ leading to the perturbation equation
\begin{equation}
	\ddot{\psi}_0 + \left( B - \frac{A^2 + 2 \dot{A}}{4} \right) \psi_0 = 0 \,.
\end{equation}
This allows us to write down an effective dispersion relation for the perturbations as
\begin{equation}
	\omega^2 = m_{\rm eff}^2 + c_s^2( {\bf k}) |{\bf{k}}|^2 \,,
\end{equation}
where $m_{\rm eff}^2$ is the effective mass square and $c_s^2( {\bf k})$ is the ${\bf k}$-dependent square sound speed. Our definition of the sound speed corresponds to a momentum-dependent phase velocity. Note that at leading order in $|{\bf{k}}|$, the notions of phase velocity and group velocity are equivalent when the mass is negligible, which is the case under consideration. Considering only the $g_8, g_{10},$ and $g_{12}$ operators, we find at order $\delta_1^2\delta_2^2$, (where the expansion parameters $\delta_1$ and $\delta_2$ are defined in \eqref{eq:validity}),
\begin{align}
	m_{\rm eff}^2&=-\frac{12}{\Lambda^4}g_8\partial_t \left( \ddot{\bar{\phi}} \dot{\bar{\phi}} \right) \ , \label{eq:HomMeff} \\
c_s^2( {\bf k}) &=1- \frac{8}{\Lambda^4} g_8 \dot{\bar{\phi}}^2 - \frac{8}{\Lambda^8}g_{12}|{\bf{k}}|^2 \ddot{\bar{\phi}}^2 +\frac{96}{\Lambda^8} g_8^2 \dot{\bar{\phi}}^4   \, , \label{eq:HomSpeed}
\end{align}
up to terms that are more suppressed.
Note that the $g_{10}$ contribution to the square sound speed vanishes as expected as the quartic Galileon\footnote{As expected, if one were to choose the parametrization for the $g_{10}$ operator in \eqref{eq:L} where the term $\Box \phi (\p \phi)^2$ is removed by field redefinition, the $g_{10}$ contribution to $c_s^2({\bf k})$ would be a total derivative and would also vanish at the level of the time delay so long as one considers background profiles with vanishing boundary terms, as is done in our analysis.} vanishes on an effectively one-dimensional background. To analyze this term, one needs to explore backgrounds that are effectively at least two-dimensional and in the next section we will consider a spherically-symmetric background\footnote{Cylindrically-symmetric background were also considered and lead to no additional insights, we shall therefore not present them in this work.}.

Note that to stay within the regime of validity of the EFT we require that
\begin{equation}
\frac{H\bar{\Phi}_0}{\Lambda^2} \ll 1 \, , \quad  \frac{H}{\Lambda}\ll 1 \, , \quad \text{and} \quad  \frac{\omega H}{\Lambda^2} \ll 1 \ ,
\end{equation}
where $\bar{\Phi}_0$ is the overall scale of the background field, or one can take $\bar{\Phi}_0={\rm max}(|\bar\phi(t)|)$.
As a consequence, this ensures that $c_s \sim 1$, up to small perturbative corrections. Furthermore, the validity of the WKB regime where the perturbations vary much faster than the background implies that $| {\bf k}|H^{-1}\gg 1$. These requirements imply that the speed \eqref{eq:HomSpeed} is subluminal for $g_8>0, g_{12}=0$ and for $ g_{8}=0, g_{12}>0$. Even though the departure from the speed of light will be small, this effect may pile-up when dealing with large observation times and lead to macroscopic effects. To understand how this could occur, we establish the amount of support $\Delta x$ the field would be able to gain outside the standard Minkowski light-cone
\begin{equation}
\Delta x=H^{-1} \int_{\tau_i}^{\tau_f} (1-c_s(\tau)) \mathrm{d} \tau \ ,
\end{equation}
where we introduced the dimensionless time $\tau=H t$. The light-cone observed by the perturbation is smaller than the Minkowski one by $\Delta x$. Hence, violations of causality arise for waves with three-momentum $\bf{k}$  enjoying $|{\bf{k}}|\Delta x<-1$ for any $\Delta \tau=\tau_f-\tau_i>0$ while remaining within the regime of validity of the EFT.  That is, if the distance that the perturbations can propagate outside of the Minkowski light-cone becomes larger than the wavelength of the perturbation. As is well known, in this setup, there is no risk of causality violation  if $g_8>0$ and $g_{12}>0$.
However if $g_8<0$, one can easily find solutions on which $|{\bf{k}}|\Delta x<-1$.

Consider for instance a time-localized profile of the form $\bar \phi(\tau)=\bar{\Phi}_0 e^{-\tau^2}$. The resulting support outside the  light-cone will then be
\begin{equation}
|{\bf k}|\Delta x=\frac{|{\bf k}|}{H} \int_{-\infty}^{\infty} (1-c_s(\tau)) \mathrm{d} \tau=4\sqrt{\pi}\frac{|{\bf k}|}{H}\left[\frac{H \bar{\Phi}_0 }{\Lambda^2}\right]^2\(g_8+3g_{12}\frac{|{\bf k}|^2H^2}{\Lambda^4}-\frac{9g_8^2}{\sqrt{2}}\frac{H^2 \bar{\Phi}_0^2}{\Lambda^4}\) \ .
\end{equation}
In the regime of validity of the EFT, $H \bar{\Phi}_0\ll \Lambda^2$  and the terms quadratic in $g_8$ are naturally negligible. While the prefactor in square brackets should be small, this can always be compensated by a sufficiently large $|{\bf k}|H^{-1}\gg 1$ as required from the validity of the WKB approximation. For those solutions the term linear in $g_{12}$ is always subdominant as $|{\bf k}| H\sim \omega H \ll \Lambda^2$.  Hence as $g_8<0$, there are solutions within the regime of validity of the EFT for which the time advance is resolvable $|{\bf k}|\Delta x<-1$ signalling a violation of causality. This result complements that derived in \cite{deRham:2020zyh}.

On the other hand for more involved profiles, the term linear in $g_{12}$ can be sufficiently enhanced so that it dominates over the term linear in $g_{8}$ despite the  $|{\bf k}|^2H^2 \Lambda^{-4}$ suppression. As a simple proof of principle, we could consider for instance a profile of the form $\bar \phi(\tau)=\bar{\Phi}_0 \tau^{2} e^{-\tau^2}$ for which the support then becomes
\ba
|{\bf k}|\Delta x=\frac{7\sqrt{\pi}}{2\sqrt{2}}\frac{|{\bf k}|}{H} \left[\frac{H \bar{\Phi}_0 }{\Lambda^2}\right]^2\(g_8+\frac{57}{7}g_{12}\frac{|{\bf k}|^2H^2}{\Lambda^4}-\frac{6129}{1792}\frac{g_8^2}{\sqrt{2}}\frac{H^2 \bar{\Phi}_0^2}{\Lambda^4}\)\,,
\ea
hence taking $|{\bf k}| H/\Lambda^2 \sim 0.2$ ensures validity of the EFT, while the $g_{12}$ dominates over the $g_8$ term and hence a resolvable time advance will be possible within the regime of validity of the EFT for negative $g_{12}\sim -1$ even if $g_8\sim 1$. One could push the analysis to more generic profiles and derive a more systematic resolvable support outside the light-cone whenever $g_{12}$ is negative however at this stage moving on to  spherically symmetric profiles will prove more instructive and positivity of $g_{12}$ from pure causality considerations will be proven in that context (see the summary of the causality constraints depicted in Fig.~\ref{fig:bounds} where it is clear that even in the presence of a generic positive $g_8$, $g_{12}$ still ought to be positive to ensure causality on generic configurations that remain in the regime of validity of the EFT). No additional information can be obtained from this analysis since $g_{10}$ has no effects on the dispersion relation as explained under Eq.~\eqref{eq:HomSpeed} and, as argued previously, the $g_8$ and $g_{12}$ contributions to the speed are sign definite so we cannot bound these coefficients from above.

\section{Spherically-symmetric background} \label{sec:Spherical}
We proceed to explore causality constraints on a static and spherically-symmetric background $\bar{\phi}(r)$ for which the operator $g_{10}$ is relevant. This  allows us to establish to which extent the non-linear positivity bounds in Eq.~\eqref{eq:pos} can be reproduced using causality considerations at low energy without other information from its UV completion. Given the symmetries of the background, we perform an expansion in spherical harmonics\footnote{Due to the azimuthal symmetry, we can neglect the $\varphi$ dependence of the spherical harmonics and work with the Legendre polynomials.} (partial waves) and write our perturbation as $\psi=\sum_{\ell} e^{i\omega t}Y_{\ell}(\theta) \delta \rho_{\ell}(r)$. We obtain an equation of motion for the $\ell$-mode radial perturbation, $\rho_{\ell}$, which schematically is
\begin{equation}
	\delta\rho''_{\ell}(r)+A(\omega^2,r)\delta\rho'_{\ell}(r)+\left(\omega^2 C(\omega^2,r)-\frac{\ell(\ell+1)}{r^2}+B(r,\ell) \right)\delta\rho_{\ell}(r)=0 \, . \label{eq:radialeomFriction}
\end{equation}
In the absence of interactions we have $B(r,\ell)=0$ (up to the mass of the scalar field which we treat as negligible). We perform a field redefinition, $\delta\rho_{\ell}(r)=e^{-\int A(\omega^2,r)/2 \mathrm{d}r}\chi_{\ell}(r)$, to remove the friction term and get
\begin{equation}
	\chi''_{\ell}(r)+\frac{1}{c_s^2(\omega^2,r)}\left(\omega^2-V_{\text{eff}}\right)\chi_{\ell}(r)=0 \ , \quad {\rm with}\quad  V_{\text{eff}}\equiv \frac{\ell(\ell+1)}{r^2}+\tilde{B}(r,\ell) \, . \label{eq:radialeom}
\end{equation}
We can obtain this equation in an exact form, but for our purposes, we will consider an expansion in the parameters $\delta_1$ and $\delta_2$ defined in Eq.~\eqref{eq:validity}. Let us consider a spherically-symmetric background of the form $\bar\phi(r)=\bar{\Phi}_0 f(r)$ and change coordinates to $R=r/r_0$, where $r_0$ is an arbitrary length scale that measures the variation of the background and $\bar{\Phi}_0$ has dimensions of mass,
\begin{equation}
	\bar\phi(r/r_0) = \bar{\Phi}_0 f(R) \,.
	\label{eq:deff}
\end{equation}
With these definitions, both the profile $f$ and the radius $R$ are dimensionless. Note that the definitions of all dimensionless parameters and functions are reported in Table~\ref{tab:dictionaryDimensionless} of Appendix \ref{ap:DefDimLess}. Moreover, we expect $f$ and its derivatives $f^{(n)}$ (where the differentiation is taken with respect to $R$) to be at most of $\mathcal{O}(1)$. The validity of the EFT implies that:
\begin{equation}
	\epsilon_1\equiv\frac{\bar{\Phi}_0}{r_0\Lambda^2}\ll1 \, , \quad  \epsilon_2\equiv\frac{1}{r_0 \Lambda}\ll1 \, , \quad  \text{and} \quad \Omega \epsilon_2\equiv \frac{\omega}{r_0\Lambda^2} \ll 1 \ . \label{eq:eps}
\end{equation}
At the level of the phase shift, each contribution from the $g_i$ terms would scale as follows
\begin{equation}
	g_8 : \mathcal{O}\left( \epsilon_1^2\right), \qquad g_{10} : \mathcal{O} \left(\epsilon_1^2 \epsilon_2^2\right), \qquad g_{12} : \mathcal{O} \left(\epsilon_1^2\epsilon_2^2\Omega^2\right) \,.
\end{equation}
More generally, any term coming from $g_8^{n_1} g_{10}^{n_2} g_{12}^{n_3}$ will be suppressed by at least a power
\begin{equation*}
	\epsilon_1^{2(n_1+n_2+n_3)} \epsilon_2^{2(n_2+n_3)}\Omega^{2n_3} \ .
\end{equation*}
We write our expressions in terms of $\epsilon_1$, $\epsilon_2$, and $\Omega$ and then perform an expansion up to order $\epsilon_1^2\epsilon_2^2$ and  $\epsilon_1^4$, which requires the assumptions $\epsilon_1^2\ll\epsilon_2$ and $\epsilon_2^2\ll\epsilon_1$. In fact, we will consider $\epsilon_1$ and $\epsilon_2$ of the same order, but allow the freedom of the exact value of these scales to be different. We will refer to these contributions as the leading order (LO) or $\mathcal{O}(\epsilon^4)$. Thus, we need to keep track of the contributions coming from the $g_8, \ g_{10}, \ g_8^2, \ g_{12}$ terms. Expanding up to order $\epsilon_1^2\epsilon_2^2$ we find
\begin{equation}
	\chi''_{\ell}(R)+W_{\ell}\chi_{\ell}(R)=0 \, , \quad W_{\ell}=\frac{(\omega r_0)^2}{c_s^2(\omega^2,R)}\left(1-\frac{V_{\text{eff}}(R)}{(\omega r_0)^2}\right) \, ,  \label{eq:WKBR}
\end{equation}
where prime now denotes a derivative with respect to $R$. Note that the expansion is not just in $\epsilon$, but in $g_i\epsilon$. For this perturbative result to be correct, we need to make sure higher-order corrections to the series expansion in Eq.~\eqref{eq:WKBR} above are small. Thus, we require schematically that $g_i \epsilon \ll 1$, which as expected, simply tells us that we should not consider too large values for the Wilson couplings. Wilson coefficients much larger than unity should be rescaled appropriately in the cutoff $\Lambda$ resulting in a lower cutoff scale. We now  solve Eq.~\eqref{eq:radialeom} using the WKB approximation, first analyzing how far in the WKB approximation one needs to include contributions to be consistent with the EFT expansion. Once the consistency of the WKB expansion with the EFT expansion is established, we can then explore the parameter space in which causality violations can arise. We will do so for the cases $\ell=0$ and $\ell\neq 0 $ separately.

\subsection{Regime of validity of the WKB approximation}
We start by considering Eq.~\eqref{eq:WKBR} given in terms of dimensionless variables
\begin{equation}
	\chi_{\ell}''(R) + (\omega r_0)^2 \hat{W}_{\ell}(R) \chi_{\ell}(R) = 0 \,, \qquad \hat{W}_{\ell}(R) = \frac{W_{\ell}(R)}{(\omega r_0)^2} = \frac{1}{c_s^2(\omega^2,R)}\left(1-\frac{V_{\text{eff}}(R)}{(\omega r_0)^2}\right) \,. \label{eq:eomDim}
\end{equation}
Since we assume that the perturbation fluctuates faster than the background, namely,
\begin{equation}
\frac{	\lambda_{\text{perturbation}} }{\lambda_{\text{background}}}=\frac{1}{\omega r_0} =\frac{\epsilon_2}{\Omega}\ll 1 \ , \label{eq:WKBreq}
\end{equation}
we can solve the equation above using the WKB method. In this approach, the solution to the equation of motion up to $n$th-order correction in the WKB formula is given by
\begin{equation}
	\chi_{\ell}^{(n)}(R) \propto \left( e^{i (\omega r_0) \int_0^R \sum_{j \geq 0}^n \delta_{\rm WKB}^{(j)} \mathrm{d}R} - e^{- i (\omega r_0) \int_0^R \sum_{j \geq 0}^n \delta_{\rm WKB}^{(j)} \mathrm{d}R} \right) \,,
\end{equation}
where the boundary conditions were chosen such that $\chi_{\ell}^{(n)}(R=0)=0$ and  $\delta_{\rm WKB}^{(j)}$ is the $j$th-order term in the WKB series expansion whose explicit expressions can be found in \cite{WKBbook} and we list the relevant ones for our analysis below. Noting that $\hat{W}_{\ell}>0$, it is easy to realize that $\delta_{\rm WKB}^{(j)}$ are purely imaginary total derivatives when $j$ is odd, meaning that they do not contribute to the phase but simply to the amplitude. In the end, we have that the phase is proportional to
\begin{equation}
	\sum_{j \geq 0} \delta_{\rm WKB}^{(2j)} = \delta_{\rm WKB}^{(0)} + \delta_{\rm WKB}^{(2)} + \delta_{\rm WKB}^{(4)} + \cdots \,,
\end{equation}
where the first three contributions are
\begin{align}
	\delta_{\rm WKB}^{(0)} &= \sqrt{\hat{W}_{\ell}} \,, \\
	\delta_{\rm WKB}^{(2)} &= - \frac{1}{(\omega r_0)^2} \frac{1}{8\sqrt{\hat{W}_{\ell}}} \left( \frac{\hat{W}''_{\ell}}{\hat{W}_{\ell}} - \frac{5}{4} \left( \frac{\hat{W}'_{\ell}}{\hat{W}_{\ell}} \right)^2 \right) \,, \\
	\delta_{\rm WKB}^{(4)} &= \frac{1}{(\omega r_0)^4} \frac{1}{32 \hat{W}_{\ell}^{3/2}} \left[ \frac{\hat{W}^{(4)}_{\ell}}{\hat{W}_{\ell}} - 7 \frac{\hat{W}'_{\ell} \hat{W}^{(3)}_{\ell}}{\hat{W}^2_{\ell}} - \frac{19}{4} \left( \frac{\hat{W}''_{\ell}}{\hat{W}_{\ell}} \right)^2 + \frac{221}{8} \frac{\hat{W}''_{\ell} \hat{W}'_{\ell}{}^2}{\hat{W}^3_{\ell}} - \frac{1105}{64} \left( \frac{\hat{W}'_{\ell}}{\hat{W}_{\ell}} \right)^4 \right] \,.
	\label{eq:WKBcorrec}
\end{align}
For our purposes, we need to consider terms up to order $\mathcal{O}(\epsilon^4)$. Below, we will see that $\delta_{\rm WKB}^{(2)}$ has contributions of order $\epsilon_1^2/(\omega r_0)^2=\epsilon_1^2\epsilon_2^2/\Omega^2$ that should be taken into account in order to have a consistent expansion at leading-order (LO) for the phase shift, and hence the time delay. In some cases, we would be interested in computing the next EFT contribution to ensure that it is a small effect that does not change our bounds. These next-to-leading (NLO) corrections include terms of the following orders $\mathcal{O}(\epsilon_1^6, \epsilon_1^2 \epsilon_2^4, \epsilon_1^4 \epsilon_2^2)$. In this case, we will need to include $\delta_{\rm WKB}^{(4)}$ corrections to the WKB formula.

We can establish the validity of the WKB approximation by looking at the relative error between the exact solution $\chi_{\ell}$ and the WKB approximation up to $n$th-order corrections $\chi_{\ell}^{(n)}$. Thus we require that
\begin{equation}
\frac{\chi_{\ell}(R)-\chi_{\ell}^{(n)}(R)}{\chi_{\ell}(R)}\sim\frac{1}{(\omega r_0)^n}\int_0^R \delta_{\rm WKB}^{(n+1)} \mathrm{d}R \ll 1 \,,
\end{equation}
as well as
\begin{equation}
\frac{1}{(\omega r_0)^n}\int_0^R \delta_{\rm WKB}^{(n+1)} \mathrm{d}R	\ll \frac{1}{(\omega r_0)^{n-1}}\int_0^R \delta_{\rm WKB}^{(n)} \mathrm{d}R \,,
\end{equation}
in order for the WKB to be a useful approximation given by an asymptotic series in $(\omega r_0)^{-1}$ \cite{WKBbook}. Similarly, we want to ensure that the next order WKB terms are indeed negligible at the order in the perturbative expansion that we are working on. This can be checked by computing the next order in Eq.~\eqref{eq:eomDim} inferring that,
\begin{equation}
{\chi^{(n)}_{\ell}}''(R) + (\omega r_0)^2 \hat{W}_{\ell}(R) \chi^{(n)}_{\ell}(R) = 	\mathcal{E}_{\ell}^{(n+1)}\sim  \frac{\delta_{\rm WKB}^{(n+1)}}{\delta_{\rm WKB}^{(0)}} \ .
\end{equation}
From which we can see that the leftover is of order $ \mathcal{O}((\omega r_0)^{-(n+1)})$ which is small provided $(\omega r_0) \gg 1$, as postulated earlier. In practice, we compute carefully the order of these leftover and make sure that it vanishes at LO.

\subsection{Case 1: Monopole} \label{sec:monopole}
In this section, we analyze the causality bounds on the EFT Wilson coefficients that arise when scattering the monopole mode. To do so, we consider Eq.~\eqref{eq:WKBR} with $\ell=0$. At leading order, the function $\hat{W}_0$ that appears in the equation of motion reads
\begin{align}
	\left. \hat{W}_0(R) \right|_{\rm LO} =& 1+8 g_8 \epsilon _1^2 f'(R)^2+96 g_8^2 \epsilon _1^4 f'(R)^4+8 g_{12} \Omega ^2 \epsilon _1^2 \epsilon_2^2 f''(R)^2 +24 g_{10} \epsilon _1^2 \epsilon _2^2 \frac{f'(R) f''(R)}{R}\nonumber \\
	&+12 \frac{g_8}{\Omega ^2} \epsilon _1^2 \epsilon _2^2 \left(2 \frac{f'(R) f''(R)}{R}+\frac 12 \p_R^2 f'(R)^2\right) \,,
\end{align}
where $f$ and $R$ are respectively the dimensionless spherically-symmetric background and radius defined in \eqref{eq:deff}. As explained earlier we have performed an expansion in the small dimensionless parameters $\epsilon_1$, $\epsilon_2$, $\Omega\epsilon_2$ that measure the validity of the EFT. There is another dimensionless parameter which measures the validity of the WKB expansion and can be written in terms of the previous parameters, namely, $(\omega r_0)^{-1}=\epsilon_2/\Omega$. In order to obtain tight bounds for the Wilsonian coefficients one needs to consider the extreme situation where these small parameters are as large as possible while maintaining the EFT under control and being able to compute the necessary WKB corrections at this order. We will be computing the time delay at LO while ensuring validity of the EFT  by imposing Eq.~\eqref{eq:eps}. These requirements together with the validity of the WKB approximation lead to $\epsilon_2\ll\Omega \ll 1/ \epsilon_2$, but in practice, we require slightly tighter bounds given by $\sqrt{\epsilon_2}<\Omega < 1/\sqrt{ \epsilon_2}$,
together with $\epsilon_1^2<\epsilon_2$ and $\epsilon_2^2<\epsilon_1$ in order to have a well-defined expansion truncated at $\mathcal{O}(\epsilon^4)$. For example, this means that we keep corrections of order $(\epsilon_1/(\omega r_0))^2$ but neglect $(\epsilon_1/(\omega r_0)^2)^2$. The latter type of corrections arise in the effective potential, but not in the speed.

It is instructive to look at the sound speed and effective potential which, at leading order, are given  by
\begin{align}
	&\left. c_s^2(\omega^2,R) \right|_{\rm LO}=1-8 g_8 \epsilon _1^2 f'(R)^2-32 g_8^2 \epsilon _1^4 f'(R)^4 - 8 g_{12} \epsilon _1^2 \epsilon _2^2 \frac{\omega^2}{\Lambda ^2} f''(R)^2 -24 g_{10} \epsilon _1^2 \epsilon _2^2 \frac{f'(R) f''(R)}{R}  \, ,\nn \\
	&\left. V_{\text{eff}}(R) \right|_{\rm LO}= -12 g_8 \epsilon _1^2 \left( 2 \frac{f'(R)f''(R)}{R} + \frac 12 \p_R^2 f'(R)^2  \right)   \, .
\end{align}
One should note here that the effective potential term is suppressed by $(\omega r_0)^{-2}=\epsilon_2^2 \Omega^{-2}$ with respect to the sound speed term. The corrections at NLO to the speed of sound and to the effective potential are listed in Appendix \ref{ap:NLO}. From the sound speed expression, we can understand whether we should expect to be able to reproduce any of the positivity bounds in Eq.~\eqref{eq:pos}. Firstly, as already argued in Section \ref{sec:speed}, the $g_8$ and $g_{12}$ contributions to the speed are clearly sign definite. Next, we analyze the $g_{10}$ contribution. This term appears to be sign indefinite, but under the integral, it is equivalent to a sign definite contribution up to total derivatives that will vanish at the boundaries. Hence, with use of the monopole, we can only expect to be able to bound the $g_{10}$ coefficient from below and we will need to resort to higher multipoles to bound $g_{10}$ from above. 

We can now determine the phase shift experienced by the perturbation travelling in the spherically-symmetric background. For that we first rewrite the solution to the perturbed equation of motion in the following way
\begin{equation}
	\chi^{(n)}_0(R) \propto e^{-i (\omega r_0) \int_0^R \left( \sum_{j \geq 0}^n \delta_{\rm WKB}^{(j)} - 1 \right) \mathrm{d}R} \left( e^{2 i (\omega r_0) \int_0^R \left( \sum_{j \geq 0}^n \delta_{\rm WKB}^{(j)} -1 \right) \mathrm{d}R} e^{i (\omega r_0) R} - e^{-i (\omega r_0) R} \right) \,,
\end{equation}
so that it can be compared to the asymptotic solution $\chi_0(R) \propto \left( e^{2i \delta_0} e^{i (\omega r_0) R} - e^{-i (\omega r_0) R} \right)$
to find that the expression for the phase shift at $\ell=0$ reads
\begin{equation}
	\delta_0(\omega) = \omega r_0\int_0^{\infty} \left( \sum_{j \geq 0} \delta_{\rm WKB}^{(j)}- 1 \right) \mathrm{d}R  \,, \label{eq:phaseshift}
\end{equation}
which is positive for $0 < c_s < 1$ and large enough $\omega r_0$ as seen when using Eqs.~\eqref{eq:eomDim} and \eqref{eq:WKBcorrec}:
\begin{equation}
	\delta_0(\omega)\sim \omega r_0\int_0^{\infty} \left( \frac{1}{c_s}- 1 \right) \mathrm{d}R  \, .
\end{equation}
From Eq.~\eqref{eq:phaseshift}, we see that the dimensionless time delay of a partial wave with zero angular momentum is given by
\begin{equation}
	\omega \Delta T_0(\omega) = 2 \omega \frac{\p\delta_0(\omega)}{\p \omega} = 2\omega \int_0^{\infty}\frac{\p}{\p \omega}\left( (\omega r_0) \left( \sum_{j \geq 0} \delta_{\rm WKB}^{(j)} - 1 \right)\right)  \mathrm{d}R \equiv  \int_0^{\infty}\mathcal{I}_0(\omega,R) \mathrm{d}R \, ,
\end{equation}
where up to $\mathcal{O}(\epsilon^4)$ we have
\begin{align}
	\left. \mathcal{I}_0(\omega^2,R) \right|_{\rm LO}= &8 (\omega r_0) \epsilon_1^2 \left[ g_8 f'(R)^2 + 10 g_8^2 \epsilon_1^2 f'(R)^4 +3 g_{12} \Omega^2 \epsilon_2^2 f''(R)^2 \vphantom{\frac{1}{2}} \right. \nonumber \\
	& \left. - \frac{g_8}{\Omega^2} \epsilon_2^2 \left( 3 \frac{f'(R)f''(R)}{R} + \frac 12 \p_R^2(f'(R)^2) \right) + 3 g_{10} \epsilon_2^2 \frac{f'(R) f''(R)}{R} \right] \, .
\end{align}
As appropriate for a scattering regime which is intrinsically wave-like such as the $\ell=0$ case,  we are computing the time delay at fixed $\ell$. As mentioned earlier, we will consider background profiles giving null boundary terms so that we can neglect any contribution from total derivative terms. Taking this into consideration and performing integration by parts we find that the above equation can be written as
\begin{align}
	\left. \mathcal{I}_0(\omega^2,R) \right|_{\rm LO}=&\ 8 (\omega r_0) \epsilon_1^2 \left[ g_8 f'(R)^2 + 10 g_8^2 \epsilon_1^2 f'(R)^4 + 3 \epsilon_2^2 \left( g_{12} \Omega^2 f''(R)^2 + \frac{1}{2} \left( g_{10} - \frac{g_8}{\Omega^2} \right) \frac{f'(R)^2}{R^2} \right) \right] \nonumber \\
	&+ \text{total derivatives}\, . \label{eq:TDintegrand}
\end{align}
We can now explicitly see that the contribution from each term in the EFT expansion is sign definite when looking at the scattering of $\ell=0$ modes. From these expressions and the constraints from the validity of the EFT and the WKB approximation in Eqs.~\eqref{eq:eps},~\eqref{eq:WKBreq}, we can easily see that the $g_8$ and $g_{12}$ terms can give rise to resolvable time delays. In fact, the time delay is positive for $g_8>0$ when $g_{10}=g_{12}=0$ and for $g_{12}>0$ when $ g_{8}=g_{10}=0$, however we will soon be able to make more general statements. For the $g_{10}$ terms, one can also obtain a resolvable time delay, but this requires tuning of the function $f$ to make the time delay large while satisfying Eq.~\eqref{eq:eps}. After considering the high $\ell$ case in the following section, we will analyze the situations when a resolvable time advance can occur in section~\ref{sec:theory}.

\subsection{Case 2: Higher-order multipoles} \label{sec:multipole}
We will now consider the case of partial waves with $\ell>0$. As first noted by Langer \cite{PhysRev.51.669}, the standard WKB approach fails to be useful when considering low multipole contributions since the approximation fails to reproduce the behavior of the solutions near $r=0$. To deal with this, one can perform a change of variable, $r=e^{\rho}$, in order to map the singularity $r=0$ to $\rho=-\infty$. Then, the exponentially decaying WKB solution reproduces the correct asymptotics at $\rho=-\infty$. We proceed to change the variables in Eq.~\eqref{eq:radialeomFriction} as described above and obtain an equation of motion that contains a friction term which we remove with a field redefinition to get,
\begin{equation}
	\p_{\rho}^2 \delta \rho_{\ell}(\rho)= - \widehat{W}_{\ell}(\rho) \delta \rho_{\ell}(\rho)\,.
\end{equation}
Then, we solve this equation using the WKB approximation. To find the phase shift, we want to express $\widehat{W}_{\ell}(\rho)$ back in terms of the dimensionless radial coordinate $R=r/r_0$. For generic multipole we define the dimensionless quantity
\begin{equation}
W_{\ell}(R)\ \equiv \frac{1}{(\omega r)^2} \widehat{W}_{\ell}(\rho(r)) \,,
	\label{eq:defWlr}
\end{equation}
 note that this is not precisely the same definition as what was performed in \eqref{eq:eomDim}. Here
 the factor $1/r^2$ captures the Jacobian of the transformation:
  \be
  \int \sqrt{\widehat{W}_{\ell}(\rho)} \mathrm{d}\rho  = \omega r_0 \int \sqrt{W_{\ell}(r)} \mathrm{d}R \, .
  \ee

Before moving on, we note that within the present formalism we cannot compute the time delay beyond the leading-order WKB approximation for the $\ell>0$ case. It is well known that higher-order ($n>0$) WKB corrections are divergent at the turning point. This simply signals the breaking of the approximation in this region and the WKB solution can be improved by matching to an asymptotic solution near the turning point. Nevertheless, this does not modify the asymptotic behavior of the WKB solution and thus does not change the inferred time delay. While these subleading contributions seem to involve infinities at finite order in the WKB series expansion, the physical phase shift is finite so that upon appropriate re-organization or  resummation of the series the result will end up being finite. We could in principle carry out this resummation or re-organization of the series, however  for simplicity  we focus here instead in the regime where $(\omega r_0) \gg 1$ so that all WKB corrections can safely be ignored.

At first order in the WKB approximation, the phase shift of the partial wave with $\ell>0$ can be identified as (see Appendix B of \cite{deRham:2020zyh})
\begin{equation}
	\delta_{\ell} = (\omega r_0) \left[ \int_{R_t}^{\infty} \mathrm{d}R \left( \sqrt{W_{\ell}(R)} - 1 \right) - R_t + \frac{1}{2} B \pi \right] \, , \label{eq:PSl}
\end{equation}
where the  $R_t=r_t/r_0$ is the dimensionless turning point such that $W_{\ell}(r_t)=0$ and we have introduced the dimensionless impact parameter $B = b/r_0$, where $b = (\ell + 1/2)/ \omega$ is the impact parameter of the free theory, \ie when $g_{i}=0$. We remind the reader that the definitions of all dimensionless parameters are reported in Table \ref{tab:dictionaryDimensionless} of Appendix~\ref{ap:DefDimLess}.

In order to perform the integral in Eq.~\eqref{eq:PSl} analytically, we will expand the integrand at LO as in the monopole case. We have to be careful when splitting the integral order by order so that each term is a converging integral. To do so, we start by writing
\begin{equation}
	W_{\ell}(R) =	W_{\ell}(R)|_{{g_i=0}}+\delta W_{\ell}(R)\ ,
\end{equation}
where $	W_{\ell}(R)|_{{g_i=0}}$ is the contribution arising purely from the angular momentum contributions but no self-interactions. Using the fact that $W_{\ell}(R_t)=0$, this can be rewritten as
\begin{equation}
	W_{\ell}(R) =\left(1 - \frac{R_t^2}{R^2}\right) + \Delta W_{\ell}(R)\ , \quad   \Delta W_{\ell}(R)=\delta W_{\ell}(R)-\frac{R_t^2}{R^2}\delta W_{\ell}(R_t)  \ ,
\end{equation}
so that each contribution is finite at the integration boundaries and the integrals at each order in $\epsilon_1$ and $\epsilon_2$ converge. Now, expanding the square root at $\mathcal{O}(\epsilon^4)$ gives
\begin{equation}
	\sqrt{W_{\ell}(R)} = \sqrt{ 1 - \frac{R_t^2}{R^2} }  +  \frac{U_{\ell}(R)}{\sqrt{ 1 - \frac{R_t^2}{R^2} }} \,,
\end{equation}
where $U_{\ell}(R_t)=0$. The LO explicit expressions for $W_{\ell}(R)$, $U_{\ell}(R)$, and $R_t$ can be found in Appendix \ref{ap:NLOmultipole}. Integrating this expression gives the phase shift at $\mathcal{O}(\epsilon^4)$. Note that $R_t = B + \mathcal{O}(\epsilon^4)$ and $U_{\ell} = \mathcal{O}(\epsilon^2)$, hence, when dealing with the $U_{\ell}$ term, the turning point $R_t$ can be replaced by $B$ since any corrections will contribute at NLO. This means that we can write
\begin{equation}
	\int_{R_t}^{\infty} \left( \sqrt{W_{\ell}(R)} -1 \right) \, \mathrm{d}R = \int_{R_t}^{\infty} \left( \sqrt{ 1 - \frac{R_t^2}{R^2} } -1 \right) \, \mathrm{d}R + \int_B^{\infty} \frac{U_{\ell}(R)}{\sqrt{ 1 - \frac{B^2}{R^2} }} \, \mathrm{d}R \,,
\end{equation}
giving
\begin{equation}
	\delta_{\ell}(\omega) = (\omega r_0) \left[ \int_B^{\infty} \frac{U_{\ell}(R)}{\sqrt{ 1 - \frac{B^2}{R^2} }} \, \mathrm{d}R + \frac{\pi}{2} \left( B - R_t \right) \right] \,. \label{eq:PSlargeL}
\end{equation}
To get the time delay, we need to differentiate the expression above with respect to $\omega$. As opposed to the monopole case where we fixed $\ell$, when going to higher multipoles it is convenenient to think of the scattering not in terms of the scattering of waves but of particles specified by a given impact parameter. That is to say, what is naturally held fixed for particle scattering is the impact parameter $b$ (or $B$). This is the time delay traditionally considered in the eikonal approximation (see for example \cite{Camanho:2014apa}). Thus, the time delay reads
\begin{equation}
	(\omega \Delta T_{b}(\omega)) = 2 \frac{\partial \delta_{\ell}(\omega)}{\partial \omega} \big|_{b}= 2 (\omega r_0) \left[ \int_{R_t}^{\infty} \left( \p_{\omega} \left( \omega \sqrt{W_{\ell}(R)} \right) - 1 \right) \mathrm{d}R - R_t + \frac12 B \pi \right] \, ,
\end{equation}
which after using Eq.~\eqref{eq:PSlargeL} can be written as
\begin{equation}
	(\omega \Delta T_{b}(\omega)) = 2 (\omega r_0) \left[ \int_{B}^{\infty} \frac{\p_{\omega} (\omega U_{\ell}(R))}{\sqrt{1 - \frac{B^2}{R^2}}} \mathrm{d}R + \frac{\pi}{2} \left( B - \p_{\omega} (\omega R_t) \right) \right] \,. \label{eq:TDlargel}
\end{equation}

In the next section, we will explore the regions in Wilson coefficient space that can lead to a resolvable time advance given by Eq.~\eqref{eq:TDlargel}. Contrary to the $\ell=0$ case, the contribution to the time delay from $g_{10}$, found in Eq.~\eqref{eq:velocityHighL}, is not sign definite when we have angular momentum. This will allow us to bound the $g_{10}$ coefficient from above and below. The tightest bounds will arise from considering the scattering of higher-order multipole modes. Note that while we can take the large-$\ell$ limit, we cannot take $\ell\rightarrow\infty$. This can be seen by writing $L=\ell+1/2$, and
\begin{equation}
	L =\omega b=\frac{B \Omega}{\epsilon_2} \,.
\end{equation}
The impact parameter $B$ cannot be taken to infinity, otherwise there would be no scattering. Meanwhile, $\Omega$ is bounded by Eq.~\eqref{eq:eps} so that we stay within the regime of validity of the EFT. Thus, at a fixed impact parameter, the angular momentum has an upper bounded given by
\begin{equation}
L\ll \frac{B}{\epsilon_2^2} \ .
\end{equation}
Note that this large angular momentum limit is related to the standard approach of computing phase shifts by looking at the eikonal limit of $2 \rightarrow 2$ scatterings.
\subsection{Causal shift-symmetric theories}
\label{sec:theory}
We now consider a specific background profile to obtain constraints on the Wilson coefficients given by causality. We use an analytic function in order to avoid any possible divergences at $R=0$. Furthermore, we require that the background vanishes at infinity to have a well-defined scattering around an asymptotically flat background, that is, the light-cones observed by the perturbation approach the Minkowski ones near infinity. Therefore, we will consider a profile of the form
\begin{equation}
	f(R^2)=\left(\sum_{n=0}^{p} a_{2n} R^{2n} \right) e^{-R^2} \ ,
	\label{eq:profilefR2}
\end{equation}
where $a_{2n}$ are arbitrary coefficients of order $1$. For a generic scalar field EFT in its own right (not coupled to gravity), one can always consider an external  source $J$ that would generate such a  profile. In more specific contexts where the scalar field is considered to be diagnosing one of the degrees of freedom of gravity (as would for instance be the case for the helicity-0 mode in \cite{Dvali:2000hr} or in massive gravity \cite{deRham:2010ik,deRham:2010kj}), one may consider more carefully how such a profile could be generated as discussed in Appendix~\ref{app:Matter_coupling}.

When considering such profiles \eqref{eq:profilefR2}, the largest contributions to the time delay (or advance) come from small powers $n$, so in practice we truncate the series by choosing $p=3$, \ie including terms up to $a_6$. Given this profile, we can explore the regions where one can obtain a resolvable time advance, that is, $\omega \Delta T_{b} \lesssim -1$ while maintaining the EFT under control and hence violate causality.
Since we have already established in the homogeneous case that $g_8$ ought to be positive we can set $g_8=1$ without loss of generality (this simply corresponds to a rescaling of all the Wilson coefficients by $g_8$). The case $g_8=0$ will be considered separately in what follows. We  use an extremization procedure to find the largest region where causality is violated following the method is explained in Appendix~\ref{ap:procedure}.

\begin{figure}[!h]
	\begin{center}
		\includegraphics[width=0.5 \textwidth]{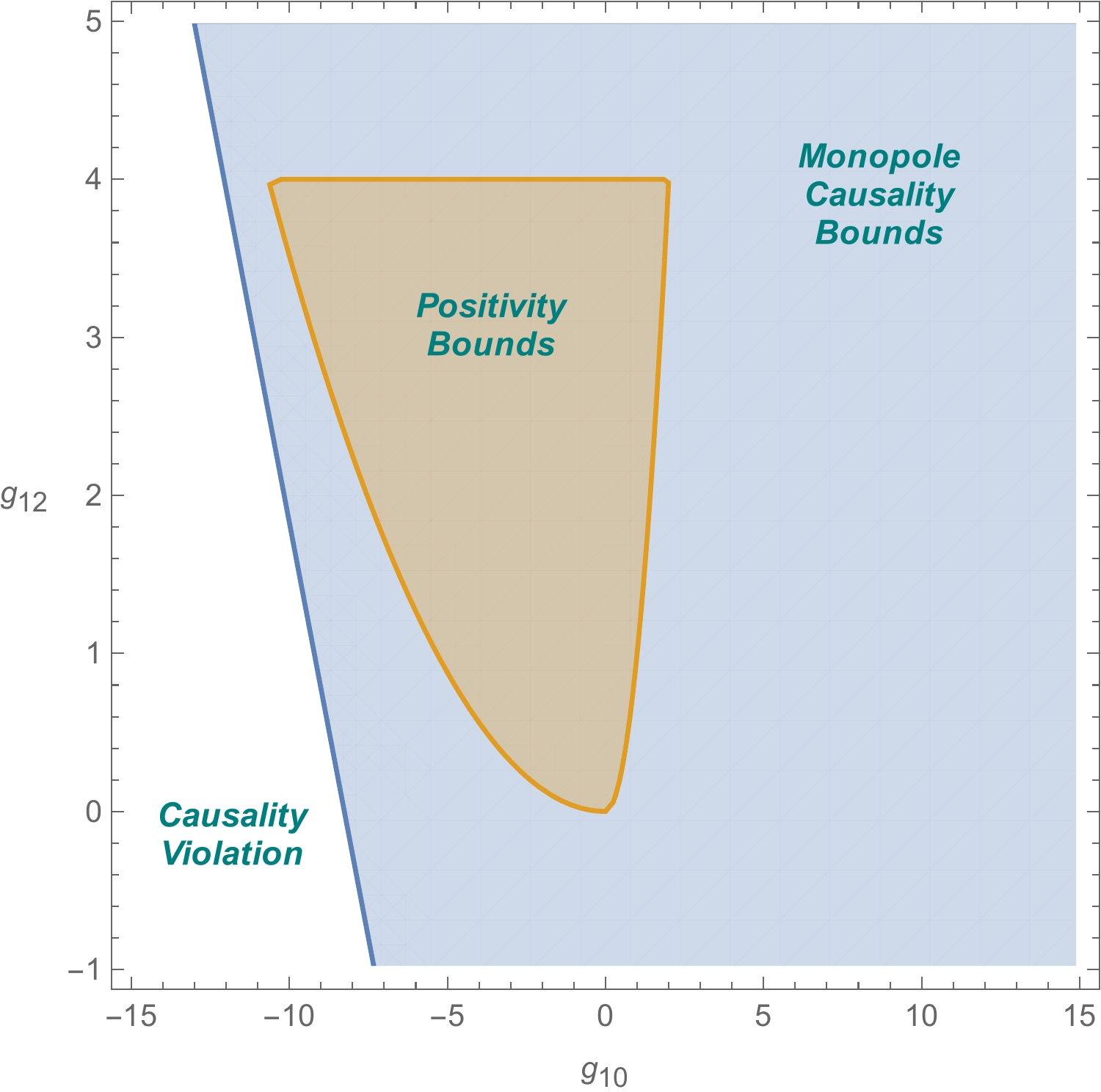}
	\end{center}
	\caption[Monopole causality bounds in shift-symmetric scalar EFT.]{Positivity and monopole causality constraints for the shift-symmetric scalar EFT considered in \eqref{eq:L}. In white, we observe a region that can lead to violations of causal propagation in the infrared, \ie where $\Delta T_0 < -1/\omega$. The blue region is its complement where there is no yet any indication of causality violation. Here, we have focused on bounds arising from monopole modes with a background profile given by Eq.~\eqref{eq:profilefR2} and the coefficient of the $(\p \phi)^4$ operator set to $g_8=1$. In orange, we observe the region that satisfies the positivity constraints in \cite{Tolley:2020gtv}, and assumes physical properties of the UV completion (such as locality and Lorentz-invariance). Please note that these positivity bounds, given by Eq.~\eqref{eq:pos}, are not the results of my work, but the causality ones are.}
	\label{fig:boundl0}
\end{figure}

\paragraph{Monopole modes:}
We first consider the $\ell=0$ case. In the extremization procedure, we include all the constraints on the dimensionless parameters arising from the validity of the EFT as in \eqref{eq:eps} and we require that the LO and NLO results differ only by a $3\%$ for $g_{14}$ of order $1$. The NLO contributions, found in Appendix \ref{ap:NLO},  include WKB corrections as well as higher-order EFT terms. Within our parametrization, we find the tightest constraints by considering
\begin{equation}
	a_0=1 \ ,\ a_2=0.72 \ , \ a_4\sim 0 \ , \ a_6=0.14 \ , \ \epsilon_1=0.36 \ ,\ \epsilon_2=0.35 \ , \ \Omega=0.70 \ , \label{eq:param}
\end{equation}
although it is likely that even tighter constraints could be derived if one considered other classes of profiles, the bounds we obtain here already serve as proof of principle.
The bounds arising from the previous choice are shown in blue in Fig.~\ref{fig:boundl0} together with the orange positivity bound from \cite{Tolley:2020gtv}. It is easy to prove, by examining Eq.~\eqref{eq:TDintegrand}, that the slope of the line delimitating the causal region from the acausal one is negative for any choice of coefficients when considering $\ell=0$. This means that when considering the monopole, we can only get a left-sided bound as argued earlier.

Note that our choice in Eq.~\eqref{eq:param} implies $\omega r_0\sim 2$ which does not suppress higher-order WKB corrections. Nevertheless, when working at $\mathcal{O}(\epsilon^4)$ we can safely consider this case since all the corrections $\delta_{\rm WKB}^{(2n)}$ for $n\geq 2$ will correspond to total derivatives that do not contribute to the phase shift. One can see that this is the case by looking at Eq.~\eqref{eq:WKBcorrec} and noting that after expanding in $\epsilon_1$ and $\epsilon_2$ up to order $\mathcal{O}(\epsilon^4)$ the WKB corrections will arise from $W^{(2n)}$.

\paragraph{Higher multipole modes:}
Moving to the higher multipoles, $\ell>0$, we consider the same profile as in Eq.~\eqref{eq:profilefR2}. By allowing finite values of $\ell$, the equation for the line $(\omega \Delta T_{b}) = -1$ separating regions of ``causality-violation"  has a new free parameter and now allows for a positive slope. This opens the possibility to constrain the causality region from both sides and below. We do not get a better lower-sided bound on $g_{10}$ but we do get an upper bound by considering the union of the constraints arising from a set of parameters as explained in Appendix~\ref{ap:procedure}. Remarkably, this method also sets a lower bound on $g_{12}$ which ought to be positive, in complete agreements with positivity bounds. In itself this is a remarkable statement as in the presence of the $g_8$ operator the speed is typically dominated by that term and little would be inferred from $g_{12}$.

\begin{figure}[!h]
	\begin{center}
		\includegraphics[width=0.5 \textwidth]{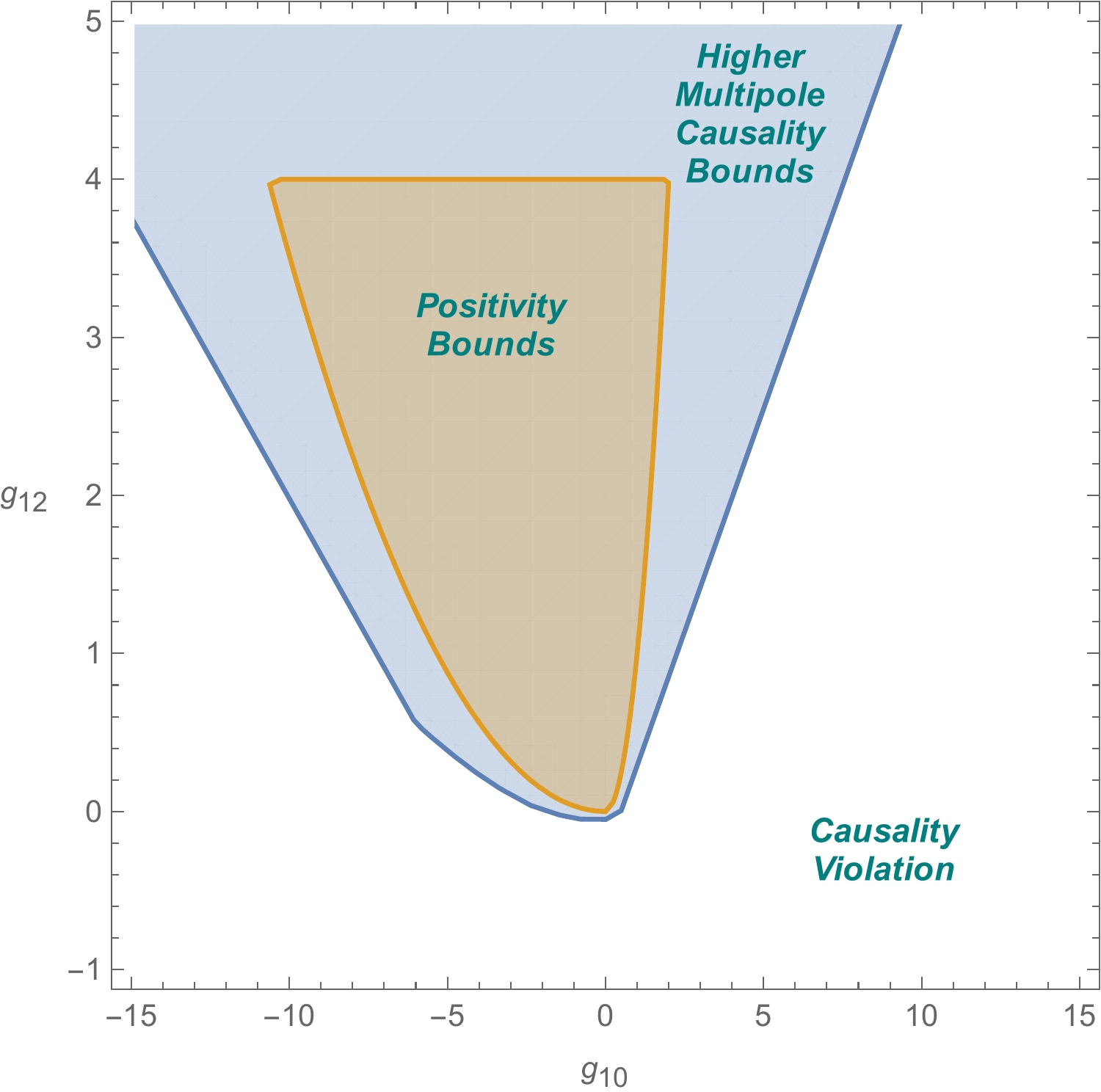}
	\end{center}
	\caption[Higher multipole causality bounds in shift-symmetric scalar EFT.]{The blue and orange regions represent the EFTs satisfying causality bounds from higher multipoles and positivity bounds respectively. The regions are computed as in Fig.~\ref{fig:boundl0} with $g_8=1$, but the causality constraints are those arising from higher-order multipole modes.}
	\label{fig:boundg81}
\end{figure}

Our results are shown in Fig.~\ref{fig:boundg81} where the causality bound region corresponds to the intersection of regions in Wilson coefficient space that do not give rise to resolvable time advances as defined in Appendix~\ref{ap:procedure}. An example of a set of parameters that we use to obtain the causality bounds is given by
\begin{equation}
	a_0=-5 \ ,\ a_2=-5 \ , \ a_4=5 \ , \ a_6=-0.91 \ , \ \epsilon_1=0.17 \ ,\ \epsilon_2=0.17 \ , \ \Omega=3 \ , \label{eq:paramLargel}
\end{equation}
leading to our tightest bound on $g_{10}$ at $g_{12}=0$. Once again we do not preclude the possibility that stronger bounds could be obtained by improved optimization methods or by considering more generic classes of profiles, however great care should be taken so as to ensure validity of the EFT and WKB approximation.  As in the previous case, we ensure that we are within the regime of validity of the EFT by satisfying Eq.~\eqref{eq:eps}. Furthermore, we only work at leading order in the WKB approximation and guarantee that higher-order corrections are negligible by taking $\omega r_0\sim\mathcal{O}(20)$. Note that, as explained earlier, odd higher-order WKB corrections only contribute to the overall amplitude and hence, corrections to the phase shift (and time delay) only come from even higher-order WKB corrections, which are then suppressed by powers of $(\omega r_0)^2\sim\mathcal{O}(400)$. In contrast to the $\ell=0$ analysis, we cannot compare to the NLO corrections since these would include WKB corrections that we cannot compute within our formalism as explained in the previous section. However, we do ensure smallness of the corrections by relying on dimension analysis given by Eq.~\eqref{eq:WKBreq}. It is interesting to note that the tightest bounds that we found come from the region where $\ell\sim\mathcal{O}(30)$, and thus are related to calculations in the eikonal limit. On the other hand, the results from the previous case ($\ell=0$) arise in the opposite regime that is less explored in the literature.

Combining monopole and higher multipoles causality bounds gives rise to the left panel of Fig.~\ref{fig:bounds} strongly constraining the viable region of the $\{g_{10}, g_{12}\}$ parameter space.
We highlight that there is room for our procedure to be further tightened (for instance by considering more generic backgrounds and more freedom in their parameterizations and their scaling). As a result the white regions ruled out in  Figs.~\ref{fig:boundl0} and \ref{fig:boundg81} are very likely not the most optimal bounds that one can obtain from causality but already provide a close contact with standard positivity bounds and new compact positivity bounds.

\subsection{Causality in Galileon theories}
Besides the shift-symmetric theory considered throughout this Chapter, one can impose a more constraining, spacetime-dependent, shift symmetry given by
\begin{equation}
	\phi\rightarrow \phi +c + b_\mu x^\mu \ ,
\end{equation}
where $c$ is a constant and $b_\mu$ a constant vector. This is the Galileon symmetry \cite{Nicolis:2008in} which arises in various contexts such as massive gravity theories, brane-world models, accelerating universes, inflationary models and alternatives to inflation \cite{deRham:2012az}. Imposing this new symmetry requires that we set $g_8=0$ in Eq.~\eqref{eq:lag}. Note that $\textit{any}$ scalar low energy EFT that enjoys a  Galileon symmetry (with no other light degrees of freedom) is forbidden by positivity bounds. Setting $g_8=0$ the positivity bounds~\eqref{eq:pos} then impose $g_{10}=g_{12}=0$. This means that when viewed as a low energy scalar EFT, a Galileon  cannot have a Wilsonian UV completion that is local, unitary, causal, and Poincar\'e invariant. Here, we would like to understand whether we can obtain similar stringent bounds from infrared causality alone with no further input on the UV completion.

The analysis follows in a similar way as the shift-symmetric case above, with the only modification arising from the requirements for the validity of the EFT which now read
\begin{equation}
	\epsilon_1 \epsilon_2\ll1 \, , \quad  \text{and} \quad \Omega \epsilon_2 \ll 1 \ . \label{eq:eftGal}
\end{equation}
The validity of the WKB approximation and the above EFT requirements imply that  $\epsilon_2\ll\Omega \ll 1/ \epsilon_2$. In order to have a well-defined $\epsilon$ expansion, we require slightly tighter lower bounds given by $\sqrt{\epsilon_2}\ll\Omega$. While in the shift-symmetric case we had $\epsilon_1\sim\epsilon_2$, here $\epsilon_1$ can in principle be larger since, thanks to the Galileon symmetry, all operators are always suppressed by some power of $\epsilon_2$. The LO or $\mathcal{O}(\epsilon^4)$ corrections simply include the $\epsilon_1^2 \epsilon_2^2$ terms.

\begin{figure}[!ht]
	\begin{center}
		\includegraphics[width=0.5 \textwidth]{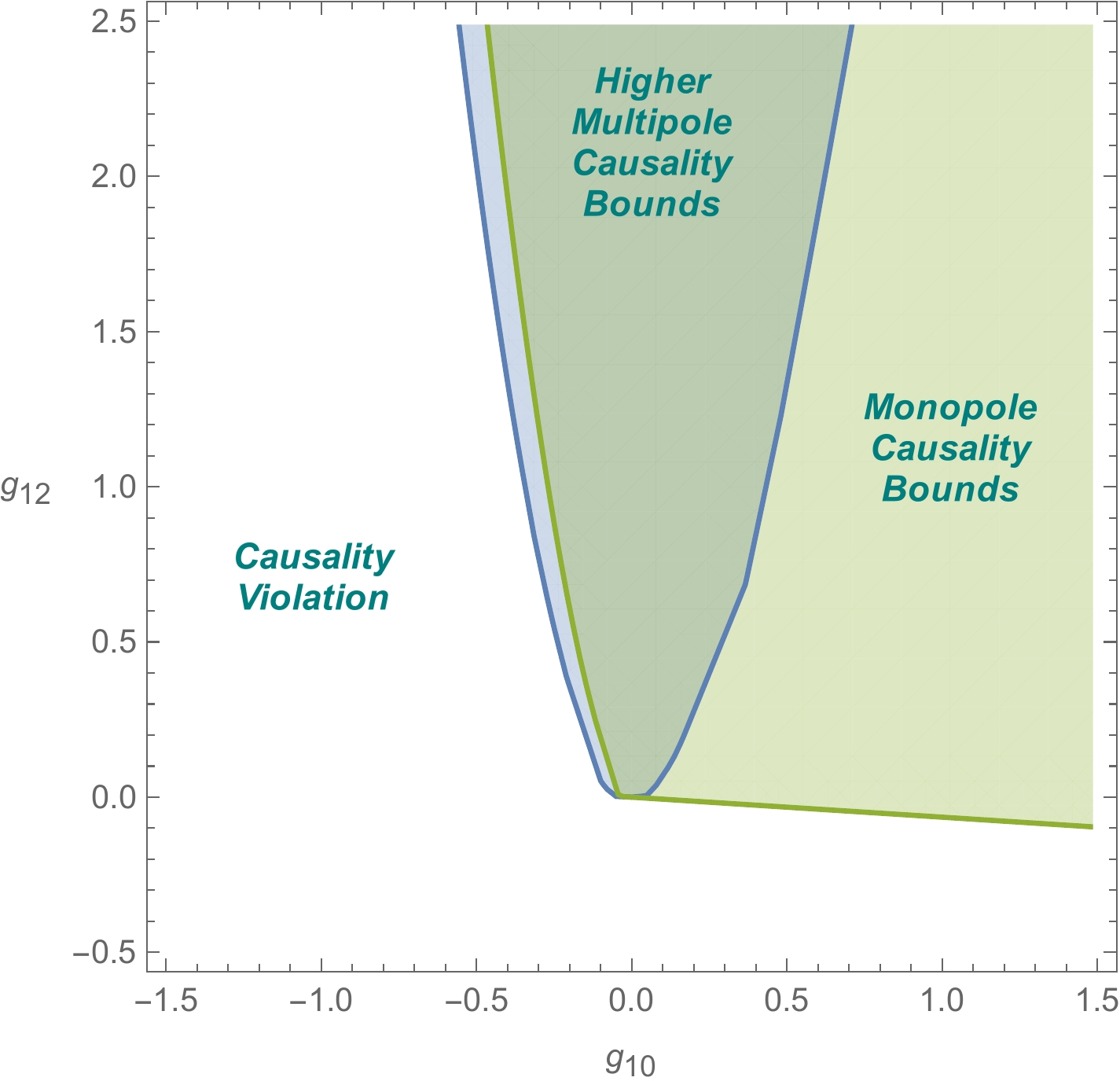}
	\end{center}
	\caption[Causality bounds for the Galileon EFT.]{Causality bounds for the Galileon EFT ($g_8=0$). The green region represents the monopole causality bounds (for the profile considered in Eq.~\eqref{eq:profilefR2}). The blue region represents causality bounds from higher multipole, leading to two-sided bounds. Only the intersection of the blue and green regions is so far causally viable.}
	\label{fig:boundGal}
\end{figure}

As in the previous case, we consider propagation around the background profile in Eq.~\eqref{eq:profilefR2}. When computing the time delay for $\ell=0$ modes, we require that the NLO result differs from the LO only by a $3\%$ for $g_{14}$ of order $1$. As in the shift-symmetric case, we can only get lower bounds on $g_{10}$ in this regime. Note that the monopole constraint for the Galileon symmetry gives $g_{10} \gtrsim 0$, which is nearly as good as it can be for a one-sided bound. Meanwhile, in the higher multipoles case, \ie $\ell>0$, we only consider the leading-order WKB results as in the previous case. For this case, we closely reproduce the $\ell=0$ left-sided bound and get a new maximal right-sided bound as seen in blue in Fig.~\ref{fig:boundGal}.

\section{Discussion and conclusions} \label{sec:Concl}

\begin{figure}[!h]
	\begin{center}
		\includegraphics[width=0.45 \textwidth]{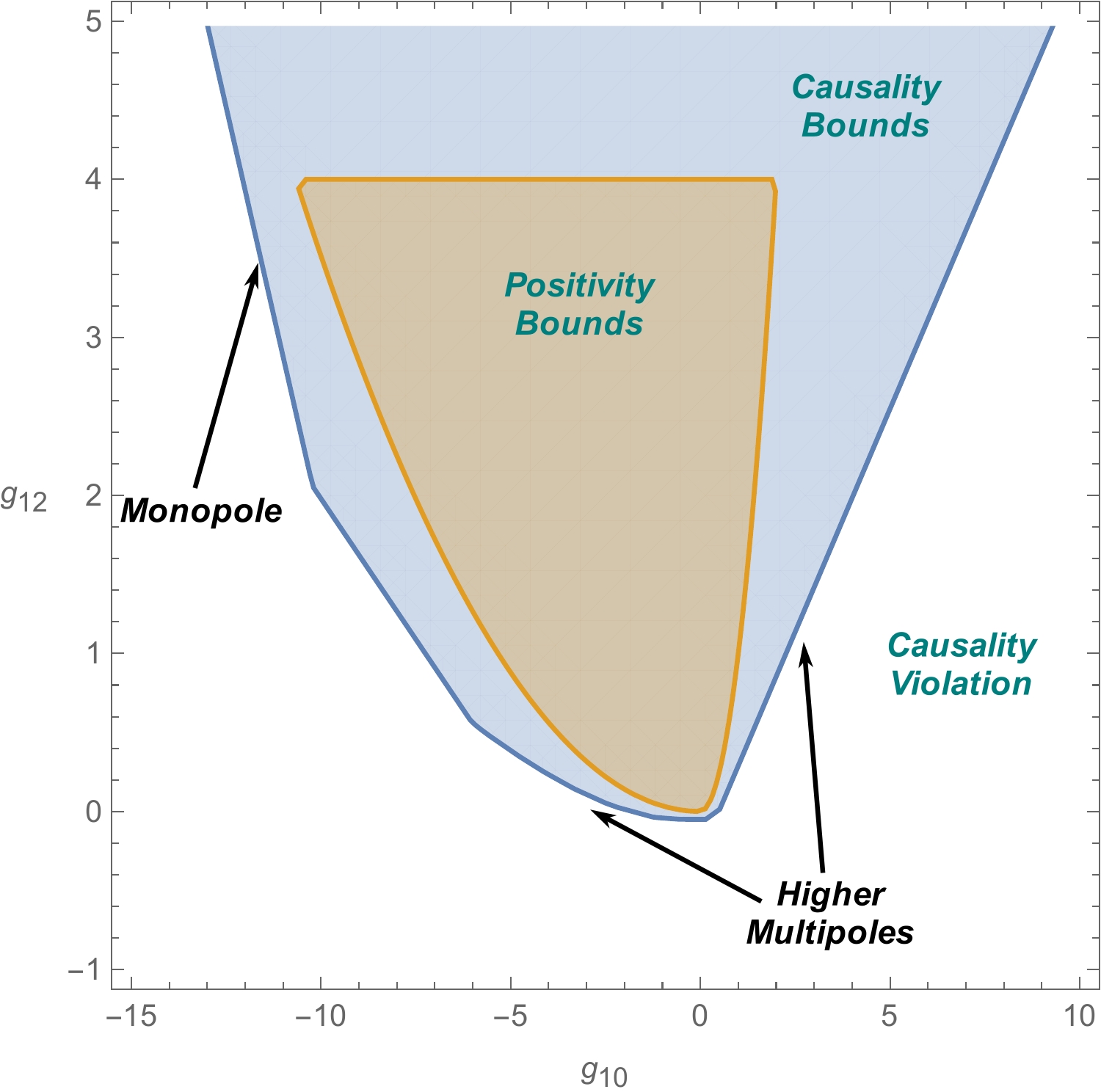}
		\hspace{0.05\textwidth}
		\includegraphics[width=0.45 \textwidth]{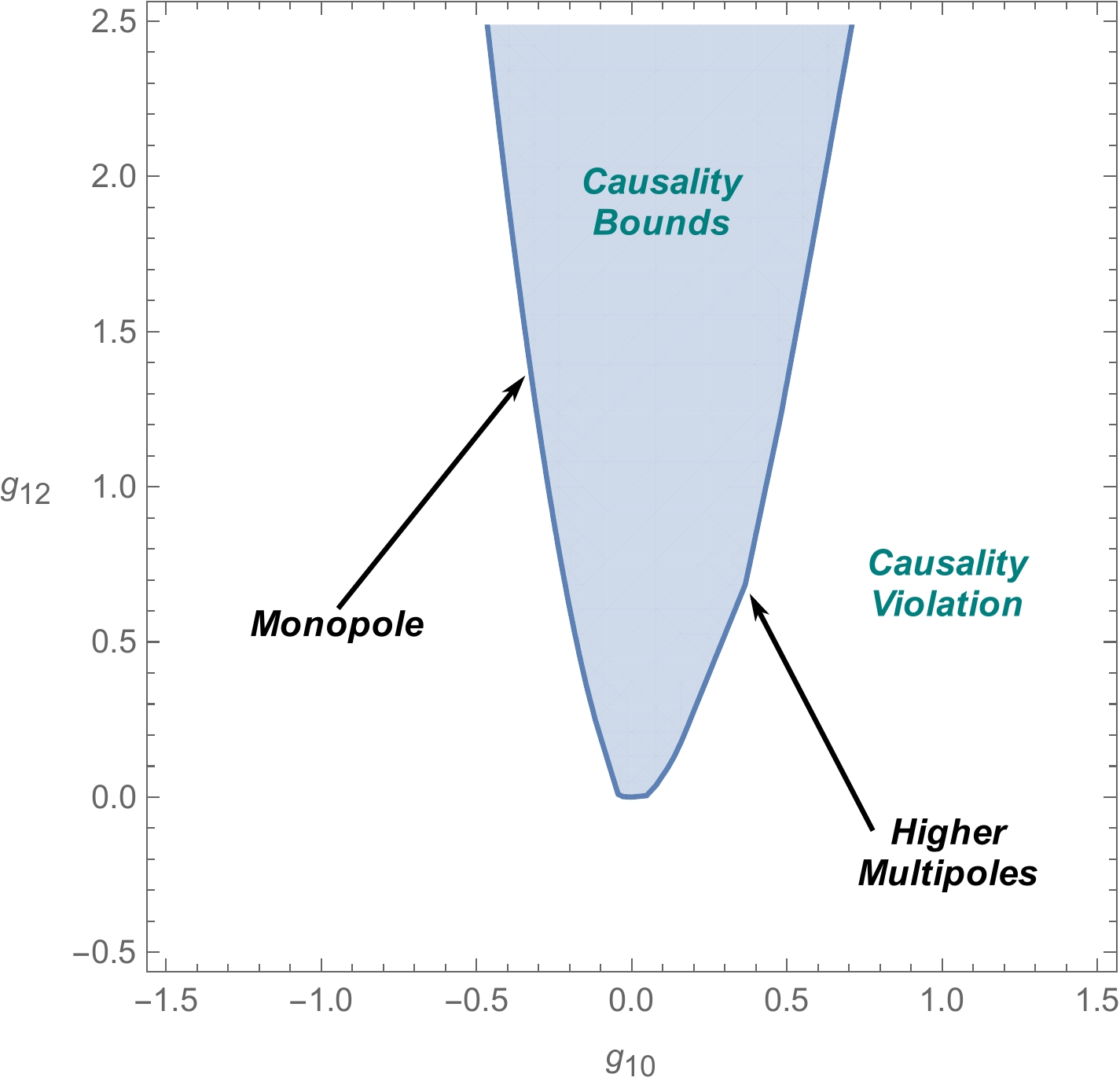}
	\end{center}
	\caption[Final causality bounds for the shift-symmetric and Galileon scalar EFT.]{Infrared Causality constraints on the Wilson coefficients of two scalar low-energy EFT, a shift-symmetric one with $g_8=1$ on the left and a Galileon-symmetric one with $g_8=0$ on the right. In both cases, the white areas are regions in the Wilson coefficients space where a violation of causality can be observed at low-energy,  whereas the orange one is derived from positivity bounds requiring assumptions in the UV.  To obtain these results, we combined lower and upper bounds derived respectively in the $\ell=0$ and $\ell>0$ cases.}
	\label{fig:bounds}
\end{figure}

We have seen that requiring that the effective field theory only leads to causal propagation around a given spherically-symmetric background allows us to put tight bounds on the Wilson coefficients of a low energy EFT, independently of its ultimate high energy completion. Remarkably, there are two physical regimes that give rise to different bounds. The propagation of zero angular momentum partial waves gives rise to lower bounds while the propagation of high $\ell$ modes imposes both lower and upper bounds, although the lower bounds are in general not competitive with those arising from $\ell=0$ modes. We can summarize our findings by combining both results from the monopole and the higher-order multipoles. This is shown in the blue causal regions depicted in Fig.~\ref{fig:bounds}.

On the left pane of Fig.~\ref{fig:bounds} we observe the causality bounds (blue) compared to the positivity bounds (orange). While our causality bounds are not as constraining as the positivity ones, we note two important points. First, contrary to the positivity bounds, causality bounds do not require any assumptions of the UV completion (including notably, unitarity and locality) they arise purely from infrared physics that is well described by the EFT. Second, positivity bounds have by now been optimized using various techniques allowing to probe features of the EFT beyond its forward limit, while ours were so far obtained using a simple static and spherically symmetric profile with a simple extremization procedure. It is likely that tighter bounds could be derived by allowing for more generic and less symmetric profiles.

More importantly, we highlight that the precise numerical values of the causality bounds should not be the main focus of our results. The fact that by simply requiring causal propagation in the infrared we can obtain such semi-compact bounds is in itself remarkable.  A naive version of the right-sided positivity bound is given by $g_{10}<2g_8$  and can be derived simply using the $s \leftrightarrow u$ dispersion relation \cite{Tolley:2020gtv}. This bound is slightly optimized when using triple crossing symmetry $s \leftrightarrow t \leftrightarrow u$. Note that in our causality bounds, we only produce an upper bound for $g_{10}$ and lower bound for $g_{12}$ when looking at higher multipoles. On the other hand, the left-sided positivity bounds are fully coming from triple crossing symmetry. In our analysis this lower bound can be reproduced by looking at both high $\ell$ and $\ell=0$ scattering, but the stronger bound comes from the monopole bound. This suggests that our analysis approximately reproduces bounds purely from $s \leftrightarrow u$ dispersion relation in the UV when looking at higher multipoles and triple crossing symmetry when looking at the monopole. However, this seems to be the opposite behaviour of the one observed in \cite{Tolley:2020gtv,Caron-Huot:2020cmc}, where the upper bound is obtained at $\ell=0$ and the lower one at $\ell \geq 2$.

Correspondingly, in the right pane  of Fig.~\ref{fig:bounds} we see that requiring infrared causality of the Galileon theory allows us to recover a very similar result to the recently derived full-crossing symmetric positivity bounds that entirely rule out the quartic Galileon by assuming properties of the UV completion. Thus, we effectively rule out the quartic Galileon as a causal low energy scalar effective field theory with no other light degrees of freedom. This does not imply that we rule out the quartic Galileon coupling that would arise in a gravitational setting.  For example, the Galileon theory is a meaningful decoupling limit of massive gravity theories, but can never be considered as a low energy description without the inclusion of other modes. Moreover the Galileon field would generically couple to the trace of the stress-energy tensor, which must obey some consistency conditions of its own. We discuss this point in Appendix~\ref{app:Matter_coupling} and leave for future work the analysis of the situation where we have a gravitational coupling in which one has to impose conditions on the sources to be physical. Instead, our analysis holds if we assume that we are dealing with a  scalar EFT in its own right that can be coupled to an arbitrary external source so that causal propagation is required for any possible external source configuration.

Over the past few years, remarkable progress has been made in deriving new sets of non-linear, compact positivity bounds that make use of full $s\leftrightarrow t \leftrightarrow u$ crossing symmetry. This work serves as proof of principle that low energy causality arguments alone can go a long way in making contact with known positivity bounds. This extends the earlier observation of \cite{Adams:2006sv} (for a more recent discussion connecting time delays and positivity bounds see Appendix A of \cite{Arkani-Hamed:2020blm}).
It would be interesting to understand how constraining low energy causality is when optimizing the bounds derived in this Chapter across more general backgrounds similar to that considered in \cite{deRham:2020zyh,Chen:2021bvg,deRham:2021bll}. One might expect that fewer symmetries could lead to stronger bounds. Similarly, one could use this approach to constrain Wilson coefficients of higher derivative terms that arise in the EFT which have been previously bounded using positivity arguments. One appeal of these constraints is that they can easily be generalizable to include operators that are higher order in the field and hence would not contribute at tree-level to known $2\to 2$ positivity bounds.  Furthermore, the requirement of low energy causality can be imposed on gravitational theories and curved backgrounds without running into problems related to the lack of an S-matrix or broken Lorentz symmetries, which would make them particularly appealing for instance for  cosmological \cite{Melville:2019wyy,deRham:2021fpu,Grall:2021xxm} or black hole gravitational bounds \cite{Chen:2021bvg,deRham:2021bll}. In future work, we will explore how causality can give rise to bounds in such situations.

\chapter{Causal vector effective field theory}
\label{chap:causalVector}

Effective field theories (EFT) are strongly constrained by such fundamental principles as unitarity, locality, causality and Lorentz invariance. Taking the example of the EFT of massless photons we compare different approaches to finding these constraints in terms of Wilson coefficients. In this Chapter, we present an analytic derivation of the constraints arising from the requirements of causality of the photon propagation on top of different backgrounds, and we compare them with implications of unitarity (linear and non-linear positivity bounds). We find that low energy causality conditions can give complementary constraints, compared to the positivity bounds. Applying both constraints together may significantly squeeze the allowed region of the photon EFT parameters.

\section{Photon Effective Field Theory}
We wish to write down the most generic effective field theory (EFT) of a massless vector field $A_{\mu}$ enjoying a $U(1)$ gauge symmetry. We define the field-strength (or Faraday) tensor $F\mn$ and its Hodge dual $\tilde{F}\mn$ in the following way
\begin{equation}
	F\mn = \p_{\mu} A_{\nu} - \p_{\nu} A_{\mu} \,, \qquad \tilde{F}\mn = \frac12 \epsilon_{\mu \nu \rho \sigma} F^{\rho \sigma} \,.
	\label{eq:defF}
\end{equation}
We make the hypothesis that the $U(1)$-symmetric Lagrangian only depends on gauge-invariant quantities, \ie
\begin{equation}
	\L_{F} = \L(F,\tilde{F},\p F, \p \tilde{F}, \dots) \,.
	\label{eq:gaugeinvariantquantities}
\end{equation}
Note that we will forbid any operator with an odd number of $\tilde{F}$ to avoid any parity breaking. We will assume that our EFT breaks down at a scale $\Lambda$ and we will consider operators of at most $4$ fields and dimension $12$. We assume that operators with more than $4$ fields are suppressed at the EFT order that we work at, which allows us to compare the causality bounds to positivity bounds. After neglecting redundant terms that can be removed through field redefinitions, we write a set of independent operators up to dimension $10$ which reads
\begin{align}
	\label{eq:Lagrangian}
	\L =& - \frac14 F\mn F\mnup \nonumber \\
	&+\frac{c_1}{\Lambda^4} F\mnup F\mn F^{\alpha \beta} F_{\alpha \beta} + \frac{c_2}{\Lambda^4} F\mnup F^{\alpha \beta} F_{\mu \alpha} F_{\nu \beta} \nonumber \\
	&+ \frac{c_3}{\Lambda^6} F^{\alpha \mu} F^{\nu \beta} \p_{\mu} F_{\beta\gamma} \p_{\nu} F_{\alpha}^{\phantom{\alpha}\gamma} + \frac{c_4}{\Lambda^6} F^{\alpha \mu} F^{\nu \beta} \p_{\beta} F_{\mu\gamma} \p^{\gamma} F_{\alpha \nu} + \frac{c_5}{\Lambda^6} F^{\alpha \mu} F^{\nu \beta} \p_{\beta} F_{\nu\gamma} \p^{\gamma} F_{\alpha \mu} \nonumber \\
	&+ \frac{c_6}{\Lambda^8} F\mnup \p_{\mu} F_{\nu \rho} \p^{\rho} \p^{\alpha} F^{\beta\gamma} \p_{\alpha} F_{\beta\gamma} + \frac{c_7}{\Lambda^8} F\ud{\mu}{\gamma} \p_{\mu} F_{\nu \rho} \p^{\nu} F_{\alpha\beta} \p^{\rho} \p^{\gamma} F^{\alpha\beta} \nonumber \\
	&+ \frac{c_8}{\Lambda^8} F^{\mu\gamma} \p_{\mu} F_{\nu\rho} \p^{\rho} \p^{\beta} F_{\alpha\gamma} \p^{\alpha} F\ud{\nu}{\beta} \,.
\end{align}
The $4$-point tree-level scattering amplitude arising from this theory can be parametrized as 
\begin{subequations}
	\begin{align}
		\mathcal{A}_{++++}&=f_2\left(s^2+t^2+u^2\right)+f_3 s t u+f_4\left(s^2+t^2+u^2\right)^2\\
		\mathcal{A}_{++--}&=g_2 s^2+g_3 s^3+g_4 s^4+g_4^{\prime} s^2 t u\\
		\mathcal{A}_{+++-}&=h_3 s t u
	\end{align}
\label{eq:AmplParam}
\end{subequations}
where all other helicity configurations can be obtained by symmetry considerations (parity, time-reversal, boson exchange, crossing symmetry) and we consider all particles incoming. The scattering amplitude parameters above are related to the Wilson coefficients in Eq.~\eqref{eq:Lagrangian} as
\begin{align}
f_2=2\left(4 c_1+c_2\right)	\ , \quad g_2=2\left(4 c_1+3 c_2\right) \, \\
f_3=-3\left(c_3+c_4+c_5\right) \ , \quad	g_3=-c_5 \ ,  \quad h_3=-\frac{3}{2} c_3 \ , \\
f_4=\frac{1}{4}c_6 \ , \quad g_4 =\frac{1}{2} (c_6 - c_8)+c_7 \ , \quad  g_4'=-\frac{1}{2} (c_7+ c_8) \ .
\end{align}

\section{Causality bounds}
In this section, we analyze bounds arising on the Wilson coefficients of the photon EFT from a related, but different perspective than in the previous sections. Here, we will obtain bounds by imposing causal propagation of the two physical photon modes around a non-trivial electromagnetic background. This is a purely low-energy calculation and does not require any assumptions on the UV completion of the theory. We will follow the analysis performed in \cite{CarrilloGonzalez:2022fwg} for a scalar field theory. More specifically, we will consider the propagation of a linearized mode around a spherically symmetric electromagnetic background in the regime where the scale measuring the variations of the background is much larger than the scale at which the perturbative mode varies. This will allow us to compute the time delay experienced by the mode traveling on a non-trivial background compared to that of a mode traveling in a background with $F_{\mu\nu}=0$. Causal propagation dictates that the time delay is bounded as
\begin{equation}
	\Delta T > -1/\omega \ , \label{eq:CausalBound}
\end{equation}
where $\omega$ is the frequency of the mode. This simply indicates that one should not have a measurable time-advance. 

\subsection{Spherically symmetric backgrounds} \label{sec:spherical}
To proceed we compute the equations of motion (explicitly shown in Appendix \ref{ap:eom}) and consider small fluctuations $\mathcal{A}_{\mu}$ on top of a background $\bar{A}_{\mu}(r)$ such that
\begin{equation}
	A_{\mu}(t,r,\theta,\varphi) = \bar{A}_{\mu}(r) + \mathcal{A}_{\mu}(t,r,\theta,\varphi) \,,
	\label{eq:Aspherical}
\end{equation} 
where the spherically-symmetric background is given by\footnote{Note that a radial component dependent only on $r$ could be included but this is just a gauge mode that does not contribute to the field strength.}
\begin{equation}
	\bar{A}_{\mu}(r) dx^{\mu} = \bar{A}_0(r) dt \,.
\end{equation}
We assume that this background is sourced by an arbitrary, spherically symmetric, external current $J^\nu$. One is then left with the task to expand the vector perturbations in spherical harmonics. In order to do so, we will assume azimuthal symmetry and thus neglect any $\varphi$ dependence, meaning that we effectively take the quantum number $m$ to be vanishing and then the spherical harmonics reduce to Legendre polynomials instead. We will also assume that the time dependence factorizes out in the form $e^{i \omega t}$ where $\omega$ is the frequency of the wave. We parametrize the gauge field as
\begin{equation}
	\mathcal{A}_{\mu} (t,r,\theta) = \frac{1}{r} \sum_{I=1}^4 \sum_{\ell \geq 0} D_I^{\ell}(r) Z^{(I)\ell}_{\mu}(\theta) e^{i \omega t} \,,
\end{equation}
where we take the basis
\begin{align}
	Z^{(1)\ell}_{\mu} &= \delta_{\mu}^t Y_{\ell}(\theta) \,, \nonumber \\
	Z^{(2)\ell}_{\mu} &= \delta_{\mu}^r Y_{\ell}(\theta) \,, \nonumber \\
	Z^{(3)\ell}_{\mu} &= \frac{r}{\sqrt{\ell(\ell+1)}} \delta_{\mu}^{\theta} \p_{\theta} Y_{\ell}(\theta) \,, \nonumber \\
	Z^{(4)\ell}_{\mu} &= \frac{r}{\sqrt{\ell(\ell+1)}} \sin\theta \delta_{\mu}^{\varphi} \p_{\theta} Y_{\ell}(\theta) \,.
\end{align}
The functions $Y_{\ell}(\theta)$ are the usual Legendre polynomials and $\ell$ denotes the order of a given partial wave. The functions $Z^{(I)\ell}_{\mu}$ with $I=1,2,3$ have polar or even parity, whereas $I=4$ has axial or odd parity\footnote{The parity of each mode can be understood by their behavior under a space inversion $\theta \rightarrow \pi - \theta$. Under this transformation we have $\left(Z^{(1,2,3)\ell}_{\mu}, Z^{(4)\ell}_{\mu} \right) \rightarrow \left((-1)^{\ell} Z^{(1,2,3)\ell}_{\mu}, (-1)^{\ell + 1} Z^{(4)\ell}_{\mu} \right)$.}, It follows that the perturbation modes $D_{I}^{\ell}(r)$ are also real and inherit polar parity for $I=1,2,3$ and axial parity for $I=4$, meaning that those two sectors decouple at leading order. One of the $D_I^\ell$ modes can be removed straightforwardly via a gauge transformation of the form $A_{\mu} \rightarrow A_{\mu} + \p_{\mu} \chi$ with 
\begin{equation}
	\chi(t,r,\theta) = - \frac{D_3^{\ell}(r)}{\sqrt{\ell(\ell+1)}} Y_l(\theta) e^{i \omega t} \,. 
\end{equation}
We can now consider the redefinition 
\begin{align}
	u_1^{\ell} &\equiv D_1^{\ell} - \frac{i \omega r}{\sqrt{\ell (\ell + 1)}} D_3^{\ell} \,, \nonumber \\
	u_2^{\ell} &\equiv \frac{1}{r} \left( D_2^{\ell} - \frac{r}{\sqrt{\ell (\ell + 1)}} (D_3^{\ell})' \right) \,, \nonumber \\
	u_4^{\ell} &\equiv D_4^{\ell} \,,
\end{align}
to obtain that  the vector perturbations take the following form
\begin{equation}
	\mathcal{A}_{\mu} = \left( \frac{u_1^{\ell}}{r} Y_{\ell}, u_2^{\ell} Y_{\ell}, 0, \frac{u_4^{\ell}}{\sqrt{\ell( \ell +1)}} \sin \theta  Y_l' \right) e^{i \omega t} \,.
\end{equation}
Note that we have introduced a $1/r$ factor in the definition of $u_2^{\ell}$ which makes this function dimensionful, however, this will simplify further analysis. By analyzing the equations of motion, one can further notice that one of them corresponds to a constraint that removes the remaining unphysical degree of freedom and we are left only with two propagating modes (given by $u_2^{\ell}$ and $u_4^{\ell}$) as expected for a massless vector field in $4$ dimensions. This constraint equation sets a linear combination of $u_1^{\ell}$, $u_2^{\ell}$, and derivatives of $u_2^{\ell}$ to zero and can be found at any desired order in the EFT expansion. The explicit expression is given in Eq.~\eqref{eq:constraint}. 

We can solve the equations of motion for the 2 remaining physical degrees of freedom, $u_2^{\ell}$ and $u_4^{\ell}$, by removing higher-order radial derivatives iteratively in order to obtain a second-order differential equation for each mode and using the WKB approximation. We work with $\ell>0$ since a vector field does not propagate a monopole mode\footnote{While this is a well known fact, it can be seen explicitly in our setting by setting $\ell=0$ in the equations of motion in which case we are left with only two modes, namely $D^{0}_1$ and $D^0_2$. One of this modes can be removed by a gauge transformation $\chi(t,r,\theta) = -D_1^{0}(r)Y_0 e^{i \omega t}/(i \omega r) \,$. Meanwhile, the second one is set to zero by Maxwell's equations. Thus there are no propagating modes.
}. In this case, it is well-known that we need to take $r \rightarrow e^{\rho}$ to better describe the problem at low $\ell$ \cite{Langer:1937qr}. Once the equations of motion are expressed in this variable, we can remove the friction term and then change variables back to $r$. For more detail on how to perform this calculation see \cite{CarrilloGonzalez:2022fwg,deRham:2020zyh}. We now proceed to solve the equations of motion for the two physical degrees of freedom by using the WKB approximation to find the phase shift experienced by these modes when propagating around non-trivial backgrounds.

\paragraph{Regime of validity of the EFT and WKB approximation}
In the following, we will follow the notation in \cite{CarrilloGonzalez:2022fwg}, \ie we will denote by $r_0$ the typical oscillation length of the background $\bar{A}_0(r)$ and by $\bar{\Phi}_0$ its typical amplitude, which carries mass dimensions. Note that we have the same notations as the ones used in Section \ref{sec:Spherical} with the only difference that $\bar{\phi}(r)$ is here replaced by $\bar{A}_0(r)$. We will introduce the reduced (dimensionless) radial coordinate $R$ such that
\begin{equation}
	\bar{A}_0(r) = \bar{\Phi}_0 f(R) \, , \quad R\equiv\frac{r}{r_0} \ ,
\end{equation}
and we also recall the impact parameter of the free theory, $b$, and its reduced partner $B$
\begin{equation}
	\omega b = l + 1/2 \,, \qquad B \equiv \frac{b}{r_0} \,.
\end{equation}
In order to stay within the regime of validity of the EFT, we require the constraints given in Eq.~\eqref{eq:eps} to be satisfied. Furthermore, we will take $\epsilon_1 \ll \epsilon_2$ to neglect the $\epsilon_1^4$ contributions and $\omega r_0\gg1$ as well as $\Omega\gg1$ to ensure the consistency of our expansion within the WKB approximation. The former is simply the requirement for the WKB approximation to hold and tells us that the the typical scale of variation of the perturbation is shorter than that of the background. The latter ($\Omega\gg1$) is required so that higher WKB corrections which cannot be included consistently in the phase shift calculation can be ignored at the order we work at. Overall, this requires the following hierarchy for our parameters
\begin{equation}
	\epsilon_1 \ll \epsilon_2  \ll 1 \ll \Omega \ll \frac{1}{\epsilon_2}  \ . \label{eq:HierarchyParam}
\end{equation}

\paragraph{Calculation of the time delay}
Now that we have all the ingredients, we can apply the machinery developed in Chapter \ref{chap:causalScalar} to get the phase shift and then the time delay. The phase shift experienced by the propagating modes can be found by using the WKB approximation to solve their equations of motion, which are given by
\begin{equation}
	u_I^{\ell}(R)'' = - W_{I,\ell}(R) u_I^{\ell} \,, \qquad \text{for } I=2,4 \,, \label{eq:ModesEOM}
\end{equation}
where the explicit expressions for $W_{I,\ell}(R)$ can be found in Eq.~\eqref{eq:W2} and Eq.~\eqref{eq:W4}. Note that we write $u_I^{\ell}$ as an abuse of notation since we are now describing the evolution of the field-redefined $u_I^{\ell}$ such that their respective equations of motion are free of any friction terms. The phase shift is found by looking at the behavior of the solution at infinity, that is,  $\lim\limits_{r\rightarrow\infty} u_I^{\ell} \propto\left(e^{2 i \delta_{I,\ell}} e^{i \omega r}- e^{i \pi \ell} e^{-i \omega r}\right)$. Thus, the phase-shift takes the following form
\begin{equation}
	\delta_{I,\ell}(\omega) = (\omega r_0) \left[ \int_{R^{t}_{I,\ell}}^{\infty} \frac{U_{I,\ell}(R)}{\sqrt{1- \frac{(R^{t}_{I,\ell})^2}{R^2}}} dR + \frac{\pi}{2} \left( B - R^{t}_{I,\ell} \right) \right] \,,
\end{equation}
where $R^{t}_{I,\ell}$ is the turning point for the degree of freedom $u_I^{\ell}$ defined by $	W_{I,\ell}(R^{t}_{I,\ell}) = 0$ and $U_{I,\ell}$ is such that
\begin{equation}
	\sqrt{W_{I,\ell}(R)} = \sqrt{1- \frac{(R^{t}_{I,\ell})^2}{R^2}} + \frac{U_{I,\ell}(R)}{\sqrt{1- \frac{(R^{t}_{I,\ell})^2}{R^2}}} \,.
\end{equation}
Note that all the functions introduced here have already been introduced in Chapter \ref{chap:causalScalar} for the scalar case. However, in the vector case there is a time delay for each of the two propagating modes labelled by $I$ and hence we prefer clarifying the notation to avoid any confusion. The explicit expression for all the functions appearing in this section can be found in Appendix~\ref{ap:SphericalExpr}. Note that we perform such an expansion around the turning point to ensure the convergence of the integral. In particular, we have $U_{I,\ell}(R^{t}_{I,\ell}) = 0$. Furthermore, it is easy to show that $U_{I,\ell} = \mathcal{O}(\epsilon_1^2)$ and $R^{t}_{I,\ell}=B+\mathcal{O}(\epsilon_1^2)$, hence the turning point can safely be replaced by its leading-order value $B$ since any corrections would contribute to $\mathcal{O}(\epsilon_1^4)$ which can be neglected in the expansion scheme we have chosen. The phase shift then reads
\begin{equation}
	\delta_{I,\ell}(\omega) = (\omega r_0) \left[ \int_{B}^{\infty} \frac{U_{I,\ell}(R)}{\sqrt{1- \frac{B^2}{R^2}}} dR + \frac{\pi}{2} \left( B - R^{t}_{I,\ell} \right) \right] \,,
\end{equation}
Obtaining the time delay from the phase-shift is straightforward,
\begin{equation}
	(\omega \Delta T_{b,I,\ell}(\omega)) = 2 \left. \frac{\p \delta_{I,\ell}(\omega)}{\p \omega} \right|_b = 2(\omega r_0) \left[ \int_{B}^{\infty} \frac{\p_{\omega} \left( \omega U_{I,\ell}(R) \right)}{\sqrt{1- \frac{B^2}{R^2}}} dR + \frac{\pi}{2} \left( B - \p_{\omega} \left( \omega R^{t}_{I,\ell} \right) \right) \right] \, , \label{eq:TimeDelay}
\end{equation}
where $|_b$ means that we perform the derivative at fixed impact parameter $b$.

\subsection{Causality Bounds}
\label{ssec:CausBoundsVector}
In this section, we will work with the scattering amplitudes parameters in Eqs.~\eqref{eq:AmplParam}, instead of the Wilson coefficients of the Lagrangian in Eq.\eqref{eq:Lagrangian}. This will allow for a straightforward comparison with the positivity bounds in the previous section.

In order to find the causality bounds, we impose the requirement that we cannot get a resolvable time advance, that is, that Eq.~\eqref{eq:CausalBound} is satisfied for both even and odd sectors with the time delay given by Eq.~\eqref{eq:TimeDelay}. The precise method has been described previously in \cite{CarrilloGonzalez:2022fwg} and is summarized for completion in Appendix~\ref{ap:method}.

\paragraph{Sign-definite contributions}
We will now investigate the contribution to the time delay from the different scattering amplitude parameters, in both the even and odd sector. In particular, we identify the ones that are sign-definite as this will lead us to predict whether the causality bounds will be one-sided or compact. The results can be found in Table \ref{tab:signdef}. First, we define the following positive integrals
\begin{align}
	\mathcal{A}^+ &= 2(\omega r_0) B^2 \int_B^{\infty} \frac{(f'(R)/R)^2}{\sqrt{1-B^2/R^2}} > 0 \,, \\
	\mathcal{B}^+ &= 2(\omega r_0) \int_B^{\infty} \frac{B^2}{R^2} \frac{(f'(R)/R - f''(R))^2}{\sqrt{1-B^2/R^2}} > 0 \,, \\
	\mathcal{C}^+ &= 2(\omega r_0) \int_B^{\infty} \frac{B^2}{R^2} \sqrt{1-B^2/R^2} \left( \frac{f'(R)}{R} - f''(R) \right)^2 > 0 \,.
	\label{eq:positiveInt}
\end{align}
We also introduce the following (non sign-definite) integral for convenience,
\begin{equation}
	(\omega \Delta T_{b,4,\ell}^{(f_3)}(\omega)) = - \frac23 (\omega r_0) \epsilon_1^2 \epsilon_2^2 \int_B^{\infty} \frac{B^2}{R^2} \frac{\left[ \frac{f'(R)^2}{R^2} -(f''(R)^2 + f'(R) f^{(3)}(R)) \right]}{\sqrt{1-B^2/R^2}} \,.
\end{equation}
We will work with a localized background of the form
\begin{equation}
    f(R)=\left(\sum_{n=0}^p a_{2 n} R^{2 n}\right) e^{-R^2} \ ,
\end{equation}
where the coefficients $a_{2n}$ are of order $1$ and we will take $p=3$. Furthermore, note that the expansion scheme we have chosen is such that the final time delay is linear in \textit{all} scattering amplitude parameters and hence we have
\begin{equation}
	(\omega \Delta T_{b,I,\ell}(\omega)) = \sum_{J} \mathcal{W}_J (\omega \Delta T_{b,I,\ell}^{(J)}(\omega)) \,,
	\label{eq:WJ}
\end{equation}
where the Wilson coefficients are denoted by $\mathcal{W}_J$ and the index $J$ runs from $1$ to $8$ such that $\mathcal{W}_J = \left\lbrace f_2 , g_2 , f_3 ,g_3 , h_3 ,f_4 , g_{4} , g_{4}' \right\rbrace_J$ (even though we will see that the time delay doesn't depend on $g_{4}'$ in either sector), and where $\Delta T_{b,I,\ell}^{(J)}$ are numerical factors depending on $\epsilon_1, \epsilon_2, \Omega, B, a_0, a_2, a_4$, and $a_6$ but \textit{not} on the Wilson coefficients $\mathcal{W}_J$. 
\begin{table}[h!]
	\centering
	\begin{tabular}{ | c | c | c | c | }
	\hline
	Wilson coefficient & $(\omega \Delta T_{b,2,\ell}(\omega))$ & $(\omega \Delta T_{b,4,\ell}(\omega))$ & Sign-definiteness \\ \hline
	$f_2$ & $\epsilon_1^2 \mathcal{A}^+ > 0$ & $-\epsilon_1^2 \mathcal{A}^+ < 0$ & No \\
	$g_2$ & $\epsilon_1^2 \mathcal{A}^+ > 0$ & $\epsilon_1^2 \mathcal{A}^+ > 0$ & $(+)$ \\
	$f_3$ & $- \frac13 \epsilon_1^2 \epsilon_2^2 \mathcal{B}^+ < 0$ & $ \omega \Delta T_{b,4,\ell}^{(f_3)}(\omega)$ & No \\
	$g_3$ & $- \epsilon_1^2 \epsilon_2^2 \mathcal{B}^+ < 0$ & $-3 (\omega \Delta T_{b,4,\ell}^{(f_3)}(\omega))$ & No \\
	$h_3$ & non sign-definite & $4 (\omega \Delta T_{b,4,\ell}^{(f_3)}(\omega)) + 2 \epsilon_1^2 \epsilon_2^2 \frac{\mathcal{A}^+}{B^2}$ & No \\
	$f_4$ & $24 \epsilon_1^2 \epsilon_2^2 \Omega^2 \mathcal{C}^+ > 0$ & $-24 \epsilon_1^2 \epsilon_2^2 \Omega^2 \mathcal{C}^+ < 0$ & No \\
	$g_{4}$ & $12 \epsilon_1^2 \epsilon_2^2 \Omega^2 \mathcal{C}^+ > 0$ & $12 \epsilon_1^2 \epsilon_2^2 \Omega^2 \mathcal{C}^+ > 0$ & $(+)$ \\
	$g_{4}'$ & $0$ & $0$ & $0$ \\ \hline
\end{tabular}
	\caption[Wilson coefficients' contributions to the time delay in the odd and even sectors.]{Contributions to the time delay in the odd and even sectors from various Wilson coefficients.}
	\label{tab:signdef}
\end{table}

 We can see from Table \ref{tab:signdef} that the time delay is \textit{independent} of $g_4'$ and that the coefficients $g_2$ and $g_4$ produce sign-definite contributions to the time delay across both even and odd sectors. Any $2$-dimensional (2d) bound that does not involve $g_2$ and $g_4$ will be a two-sided bound that can lead to a compact causal region. In the following we show several representative $2d$ bounds. First we analyze the $f_2$ and $g_2$ case which will allows us to fix the value of $g_2$. For the other bounds, we plot a $2d$ slice of whole $6d$ parameter space ${f_2,f_3,g_3,h_3,f_4,g_4}$, that is, we fix the value of four of those coefficients and plot the bounds on the other two.

\subsubsection{2-dimensional bounds}

\paragraph{Bounds on $f_2$ and $g_2$}
It is possible to obtain the strongest bounds on the leading order Wilson coefficients by considering a regime of validity of the EFT and WKB approximation in which all higher order corrections are highly suppressed. Instead of the hierarchy for the parameters in Eq.~\eqref{eq:HierarchyParam}, we can take
$\epsilon_2 \ll \epsilon_1  \ll 1$. Within this setting, one can show that
\begin{equation}
	(\omega \Delta T_{b,I,\ell}(\omega)) = \epsilon_1^2 X_I\mathcal{A}^+ + \mathcal{O}(\epsilon_1^2 \epsilon_2^2, \epsilon_1^2 \epsilon_2^2 \Omega^2)  \,,
\end{equation}
where 
\begin{equation}
	X_I = \begin{cases}
		f_2+g_2 \qquad &\text{for } I=2 \,, \\
		-f_2+g_2 \qquad &\text{for } I=4 \,.
	\end{cases}
\end{equation}
Thus one can easily see that the time delay is sign-definite for both modes at leading order. The causality constraint  $(\omega \Delta T_{b,I,\ell}(\omega)) > -1$ translates into $X_I > - 1/(\epsilon_1^2 \mathcal{A}^+) \sim - 1/((\omega r_0) \epsilon_1^2)$ where $(\omega r_0) \epsilon_1^2 = \Omega \epsilon_2 (\epsilon_1/\epsilon_2)^2$ can be made arbitrarily large for $\epsilon_2 \ll \epsilon_1$ and hence we get
\begin{equation}
	f_2+g_2 > 0\,, \qquad -f_2+g_2 > 0 \,.   \label{eq:boundg2f2}
\end{equation}
Together, this imply that $g_2$ is positive. Note that this constraint includes the predictions from the low energy limit of string theory in $4$ dimensions \cite{TSEYTLIN1986391} in which $f_2=0$ since all the EFT corrections are proportional to powers of  $T_{\mu\nu}T^{\mu\nu}$, where $T^{\mu\nu}$ is the stress-energy tensor of the photon.

From now on, we will set $g_2=1$ which simply fixes the cutoff scale of the EFT to be determined solely by $\Lambda$. The $g_2=0$ case will be treated separately at the end of this section.

\paragraph{Bounds on $f_2$ and $g_3$}
There won't be any sign-definite contributions across both even and odd sectors in the $(f_2,g_3)$ plane, which allows for compact bounds, but we get one-sided bounds from the even sector, which can easily be explained.
\begin{figure}[!h]
	\begin{center}
		\includegraphics[width=0.5\textwidth]{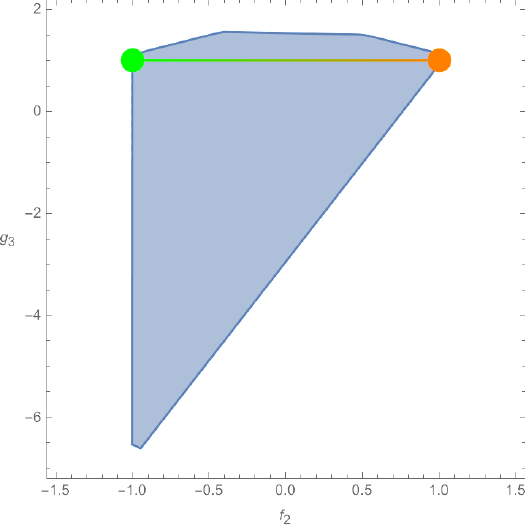}
	\end{center}
	\caption[$f_2 - g_3$ causality bounds, compared to scalar and axion partial UV completions.]{Causality bounds in the $(f_2, g_3)$ plane for $f_3=3f_2$, $h_3=0$, $f_4=f_2/2$, and $g_4=1$ which are consistent with the values of the scalar and axion partial UV completions. Once again, one can see that the points denoting the values of the partial UV completions, scalar in green and axion in orange, lie in the boundary of the causal region. The line connecting them involves values of a partial UV completion involving both the scalar and axion.}
	\label{fig:plotf2g3SA}
\end{figure}

Starting with the scalar and axion UV completion as depicted in Fig.~\ref{fig:plotf2g3SA}, we have both $f_3$ and $f_4$ that are proportional to $f_2$ to allow for a smooth transition between the scalar and axion UV completions. The line corresponds to a partial UV completion involving an axion and a scalar whose coefficients are simply the sum of the axion and scalar cases with a parameter (namely $f_2$ in this case) tuning the contribution from each. More precisely, the coefficients in this line are given by 
\begin{equation}
	\mathcal{W}_J= \cos{\theta} \  \mathcal{W}_J^{\text{scalar}} + \sin{\theta} \  \mathcal{W}_J^{\text{axion}} \ , \quad  \theta\in [0,\pi/2] \ ,
    \label{eq:UVLineWJ}
\end{equation}
so that at $\theta=0$ we have the purely scalar case and at $\theta=\pi/2$ we have the purely axionic case. The line precisely lies within our causality bounds whereas the scalar and axion UV completions exactly sit on the boundary.

Along this line joining the scalar and axion UV completions, we set $f_3$ and $f_4$ to be proportional to $f_2$ and hence we have a time delay in the form of $(\omega \Delta T^{(f_2)} + 3\omega \Delta T^{(f_3)} + \omega \Delta T^{(f_4)}/2) f_2 + \omega \Delta T^{(fg_3)} g_3 + ({\rm constant})$. Interestingly, the term multiplying $f_2$ is strictly positive in the even sector, whereas the one in front of $g_3$ is negative. When imposing the causality constraint we end up with an equation of the form $g_3 < {\rm (positive)} f_2 + {\rm (constant)}$ and hence the even sector is responsible for the top and/or left bounds (the same configuration will arise in the $(f_2,f_3)$ plane below), whereas we obtain the rest of the compact region thanks to the less rigid structure in the odd sector. Both UV partial completions and the whole line joining them are exactly contained within our causality bounds.

\begin{figure}[!h]
	\begin{center}
		\includegraphics[width=0.5\textwidth]{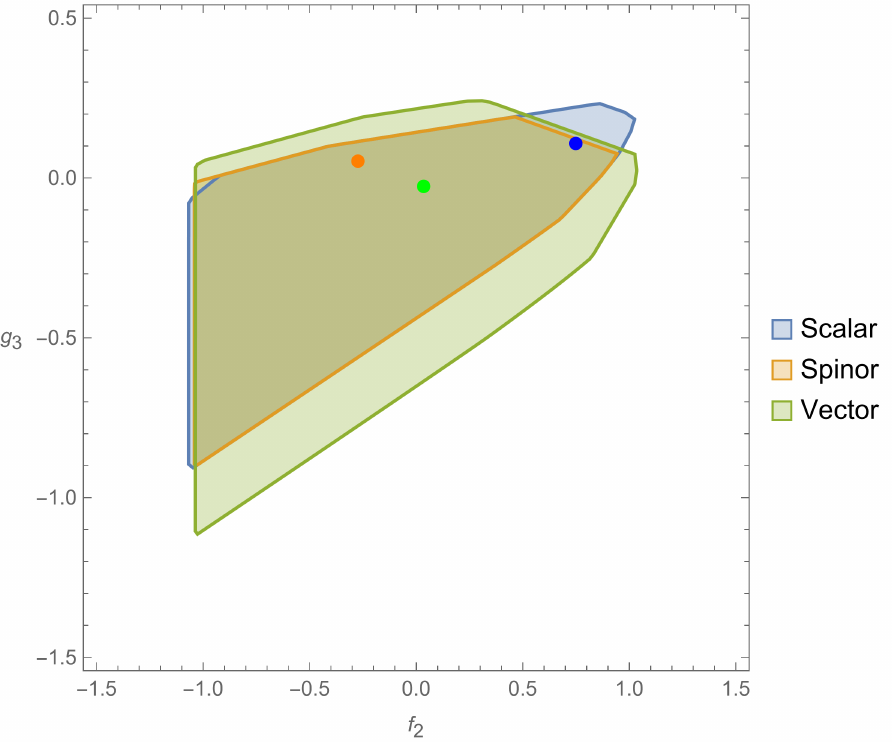}
	\end{center}
	\caption[$f_2 - g_3$ causality bounds, compared to QED partial UV completions.]{Causality bounds in the $(f_2, g_3)$ plane for various values of $(f_3,h_3,f_4,g_4)$ that are consistent with the scalar, spinor and vector QED partial UV completions, respectively in blue, orange, and green, and are given in Table \ref{tab:UVcomp}. It is important to note that the three bounds are superimposed but each are derived for different values of $(f_3,h_3,f_4,g_4)$.}
	\label{fig:plotf2g3QED}
\end{figure}
Let us now turn to 1-loop QED where three different set of coefficients are known to correspond to the scalar, spinor, and vector UV completions, as given in Table \ref{tab:UVcomp}. For each case independently, we leave $f_2$ and $g_3$ arbitrary and set the other values so that they match the ones of the relevant UV completion. It is clear that the structure is the same, as in the left and/or top bounds seen in Fig.~\ref{fig:plotf2g3QED} come from the even sector whereas the right and bottom ones arise from the odd sector.

It is interesting to note that the causality bounds for the three sets of different $(f_3,h_3,f_4,g_4)$ coefficients corresponding to the values they take in the scalar, spinor and vector one-loop QED partial UV completions are very similar. One main difference however is the lower bound, coming from the odd sector. It has the same slope for all three but is noticeably less constraining in the vector case. To understand this, let's turn to the generic equation governing such constraints. The slope is given by the coefficients of $f_2$ and $g_3$, but the vertical displacement of the bound is linear in $\sum_{J} \mathcal{W}_J (\omega \Delta T_{b,I,\ell}^{(J)}(\omega))$ where $J$ runs over the fixed values of $(f_3,h_3,f_4,g_4)$. From our analysis, we see that it is the $h_3$ contribution that is responsible for the discrepancy between the vector case and the other two. Let's focus on this and forget about the other Wilson coefficients. Then, the equation for the lower bound is given by

\begin{equation}
	g_3 > {\rm (positive)} f_2 + \mathcal{H}_3 h_3 + {\rm (constant)} \,,
\end{equation}
where the numerical factor $\mathcal{H}_3$ is optimized such that the combination $\mathcal{H}_3 h_3$ is maximized for a tighter bound. This in turn means that the vertical position of the lower bound only depends on the absolute value $|\mathcal{H}_3 h_3|$. Turning now to Table \ref{tab:UVcomp}, it is easy to see that the value for $|h_3|$ is an order of magnitude smaller in the vector case than both the scalar and spinor and hence, this explains the fact that causality is less constraining in the vector case.

Lastly, we analyze tree-level UV completions involving spin-2 fields. More precisely, we will look into the partial UV completions constructed in \cite{Henriksson:2021ymi,Henriksson:2022oeu,Haring:2022sdp}. The spin-2 partial UV completion in \cite{Henriksson:2021ymi} is constructed by integrating out a massive spin-2 with a minimal coupling to the photon, that is, a coupling $h_{\mu\nu}T^{\mu\nu}$ where $T^{\mu\nu}$ is the stress-energy tensor of the photon. Meanwhile, the partial UV completions in \cite{Haring:2022sdp} are constructed by on-shell amplitude methods. First, they construct the residue at the spin-2 pole and split this in two cases: parity even and odd. They allow for the freedom of adding arbitrary contact terms which are then fixed so that the amplitude has the desired Regge limit (growing as $\mathcal{O}\left(s^2, u^2, t^2\right)$). Without taking into account these additional contributions from contact terms, the even and odd partial UV completions indeed only propagate even or odd modes, but these additional contact terms do not respect the parity and the partial UV completions in \cite{Haring:2022sdp} propagate both even and odd modes. From a Lagrangian perspective, the construction of \cite{Haring:2022sdp} includes higher derivative couplings between the photon and the massive spin-2 field. 

\begin{figure}[!h]
	\begin{center}
		\includegraphics[width=0.46\textwidth]{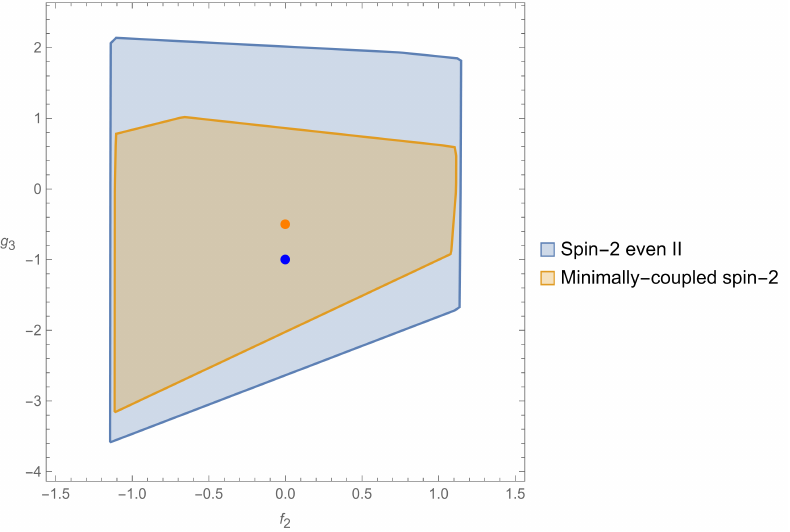}
  \includegraphics[width=0.4\textwidth]{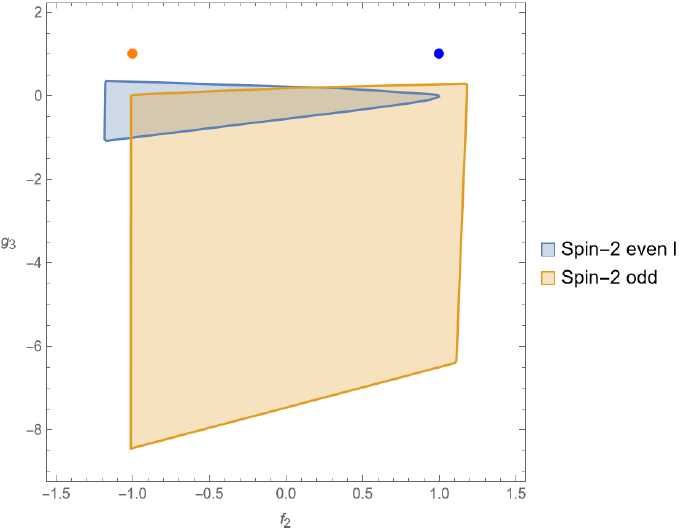}
	\end{center}
	\caption[$f_2 - g_3$ causality bounds, compared to spin-2 partial UV completions.]{Causality bounds in the $(f_2, g_3)$ plane with all other coefficients set to the values of the corresponding partial UV completions as in Table \ref{tab:UVcomp}. On the left side, we see the partial UV completions with causal propagation. Note that the spin-2 even II partial UV completion is related to the massive graviton one by rescaling $g_2$ by a factor of $2$ and keeping all other parameters unchanged. On the right side, we plot the case of the partial UV completions of \cite{Haring:2022sdp} that have acausal propagation. If we considered those partial UV completions, but removed the addition of the contact term that gives a $g_3$ contribution, i.e. setting $g_3=0$, then both would lie in the boundary of the causal region.}
	\label{fig:plotf2g3spin2}
\end{figure}

It was already noticed in \cite{Haring:2022sdp} that these partial UV completions lie outside the bounds found in \cite{Henriksson:2021ymi,Henriksson:2022oeu}. While those bounds require that the Regge limit is strictly softer than $s^2$ this is a requirement on the UV theory not on the EFT or the partial UV completion. One can assume that at very high energies there is a UV completion with the desired behavior. From our analysis we see in Fig. \ref{fig:plotf2g3spin2}. that some of these partial UV completions also lie outside the region of causal IR propagation. We highlight again that this does not require any UV assumptions, thus the EFTs constructed in this way do not have causal propagation. In other words, the non-minimal coupling between the photon and the spin-2 in the partial UV completion lead to acausal propagation.

\begin{figure}[!h]
	\begin{center}
		\includegraphics[width=0.5\textwidth]{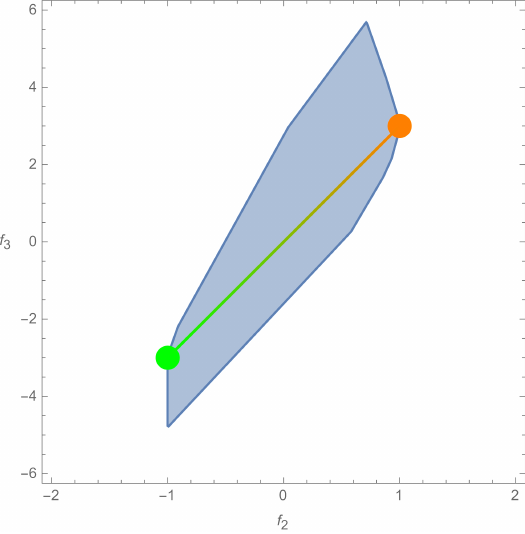}
	\end{center}
	\caption[$f_2 - f_3$ causality bounds, compared to scalar and axion partial UV completions.]{Causality bounds in the $(f_2, f_3)$ plane for $g_3=1$, $h_3=0$, $f_4=f_2/2$, and $g_4=1$ which are consistent with the values of the scalar and axion partial UV completions. One can see that the points denoting the values of the partial UV completions, scalar in green and axion in orange, lie in the boundary of the causal region. The line connecting them corresponds to a partial UV completion involving both a scalar and an axion.}
	\label{fig:plotf2f3}
\end{figure}

\paragraph{Bounds on $f_2$ and $f_3$} 
Neither $f_2$ nor $f_3$ contribute in a sign-definite way to the total time delay (see Table \ref{tab:signdef}) over both even and odd sectors and hence there are no obvious one-sided bounds. This freedom is welcome as it allows for compact causality regions in the $(f_2, f_3)$ plane as can be seen in Fig.~\ref{fig:plotf2f3}.

To better understand these bounds, we can analyze the even and odd sectors separately. Note that we have set $f_4=f_2/2$ so that one can use $f_2$ as a dialing parameter to extrapolate between the scalar and axion UV completions, as can be seen in Fig.~\ref{fig:plotf2f3}. The time delay now reads $(\omega \Delta T^{(f_2)} + \omega \Delta T^{(f_4)}/2)f_2 + \omega \Delta T^{(f_3)} f_3 + {\rm (constant)}$ with the terms multiplying $f_2$ and $f_3$ in this expression being respectively strictly positive and negative in the even sector, similarly to the previous $(f_2,g_3)$ analysis. This means that the even sector will provide upper and/or left bounds depending on the magnitude of the positive slope, $f_3 < {\rm (positive)} f_2 + {\rm (constant)}$. When carefully processing the bounds, we confirm that the left and top bounds come from the even sector, whereas the odd sector has a richer structure and produces the remaining constraints, hence explaining the hard changes of slopes on the right side.

Once again, we plot the whole line joining the scalar and axion UV completion points, parametrized by $f_2$ going from $1$ to $-1$, in Fig. \ref{fig:plotf2f3}. We have the same qualitative behavior, i.e. the end points corresponding to the scalar and axion UV completions are on the boundary of the causality bounds whereas the rest of the line is within.

\paragraph{Bounds on $f_2$ and $h_3$}
\begin{figure}[!h]
	\begin{center}
		\includegraphics[width=0.5\textwidth]{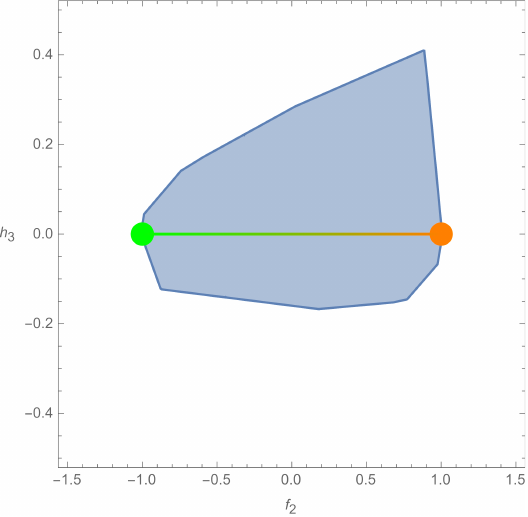}
	\end{center}
	\caption[$f_2 - h_3$ causality bounds, compared to scalar and axion partial UV completions.]{Causality bounds in the $(f_2, h_3)$ plane for $f_3=3f_2$, $g_3=1$, $f_4=f_2/2$, and $g_4=1$ which are consistent with the values of the scalar and axion partial UV completions. The points denoting the values of the partial UV completions, scalar in green and axion in orange, lie in the boundary of the causal region as in the previous cases. Similarly, the line denoting the mixed partial UV completion of a scalar and axion also lies within the causal region.}
	\label{fig:plotf2h3SA}
\end{figure}
When going to the $(f_2,h_3)$ plane, we choose to set all remaining coefficients in such a way that they can satisfy either the scalar or the axion partial UV completions. This is done by allowing the remaining Wilson coefficients to depend on $f_2$, which will then once again act as a dialing parameter. One can extrapolate between both end points and obtain a full segment of partial UV completions as shown in Fig.~\ref{fig:plotf2h3SA}. The segment is fully contained within our causality bounds, with the scalar and axion points exactly lying on the boundary, as was previously the case in the $(f_2,f_3)$ and $(f_2,g_3)$ planes, respectively plotted in Figs.~\ref{fig:plotf2f3} and \ref{fig:plotf2g3SA}.

In this example, neither $f_2$ nor $h_3$ enjoy a sign-definite contribution to the time delay in either of the even and odd sectors. This way, the latter contributed respectively to left and right-sided bounds.

\begin{figure}[!h]
	\center
	\includegraphics[height=5cm]{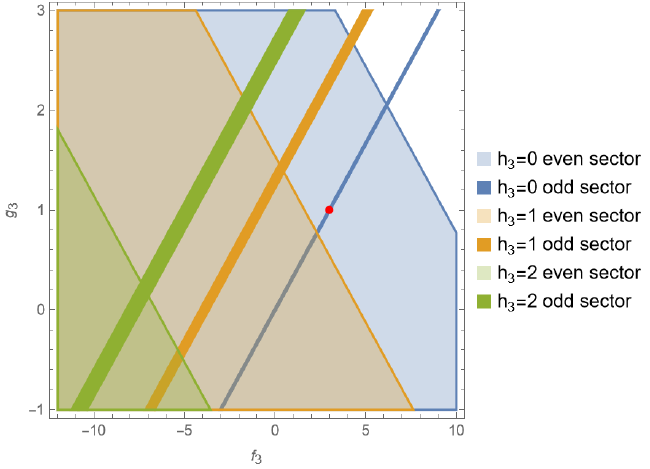}
	\hspace{0.3cm} \includegraphics[height=5cm]{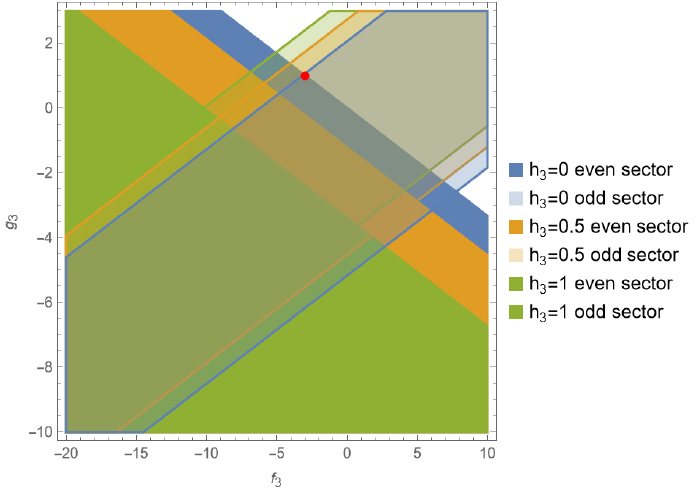}
	\caption[$f_3 - g_3$ causality bounds for several values of $h_3$.]{Causality bounds on $(f_3,g_3)$ plane for the odd and even sectors with the values of the other coefficients fixed as in the scalar partial UV completion on the left and the axion one on the right. The red dot indicates the value of the coefficients for the corresponding partial UV completion. The final causal region is obtained as the intersection of the causal regions for both the odd and even sectors. This region won't be compact in the negative $(f_3,g_3)$ direction. Note that for the scalar (left plot) the even sector is transparent while the odd is solid and for the axion (right plot) this is reversed for clarity of the plots.}
	\label{fig:plotF3G3}
\end{figure}

\begin{figure}[!h]
	\center
	\includegraphics[height=5cm]{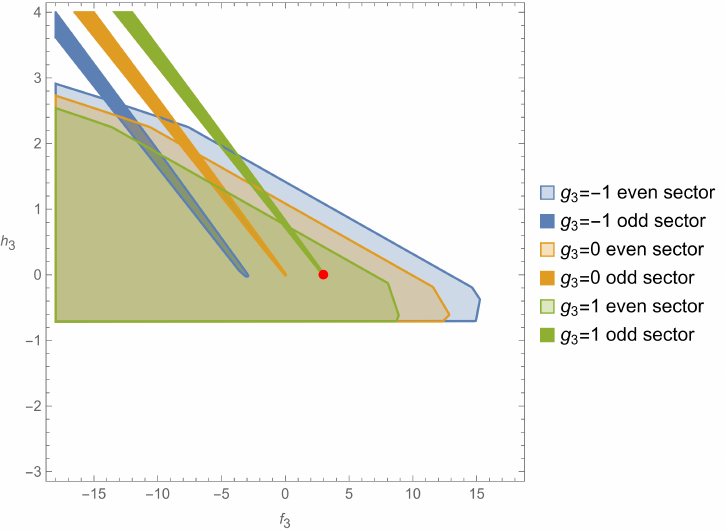}
	\hspace{0.3cm} \includegraphics[height=5cm]{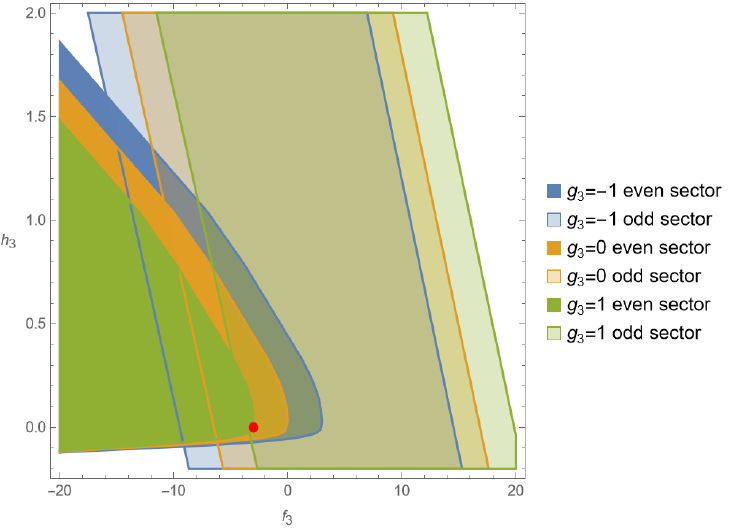}
	\caption[$f_3 - h_3$ causality bounds for several values of $g_3$.]{Causality bounds on $(f_3,h_3)$ plane for the odd and even sectors with all other parameters set by the values of the scalar (left) or axion (right) partial UV completions. The final causal region is obtained as the intersection of the causal regions for both the odd and even sectors. The red dot represents the corresponding values of the partial UV completions and lies in the boundary of the causal region.}
	\label{fig:plotF3H3}
\end{figure}

\begin{figure}[!h]
	\center
	\includegraphics[height=5cm]{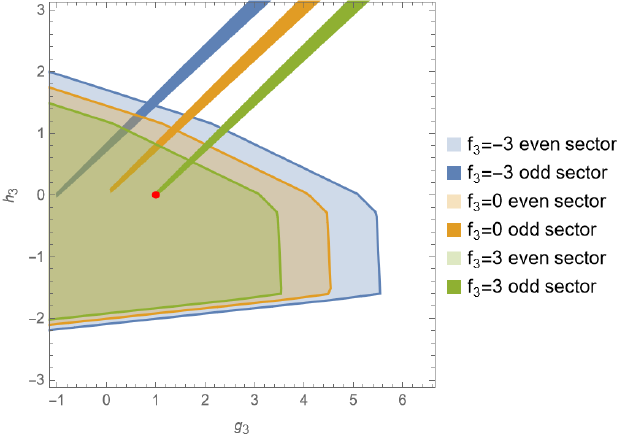}
	\hspace{0.3cm} \includegraphics[height=5cm]{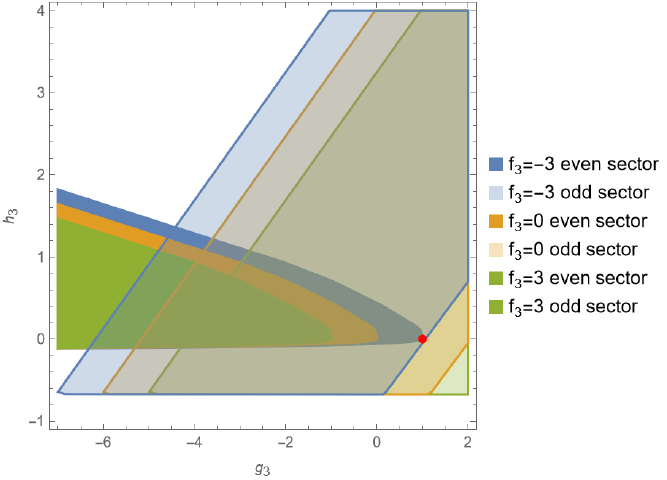}
	\caption[$g_3 - h_3$ causality bounds for several values of $f_3$.]{Causality bounds on $(g_3,h_3)$ plane for the odd and even sectors with all other parameters set by the values of the scalar (left) or axion (right) partial UV completions. As in the previous cases, the final causal region is obtained as the intersection of the causal regions for both the odd and even sectors and the red dot represents the corresponding partial UV completion.}
	\label{fig:plotG3H3}
\end{figure}

\paragraph{Bounds on $f_3$, $g_3$, and $h_3$}
We now turn to analyze the bounds for the dimension $10$ coefficients. For these parameters we obtain bounds of the following form
\begin{align}
	f_3+3g_{3}&<X_{u^l_{2}}+Y_{u^l_{2}}h_3 \ , \\
 -X_{u^l_{4}}-Y_{u^l_{4}}h_3<f_3-3g_{3}+4h_3&<X_{u^l_{4}}+Y_{u^l_{4}}h_3\ ,
\end{align}
where $X_{u^l_{2,4}}$ depends on the other amplitude parameters as observed in Figs.~\ref{fig:plotF3G3}, \ref{fig:plotF3H3}, and \ref{fig:plotG3H3} and on the specific background. Thus, the value of $X_{u^l_{4}}$ and $Y_{u^l_{4}}$ can be different for the upper and lower bounds since it will be optimized to have the tightest bounds. This analysis simplifies in the even sector for the choice of the amplitude parameters as in the axion partial UV completion giving $X_{u^l_{2}}=1/(\epsilon_1^2 \epsilon_2^2 \mathcal{B}^+)$ and in the odd sector for the scalar partial UV completion giving $X_{u^l_{4}}=1/|\omega \Delta T_{b,4,\ell}^{(f_3)}(\omega)|$. 

On the $(f_3,g_3)$ plane, we see that the even sector only gives rise to an upper bound due to the sign definite contribution from both $g_3$ and $f_3$. This upper bound becomes stronger as we increase $h_3$ since this can contribute with a negative time delay that needs to be balanced by a more positive contribution from $f_3$ and $g_3$. Meanwhile, the odd sector contributions are not sign definite and give both upper and lower bounds. The strength of these two-sided bounds does not change drastically when varying $h_3$ since the background is optimized so that the width of the bound on the $f_3$ direction is the smallest possible bound. While the strength of the bounds in the odd sector does not vary drastically with $h_3$, the location does change and moves towards the negative $f_3$ direction as $h_3$ is increased.  Note that the values of $h_3$ considered here are chosen so that we have a non-vanishing causal region, which in these cases requires $h_3\gtrsim0$ as can be seen below.

In the $(f_3,h_3)$ plane, we should expect upper and lower bounds. This is observed in Fig.~\ref{fig:plotF3H3} where the bounds arising from the even and odd sectors are plotted separately. We can appreciate that each sector separately does not give a compact region; it is the combination of both that gives rise to the compact causal region. On the even sector, the contribution of $f_3$ is sign definite, so we can only get an upper found, while in the odd sector, we have both an upper and lower bound. As in the $(f_3,g_3)$ bounds, we have tight bounds on the odd sector for the scalar partial UV completion case since the contributions from all the other amplitude parameters vanish and the actual values of the UV completion lie on the boundary. Similarly, we observe that the axion partial UV completion is in the boundary of the even sector causal region. The bounds are not as strong as in the scalar case due to the sign-definite contribution of $f_3$. If we were to change the contributions from the dimension $12$ coefficients, we will be able to tune the background to obtain a negative contribution to the odd sector time delay which leads to a stronger bound. The effect of this is to move the odd sector causal region further up so that it no longer intersects with the even sector causal region. 

Last, we analyze the bounds on the $(g_3,h_3)$ plane. The analysis performed in the $(f_3,h_3)$ case is identical for the $(g_3,h_3)$ bounds since the contribution of $g_3$ is degenerate with the one in $f_3$. In the even sector, the $f_3$ and $g_3$ contributions have the same sign, but in the odd sector, they have opposite signs, so the only difference will be the change in the sign of the slope of the odd sector bounds as seen in Fig. \ref{fig:plotG3H3}.

\paragraph{Bounds on $f_4$, $g_4$, and $g^'_4$}
The leading order contribution to the time delay from the dimension $12$ operators is
\begin{equation}
	\omega \Delta T_{b,\{2,4\},\ell} \supset 12 \left(\pm 2f_4+g_{4}\right)\epsilon_1^2 \epsilon_2^2 \Omega^2 \mathcal{C}^+  \ , \label{eq:Dim12Contrib}
\end{equation}
where $\mathcal{C}^+>0$ is defined in Eq.~\eqref{eq:positiveInt} and the $+$($-$) sign comes from the even (odd) sector. Thus, we expect to get bounds of the form:
\begin{equation}
	2f_4+g_{4}>Z_{u^l_{2}} \ , \quad -2f_4+g_{4}>Z_{u^l_{4}} \ ,
\end{equation}
where $Z_{u^l_{2,4}}$ are numbers determined by the optimization of the background which also depend on the other scattering amplitude parameters. They depend very weakly on $f_2$ and only become slightly tighter as we approach $f_2=\pm 1$ boundaries. Together this give rise to a lower bound $g_4>(Z_{u^l_{2}}+Z_{u^l_{4}})/2$ as predicted in Table~\ref{tab:signdef}. By optimizing our bounds, we are able to find 
\begin{equation}
	g_4>0 \ .
\end{equation}
this lower bound can be stronger for specific $f_3, g_3, h_3$ values.

\begin{figure}[!h]
	\center
	\includegraphics[height=5cm]{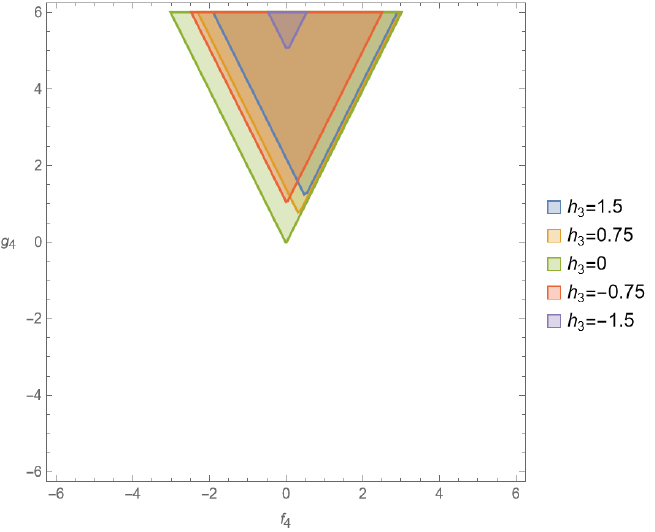}
	\includegraphics[height=5cm]{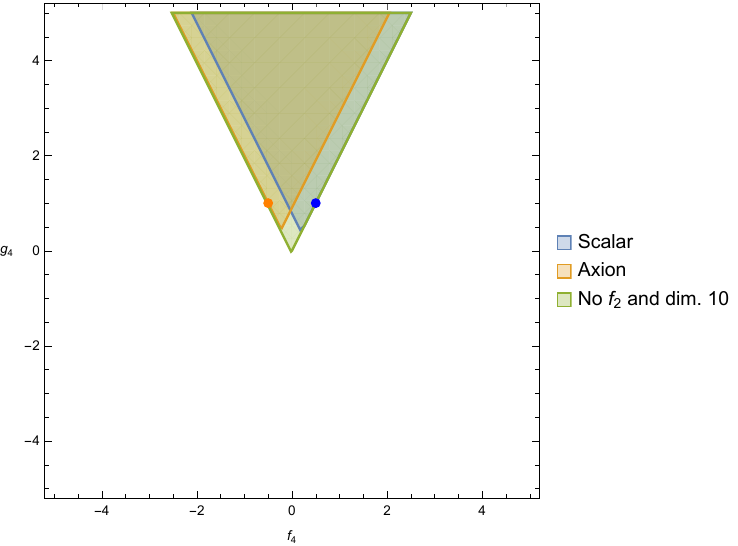}
	\caption[$f_4 - g_4$ causality bounds, compared to scalar and axion partial UV completions.]{Bound on the $(f_4,g_4)$ plane. The right-hand side has $f_2=f_3=g_3=0$ and varying $h_3$. On the left-hand side, we choose the parameters not being plotted to have the corresponding values of the partial UV completion as in Table \ref{tab:UVcomp} for the scalar (blue) and axion (orange) and $f_2=f_3=g_3=h_3=0$ for the green region. Note that the plot extends infinitely to the top since we cannot bound $g_4$ from above.}
	\label{fig:Boundsf4g4}
\end{figure}

In Fig.~\ref{fig:Boundsf4g4}, we observe the bounds on the $(f_4,g_4)$ plane. The lower bound on $f_4$ becomes stronger when $f_3+3g_3>0$ since these terms can give a negative contribution to the time delay of $u_2^\ell$ which in turn requires a larger positive value of $f_4$ to not obtain an observable time advance. Meanwhile, a non-zero $f_3-3g_3$ contribution can be tuned to give a negative time delay for $u_2^\ell$ so that the upper bound on $f_4$ becomes stronger. Both the lower and upper bounds on $f_4$ are improved for a non-zero $h_3$. This is a large effect compared to that of $f_3$ and $g_3$ since these parameters have a suppression of $B^2/R^2$ in the integrand of the time delay with respect to a part of the $h_3$ contribution (See Table \ref{tab:signdef}). In the odd sector, the non-suppressed $h_3$ contribution is positive definite so a positive $h_3$ does not improve largely the upper bounds on $f_4$. This is not the case in the even sector so a positive $h_3$ does improve significantly the lower bounds. For a negative $h_3$ the non-suppressed contributions from both the even and odd sectors can be made negative and largely improve both upper and lower bounds.

Note that up to the EFT order we are working on, we have no contribution from the Wilson coefficient $c_7$, or equivalently from the scattering amplitude parameter $g'_4$. The contribution of $g'_4$ ($c_7$) starts at order $\epsilon_1^2 \epsilon_2^4$. If we choose to include this contribution, but still neglect all the WKB corrections (since they are not calculable in our setting), it would require that we neglected $\epsilon_1^2\epsilon_2^2/\Omega^2$ corrections, but include $\epsilon_1^2 \epsilon_2^4$, $\epsilon_1^2 \epsilon_2^4\Omega^2$, $\epsilon_1^2 \epsilon_2^6\Omega^2$ terms. The latter contributions arise from operators with dimension $14$ and $16$. In this new setting, the validity of the EFT will require a large $\Omega$ bounded as $\epsilon_2^{-1}\ll\Omega \ll \epsilon_2^2$ which will naturally enhance the new contributions from the operators of dimension $14$ and $16$. While we can tune the background solution to decrease the effect of these operators and enhance that of $g'_4$, these explorations are beyond the scope of this work. Instead, we can parametrize the amplitude as in \cite{Henriksson:2021ymi,Henriksson:2022oeu}, in which case we can obtain bounds for both $g_{4,1}$ and $g_{4,2}$ parameters defined in Table \ref{tab:conversion}. This simply follows from their leading order contribution to the time delay which will give bounds of the form
\begin{equation}
	2f_4+g_{4,1}+2g_{4,2}>A_{u^l_{2}} \ , \quad -2f_4+g_{4,1}+2g_{4,2}>A_{u^l_{4}} \ .
\end{equation}

\subsubsection{Case $g_2=0$}
\begin{figure}[!h]
	\center
	\includegraphics[width=0.5 \textwidth]{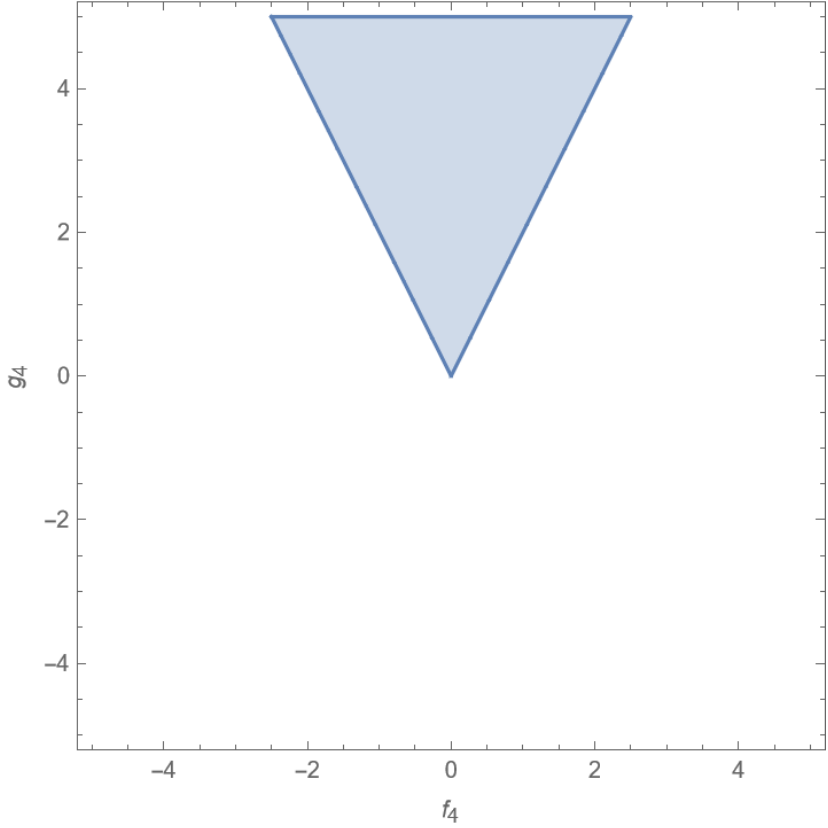}
	\caption[$f_4 - g_4$ causality bounds in the case $g_2=0$.]{Bounds on the $(f_4,g_4)$ plane for the case $g_2=0$. These bounds are independent on the value of all the other scattering amplitude parameters.}
	\label{fig:Boundsf4g4G2isZero}
\end{figure}

One can show that when $g_2=0$ causal propagation implies\footnote{Strictly speaking we obtain bounds of the form $-\epsilon<\mathcal{W}_J<\epsilon$, where $\epsilon$ is smaller than the WKB and EFT contributions that we are neglecting so we can effectively take $\epsilon=0$.} 
\begin{equation}
	f_2=f_3=g_3=h_3=0 \ ,
\end{equation}
Once both $f_2$ and $g_2$ vanish, the requirements for the validity of the EFT in Eq.~\eqref{eq:HierarchyParam} change. Instead, we have the less restrictive situation where:
\begin{equation}
	\epsilon_1,\epsilon_2   \ll 1\ , \quad  \epsilon_2^{1/2}\ll\Omega \ll \frac{1}{\epsilon_2}  \ . \label{eq:HierarchyParamG2Zero}
\end{equation}
Using this new setup, we can find the causality bounds on the $(f_4,g_4)$ plane as shown in Fig.~\ref{fig:Boundsf4g4G2isZero}. We can see that the bounds are equal to the $g_2\neq0$ case with all other amplitude parameters set to zero, that is, they are given by
\begin{equation}
	2f_4+g_{4}>0 \ , \quad -2f_4+g_{4}>0 \ ,
\end{equation}
which implies that $g_4$ is positive. Note that since we can only obtain a one sided bound for the $g_4$ coefficient within our EFT setup, we are only able to constrain its sign. One should note that $f_4$ will be required to vanish if $g_4$ were to vanish.

\section{Comparison with positivity bounds}
\label{sec:CompPosBounds}

Now that we have derived causality bounds for a variety of interesting $2d$ slices of the low-energy parameter space of the massless photons EFT \eqref{eq:Lagrangian}, we wish to compare them to positivity bounds. To do so, we will use the most recent set of positivity bounds computed in \cite{CarrilloGonzalez:2023cbf}. Note that I haven't been involved in the derivation of the aforementioned bounds and that they are purely the work of my collaborators. We will not enter into details on the method behind them and refer the reader to Section \ref{sec:IntroPosBounds} for a brief introduction, the full derivation being explained in \cite{CarrilloGonzalez:2023cbf}.

Most of the features of the causality bounds reproduced in Figures \ref{fig:plotf2g3SA}--\ref{fig:Boundsf4g4G2isZero} have already been explained in Section \ref{ssec:CausBoundsVector} so we will focus on their comparison with positivity bounds and discuss their complementarity. In all the following plots, the regions constrained by IR causality and positivity will be bounded by a thick line and a dashed line respectively.

\begin{figure}[!h]
	\begin{center}
		\includegraphics[width=0.5\textwidth]{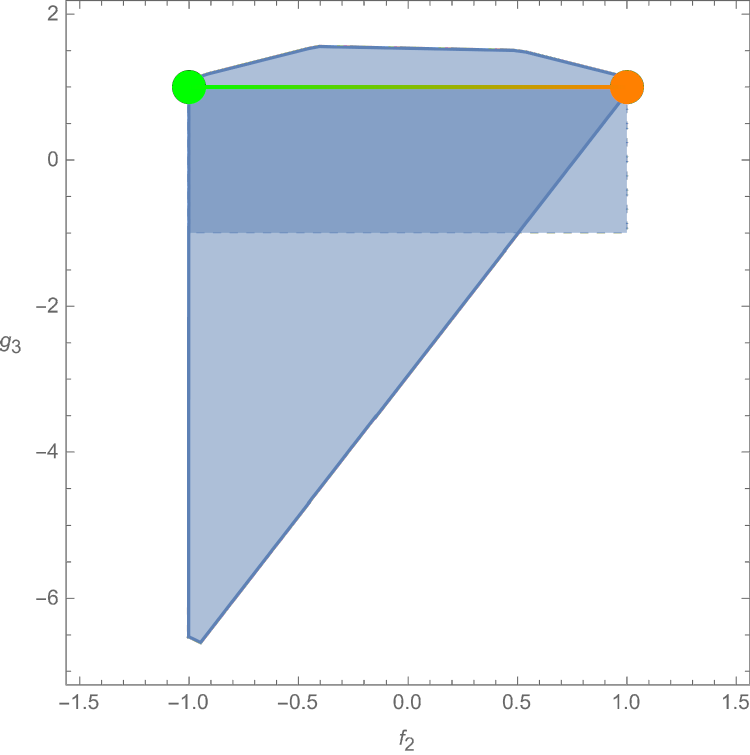}
	\end{center}
	\caption[$f_2 - g_3$ causality and positivity bounds, scalar and axion partial UV completions.]{Causality bounds in the $(f_2, g_3)$ plane for $f_3=3f_2$, $h_3=0$, $f_4=f_2/2$, and $g_4=1$ which are consistent with the values of the scalar and axion partial UV completions. The causality bounds are within the thick boundary whereas the positivity ones are within the dashed one.}
	\label{fig:f2g3UVLineComparison}
\end{figure}

\paragraph{Bounds on $f_2$ and $g_3$}
Starting with values extrapolating between the scalar and axion UV completion, the bounds on $f_2$ and $g_3$ are plotted in Fig.~\ref{fig:f2g3UVLineComparison}. Both partial UV completions and the line joining them are allowed by both causality and positivity bounds. The positivity bounds, delimited by the dashed boundary, are more constraining in this case but get refined when also considering the causality bounds.

\begin{figure}[!h]
	\begin{center}
		\includegraphics[width=0.5\textwidth]{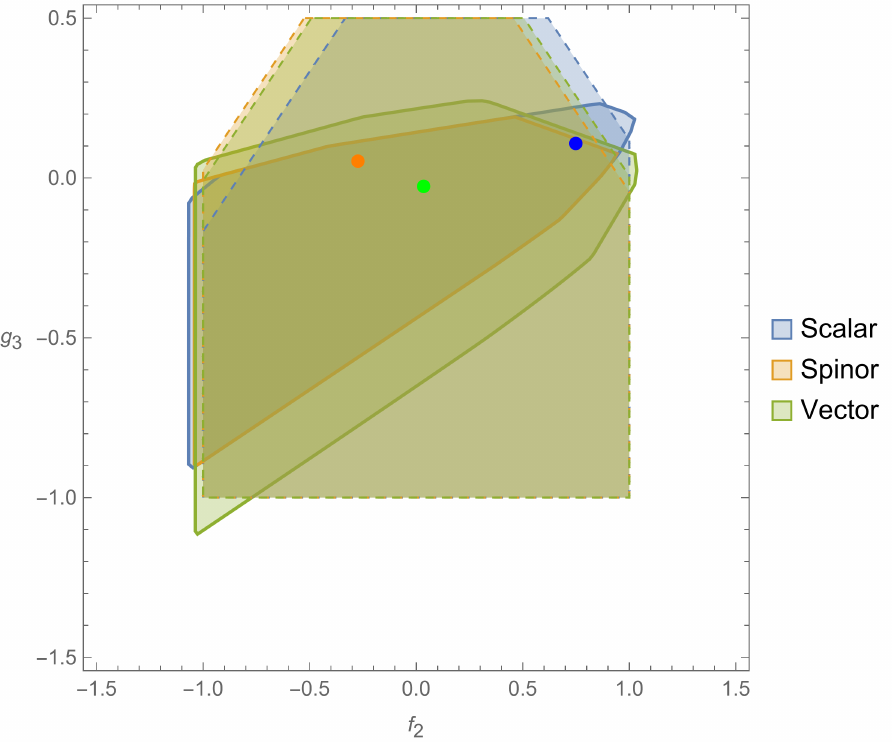}
	\end{center}
	\caption[$f_2 - g_3$ causality and positivity bounds, scalar and axion partial UV completions.]{Causality (thick line) and positivity (dashed line) bounds in the $(f_2, g_3)$ plane for various values of $(f_3,h_3,f_4,g_4)$ that are consistent with the scalar, spinor and vector QED partial UV completions.}
	\label{fig:f2g3QEDComparison}
\end{figure}

Turning now to the QED partial UV completions, the causality bounds are compared to the region compatible with positivity in Fig.~\ref{fig:f2g3QEDComparison}. It is interesting to note that the causality bounds lie nearly entirely in the positivity bounds and hence this is a case where IR causality is stronger than positivity. This is the first such example and it proves that causality in the IR and in the UV are different requirements and that, a priori, one is not necessarily more constraining than the other in practice. Note finally that all QED partial UV completions lie within both causality and positivity bounds, as expected for consistency.

\begin{figure}[!h]
	\begin{center}
		\includegraphics[width=0.46\textwidth]{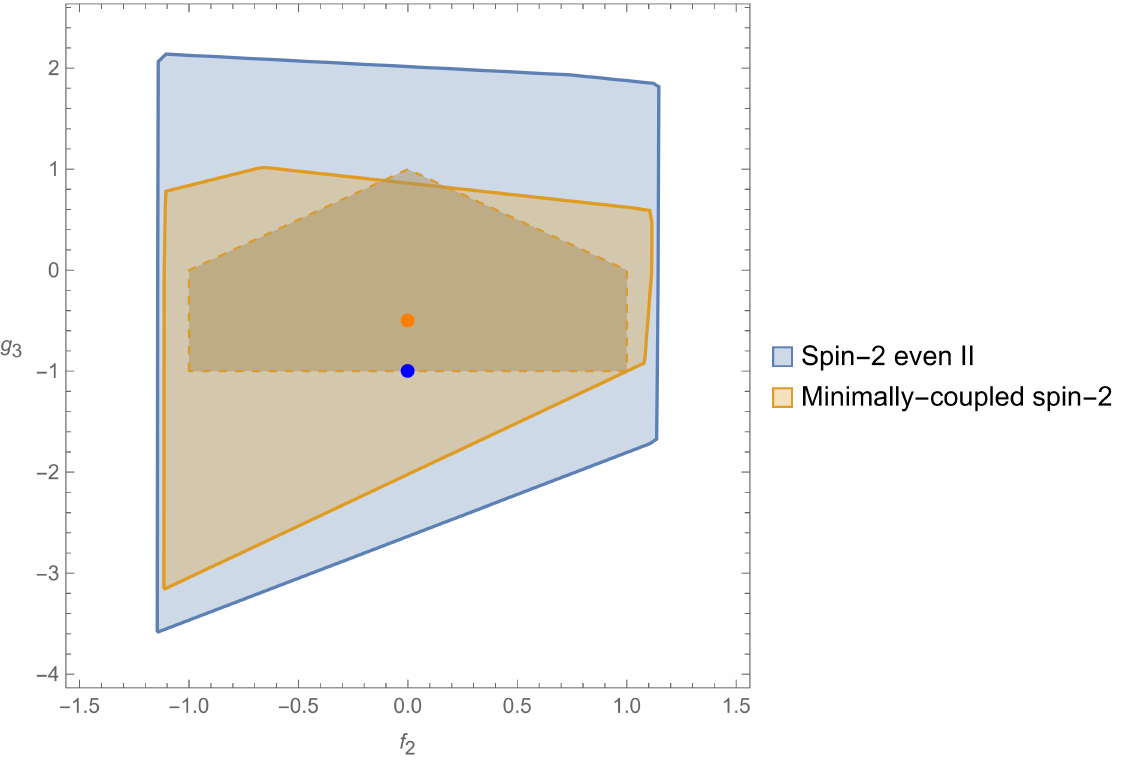}
  \includegraphics[width=0.4\textwidth]{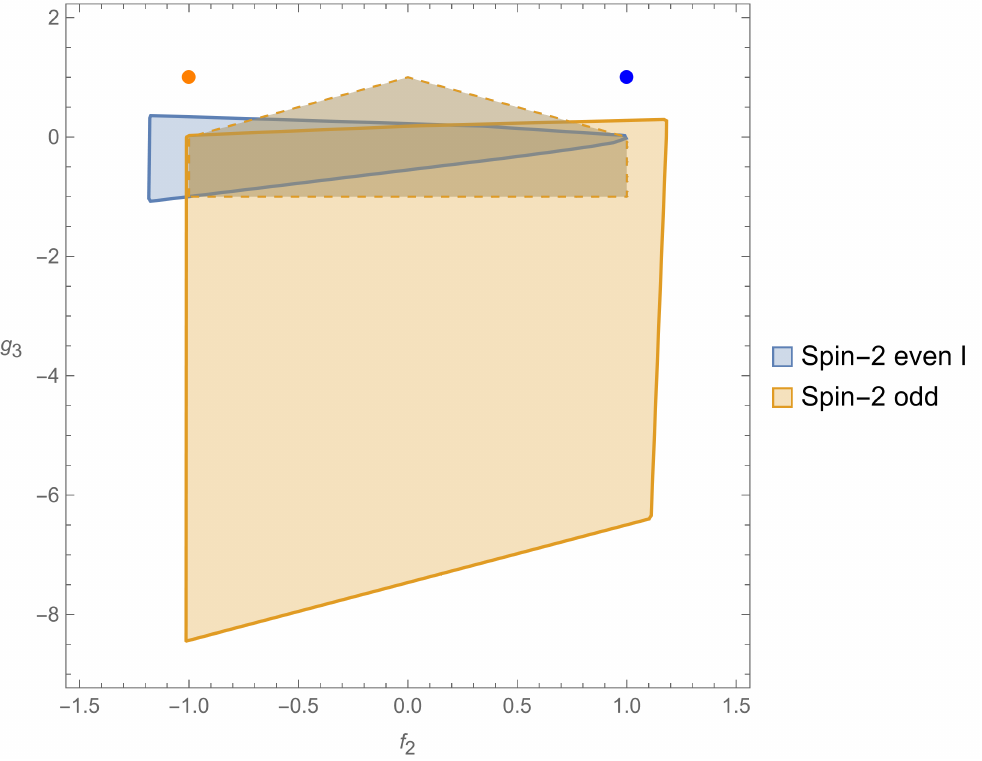}
	\end{center}
	\caption[$f_2 - g_3$ causality and positivity bounds, spin-2 partial UV completions.]{Causality (thick line) and positivity (dashed line) bounds in the $(f_2, g_3)$ plane with all other coefficients set to the values of the corresponding partial UV completions as in Table \ref{tab:UVcomp}. The partial UV completions with causal propagation appear on the left whereas the ones on the right do not agree with either of the causality and positivity bounds. Note that the positivity bounds for both cases are the same, hence why only one region appears with a dashed contour.}
	\label{fig:f2g3Spin2Comparison}
\end{figure}

Finally, the causality and positivity bounds for the multiple spin-2 partial UV completions are presented in Fig.~\ref{fig:f2g3Spin2Comparison}. The positivity bounds are obtained through a different procedure than the causality ones and the result is independent of the dimension-$12$ operators. According to Table \ref{tab:UVcomp}, all dimension-$8$ and $10$ free coefficients (i.e. with the exception of $f_2$ and $g_3$ in this specific case) are identical in each of the four proposed partial UV completions. Hence, there is one and only one region given by positivity for each of the four cases. For the two that are causal, they also lie within the positivity bounds and the latter are more constraining. The remaining two `UV completions' (on the right side) are also disfavoured by positivity as they lie outside the allowed regions. If these two points were only ruled out by the causality bounds, one could have had doubts about the validity of the result, but it is now confirmed by the well-established positivity method. Ruling out potential UV completions is thus another powerful application of both causality and positivity bounds, especially when taken together.

\paragraph{Bounds on $f_2$ and $f_3$} 
In the $(f_2,f_3)$ plane the positivity bounds exactly reduce to the line extrapolating between the scalar and axion UV completions as seen in Fig.~\ref{fig:plotf2f3}. In this case, the positivity bounds coincide with a one-dimensional region of the parameter space that admits a partial UV completion and is thus stronger than the causality bounds.

\begin{figure}[!h]
	\begin{center}
		\includegraphics[width=0.5\textwidth]{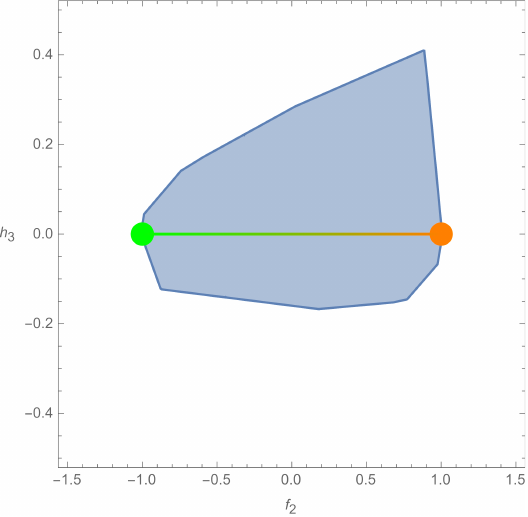}
	\end{center}
	\caption[$f_2 - h_3$ causality and positivity bounds, scalar and axion partial UV completions.]{Causality (thick line) and positivity (dashed line) bounds in the $(f_2, h_3)$ plane for $f_3=3f_2$, $g_3=1$, $f_4=f_2/2$, and $g_4=1$ which are consistent with the values of the scalar and axion partial UV completions.}
	\label{fig:f2h3UVLineComparison}
\end{figure}

\paragraph{Bounds on $f_2$ and $h_3$}
Similarly to the $(f_2,h_3)$ case, when setting the free Wilson coefficients to be compatible with the scalar and axion partial UV completions in the $(f_2,h_3)$ plane, the causality bounds are very close to shrinking to the single UV completion segment. The causality bounds do not improve the positivity constraints as the latter are fully contained in the former, as seen in Fig.~\ref{fig:f2h3UVLineComparison}.

\begin{figure}[!h]
	\center
	\includegraphics[height=5cm]{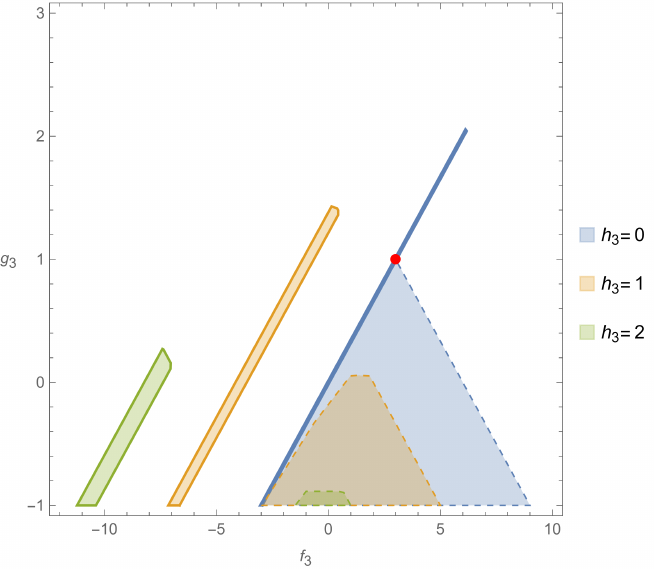}
	\hspace{0.3cm} \includegraphics[height=5cm]{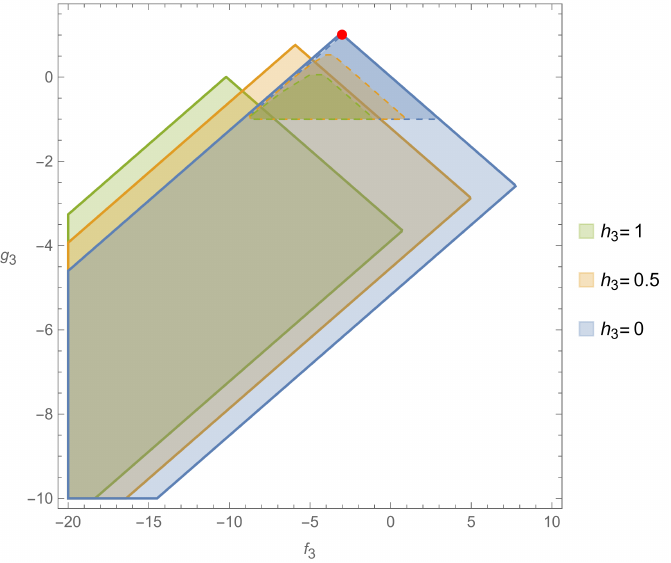}
	\caption[$f_3 - g_3$ causality and positivity bounds for several values of $h_3$.]{Causality (thick line) and positivity (dashed line) bounds on $(f_3,g_3)$ plane with the values of the other coefficients fixed as in the scalar partial UV completion on the left and the axion one on the right.}
	\label{fig:f3g3ScalAxionComparison}
\end{figure}

\paragraph{Bounds on $f_3$, $g_3$, and $h_3$}
Let's start by discussing the $(f_3,g_3)$ plane where all other coefficients are set to respect the values of either the scalar or axion UV completions as written in Table \ref{tab:UVcomp}. The bounds are reported in Fig.~\ref{fig:f3g3ScalAxionComparison}. The first thing to note is that even though some allowed regions exist both from causality and positivity bounds for $h_3=1$ and $h_3=2$, their union is empty and hence these values are discarded in the sense that they cannot correspond to theories endowed with causal propagation either in the IR or the UV. Both scalar and axion partial UV completions (corresponding to the case $h_3=0$) are allowed by the two methods. In this case, the causality bounds are much more constraining than their positivity counterparts, even though the latter provide a slightly lower upper value. 

For the right side of Fig.~\ref{fig:f3g3ScalAxionComparison}, once again the axion partial UV completion is consistent with both methods. This time the positivity bounds are more constraining than the bounds obtained from IR causality but it is still interesting to compare them since for instance, their union seems to become vanishingly small when $h_3$ gets closer to $1$. The combination of the bounds, even though the positivity ones are not that strong in this case, is still a powerful discriminating tool.

\begin{figure}[!h]
	\center
	\includegraphics[height=5cm]{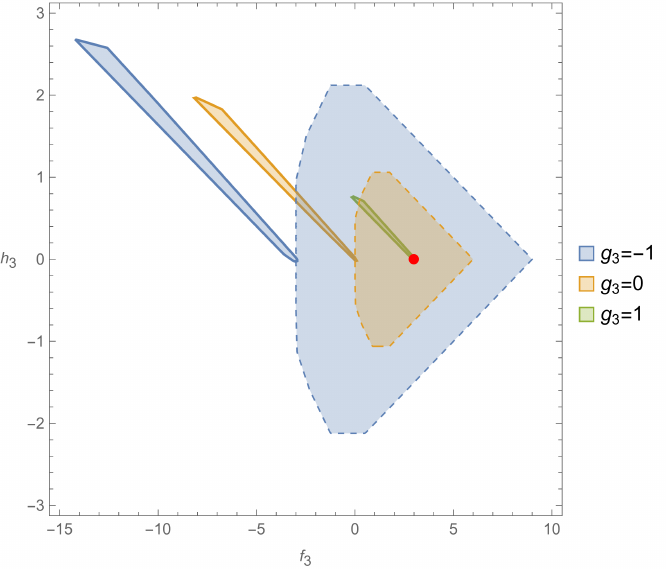}
	\hspace{0.3cm} \includegraphics[height=5cm]{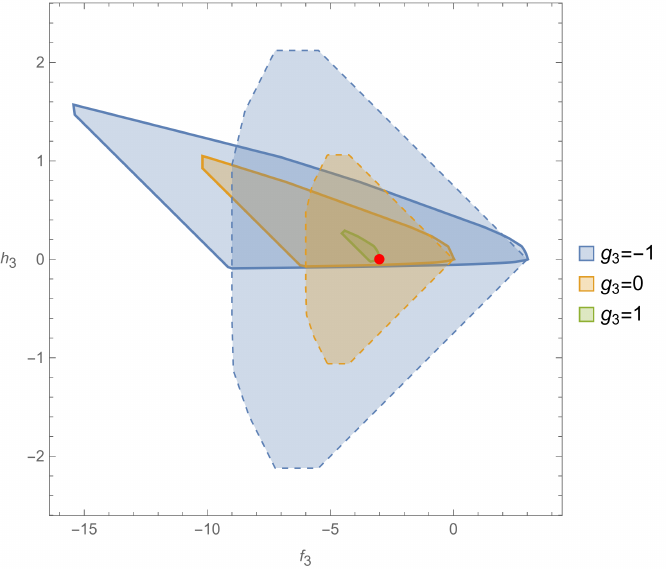}
	\caption[$f_3 - h_3$ causality and positivity bounds for several values of $g_3$.]{Causality (thick line) and positivity (dashed line) bounds on $(f_3,h_3)$ plane with all other parameters set by the values of the scalar (left) or axion (right) partial UV completions.}
	\label{fig:f3h3ScalAxionComparison}
\end{figure}

In the $(f_3,h_3)$ plane, where values are consistent with the scalar partial UV completion (left panel of Fig~\ref{fig:f3h3ScalAxionComparison}), one can see that the causality bounds are much more constraining than their positivity counterparts, except in the case $g_3=1$ where the latter reduce to the single scalar UV completion point. For the other cases, even though both sets of bounds have a finite size, their union also reduce to a single point, once again proving the power of this combined approach. Turning now to the right panel (axion), both causality and positivity bounds have similar shapes, with the causality ones being slightly more constraining this time. The scalar and axion UV partial UV completions are consistent with both methods, as it has been the case for every case analyzed so far (with the exception of the problematic spin-2 sector where the issue was addressed in Section \ref{ssec:CausBoundsVector}).

\begin{figure}[!h]
	\center
	\includegraphics[height=5cm]{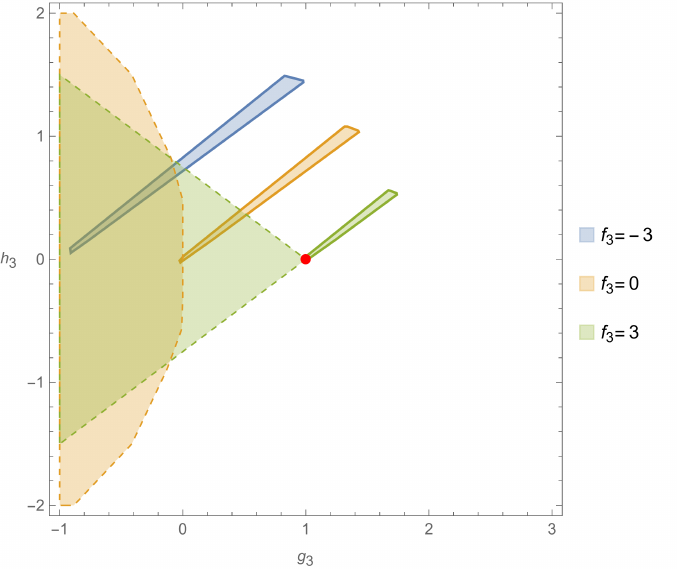}
	\hspace{0.3cm} \includegraphics[height=5cm]{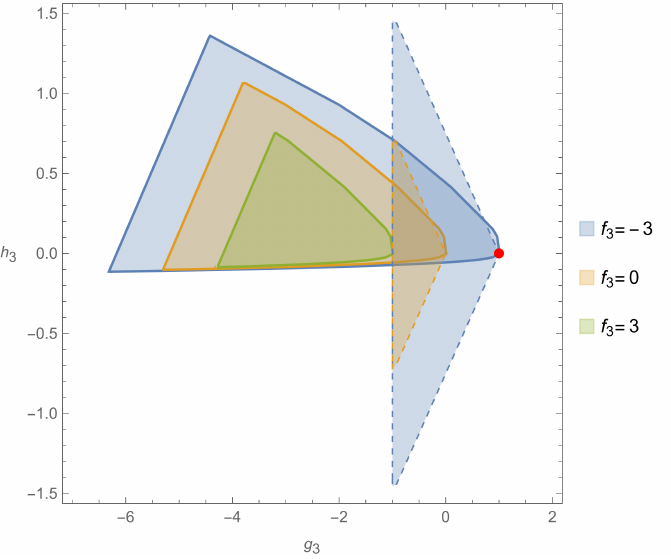}
	\caption[$g_3 - h_3$ causality and positivity bounds for several values of $f_3$.]{Causality (thick line) and positivity (dashed line) bounds on $(g_3,h_3)$ plane with all other parameters set by the values of the scalar (left) or axion (right) partial UV completions.}
	\label{fig:g3h3ScalAxionComparison}
\end{figure}

The $(g_3,h_3)$ case is extremely interesting. When focusing on the value $f_3=3$ on the left panel of Fig.~\ref{fig:g3h3ScalAxionComparison}, which is precisely the value that is compatible with the scalar partial UV completion, the intersection of the causality and positivity bounds reduce to this single point. It is the first time we get such stringent results. This seems to indicate that in this plane there exists no other UV completions for such coefficients. This example perfectly illustrates the power of associating the two methods, which can dramatically reduce the space of causal theories, to the point of converging to a point-like region.

The right panel shows the axion case. The causality bounds are slightly less constraining than the positivity ones but their union is smaller. It is worth noting that the axion partial UV completion is consistent with  both bounds and that the union of the bounds forbid any theory with $f_3=3$. 

\section{Discussion}
\label{sec:DiscVect}

Throughout this Chapter, we have proven that it was possible to apply the method derived in Chapter \ref{chap:causalScalar} for scalars to massless vectors. More precisely, we have been able to constrain the low-energy coefficients of the massless photon EFT \eqref{eq:Lagrangian} by imposing that none of the two propagating modes acquire any resolvable time advance compared to a freely propagating wave. The analysis performed here is more complicated than the scalar case as the number of Wilson coefficients grows more rapidly for the same truncating order. This enlargement of the parameter space meant that we had to commit to given $2d$ slices to be able to plot our bounds and extract meaningful information. 

The causality bounds derived in Section \ref{ssec:CausBoundsVector} are predominantly double-sided and compact, with a few exceptions, especially the $(f_4,g_4)$ case that does not admit any upper bound. This is already a great accomplishment that deserves to be noted. Furthermore, we used all partial UV completions that we knew existed (listed in Table \ref{tab:UVcomp}) to check the consistency of our results and we confirmed that all our bounds allow for these known partial UV completions, as expected. There is a little caveat here as the ones written as `spin-2 even I' and `spin-2 odd' are explicitly forbidden by our bounds but we believe that their derivation suffered from inconsistencies and cannot be trusted. This has actually been an interesting test for our bounds as it has unveiled a new way to use them: discriminate between legitimate and erroneous partial UV completions.

However, we did not stop there and decided to compare our bounds to the positivity bounds of \cite{CarrilloGonzalez:2023cbf} (once again, I need to stress that I was not involved in their computation but my collaborators were), which revealed to be extremely powerful. It is worth noting that depending on the cases, either one or the other method proved to be more constraining than the other, without any anticipation of such results. However, it was always true that taking their union, i.e. requiring that the theory admits both causal propagation in the IR and a causal UV completion, significantly reduced the region of EFTs allowed by causality and positivity in the space of parameters. In some cases, the union exactly reduced to a single point, a point that precisely coincided with a known partial UV completion. 

In conclusion, causality bounds have proven to be constraining for spin-1 models too. This method has ruled out considerable regions of the space of parameters on its own while always allowing for known partial UV completions. This proves that the first try at applying them to scalars \cite{CarrilloGonzalez:2022fwg} in Chapter \ref{chap:causalScalar} has been successfully extended to vectors and we are hopeful that we can tackle massless and massive spin-2 particles in the same way in the future. However, the strongest results of this analysis were obtained when combining our bounds with positivity bounds. This further reduced the allowed causal regions, sometimes to a single UV completion point, leaving no other viable alternative. Both causality and positivity methods are hence complementary and using them together has the potential to dramatically improve any results derived so far.

\part*{Conclusions and Summary of new results}
\label{part:Conclusions}
\addcontentsline{toc}{part}{Conclusions and Summary of new results}

\thispagestyle{plain}
\paragraph{Conclusions}

Effective Field Theories are a useful physical tool to describe physical phenomena at low energy. However, making predictions can be a difficult task if no, or too little, experimental data to compare with is available. In this context, it becomes interesting to extract as much information as possible from the theoretical consistency of the theory at hand. In this thesis, we have explored a variety of such tools and have been able to constrain EFT parameters, or Wilson coefficients, further and more efficiently than they were before. 

More than simply refining some known theories, we derived a brand new model of massive vector fields, dubbed Extended Proca-Nuevo (EPN) \cite{deRham:2020yet} in Chapter \ref{chap:PN}. We have proven that this model is inequivalent to Generalized Proca and more general in the sense that it encapsulates most of its interactions while exploring a class of highly non-linear operators based on Massive Gravity. The main difference manifests itself at the level of the null eigenvector which was used to discover the existence of a non-trivial constraint. The second constraint \cite{deRham:2023brw} was later found in Chapter \ref{chap:Constraint}, finishing the proof of the absence of ghosts in any EPN model. The requirement that the theory should be ghost-free vastly shrinks the space of allowed massive spin-1 EFTs, to a point that only GP and EPN are viable candidates to this day. The wider quest of finding the most general ghost-free EFT of a massive vector field is still an open question.

After having established the consistency of EPN on flat backgrounds, we explored the question of its covariantization in Chapter \ref{chap:CovEPN}. To begin with, we considered Generalized Massive Gravity as the natural way of covariantizing EPN given the fact that its interactions are inspired by the $\Lambda_3$ decoupling limit of the former. We then showed that there exist alternative covariantizations and that they maintain the existence of the constraint. Finally, we discussed cosmological covariantizations (with or without non-minimal couplings) and explicitly showed that the correct number of degrees of freedom propagate.

We then decided to explore the cosmological implications of using EPN to model dark energy \cite{deRham:2021efp} in Chapter \ref{chap:Cosmo}. For this, we derived the equations of motion for all (i.e. scalar, vector and tensor) linear perturbations propagating on top of an FLRW background. We used a different way to impose the consistency of our theory both at the level of the background and the perturbations. We proved that our model featured hot Big Bang solutions with successive epochs of radiation, matter and dark energy domination, with the latter corresponding to a ``self-accelerating'' phase, being driven by the vector field condensate and not a cosmological constant. We have then been able to show that it was possible to find parameters such that the perturbations are sub-luminal, stable and ghost-free. More, we have also shown that we can restrict our model to have exact luminal propagation of the tensor mode, in agreement with the tightest bounds on the speed of propagation of gravitational waves. Finally, we explicitly provided a signature of our cosmological model in the fact that we predict a possible deviation from the speed of propagation of the phonon (the longitudinal fluctuation of the perfect fluid) as expected in GR (and also GP). This is interesting as it provides us with an experimental way of distinguishing EPN from both GR and GP.
\thispagestyle{plain}

In Part \ref{part:CausalityBounds}, we decided to focus our efforts on finding consistent EFTs by requiring infrared causality. This led us to the derivation of so-called causality bounds for shift-symmetric scalars \cite{CarrilloGonzalez:2022fwg} and massless photons \cite{CarrilloGonzalez:2023cbf} respectively in Chapters \ref{chap:causalScalar} and \ref{chap:causalVector}. In both cases, we have used the freedom of requiring the absence of any resolvable time advance for perturbations propagating on top of any arbitrary background. Such a constraint attached to a given background translates into a trivial one-sided bound in the space of parameters of the EFT, but varying over the backgrounds allowed us to extract a large number of such bounds that formed tight and compact bounds for the region of allowed parameters once taken together. Applying causality bounds to vector fields bridges the gap with the focus of Part \ref{part:PN} and paves the way to a natural extension, i.e. constraining the EFT parameters of EPN, which would be interesting to pursue in the future.

Such causality bounds can be compared to the corresponding positivity bounds (in the case that the latter exist) and both can be used in a common effort to get the tightest possible regions in the space of the Wilson coefficients that allow for a causal low-energy theory with a causal UV completion. We have proven that the combination of both methods can be as powerful as to reduce some $2d$ regions of the space of parameters to a single point, featuring a viable UV completion. However, it is worth noting that positivity bounds can only be derived for asymptotically-flat backgrounds and hence cannot predict anything for theories living on an arbitrarily curved background. This is not true for causality bounds and the hope is that this thesis has provided proof of principle that the latter is a robust and consistent method and that it should now be applied to configurations that are out of reach of the positivity bounds to extract all its predictive power. 
\thispagestyle{plain}

\paragraph{Future directions}
On top of the achievements of this thesis, it is also worth highlighting the fact that this work could be extended in a variety of promising directions.

\thispagestyle{plain}
\begin{itemize}
	\item This thesis has enlarged the space of viable massive ghost-free spin-1 EFTs. In reference to Fig.~\ref{fig:proca theory space}, the space of Proca theories is now comprised of GP-like and EPN-like interactions with some overlap between the two. However, the real question is to know whether this is all there is or if there exist other viable interactions yet to be discovered. And if so, how would they realize the Hessian constraint?
	\item A natural step that would fully unify Part \ref{part:PN} and Part \ref{part:CausalityBounds} would be to apply causality bounds to Extended Proca-Nuevo and constrain the space of low-energy parameters of the theory. The EFT has been proven to be stable, consistent, ghost-free, etc. and we understand how to compute time delays for vector modes. The main difference would be the propagation of a third mode. Solving the $3$ decoupled equations of motion (one for each linear perturbation) could be a more difficult task but can be done in principle. Then the same procedure would apply and the first theoretical bounds from causality requirements could be obtained for EPN.
	\item After a thorough analysis of scalar and vector EFTs, it would be interesting to go on and apply causality bounds to spin-2 particles, whether massless or massive. This could be combined with positivity bounds in the same fashion as in Chapter \ref{chap:causalVector} to get tighter bounds and extract as much information as possible from the requirements of causality.
	\item Finally, as we have already mentioned, positivity bounds have their limitations, especially in the fact that they rely on being able to define an S-matrix, and that this very definition requires asymptotic ingoing and outgoing states to propagate on a flat background. In other words, in the current state of the art, positivity can only be imposed on asymptotically-flat backgrounds. However, one can easily get around this issue when turning to causality bounds as there are no issues in computing the classical time delay in any arbitrary background. Deriving causality bounds on such backgrounds would be an important step towards showing their greater versatility.
\end{itemize}
\thispagestyle{plain}

In conclusion, we have explored a range of scalar and vector Effective Field Theories and have studied several theoretical tools to constrain their low-energy coefficients and ensure physical consistency. I believe this thesis compiles new results and innovative methods while also shining a light on interesting avenues to explore in the future.

\bibliographystyle{JHEP}
\bibliography{references_thesis}

\begin{appendices}

\chapter{Proca-Nuevo}

\section{Null Eigenvector for generic \Pros theories}
\label{app:Null}

In subsection \ref{sssec:Null4d} we proved explicitly that the vector $V_a$ defined in \eqref{eq:Va} as $V_a = \Z^{0 \mu} \p_{\mu} \phi_a$ with $\Z^{-1}=\eta \X$ and $\X=\sqrt{\eta^{-1}f}$ is a null eigenvector of the Hessian associated with the Lagrangian $\L_1[\X]$. We now proceed to prove this result for every other $\L_n[\X]$.

We will not go through the derivation of this result for each order in $\X$ or $\K$ but we provide here intermediate results, \ie the momenta and Hessian matrices. 

At any order in the Lagrangian expansion \eqref{eq:defLK}, we define
\ba
	p_a^{(n)} = \frac{\p (\Lambda_2^4 \L_n[\X])}{\p \dot{A}^a} \nn \quad {\rm and }\quad
	\mathcal{H}_{ab}^{(n)}= \frac{\p^2 (\Lambda_2^4 \L_n[\X])}{\p \dot{A}^a \p \dot{A}^b} = \frac{\p p_a^{(n)}}{\p \dot{A}^b} \,. \nn
\ea
Since the Lagrangians $\L_n[\X]$ and $\L_n[\K]$ are related by linear relations,
\ba
	\L_n[\K] = \sum_k c_{n,k} \L_k[\X] \quad \Rightarrow \quad
	\mathcal{H[\K]}_{ab}^{(n)} &= \sum_k c_{n,k} \mathcal{H[\X]}_{ab}^{(k)}\,. \nn
\ea
$\bullet$ For the Lagrangian $\L_2[\X]$, we have an associated contribution to the conjugate momentum given by
\begin{equation}
	\Lambda_2^{-2}p_a^{(2)} = 4([\X]V_a + \dot{\phi}_a)
\label{eq:pa2}
\end{equation}
resulting in a contribution to the Hessian given by
\begin{equation}
	\mathcal{H}_{ab}^{(2)} = 4\left(\Lambda_2^2[\X]\frac{\partial V_a}{\partial \dot{A}^b} + V_b V_a + \eta_{ab}\right)\,.
\label{eq:H2}
\end{equation}
Given the Hessian \eqref{eq:H2}, it is straightforward to see that $V^a$ is indeed a null eigenvector, meaning that $\mathcal{H}_{ab}^{(2)} V^a=0$
\ba
	\mathcal{H}_{ab}^{(2)} V^a = 4\left(\Lambda_2^2[\X]\frac{1}{2}\frac{\partial V_a V^a}{\partial \dot{A}^b} + V_b (V_a V^a) + V_b\right)
	                           = 0\,.
\label{eq:HV2}
\ea
$\bullet$ For the Lagrangian $\L_3[\X]$, we have an associated contribution to the conjugate momentum given by
\begin{equation}
\Lambda_2^{-2}	p_a^{(3)} = 3\left([\X]^2-[\X^2]\right)V_a + 6[\X]\dot{\phi}_a + 6 X^{0 \mu}\partial_{\mu} \phi_a
\label{eq:pa3}
\end{equation}
leading to a Hessian
\begin{equation}
	\mathcal{H}_{ab}^{(3)} = 6\left([\X](V_a V_b + \eta_{ab}) + V_a\dot{\phi}_b + V_b\dot{\phi}_a + \X^{00}\eta_{ab}\right) + 3\Lambda_2^2\left([\X]^2-[\X^2]\right)\frac{\partial V_a}{\partial \dot{A}^b} + 6\Lambda_2^2\frac{\partial X^{0 \mu}}{\partial \dot{A}^b} \partial_{\mu} \phi_a
\label{eq:H3}
\end{equation}
for which we can again explicitly check that $V^a$ is a null vector,
\begin{align}
	\mathcal{H}_{ab}^{(3)}V^a &= 6\left([\X](- V_b + V_b) -\dot{\phi}_b + V_b\dot{\phi}_a V^a + \X^{00} V_b\right) + \frac{3}{2}\Lambda_2^2\left([\X]^2-[\X^2]\right)\frac{\partial (-1)}{\partial \dot{A}^b} + 6\Lambda_2^2\frac{\partial \X^{0 \mu}}{\partial \dot{A}^b} \X^0_{\phantom{0}\mu} \nn \\
	&= 6(\X^{00} + \dot{\phi}_a V^a)V_b + 3\left(\Lambda_2^2\frac{\p f^{00}}{\partial \dot{A}^b}-2\dot{\phi}_b\right) \nn \\
	&= 0
\label{eq:VH3}
\end{align}
$\bullet$ Finally, for the Lagrangian $\L_4[\X]$, the associated  conjugate momentum is given by
\begin{equation}
\Lambda_2^{-2}	p_a^{(4)} = 4\left([\X]^3 - 3[\X][\X^2] + 2[\X^3]\right)V_a + 12\left([\X]^2-[\X^2]\right)\dot{\phi}_a + 24\left([\X]\X^{0 \mu}-f^{0 \mu}\right)\partial_{\mu}\phi_a\,,
\label{eq:pa4}
\end{equation}
leading to the Hessian
\begin{align}
	\mathcal{H}_{ab}^{(4)} =& 12\left([\X]^2-[\X^2]\right)(V_b V_a + \eta_{ab}) + 24\left([\X]\dot{\phi}_b + \X^{0\mu}\partial_{\mu}\phi_b\right)V_a + 4\Lambda_2^2\left([\X]^3 - 3[\X][\X^2] + 2[\X^3]\right) \frac{\partial V_a}{\partial \dot{A}^b} \nonumber \\
	                        &+ 24\left([\X]V_b + \dot{\phi}_b\right)\dot{\phi}_a + 24\left(\X^{0 \mu} V_b + [\X] \Lambda_2^2 \frac{\partial \X^{0 \mu}}{\partial \dot{A}^b} + \partial^{\mu}\phi_b + \eta^{0\mu}\dot{\phi}_b\right)\partial_{\mu}\phi_a \nonumber \\
&+ 24\left([\X]\X^{00}-f^{00}\right)\eta_{ab}
\label{eq:H4}
\end{align}
for which $V^a$ is yet again a null eigenvector,
\begin{align}
	\mathcal{H}_{ab}^{(4)}V^a =& 12\left([\X]^2-[\X^2]\right)(- V_b + V_b) - 24\left([\X]\dot{\phi}_b + \X^{0\mu}\partial_{\mu}\phi_b\right) + 2\Lambda_2^2\left([\X]^3 - 3[\X][\X^2] + 2[\X^3]\right) \frac{\partial (-1)}{\partial \dot{A}^b} \nonumber \\
	                           &+ 24\left([\X]V_b + \dot{\phi}_b\right)\X^{0}_{\phantom{0}0} + 24\left(\X^{0 \mu} V_b + [\X] \Lambda_2^2 \frac{\partial \X^{0 \mu}}{\partial \dot{A}^b} + \partial^{\mu}\phi_b + \eta^{0\mu}\dot{\phi}_b\right)\X^{0}_{\phantom{0}\mu} \nonumber \\
                             &+ 24\left([\X]\X^{00}-f^{00}\right)V_b \nn \\
                            =& 24 \left\{ (\X^{0}_{\phantom{0}0} + \X^{00})([\X]V_b+\dot{\phi}_b) + (\X^{0\mu}\X^{0}_{\phantom{0}\mu}-f^{00})V_b + \left( \Lambda_2^2 \frac{\partial \X^{0 \mu}}{\partial \dot{A}^b} - \dot{\phi}_b \right)[\X] \right\}
														= 0\,.
\label{eq:HV4}
\end{align}

We can therefore conclude that for any linear combination of the \Pros vector Lagrangians,
\ba
\L_{\K}[A] = \Lambda_2^4 \sum_{n=0}^4 \alpha_n(A^2) \L_n[\K[A]]= \Lambda_2^4 \sum_{n=0}^4 \beta_n(A^2) \L_n[\X[A]]\,,
\label{eq:LKX}
\ea
where the relation between the coefficients $\alpha_n$ and $\beta_n$ is given in \cite{deRham:2014zqa}
the resulting Hessian is of the form
\ba
\mathcal{H}_{ab}= \sum_{n=0}^4 \beta_n \mathcal{H}_{ab}^{(n)}\,.
\ea
Since all the individual Hessians $\mathcal{H}_{ab}^{(n)}$ have the same null direction, with null eigenvector $V^a$, it automatically follows that $V^a$ is also a null eigenvector of the full Hessian $\mathcal{H}_{ab}$ and the full \Pros theory carries a constraint. Remarkably, it is clear from this construction that \Pros theories lie on a different branch of theories as compared to GP theories in terms of how the constraint comes to be implemented. Even though both GP and \Pros are ghost-free theories that carry a constraint, linear combinations of both theories typically break the constraint.

In two dimensions, we can check that the null eigenvector reproduces the exact analytic result  \eqref{eq:L2d}. Recalling  that
\begin{equation}
	\phi^a = x^a + \frac{1}{\Lambda_2^2} A^a\,,
\label{eq:phitoA}
\end{equation}
which then gives
\begin{equation}
V^a=\Z^{0\mu}\p_\mu \phi^a=
	\left(\begin{array}{c}
\Z^{00}(1 - \frac{1}{\Lambda_2^2} \dot{A}_0) - \frac{1}{\Lambda_2^2} \Z^{01} A_0' \\
\frac{1}{\Lambda_2^2} \Z^{00} \dot{A}_1 + \Z^{01}(1 + \frac{1}{\Lambda_2^2} A_1')
\end{array}
	\right)\,.
\label{eq:V0V1}
\end{equation}
Rearranging these terms gives the exact non-perturbative prediction for the two--dimensional eigenvector $v^a$ introduced in  \eqref{eq:eigvec} (up to an irrelevant normalization factor),
\begin{equation}
 V_a = \frac{1}{[\X]}
		\begin{pmatrix}
			2&+&\frac{A_1' - \dot{A}_0}{\Lambda_2^2} \\
			 &-&\frac{\dot{A}_1 - A_0'}{\Lambda_2^2}
		\end{pmatrix}
=  \frac{2}{[\X]}
		\begin{pmatrix}
			1 &+&x/2\\
			&-&y/2
		\end{pmatrix}= \frac{2}{[\X]}v_a
\,.
\label{eq:eigvecbis}
\end{equation}

\section{Kinematics}
\label{sec:appKin}

To perform the scattering amplitudes computations for a given set of polarizations, we need a basis for the polarization vectors $\epsilon_{\mu}^{\lambda}(k_i)$. The polarizations are labelled by $\lambda=-1,0,+1$.

First of all, we consider the center of mass frame where $k_1$ and $k_2$ are traveling in the $\hat{z}$ direction and $k_3$ forms an angle $\theta$ with the $\hat{z}$-axis. We denote the energy by $\omega$ and the norm of the 3-momentum by $k$
\begin{align}
	k_1^{\mu} &= (\omega, 0, 0, k) \label{eq:k1 }\\
	k_2^{\mu} &= (\omega, 0, 0, -k) \label{eq:k2} \\
	k_3^{\mu} &= (\omega, k \sin(\theta), 0, k \cos(\theta)) \label{eq:k3} \\
	k_4^{\mu} &= (\omega, - k \sin(\theta), 0, - k \cos(\theta))\,. \label{eq:k4}
\end{align}
In this setup the polarization vectors basis can be chosen to be
\begin{center}
\begin{tabular}{ c c c }
	$\epsilon_{\mu}^{+}(k_1) = \begin{pmatrix} 0 \\ 1 \\ 0 \\ 0 \end{pmatrix}$ & $\epsilon_{\mu}^{-}(k_1) = \begin{pmatrix} 0 \\ 0 \\ 1 \\ 0 \end{pmatrix}$ & $\epsilon_{\mu}^0(k_1) = \begin{pmatrix} -\frac{k}{m} \\ 0 \\ 0 \\ \frac{\omega}{m} \end{pmatrix}$ \\
	$\epsilon_{\mu}^{+}(k_2) = \begin{pmatrix} 0 \\ -1 \\ 0 \\ 0 \end{pmatrix}$ & $\epsilon_{\mu}^{-}(k_2) = \begin{pmatrix} 0 \\ 0 \\ 1 \\ 0 \end{pmatrix}$ & $\epsilon_{\mu}^0(k_2) = \begin{pmatrix} -\frac{k}{m} \\ 0 \\ 0 \\ -\frac{\omega}{m} \end{pmatrix}$ \\
	$\epsilon_{\mu}^{+}(k_3) = \begin{pmatrix} 0 \\ \cos(\theta) \\ 0 \\ -\sin(\theta) \end{pmatrix}$ & $\epsilon_{\mu}^{-}(k_3) = \begin{pmatrix} 0 \\ 0 \\ 1 \\ 0 \end{pmatrix}$ & $\epsilon_{\mu}^0(k_3) = \begin{pmatrix} -\frac{k}{m} \\ \frac{\omega}{m} \sin(\theta) \\ 0 \\ \frac{\omega}{m} \cos(\theta) \end{pmatrix}$ \\
	$\epsilon_{\mu}^{+}(k_4) = \begin{pmatrix} 0 \\ -\cos(\theta) \\ 0 \\ \sin(\theta) \end{pmatrix}$ & $\epsilon_{\mu}^{-}(k_4) = \begin{pmatrix} 0 \\ 0 \\ 1 \\ 0 \end{pmatrix}$ & $\epsilon_{\mu}^0(k_4) = \begin{pmatrix} -\frac{k}{m} \\ -\frac{\omega}{m} \sin(\theta) \\ 0 \\ -\frac{\omega}{m} \cos(\theta) \end{pmatrix}$\,.
\end{tabular}
\end{center}
One can verify that this basis satisfies the polarization vector properties, for a given vector $k_i$ (\ie $i=1,...,4$ fixed)
\begin{align}
	\epsilon_{\mu}^{\lambda}(k_i) k_i^{\mu} &= 0 \label{eq:keps0} \\
	\epsilon_{\mu}^{\lambda}(k_i) \epsilon^{\mu,\lambda'}(k_i) &= \delta^{\lambda \lambda'} \label{eq:epseps0} \\
	\sum_{\lambda=-1}^1 \epsilon_{\mu}^{\lambda}(k_i) \epsilon_{\nu}^{\lambda}(k_i) &= \eta\mn + \frac{{k_i}_{\mu} {k_i}_{\nu}}{m^2}\,. \label{eq:completeness}
\end{align}
We also have the following kinematical constraints
\begin{align}
	\omega &= \frac{\sqrt{s}}{2} \label{eq:omegatos} \\
	k &= \frac{1}{2}\sqrt{s-4m^2} \label{eq:ktos} \\
	t &= -\frac{1}{2}(s-4m^2)(1-\cos(\theta)) \label{eq:ttos} \\
	u &= -(s+t)+4m^2 \,, \label{eq:utos}
\end{align}
which enable us to fully specify the kinematics with the two parameters $(s,\theta)$.

\section{Resummation of the complete DL of massive gravity}
\label{sec:appDL}

In this Appendix we provide an explicit formula resumming the DL of massive gravity to all orders in $\Phi\mn = \p_{\mu} \p_{\nu} \phi$.
For convenience, we work here in the formulation of the theory in terms of the tensor $\X$ as in \eqref{eq:LKX} and we only need to focus on the contribution of the $\beta_n$ which is independent of $A$, so in what follows we may consider the $\beta_n$'s to be constant.
As derived by Ondo and Tolley in \cite{Ondo:2013wka}, the scalar--vector sector of this DL is
\begin{align}
\label{eq:DLAndrew}
	\L_{\text{DL}} = \left\{ \vphantom{\frac{1}{2}} \right. &- \frac{\beta_1}{4} \left( \frac{1}{2} F^a_{\mu} \omega^b_{\phantom{b}\nu} \delta^c_{\rho} \delta^d_{\sigma} + (\delta + \Phi)^a_{\mu} \delta^b_{\nu} \left[ \omega^c_{\phantom{c}\rho} \omega^d_{\phantom{d}\sigma} + \frac{1}{2} \delta^c_{\rho} \omega^d_{\phantom{d}\alpha} \omega^{\alpha}_{\phantom{\alpha}\sigma} \right] \right) \\
											                                &- \frac{\beta_2}{8} \left( 2 F^a_{\mu} \omega^b_{\phantom{b}\nu} (\delta+\Phi)^c_{\rho} \delta^d_{\sigma} + (\delta + \Phi)^a_{\mu} (\delta + \Phi)^b_{\nu} \left[ \omega^c_{\phantom{c}\rho} \omega^d_{\phantom{d}\sigma} + \delta^c_{\rho} \omega^d_{\phantom{d}\alpha} \omega^{\alpha}_{\phantom{\alpha}\sigma} \right] \right) \nonumber \\
														                                &- \left. \frac{\beta_3}{24} \left( 3 F^a_{\mu} \omega^b_{\phantom{b}\nu} (\delta+\Phi)^c_{\rho} (\delta+\Phi)^d_{\sigma} + (\delta + \Phi)^a_{\mu} (\delta + \Phi)^b_{\nu} (\delta + \Phi)^c_{\rho} \omega^d_{\phantom{d}\alpha} \omega^{\alpha}_{\phantom{\alpha}\sigma} \right) \right\} \epsilon^{\mu \nu \rho \sigma} \epsilon_{abcd}\,,\nn
\end{align}
where $\omega$ is a composite field defined by
\begin{equation}
	\omega = \sum_{n,m} \frac{(n+m)!}{2^{1+n+m}n!m!}(-1)^{n+m} \Phi^n F \Phi^m\,.
\label{eq:omega}
\end{equation}
The expression \eqref{eq:DLAndrew} has the advantage to be compact and complete but it is useful to rewrite it only in terms of $F$ and $\Phi$, the actual field content of the theory. It can be proven by basic binomial manipulations that the complete DL of massive gravity can be resummed to all orders in the following way
\begin{align}												\label{eq:resumfullDL}
	\L_{\text{DL}} =& - \frac{\beta_1+2\beta_2+\beta_3}{8}F\mnup F\mn + \frac{\beta_1+4\beta_2+3\beta_3}{8\Lambda_3^3}F^{\mu \alpha}F^{\nu}_{\phantom{\nu} \alpha} \Phi\mn - \frac{\beta_2+\beta_3}{8\Lambda_3^3}F\mnup F\mn[\Phi] \\
										 & + \sum_{p=2}^{\infty} \sum_{k=0}^{p} \frac{1}{\Lambda_3^{3p}} \left\{ \frac{\beta_1+4\beta_2}{8} - \frac{\beta_3}{8} \left( 8p - 23 - 4 \frac{k(p-k)(4p-9)}{p(p-1)} \right) \right\} \frac{(-1)^p}{2^p} {{p}\choose{k}} [F \Phi^{k} F \Phi^{p-k}] \nonumber \\
										 & + \sum_{p=2}^{\infty} \sum_{k=0}^{p-1} \frac{1}{\Lambda_3^{3p}} \left\{ - \frac{\beta_2}{4} + \frac{\beta_3}{4} \left( 2p - 9 + 2 \frac{(p-k-1)^2+k^2}{p-1} \right) \right\} \frac{(-1)^p}{2^p} {{p-1}\choose{k}} [F \Phi^{k} F \Phi^{p-k-1}]S_1(\Phi)
												\nonumber \\
										 & + \sum_{p=2}^{\infty} \sum_{k=0}^{p-2} \frac{1}{\Lambda_3^{3p}} \left\{ - \frac{\beta_3}{4} \left( 2p - 5 \right) \right\} \frac{(-1)^p}{2^p} {{p-2}\choose{k}} [F \Phi^{k} F \Phi^{p-k-2}]S_2(\Phi)
												\nonumber \\
										 & + \sum_{p=3}^{\infty} \sum_{k=0}^{p-3} \frac{1}{\Lambda_3^{3p}} \left\{ \frac{\beta_3}{6} \left( p - 2 \right) \right\} \frac{(-1)^p}{2^p} {{p-3}\choose{k}} [F \Phi^{k} F \Phi^{p-k-3}]S_3(\Phi)\,, \nonumber
\end{align}
where the $S_n(\Phi)$ are a short-hand notation for
\begin{align}
	S_1(\Phi) &= [\Phi] \\
	S_2(\Phi) &= [\Phi]^2 - [\Phi^2] \\
	S_3(\Phi) &= [\Phi]^3 - 3[\Phi][\Phi^2] + 2[\Phi^3]\,.
\label{eq:SnPhi}
\end{align}
Here we use brackets as a notation for the trace. Some terms of the expansion \eqref{eq:resumfullDL} might include contributions of the form $[F^2]$, which really stands for the trace of the square of the field-strength tensor $F\mn$. In this case, the convention is opposite to the one introduced in \eqref{eq:LK3}. Indeed,
\begin{equation}
	[F^2] = F\mnup F_{\nu \mu} = - F\mnup F\mn
\label{eq:newconv}
\end{equation}
Note that the coefficients $\beta_n$  are not linearly independent, indeed they satisfy
\begin{equation}
	\beta_1 + 2 \beta_2 + \beta_3 = 2\,.
\label{eq:betas}
\end{equation}
Expanding \eqref{eq:resumfullDL} up to quartic order and using \eqref{eq:betas} to eliminate $\beta_3$ gives
\begin{align}
	\L_{{\text{MG DL}}}^{(2)} =& -\frac{1}{4}F\mnup F\mn
		\label{eq:LMGDL2} \\
	\L_{{\text{MG DL}}}^{(3)} =& - \frac{2 - \beta_1 - \beta_2}{8} F\mnup F\mn \Box \phi  + \frac{3 - \beta_1 - \beta_2}{4} F^{\mu \alpha}F^{\nu}_{\phantom{\nu}\alpha} \p_{\mu} \p_{\nu} \phi
		\label{eq:LMGDL3} \\
	\L_{{\text{MG DL}}}^{(4)} =& - \frac{2 - \beta_1 - 2 \beta_2}{16} F\mn^2\left( (\Box \phi)^2 - (\p_{\alpha} \p_{\beta}\phi)^2 \right) - \frac{7 - 3 \beta_1 - 5 \beta_2}{8}F^{\mu \alpha} F^{\nu}_{\phantom{\nu} \alpha} \p_{\mu} \p^{\beta} \phi \p_{\nu} \p_{\beta} \phi \nonumber \\
								             & + \frac{6 - 3 \beta_1 - 5 \beta_2}{8} F^{\mu \alpha} F^{\nu}_{\phantom{\nu} \alpha} \p_{\mu} \p_{\nu} \phi \Box \phi - \frac{5 - 2 \beta_1 - 3 \beta_2}{8}F\mnup F\abup \p_{\mu} \p_{\alpha} \phi \p_{\nu} \p_{\beta} \phi
			\label{eq:LMGDL4}
\end{align}
Comparing \eqref{eq:LKDL2}--\eqref{eq:LKDL4} to \eqref{eq:LMGDL2}--\eqref{eq:LMGDL4}, and using the relation between the coefficients $\alpha_n$ and $\beta_n$ as provided in \cite{deRham:2014zqa} one can see that the vector-scalar sector of our new Proca interactions in the DL exactly coincides with this sector in the DL of massive gravity.
%

\section{Recovering GP from EPN}
\label{sec:GPfromEPN}

In this Appendix, we will show how one can recover most of the  GP from the Extended Proca-Nuevo Lagrangian given in \eqref{eq:LextPNflat}, in the limit where $\tilde \Lambda \to \infty$ while keeping the scale $\Lambda$ finite and the vector field mass finite.

We start with the extended PN Lagrangian \eqref{eq:LextPNflat} written in the form
\beq
 \label{eq:LextPNflatc}
	\L_{\text{EPN}} =\tilde\L_{\rm PN} + \L_{\rm GP}\,,
\eeq
with
\ba
\tilde\L_{\rm PN} &=& \tilde{\Lambda}^4 \sum_{n=0}^4 \alpha_n(\tilde{X}) \L_n[\tilde{\K}[A]] \,,
\ea
and where  $\L_{\rm GP}$ includes all the GP interactions aside from the generic function $f(F\mn,\tilde F\mn, X)$,
\ba
\L_{\rm GP}&=& \Lambda^4 \sum_{n=1}^4 d_n(X) \frac{\L_n[\p A]}{\Lambda^{2n}} \,.
\ea
Assuming analyticity of the functions $\alpha_n$ and $d_n$,
the mass term of the vector field is given by
\ba
m^2=\tilde \Lambda^2 \alpha_0'(0)\,.
\ea
Keeping the mass of the vector field finite in the $\tilde \Lambda \to \infty$ limit therefore requires scaling the coefficient $\alpha_0'(0)$ as $\alpha_0'(0)\to m^2 / \tilde \Lambda^2$. Since $\L_n(\tilde K)\sim \mathcal{O}\left( (\p A)^{n}/\tilde \Lambda^{2n} \right)$, we see that in the limit $\tilde \Lambda \to \infty$, keeping the scale $m$ fixed, the only relevant terms of PN origin are, up to irrelevant constant and total derivatives
\ba
\tilde\L_{\rm PN}
 \xrightarrow[\tilde \Lambda \rightarrow \Lambda]{} -\frac 12 m^2 A^2+\alpha_0''(0)A^4
 -\frac 12 \alpha_1'(0) A^2 \p A
 + \frac 14 \alpha_1(0) F\mn^2-\frac 12 \alpha_2(0)F\mn^2\,.
\ea
These are all of GP nature, so added to $\L_{\rm GP}$, we directly deduce that in the limit $\tilde \Lambda \to \infty$, keeping the scales $m$ and $\Lambda$ fixed, the extended PN Lagrangian given in  \eqref{eq:LextPNflatc} includes all the GP interactions aside from the generic function $f(F\mn,\tilde F\mn, X)$.

\chapter{Constraint Analysis for Proca-Nuevo}

\section{EPN constraints in the Lagrangian formalism}
\label{app:EPNLagrangian}

The analytic form of the acceleration-free part $u^{\mu}_{\rm (EPN)}$ of the equations of motion Eq.~\eqref{eq:EL} for the massive vector field $A_{\mu}$ in the two-dimensional EPN theory is given by
\begin{equation}
\begin{aligned}
u_{\rm (EPN)}^0 =& \; \frac{\tilde\alpha_1}{N_+^3} \left( x_{+} y_{+} \left( 2 \dot{A}_0' - A_1'' \right) - x_{+}^2 A_0'' + \left( x_{+}^2 + y_{+}^2 \right) \dot{A}_1' \right)  \\
	& + 2\Lambda \left( \tilde\alpha_{1,X} \frac{y_{+}}{N_{+}} + \alpha_{2,X} \left( y_+ + y_- \right) \right) \left( A_1 A_1' - A_0 A_0' \right)  \\
	& - 2\Lambda \left( \tilde\alpha_{1,X} \frac{x_{+}}{N_{+}} + \alpha_{2,X} (x_{+}+x_{-}) \right) (A_1 \dot{A}_1 - A_0 \dot{A}_0) \\
	& + 2 \Lambda^2 \left( \tilde\alpha_{0,X} + \tilde\alpha_{1,X} N_{+} + \frac12 \alpha_{2,X} \left( N_{+}^2 -N_{-}^2 \right) \right) A_0  \\
	&+  \Lambda^2 d_{1,X} \left( 2(1+\Sigma)A_0 - (y_++y_-)A_1 \right) \,, 
\end{aligned}
\end{equation}
and
\begin{equation}
\begin{aligned}
	u_{\rm (EPN)}^1 =& \; \frac{\tilde\alpha_1}{N_{+}^3} \left( x_{+} y_{+} \left( 2  \dot{A}_1' -  A_0'' \right) - y_{+}^2 A_1'' + \left( x_{+}^2 + y_{+}^2 \right)  \dot{A}_0' \right) \\
	& + 2\Lambda \left(  \tilde\alpha_{1,X} \frac{x_{+}}{N_{+}} + \alpha_{2,X} \left( x_+ - x_- \right) \right) \left( A_1 A_1' - A_0 A_0' \right)  \\
	& - 2\Lambda \left(  \tilde\alpha_{1,X} \frac{y_{+}}{N_{+}} + \alpha_{2,X} (y_{+}-y_{-}) \right) (A_1 \dot{A}_1 - A_0 \dot{A}_0)  \\
	& -2 \Lambda^2 \left( \tilde\alpha_{0,X} + \tilde\alpha_{1,X} N_{+} + \frac12 \alpha_{2,X} \left( N_{+}^2 -N_{-}^2 \right) \right) A_1  \\
	&+ \Lambda^2 d_{1,X} \left( 2\Delta A_0 - (2+x_+ -x_-)A_1 \right) \,.
\end{aligned}
\end{equation}
Furthermore, we defined the following functions entering the constraint Eq.~\eqref{eq:secondclassEVb},
\begin{equation}
\begin{aligned}
	\phi_0 =& 2 \left( \bar{x}_{+} A_0 + \bar{y}_{+} A_1 \right) \,, \\
	\phi_1 =& 2 \left( \left( 1 + \Sigma \right) A_0 + \Delta A_1  \right) \,, \\
	\phi_2 =& 4 \left( \left( \bar{x}_{+} \Sigma - \bar{y}_{+} \Delta \right) A_0 - \left( \bar{x}_{+} \Delta + \bar{y}_{+} \Sigma \right) A_1  \right) \,, \\
	\Phi_1 =& \frac{1}{\Lambda} \left( \bar{x}_{+} \p_1 \bar{y}_{+} - \bar{y}_{+} \p_1 \bar{x}_{+}  \right) \,,
\end{aligned}
\end{equation}
where we have introduced the following notation
\begin{equation}
\label{eq:notations_Sigma_Nabla}
	\Sigma = \frac12 (x_+ + x_-) = 1 + \frac{A_1'}{\Lambda} \,, \qquad \Delta = \frac12 ( y_+ - y_-) = - \frac{A_0'}{\Lambda} \,.
\end{equation}

\section{EPN Secondary constraint in the Lagrangian picture}
\label{app:EPNSecConstraint}

The secondary second-class constraint for the EPN theory is obtained by imposing the time derivative of the primary constraint $\mathcal{C}_{V1}$ presented in Eq.~\eqref{eq:secondclassEVb} to vanish. Its expression reads,
\begin{align}
\frac{\mathcal{C}_{V2}}{\Lambda^3} =& \alpha_{2,XX} \left(\frac{\dot{X}}{\Lambda} \phi_2 + 4 \left(A_0 \Delta -A_1 (1-\Sigma )\right) \left(2 A_0 \gamma _1+2 A_1 \left(\gamma _2-N_+\right)\right) \right) \\
   &+ 2 \alpha_{2,X} \left\lbrace \left(N_-^2-N_+^2\right) \bar{x}_+ +2 A_0 N_+ \Phi _1+2 N_+ - 2 \left(\left(\psi _1-\psi_2\right){}^2-A_1^2 N_+^2\right) \frac{\alpha_{2,X}}{\tilde{\alpha}_1} \right. \nonumber \\
   &\qquad \left. + \left(A_1 N_+^2 \psi _2+2 \left(\psi _2-\psi _1\right) \psi _3\right) \frac{\tilde{\alpha}_{1,X}}{\tilde{\alpha }_1} + 2 \psi _2 \left(2 A_0 \gamma _1+A_1 \left(2 \gamma _2-N_+\right)\right) \frac{\tilde{\alpha }_{0,X}}{\tilde{\alpha}_1} \right. \nonumber \\
   & \qquad \left. + 2 \left[ -4 A_0^2 \gamma _1 \left(\gamma _1-\bar{y}_+\right)-2 A_1 A_0 \left(2 \gamma _1 \left(2 \gamma_2-N_+\right)+y_-\right) \right. \right. \nonumber \\
   &\qquad \qquad \left. \left. + A_1^2 \left(2 \bar{x}_+ \left(2 \gamma _2-N_+\right)-4 \gamma _2 \left(\gamma
   _2-N_+\right)\right) \right] \frac{d_{1,X}}{\tilde{\alpha }_1} \right\rbrace \nonumber \\
   &+\tilde{\alpha }_{1,XX} \left( \frac{\dot{X}}{\Lambda} \phi_1 -4 \Delta \frac{\dot{A}_1}{\Lambda} A_0^2 +2 A_1 A_0 \left(\left(2-x_-\right) y_--x_+ y_+\right) +2 A_1^2 (1-\Sigma ) \left(-x_-+x_++2\right) \right) \nonumber \\
   &+\tilde{\alpha }_{1,X} \left\lbrace N_+ \Phi _1 \phi_0 +N_-^2-N_+^2+4 - \left(2 \psi _1 \psi _3- A_1 N_+^2 \psi _2\right)
   \frac{d_{1,X}}{\tilde{\alpha }_1} -2 \psi _2 \psi _3 \frac{\tilde{\alpha }_{0,X}}{\tilde{\alpha}_1} \right. \nonumber \\
   &\qquad \left. \vphantom{\frac{d_1}{\tilde{\alpha}_1}} + 2  \left( A_0^2 \Delta \left(y_-+y_+\right)-A_1 A_0 \left(\left(2-x_-\right) y_--x_+ y_+\right) -A_1^2 (1-\Sigma )
   \left(-x_-+x_++2\right) \right)\frac{\tilde{\alpha }_{1,X}}{\tilde{\alpha }_1} \right\rbrace \nonumber \\
   &+ \Phi_1^2 \tilde{\alpha}_1 +\tilde{\alpha }_{0,XX} \left(\frac{\dot{X}}{\Lambda}\phi_0 -2 \psi _2 \left(A_0 \Delta -A_1
   (1-\Sigma )\right)\right) \nonumber \\
   &+ \tilde{\alpha }_{0,X} \left(2 \left(2\bar{x}_+-N_+\right) - 2 \psi_1 \psi_2 \frac{d_{1,X}}{\tilde{\alpha}_1} - \psi_2^2 \frac{\tilde{\alpha}_{0,X}}{\tilde{\alpha}_1} \right) \nonumber \\
   &+ d_{1,X} \left(\left(N_-^2-N_+^2+4\right) \bar{x}_+ +2 N_+ \Phi_0 A_0 -\frac{d_{1,X}}{\tilde{\alpha }_1}\left(\psi _1-A_1 N_+\right) \left(A_1 N_++\psi _1\right) \right) \nonumber \\
   &+ d_{1,XX} \left(\frac{\dot{X}}{\Lambda} \left(\phi_0+\frac{\phi _2}{2}\right) -2 \left(A_0
   \Delta -A_1 (1-\Sigma )\right) \left(A_1 N_++\psi _1\right)\right) \,, \nonumber
\end{align}
where we defined
\begin{equation}
\begin{aligned}
	\psi_1 =& (y_- \bar{x}_+ + (2+x_-) \bar{y}_+) A_0 + ((2-x_-) \bar{x}_+ - y_- \bar{y}_+) A_1 \,, \\
	\psi_2 =& 2(\bar{y}_+ A_0 + \bar{x}_+ A_1) \,, \\
	\psi_3 =& y_- A_0 + (2-x_-) A_1 \,,
\end{aligned}
\end{equation}
and
\begin{equation}
\begin{aligned}
	\gamma_1 =& \bar{x}_+ \Delta - \bar{y}_+ \Sigma \,, \\
	\gamma_2 =& \bar{x}_+ \Sigma - \bar{y}_+ \Delta \,.
\end{aligned}
\end{equation}
It is now easy to see that, as claimed in the main text, the secondary constraint $\mathcal{C}_{V2}$ does not involve  any acceleration term.

\section{Secondary Constraint in arbitrary dimensions}
\label{app:Upsilon}
To complete the proof of the existence of a secondary constraint for the minimal model in arbitrary dimensions, we explicitly show that the vector $\Upsilon_\mu$ defined as in \eqref{eq:upsilona} by
\ba
\label{eq:UpsilonSol}
\Upsilon_\mu= \upsilon_a V^{\perp (a)}_\mu
= \sum_{a=1}^{D-1} \frac{1}{\lambda^{(a)}} (\p_i V_{\mu}) \mathcal{H}^{i0,\mu\alpha} V^{\perp (a)}_\alpha V^{\perp (a)}_\mu\,,
\ea
satisfies the relation  \eqref{eq:defUpsilon}.
First, by construction, we clearly have
\ba
\label{eq:UpsilonH}
\Upsilon_\mu \mathcal{H}^{\mu\alpha}= \upsilon_a V^{\perp (a)}_\mu \mathcal{H}^{\mu\alpha}
= \sum_{a=1}^{D-1}\upsilon_a \lambda^{(a)} V^{\perp (a)}{}^\alpha
= \sum_{a=1}^{D-1}  (\p_i V_{\mu}) \mathcal{H}^{i0,\mu\beta} V^{\perp (a)}_\beta V^{\perp (a)}{}^\alpha\,.
\ea
Now  recalling that as argued below Eq.~\eqref{eq:defUpsilon}, the  $D$ eigenvectors $\{\tilde{V}_\mu^{(\sigma)}\}_{\sigma=0,\cdots,D-1} =  \{V_\mu, V^{\perp\, (a)}_\mu\}_{a=1,\cdots,D-1}$ form a complete orthonormal basis, satisfying various properties, notably (see \eqref{eq:HW2} and \eqref{eq:norm}),
\ba
  \mathcal{H}^{i0,\mu \beta}   V_{\beta} = 0  \quad {\rm and }\quad
 V^{\perp\, (a)}_\mu V^{\perp\, (b)}{}^\mu=\delta^{ab} \,, \ \forall \ a,b=1,\cdots,D-1,
\ea
we can hence expand any vector in that complete basis. In particular we can write the vector
\ba
T^\beta=(\p_i V_\mu){\mathcal{H}}^{i0,\mu \beta}=\tau_0 V^\beta+\tau_b V^{\perp\, (b)}{}^\beta\,,
\ea
where $\tau_0=0$ since $V_\beta$ is also a NEV of  $\mathcal{H}^{i0,\mu \beta}$ and hence $T^\beta V_\beta =0=\tau_0$. With this in mind, we can therefore expand the RHS of \eqref{eq:UpsilonH} as follows
\ba
\label{eq:UpsilonH2}
\Upsilon_\mu \mathcal{H}^{\mu\alpha}&=&\sum_{a=1}^{D-1}  T^\beta V^{\perp (a)}_\beta V^{\perp (a)}{}^\alpha \nn \\
&=&\sum_{a=1}^{D-1}\sum_{b=1}^{D-1} \tau_b   V^{\perp\, (b)}{}^\beta V^{\perp (a)}_\beta V^{\perp (a)}{}^\alpha \nn \\
&=&\sum_{a=1}^{D-1} \tau_a  V^{\perp (a)}{}^\alpha \equiv T^\alpha= (\p_i V_\mu){\mathcal{H}}^{i0,\mu \alpha}\,.
\ea
This concludes the proof that the vector $\Upsilon_\mu$ given in \eqref{eq:UpsilonSol} does indeed satisfy the relation \eqref{eq:upsilona}. Interestingly, this relies non-trivially on the fact that $V_{\nu}$ is a null eigenvector of both $\mathcal{H}^{i0,\mu\nu}$ and $\mathcal{H}^{00,\mu\nu}$.

\chapter{Covariantization of Extended Proca-Nuevo}

\section{Null eigenvector on a fixed background}
\label{ssec:NEVFix}

In this section, we shall prove that the EPN theory defined in  \eqref{eq:covEPN} admits a constraint about any fixed background (no matter how curved and spacetime-dependent). To prove this, we simply need to show that the  vector $V_{\mu}$ defined in \eqref{eq:NEVcovariant} is indeed a null eigenvector for this EPN theory on any background. To be more precise, we have defined  $\mathcal{H}_{\alpha \beta}^{(n)}$ in Eq.~\eqref{eq:HessianOrderN} to be the Hessian matrix corresponding to the Lagrangian $\L_n[\X]$ for $n=1,...,4$, as expressed in \eqref{eq:L0}--\eqref{eq:L3}. In what follows, we shall show that $V^{\alpha}$ is a null eigenvector for each $\mathcal{H}_{\alpha \beta}^{(n)}$ for $n=0,...,3$. The case $n=0$ is trivial since it is purely a potential term. For the other non-trivial Lagrangians, it turns out to be easier to consider them as functions of $\X$ rather than $\K$. This change of variable is always possible since the set $\{ \L_n[\X] \}$ is linearly related the set $\{ \L_n[\K] \}$ as long as one considers $\textit{all}$ interactions, i.e. including $\L_4$. In what follows we shall thus simply prove that
\begin{equation}
	\mathcal{H}_{\alpha \beta}^{(n)} V^{\alpha} = 0, \quad \text{for } n=1,...,4 \,.
\end{equation}
The case $n=1$ was proven in the main text \eqref{eq:HV1equals0}. Let us now turn to the proof that $V$ is indeed the NEV for $\L_n[\X]$, with $n=2,3,4$. To begin with, we make use of the following identity
\begin{equation}
	\frac{\p [\X^n] }{\p \dot{A}^{\alpha}} = n \Lambda^{-2} \left( \X^{n-2} \right)^0_{\phantom{0} \mu} \left( \delta^{\mu}_{\alpha} + \Lambda^{-2} \nabla^{\mu} A_{\alpha} \right) \,.
\end{equation}
\begin{itemize}
	\item \textbf{$\L_2[\X]$}

One can start by showing that the momentum associated with this Lagrangian reads
\begin{equation}
	p^{(2)}_{\alpha} = 2 \Lambda^{2} \left\lbrace [\X] V_{\alpha} - \left( \delta^{0}_{\alpha} + \Lambda^{-2} \nabla^{0} A_{\alpha} \right) \right\rbrace \,,
\end{equation}
it is then easy to prove that $V$ is indeed the correct null eigenvector,
\ba
	\mathcal{H}^{(2)}_{\alpha \beta} V^{\alpha} = 2 \Lambda^2 \left\lbrace \frac{\p [\X]}{\p \dot{A}^{\beta}} V_{\alpha} V^{\alpha} + [\X] \frac{\p V_{\alpha}}{\p \dot{A}^{\beta}} V^{\alpha} - \frac{1}{\Lambda^{2}} g^{00} g_{\alpha \beta} V^{\alpha} \right\rbrace = 2 \Lambda^2 g^{00} \left( \frac{\p [\X]}{\p \dot{A}^{\beta}} -  \frac{V_{\beta}}{\Lambda^2} \right) = 0 \,.\qquad
\ea
	\item \textbf{$\L_3[\X]$}

The momentum associated with $\L_3[\X]$ is given by
\begin{equation}
	\Lambda^{-2} p^{(3)}_{\alpha} = 3 \left( [\X]^2 - [\X^2] \right) V_{\alpha} - 6 [\X] \left( \delta^0_{\alpha} + \Lambda^{-2} \nabla^0 A_{\alpha} \right) + 6 \X^0_{\phantom{0} \mu} \left( \delta^{\mu}_{\alpha} + \Lambda^{-2} \nabla^{\mu} A_{\alpha} \right) \,.
\end{equation}
We will make use of the following identities
\begin{align}
	&V^{\alpha} \left( \delta^{\mu}_{\alpha} + \Lambda^{-2} \nabla^{\mu} A_{\alpha} \right) = \X^{0 \mu} \,, \\
	&\frac{\p \X^{0\mu}}{\p \dot{A}^{\beta}} \X^0_{\phantom{0} \mu} = \Lambda^{-2} g^{00} \left( \delta^0_{\beta} + \Lambda^{-2} \nabla^0 A_{\beta} \right) \,,
\end{align}
so as to derive the following matrix product between the Hessian and the vector $V$,
\begin{align}
	\mathcal{H}_{\alpha \beta}^{(3)} V^{\alpha} =& 6 \left\lbrace [\X] V_{\beta} - \left( \delta^0_{\beta} + \Lambda^{-2} \nabla^0 A_{\beta} \right) \right\rbrace V_{\alpha} V^{\alpha} + \frac{3}{2} \Lambda^2 \left( [\X]^2 - [\X^2] \right) \frac{\p \left( V_{\alpha} V^{\alpha} \right)}{\p \dot{A}^{\beta}} \nonumber \\
	& - 6 V_{\beta} \left( \delta^0_{\alpha} + \Lambda^{-2} \nabla^0 A_{\alpha} \right) V^{\alpha} - 6 [\X] g^{00} V_{\beta} + 6 \Lambda^2 \frac{\p \X^0_{\phantom{0} \mu}}{\p \dot{A}^{\beta}} \left( \delta^{\mu}_{\alpha} + \Lambda^{-2} \nabla^{\mu} A_{\alpha} \right) V^{\alpha} \nonumber \\
	& + 6 \X^{00} V_{\beta} \nonumber \\
	=& 6 \left[ - g^{00} \left( \delta^0_{\beta} + \Lambda^{-2} \nabla^0 A_{\beta} \right) - \X^{00} V_{\beta} + \Lambda^2_2 \frac{\p \X^0_{\phantom{0} \mu}}{\p \dot{A}^{\beta}} \X^{0 \mu} + \X^{00} V_{\beta} \right] \nonumber \\
	=& 0  \,.
\end{align}
\item \textbf{$\L_4[\X]$}

The canonical momentum coming from the Lagrangian at order 4 reads
\begin{equation}
	\Lambda^{-2} p_{\alpha}^{(4)} = 4 \L_3[\X] V_{\alpha} - 12 (\L_2[\X] \delta^0_{\mu} -2 ([\X] \X^0_{\phantom{0} \mu} - f^0_{\phantom{0} \mu})) \left( \delta^{\mu}_{\alpha} + \Lambda^{-2} \nabla^{\mu} A_{\alpha} \right) \,,
\end{equation}
and the eigenvalue equation follows directly
\begin{align}
	\mathcal{H}_{\alpha \beta}^{(4)} V^{\alpha} =& 4 \Lambda^{-2} p_{\beta}^{(3)} g^{00} + 4 \Lambda^2 \L_3[\X] \frac{\p V_{\alpha}}{\p \dot{A}^{\beta}} V_{\alpha} \nonumber \\
	& - 12 \left[ \Lambda^{-2} p_{\beta}^{(2)} \delta^0_{\mu} - 2 \Lambda^{-2} p_{\beta}^{(1)} \X^0_{\phantom{0} \mu} - 2 \Lambda^2 [\X] \frac{\p \X^0_{\phantom{0} \mu}}{\p \dot{A}^{\beta}} + 2 g^{00}  \left( g_{\mu \beta} + \Lambda^{-2} \nabla_{\mu} A_{\beta} \right) \right. \nonumber \\
	& \qquad \quad \left. \vphantom{\frac{\p \X^{0\mu}}{\p \dot{A}^{\beta}}} + 2 \delta^0_{\mu}  \left( \delta^{0}_{\beta} + \Lambda^{-2} \nabla^{0} A_{\beta} \right) \right] \X^{0 \mu} \nonumber \\
	&- 12 (\L_2[\X] \delta^0_{\mu} -2 ([\X] \X^0_{\phantom{0} \mu} - f^0_{\phantom{0} \mu})) g^{0 \mu} V_{\beta} \nonumber \\
	=& 4 g^{00} \left[ 3 \L_2[\X] V_{\beta} - 6 [\X] \left( \delta^{0}_{\beta} + \Lambda^{-2} \nabla^{0} A_{\beta} \right) + 6 \X^0_{\phantom{0} \mu} \left( \delta^{\mu}_{\beta} + \Lambda^{-2} \nabla^{\mu} A_{\beta} \right) \right] \nonumber \\
	&-24 \left[ \left\lbrace [\X] V_{\beta} - \left( \delta^{0}_{\beta} + \Lambda^{-2} \nabla^{0} A_{\beta} \right) \right\rbrace \X^{00} - f^{00} V_{\beta} - g^{00} [\X] \left( \delta^{0}_{\beta} + \Lambda^{-2} \nabla^{0} A_{\beta} \right) \right. \nonumber \\
	& \qquad \quad \left. + g^{00} \X^{0 \mu} \left( \delta_{\mu \beta} + \Lambda^{-2} \nabla_{\mu} A_{\beta} \right) + \X^{00} \left( \delta^{0}_{\beta} + \Lambda^{-2} \nabla^{0} A_{\beta} \right) \right] \nonumber \\
	&- 12 \left( g^{00} \L_2[\X] - 2 [\X] \X^{00} + 2 f^{00} \right) V_{\beta} \nonumber \\
	=& 0 \,.
\end{align}
This concludes the proof that $V$ is the common null eigenvector to $\L_1$, $\L_2$, $\L_3$ and $\L_4$,
\begin{equation}
	\Rightarrow \mathcal{H}^{(n)}_{\alpha \beta} V^{\alpha} = 0, \quad \text{for} \quad n=1,2,3,4 \,.
\end{equation}
\end{itemize}

\section{Null eigenvector on a dynamical background}
\label{ssec:NEVDyn}

In this section, we will consider the background to be dynamical, and hence extend the dynamical phase space including those contained in the gravitational sector. The NEV $V_{\mu}^{\ast}$ defined in \eqref{eq:Vast} is now embedded in a higher-dimensional vector $\bm{\mathcal{V}}=(V_{\mu}^{\ast},0)$ where the null entries run through the metric components. We have proven that $\mathcal{H}^{(n)}_{\alpha \beta} V^{\alpha} = 0$ for $n=1,...,3$ on a non-dynamical background. When coupling EPN to gravitational degrees of freedom, the vector $V_{\mu}^{\ast}$ is related to $V_{\mu}$ by the linear transformation $V_{\mu}^{\ast}=(M^{-1})_{\mu}^{\phantom{\mu}\nu} V_{\nu}$ and thus it is immediate to see that $\mathcal{H}^{\ast,(n)}_{\alpha \beta} V^{\ast,\alpha} = 0$ for $n=1,...,3$, i.e. $\bm{\mathcal{V}}$ annihilates the pure vector sector. This is trivially true for the pure metric sector. In order to prove that the higher-dimensional vector $\bm{\mathcal{V}}$ is the correct NEV on a dynamical background, one has to check that it also annihilates the mixed vector-metric sector. The Hessian matrix for the mixed vector-metric sector is defined to be
\ba
\mathcal{H}^{*(n)}_{\mu,ij} = \frac{\p p^{*(n)}_{\mu}}{\p \dot{\gamma}^{ij}}=
\Lambda^4 \frac{\p^2 \L_n[\X]}{\p \dot{A}^{\ast\mu}\p \dot{\gamma}^{ij}}\,.
\ea
We have previously shown that $V_{\mu}^{\ast}$ is indeed a NEV for the Hessian $\mathcal{H}^{*(1)}_{\mu,ij}$ and hence   $\bm{\mathcal{V}}$ is a NEV of the full Hessian associated with $\L_1[\X]$.
We will now prove that even though $\bm{\mathcal{V}}$ fails to remain a NEV for $\L_2$ (and $\L_3$), it is possible to add non-minimal couplings to $\L_2$ such that symbolically $\mathcal{H} \bm{\mathcal{V}}$ vanishes in all sectors at leading order in $(\nabla A) / \Lambda^2$. This seems to indicate that one could possibly add further non-minimal couplings to push the constraint to the next order and so on in an infinite series. However, this is only  postulated at this stage and proving such a statement in generality is beyond the scope of this work. Nevertheless, our results are  interesting in their own right and we will further show that it immediately follows that $\bm{\mathcal{V}}$ is the NEV of $\L_2^{(\text{non-min})}$ on any background such that $\nabla_{\mu} A_{\nu}$ is symmetric, e.g. FLRW. Even though the covariantization fails on a generic dynamical background, this is a proof that EPN can be considered for cosmology. An estimation for the mass of the resulting ghost on background where the field strength tensor acquired a non-vanishing vev is given in the main text.

\paragraph{$\L_2$ without non-minimal couplings.}

We start by computing the Hessian matrix associated with $\L_2$ in the mixed vector-metric sector. First note that with the covariantization introduced in \eqref{eq:SextPN}, the time-derivatives of the spatial metric do not only enter through the curvature, but also through the covariant derivative of the vector field. To include their contributions, we first consider the following derivatives
\begin{equation}
 \frac{\p \Gamma^{\beta}_{\mu \alpha}}{\p \dot{\gamma}_{kl}} = \frac14 g^{\beta \lambda} \left( \delta^0_{\mu} \delta^k_{\lambda} \delta^l_{\alpha} + \delta^0_{\alpha} \delta^k_{\lambda} \delta^l_{\mu} - \delta^0_{\lambda} \delta^k_{\mu} \delta^l_{\alpha} \right) + \left\{ k \leftrightarrow l \right\} \,.
\end{equation}
Throughout the rest of this appendix, we will consider $A^{\mu}$ (with upper index) to be constant with respect to $\gamma^{ij}$ and as a result $\p_{\mu} A_{\nu}$ will also contribute when differentiating with respect to $\dot{\gamma}^{ij}$. The derivative of $\nabla_\mu A^\nu$ is hence given by
\begin{align}
	\frac{\p (\nabla_{\mu} A^{\nu})}{\p \dot{\gamma}^{ij}} &= - \frac14 \gamma_{ik} \gamma_{jl} g^{\nu \lambda} \left( ( \delta^0_{\mu} \delta^k_{\lambda} - \delta^0_{\lambda} \delta^k_{\mu} ) A^l + \delta^k_{\mu} \delta^l_{\lambda} A^0 \right) + \left\lbrace i \leftrightarrow j \right\rbrace \nonumber \\
	&= - \frac14 \left[ (\delta^0_{\mu} \delta^{\nu}_i - g_{\mu i} g^{0 \nu}) A_j + g_{\mu i} \delta^{\nu}_j A^0 - 2 \delta^0_{\mu} \delta^{\nu}_i N_j A^0 + \delta^0_{\mu} g^{0 \nu} N_i N_j A^0 \right] + \left\lbrace i \leftrightarrow j \right\rbrace \,,
\end{align}
while that of $\nabla_{\mu} A_{\nu}$ follows trivially. Now, in order to compute the derivative of the momentum $p^{(2)}$ with respect to $\dot{\gamma}_{ij}$, we need
\begin{align}
	\frac{\p [\X]}{\p \dot{\gamma}^{ij}} &= \frac12 \frac{\p f\mn}{\p \dot{\gamma}^{ij}} \left(\X^{-1}\right)\mnup
	= \left[ \frac{\p(\nabla_{\mu} A_{\alpha})}{\p \dot{\gamma^{ij}}} + \Lambda^{-2} \frac{\p(\nabla_{\mu} A_{\nu})}{\p \dot{\gamma^{ij}}} \nabla_{\alpha} A^{\nu} \right] \left(\X^{-1}\right)^{\mu\alpha} \nonumber \\
	&= \Lambda^{-2} \frac{\p(\nabla_{\mu} A_{\nu})}{\p \dot{\gamma^{ij}}} \left(\X^{-1}\right)^{\mu\alpha} \left( \delta^{\nu}_{\alpha} + \Lambda^{-2} \nabla_{\alpha} A^{\nu} \right)
	= \Lambda^{-2} \frac{\p(\nabla_{\mu} A_{\nu})}{\p \dot{\gamma^{ij}}} V\mnup \nonumber \\
	&= \frac14 \Lambda^{-2} \left\lbrace A_i \left( V_j^{\phantom{j} 0} - V_j \right) + A^0 \left( 2 V_i N_j - V_{ij} - N_i N_j V^0 \right) \right\rbrace + \left\{ i \leftrightarrow j \right\}  \,,
\end{align}
where we have introduced
\begin{equation}
	V^{\mu \nu} = \left( \X^{-1} \right)^{\mu \alpha} \left( \delta^{\nu}_{\alpha} + \Lambda^{-2} \nabla_{\alpha} A^{\nu} \right) \,,
\end{equation}
such that
\begin{equation}
	V^{\mu} = V^{0 \mu} \,.
\end{equation}
On the hand, we have
\begin{equation}
	\frac{\p \left(\nabla^0 A_{\alpha} \right)}{\p \dot{\gamma}^{ij}} V^{\alpha} = \frac14 g^{00} \left( A^0 ( 2 V_i N_j - N_i N_j V^0 ) - A_i V_j \right) + \left\{ i \leftrightarrow j \right\} \,.
\end{equation}
Putting everything together we find that the contraction of the Hessian of $\alpha_{2,X} \L_2$ with the null eigenvector for $\L_1$ is now
\begin{align}
	\mathcal{H}^{(2)}_{\alpha,ij} V^{\alpha} &= \frac{\p p^{(2)}_{\alpha}}{\p \dot{\gamma}^{ij}} V^{\alpha} \nonumber \\
	&= 2 \alpha_{2,X} \Lambda^2 \left\lbrace \frac{\p [\X]}{\p \dot{\gamma}^{ij}} V_{\alpha} V^{\alpha} + [\X] \frac{\p V_{\alpha}}{\p \dot{\gamma}^{ij}} V^{\alpha} - \Lambda^{-2} \frac{\p \left(\nabla^0 A_{\alpha} \right)}{\p \dot{\gamma}^{ij}} V^{\alpha} \right\rbrace \nonumber \\
	&= 2 \alpha_{2,X} \Lambda^2 \left\lbrace g^{00} \frac{\p [\X]}{\p \dot{\gamma}^{ij}} - \Lambda^{-2} \frac{\p \left(\nabla^0 A_{\alpha} \right)}{\p \dot{\gamma}^{ij}} V^{\alpha} \right\rbrace \nonumber \\
	&= \frac12 \alpha_{2,X} g^{00} \left( V_i^{\phantom{i} 0} A_j + V_j^{\phantom{j} 0}  A_i - V_{ij} A^0 - V_{ji} A^0 \right) \nonumber \\
	&= - \alpha_{2,X} g^{00} \gamma_{ij} A^0 + \frac{1}{2\Lambda^2} \alpha_{2,X} g^{00} F_{(i}^{\phantom{(i}0}A_{j)} + \mathcal{O}((\nabla A)^{2}/\Lambda^{4}) \,,
	\label{eq:HV2}
\end{align}
which does not generically vanish. Considering this result in an operator expansion, or power expansion in $\nabla A/\Lambda^2$, we see that to leading order in that expansion, we get
\begin{equation}
	V\mn = g\mn + \frac{F\mn}{2\Lambda^2} - \frac{1}{8\Lambda^4} \left( F_{\mu}^{\phantom{\mu} \alpha} F_{\nu \alpha} - 4 \nabla_{\alpha} A_{[\mu} \nabla_{\nu]} A^{\alpha} \right) + \mathcal{O}((\nabla A)^{3}/\Lambda^{6}) \,,
\end{equation}
and it is clear that \eqref{eq:HV2} does not vanish at leading order in the operator expansion.
The previous result is surprising in itself and indeed the same occurs for GP at precisely the same level. The resolution in that case is the introduction of non-minimal couplings to gravity as already provided in \cite{Heisenberg:2014rta}.

\paragraph{Addition of non-minimal couplings}

In the context of EPN, the generalization of those non-minimal couplings is however much more challenging to find, particularly due to the fact that the constraint has to be satisfied non-linearly through mixing of orders. At this stage, there is no candidate for a straightforward and natural non-minimal coupling, however for lack of a better insight, we consider the inclusion of the following non-minimal coupling:
\begin{align}
	\sqrt{-g} \Lambda^2 \alpha_2[X] R &= \sqrt{-g} \Lambda^2 \alpha_2[X] g\mnup \left( \p_{\alpha} \Gamma^{\alpha}\mn - \p_{\nu} \Gamma^{\alpha}_{\alpha \mu} + \Gamma^{\alpha}\mn \Gamma^{\beta}_{\alpha \beta} - \Gamma^{\beta}_{\mu \alpha} \Gamma^{\alpha}_{\nu \beta} \right) \nonumber \\
	&= \sqrt{-g} \Lambda^2 \alpha_2[X] g\mnup \left( \nabla_{\alpha} \Gamma^{\alpha}\mn - \nabla_{\nu} \Gamma^{\alpha}_{\alpha \mu} + \mathcal{O}(\Gamma^2) \right) \nonumber \\
	&= \sqrt{-g} \Lambda^2 \alpha_2[X] \left( \nabla_{\alpha} \left( g\mnup \Gamma^{\alpha}\mn \right) - \nabla^{\mu} \Gamma^{\alpha}_{\alpha \mu} + \mathcal{O}(\Gamma^2) \right) \nonumber \\
	&= \sqrt{-g} \Lambda^2 \left( \Gamma^{\alpha}_{\alpha \mu} \nabla^{\mu} \alpha_2[X] - g\mnup \Gamma^{\alpha}\mn \nabla_{\alpha} \alpha_2[X] + \mathcal{O}(\Gamma^2) \right) \nonumber \\
	&= \sqrt{-g} \alpha_{2,X}[X] A_{\beta} \left( g\mnup \Gamma^{\alpha}\mn \nabla_{\alpha} A^{\beta} - \Gamma^{\alpha}_{\alpha \mu} \nabla^{\mu}A^{\beta} + \mathcal{O}(\Gamma^2) \right) \,.
\end{align}
Now defining $p^{(2,R)}$ as the momentum conjugate to $A$ with respect to the non-minimal coupling part of the Lagrangian at order 2,
\begin{equation}
	\Rightarrow p^{(2,R)}_{\alpha} = \frac{\p \left( \Lambda^2 \alpha_2[X] R \right)}{\p \dot{A}^{\alpha}} = \alpha_{2,X} A_{\alpha} \left( g\mnup \Gamma^0\mn - g^{\mu 0} \Gamma^{\nu}\mn \right) \,,
\end{equation}
and $\mathcal{H}^{(2,R)}_{\alpha,ij}$ as the contribution to the second-order Hamiltonian purely coming from the second-order non-minimal coupling to gravity,
\begin{align}
	\mathcal{H}^{(2,R)}_{\alpha,ij} &= \frac{\p p^{(2,R)}_{\alpha}}{\p \dot{\gamma}^{ij}} \nonumber \\
	&= - \gamma_{ik} \gamma_{jl} \alpha_{2,X} A_{\alpha} \left( g\mnup \frac{\p \Gamma^0\mn}{\p \dot{\gamma}_{kl}} - g^{\mu 0} \frac{\p \Gamma^{\nu}\mn}{\p \dot{\gamma}_{kl}} \right) \nonumber \\
	&= - \frac14 \gamma_{ik} \gamma_{jl} \alpha_{2,X} A_{\alpha} \left( \left(4 g^{0k} g^{0l} - 2 g^{00} g^{kl} \right) - 2 g^{00} g^{kl} \right) \nonumber \\
	&= \gamma_{ik} \gamma_{jl} \alpha_{2,X} A_{\alpha} \left( g^{00} g^{kl} - g^{0k} g^{0l} \right) \nonumber \\
	&= \alpha_{2,X} A_{\alpha} \left( g^{00} \left( \gamma_{ij} + g^{00} N_i N_j \right) - \left( - g^{00} N_i \right) \left( - g^{00} N_j \right) \right) \nonumber \\
	&= g^{00} \gamma_{ij} \alpha_{2,X} A_{\alpha} \,,
\end{align}
then, we get the following eigenstate equation
\begin{align}
	\Rightarrow \mathcal{H}^{(2,R)}_{\alpha,ij} V^{\alpha} &= g^{00} \gamma_{ij} \alpha_{2,X} A_{\alpha} V^{\alpha} \nonumber \\
	&= \alpha_{2,X} g^{00} \gamma_{ij} \left( A^0 + \frac{1}{2\Lambda^2} F^{0 \alpha} A_{\alpha} + \mathcal{O}((\nabla A)^{2}/\Lambda^{4}) \right) \,.
	\label{eq:HV2R}
\end{align}
Separately, neither \eqref{eq:HV2} nor \eqref{eq:HV2R} vanish at leading order in $(\nabla A)/\Lambda^{2}$. However, when adding these two contributions, we get a cancellation at leading order,
\begin{align}
	\Rightarrow \left( \mathcal{H}^{(2)}_{\alpha,ij} + \mathcal{H}^{(2,R)}_{\alpha,ij} \right) V^{\alpha} &= \frac12 g^{00} \alpha_{2,X} \left( 2 \gamma_{ij} A_{\alpha} V^{\alpha} + V_i^{\phantom{i} 0} A_j + V_j^{\phantom{j} 0}  A_i - V_{ij} A^0 - V_{ji} A^0 \right) \nonumber \\
	&= 0 + \frac{1}{2\Lambda^2} \alpha_{2,X} g^{00} \left( F_{(i}^{\phantom{(i}0} A_{j)} + \gamma_{ij} F^{0 \alpha} A_{\alpha} \right) + \mathcal{O}((\nabla A)^{2}/\Lambda^{4}) \,.
	\label{eq:sumHcoupl}
\end{align}
Now, this equation is vanishing at leading order in $(\nabla A)/\Lambda^{2}$ but not to higher order. From this we  conclude that by itself the minimal coupling $\alpha_2[X] R$ does help with pushing the breaking of the constraint to a higher order but is not sufficient to ensure that the constraint will be satisfied to all orders. Other more general non-minimal couplings are currently under investigations but those are kept to another study since for what interests us in the context of cosmology is to ensure the absence of ghosts on cosmological backgrounds. In this context, the tensor $\nabla_{\mu}A_\nu$ is symmetric and the right hand side of \eqref{eq:sumHcoupl} then vanishes.  Indeed, if $\nabla A$ is symmetric then $f$ is nothing other than $(1 + \nabla A / \Lambda^2)^2$, meaning that $\chi$ reduces to the simple form $1 + \nabla A / \Lambda^2$. Finally, we have
\begin{equation}
	V\mnup = \left[ \left( 1 + \nabla A / \Lambda^2 \right)^{-1} \right]^{\mu \alpha} \left[ 1 + \nabla A / \Lambda^2 \right]_{\alpha}^{\phantom{\alpha}\nu} = g\mnup \,,
\end{equation}
which is simply the zeroth-order of the general formula, proving that the right hand side of \eqref{eq:sumHcoupl} vanishes for any configurations where the field strength tensor $F\mn$ vanishes.

\section{Special example with no non-minimal coupling}
\label{ssec:NEVspec}

We now establish whether the special example considered in Section \ref{sec:SpecEx} enjoys a constraint when coupled to gravity. We start by defining the momenta associated with the GP operators as
\begin{equation}
	p^{(n,{\rm GP})}_{\alpha} =\frac{\p \L_n[\nabla A]}{\p \dot{A}^{\alpha}} \,,
\end{equation}
so that we have
\begin{align}
	p^{(1,{\rm GP})}_{\alpha} &= \delta^0_{\alpha} \,, \\
	p^{(2,{\rm GP})}_{\alpha} &= 2 \left( \delta^0_{\alpha} \nabla_{\mu} A^{\mu} - \nabla_{\alpha} A^0 \right) \,, \\
	p^{(3,{\rm GP})}_{\alpha} &= 3 \delta^0_{\alpha} \left( (\nabla_{\mu} A^{\mu})^2 - \nabla_{\mu} A^{\nu} \nabla_{\nu} A^{\mu} \right) + 6 \left( \nabla_{\alpha} A^{\mu} \nabla_{\mu} A^{0} - \nabla_{\alpha} A^{0} \nabla_{\mu} A^{\mu} \right) \,.
\end{align}
From there we can immediately check that by themselves, the GP type of terms constructed out of symmetric polynomials of $(\nabla A)$  do not contribute to the Hessian matrix, namely
\begin{equation}
	\mathcal{H}^{(n,{\rm GP})}_{\alpha \beta} = 0, \qquad n=1,2,3 \,.
\end{equation}
Is is then clear that $V^{\alpha}$ (the covariantization of the Minkowski NEV for EPN) is still satisfying
\begin{equation}
	\mathcal{H}_{\alpha \beta} ( \hat{\L} ) V^{\alpha} = 0 \,.
\end{equation}
Now focusing on the part of the Hessian matrix that probes the mixing between $\dot{A}$ and $\dot{\gamma}$ we find
\begin{equation}
	\mathcal{H}_{\alpha,ij}^{(2,{\rm GP})} V^{\alpha} = - \frac12 \left[ \gamma_{ij} A^0 V^0 + g^{00} A_i V_j \right] + \left\{ i \leftrightarrow j \right\} \,,
\end{equation}
leading to
\begin{align}
	\mathcal{H}_{\alpha,ij}(\hat{\L}^{(2)}) V^{\alpha} &= \frac12 \alpha_{2,X} \left[ A^0 (\gamma_{ij} V^0 - V_{ij} g^{00} ) + g^{00} A_i ( V_j + V_j^{\phantom{j} 0} ) \right] + \left\{ i \leftrightarrow j \right\}  \\
	&= 0 + \frac{0}{\Lambda^2} - \frac{1}{8\Lambda^4} \left[ \gamma_{ij} A^0 F^{0 \alpha} F_{0 \alpha} + g^{00} \left( 2 F^{0 \alpha} A_{(i} F_{j) \alpha} - A^0 F_{i}^{\phantom{i}\alpha} F_{j \alpha} \right) \right] + \mathcal{O}((\nabla A)^{3}/\Lambda^{6}) \,,\nonumber
\end{align}
which again fails to vanish at all orders but vanishes at leading and next-to-leading order in the operator expansion and vanishes on any background for which the field strength tensor vanishes, $F\mn=0$, as is the case for cosmology.

\chapter{Cosmology of Proca-Nuevo}

\section{Definition of some coefficients in the perturbed quadratic actions}
\label{sec:DefCoefs}

\subsection{Coefficients of the scalar perturbations in the special model}
\label{ssec:DefCoefsExScal}

We define here the $7$ coefficients entering the quadratic scalar action of the the special model \eqref{eq:SSnocoupl2},
\begin{align}
	\hat{\omega}_1 &= - 2 M_{\text{Pl}}^2 H - \phi^3 \left( \alpha_{1,X} + d_{1,X} \right) \,, \nonumber \\
	\hat{\omega}_2 &= \hat{\omega}_1 + 2 M_{\text{Pl}}^2 H \,, \nonumber \\
	\hat{\omega}_3 &= -2 \phi^2 \hat{q}_V \,, \nonumber \\
	\hat{\omega}_4 &=  -3 M_{\text{Pl}}^2 H^2 + \frac{1}{2} \phi^4 \alpha_{0,XX} - \frac{3}{2} H \phi^3 \left[ ( \alpha_{1,X} +  d_{1,X} ) - \frac{\phi^2}{\Lambda^2} \left( \alpha_{1,XX} +  d_{1,XX} \right) \right] \,, \nonumber \\
	\hat{\omega}_5 &= \hat{\omega}_4 - \frac32 H ( \hat{\omega}_1 + \hat{\omega}_2 ) \,, \nonumber \\
	\hat{\omega}_6 &= - \phi^2 \left( \alpha_{1,X} + d_{1,X} \right) \,, \nonumber \\
	\hat{\omega}_7 &= - \dot{\phi} \left( \alpha_{1,X} + d_{1,X} \right) \,.
	\label{eq:coefsomegaspec}
\end{align}

\subsection{Masses of the scalar modes in the special model}
\label{ssec:MassesExScal}

In this Appendix, we present the results for the square masses of both scalar modes, the matter perturbation $\delta \rho_M / k$ and the scalar $\psi$. These masses are inferred from the dispersion relation \eqref{eq:SpecExdispSM} and hence are canonically normalized. To begin with, the mass of the matter field is given by
\begin{equation}
	\hat{m}_{S,M}^2 = \frac{\hat{\Theta}}{2 \phi^2 \hat{q}_V (\hat{\omega}_1 - 2 \hat{\omega}_2)^2} \,,
\end{equation}
where
\begin{align}
	\hat{\Theta} = 2 (\rho_M + P_M) \hat{\omega}_2^2 + \hat{\omega}_3 &\left[ (\rho_M + P_M)^2 - H (\rho_M + P_M) \left\lbrace (\hat{\omega}_1 - 2 \hat{\omega}_2 ) \left( 1 + 6 c_M^2 \right) + \frac{\dot{\hat{\omega}}_1 - 2 \dot{\hat{\omega}}_2}{H} \right\rbrace \right.  \\
	& \quad \left. + 3 H^2 (\hat{\omega}_1 - 2 \hat{\omega}_2 )^2 \left\lbrace 3 c_M^4 + c_M^2 - 2 - (1 + c_M^2) \left( \frac{\dot{H}}{H^2} - \frac{\p_t ( \rho_M + P_M)}{H ( \rho_M + P_M )} \right\rbrace \right) \right] \,.\nonumber
\end{align}
Avoiding a tachyonic instability is achieved by requiring
\begin{equation}
	\hat{\Theta} > 0 \,.
\end{equation}
On another hand, one can also derive the mass of the scalar field $\psi$,
\begin{equation}
	\hat{m}_{S,\psi}^2 = \frac{1}{(\hat{\omega}_1-2\hat{\omega}_2)^2 \hat{\omega}_3^2} \frac{\hat{\Xi}_1}{\hat{\Xi}_2} \,,
\end{equation}
where
\begin{align}
	\hat{\Xi}_1 =& - (\hat{\omega}_1-2\hat{\omega}_2)^2 \hat{\omega}_3 \hat{\Xi}_2 \phi \dot{\hat{\omega}}_6 \nonumber \\
	& + 2 (\hat{\omega}_1-2\hat{\omega}_2)(\hat{\omega}_1-\hat{\omega}_2) \hat{\omega}_3 \phi \left[ \hat{\omega}_2 (\hat{\omega}_1^2 - \hat{\omega}_1 \hat{\omega}_2 + (P_M + \rho_M) \hat{\omega}_3) + (\hat{\omega}_1 - 2 \hat{\omega}_2)(\hat{\omega}_1-\hat{\omega}_2) \hat{\omega}_6 \phi \right] \dot{\hat{\omega}}_4 \nonumber \\
	& + (\hat{\omega}_1 - 2 \hat{\omega}_2) ( \hat{\omega}_1 \hat{\omega}_2 + (\hat{\omega}_1 - 2 \hat{\omega}_2) \hat{\omega}_6 \phi) \hat{\Xi}_2 \dot{\hat{\omega}}_3 \nonumber \\
	&- \hat{\omega}_3 \phi \left[ 3 H \hat{\omega}_1^2 \left\lbrace \hat{\omega}_1 (\hat{\omega}_1 (\hat{\omega}_1 +2 \hat{\omega}_2) - 2\hat{\omega}_2^2) + (\hat{\omega}_1-2\hat{\omega}_2)(3\hat{\omega}_1-2\hat{\omega}_2)\hat{\omega}_6 \phi) \right. \right. \nonumber \\
	&\phantom{- \hat{\omega}_3 \phi} \qquad \qquad \left. \left. + (P_M + \rho_M)(\hat{\omega}_1 + 4 \hat{\omega}_2) \hat{\omega}_3 \right\rbrace \right. \nonumber \\
	&\phantom{- \hat{\omega}_3 \phi} \quad \left. - 2 \hat{\omega}_4 \left\lbrace \hat{\omega}_1 (\hat{\omega}_1 - \hat{\omega}_2) (\hat{\omega}_1 (\hat{\omega}_1 + \hat{\omega}_2) + 2(\hat{\omega}_1-2\hat{\omega}_2)\hat{\omega}_6 \phi) \right. \right. \nonumber \\
	&\phantom{- \hat{\omega}_3 \phi} \qquad \qquad \left. \left. + (P_M + \rho_M) (\hat{\omega}_1 (\hat{\omega}_1 +2 \hat{\omega}_2) - 2\hat{\omega}_2^2) \hat{\omega}_3 \right\rbrace \right] \dot{\hat{\omega}}_2 \nonumber \\
	&- \hat{\omega}_3 \phi \left[ 3 H \hat{\omega}_1 \left\lbrace \hat{\omega}_1 \hat{\omega}_2 (\hat{\omega}_1^2 - 8 \hat{\omega}_1 \hat{\omega}_2 + 6\hat{\omega}_2^2) + (\hat{\omega}_1 - 2 \hat{\omega}_2) (\hat{\omega}_1^2 - 6\hat{\omega}_1 \hat{\omega}_2 + 4\hat{\omega}_2^2) \hat{\omega}_6 \phi \right. \right. \nonumber \\
	&\phantom{- \hat{\omega}_3 \phi} \qquad \qquad \left. - (P_M + \rho_M) (\hat{\omega}_1 + 4\hat{\omega}_2) \hat{\omega}_2 \hat{\omega}_3 \right\rbrace \nonumber \\
	&\phantom{- \hat{\omega}_3 \phi} \quad \left. + 2 \hat{\omega}_2 \hat{\omega}_4 \left\lbrace (2\hat{\omega}_1- \hat{\omega}_2)(2(\hat{\omega}_1-\hat{\omega}_2)\hat{\omega}_2 + (P_M + \rho_M)\hat{\omega}_3) + 2(\hat{\omega}_1-2\hat{\omega}_2) (\hat{\omega}_1-\hat{\omega}_2) \hat{\omega}_6 \phi \right\rbrace \right] \dot{\hat{\omega}}_1 \nonumber \\
	& + 3\hat{\omega}_1^2 (\hat{\omega}_1 - 2\hat{\omega}_2) \hat{\omega}_3 \left[ \hat{\omega}_2 (\hat{\omega}_1 (\hat{\omega}_1-\hat{\omega}_2) + \hat{\omega}_3 (P_M + \rho_M)) + (\hat{\omega}_1-2\hat{\omega}_2)(\hat{\omega}_1-\hat{\omega}_2)\hat{\omega}_6 \phi \right] \phi \dot{H} \nonumber \\
	& - (\hat{\omega}_1 - 2\hat{\omega}_2) \hat{\omega}_2 \hat{\omega}_3^2 \left[ 3H \hat{\omega}_1^2 - 2 (\hat{\omega}_1-\hat{\omega}_2)\hat{\omega}_4 \right] \phi \p_t (P_M + \rho_M) \nonumber \\
	& + (\hat{\omega}_1-2\hat{\omega}_2)\hat{\omega}_3 (3 H \hat{\omega}_1^2 -2(\hat{\omega}_1-\hat{\omega}_2)\hat{\omega}_4) \nonumber \\
	& \qquad \qquad \times \left[ 2 \hat{\omega}_2(\hat{\omega}_1(\hat{\omega}_1-\hat{\omega}_2)+ \hat{\omega}_3(P_M+\rho_M)) + (\hat{\omega}_1-2\hat{\omega}_2)(\hat{\omega}_1-\hat{\omega}_2)\hat{\omega}_6 \phi \right] \dot{\phi} \nonumber \\
	& - (\hat{\omega}_1 - 2\hat{\omega}_2) \left[ 2 \hat{\omega}_2 \hat{\omega}_3 (3 H \hat{\omega}_2 + (P_M + \rho_M)) + \hat{\omega}_1^2 (2 \hat{\omega}_2 + 3 H \hat{\omega}_3) - \hat{\omega}_1 \hat{\omega}_2 (2 \hat{\omega}_2 + 9H \hat{\omega}_3) \right] \nonumber \\
	& \qquad \qquad \times (3 H \hat{\omega}_1^2 -2 (\hat{\omega}_1 -\hat{\omega}_2) \hat{\omega}_4) \phi^2 \hat{\omega}_6 \nonumber \\
	&+ \left[ 2 (\hat{\omega}_1-2\hat{\omega}_2)^2(\hat{\omega}_1-\hat{\omega}_2)^2 \hat{\omega}_6^2 \phi^3 + 2 \hat{\omega}_2 ( \hat{\omega}_1(\hat{\omega}_1-\hat{\omega}_2) + \hat{\omega}_3 (P_M + \rho_M) ) \right. \nonumber \\
	&\phantom{+} \qquad \qquad \qquad \left. \times \left\lbrace \hat{\omega}_1 \hat{\omega}_2 (\hat{\omega}_1 - \hat{\omega}_2) + \hat{\omega}_3 ( 3H(\hat{\omega}_1-2\hat{\omega}_2)(\hat{\omega}_1-\hat{\omega}_2) + \hat{\omega}_2 (P_M + \rho_M)) \right\rbrace \phi \right] \hat{\omega}_4 \nonumber \\
	&- 12 H \phi \hat{\omega}_2^3 \hat{\omega}_3^2 (P_M + \rho_M)^2 \nonumber \\
	&-3 H \hat{\omega}_1^2 \phi \left[ \hat{\omega}_2 \hat{\omega}_3 (2\hat{\omega}_1 \hat{\omega}_2 + 3H (\hat{\omega}_1 -2\hat{\omega}_2) \hat{\omega}_3) (P_M + \rho_M) \right. \nonumber \\
	&\phantom{-3 H \hat{\omega}_1^2 \phi} \quad \left.+ (\hat{\omega}_1-\hat{\omega}_2) \left\lbrace 3 H \hat{\omega}_1 (\hat{\omega}_1 - 2 \hat{\omega}_2) \hat{\omega}_2 \hat{\omega}_3 + \hat{\omega}_1^2 \hat{\omega}_2^2 + (\hat{\omega}_1 - 2 \hat{\omega}_2)^2 \hat{\omega}_6^2 \phi^2 \right\rbrace \right] \,,
\end{align}
and
\begin{equation}
	\hat{\Xi}_2 = (\hat{\omega}_1- \hat{\omega}_2) (3H \hat{\omega}_1^2 - 2(\hat{\omega}_1-\hat{\omega}_2)\hat{\omega}_4) \phi \,.
\end{equation}
The absence of tachyonic instabilities for the scalar mode $\psi$ is ensured by the positivity of $\hat{m}_{S,\psi}^2$, which is equivalent to require $\hat{\Xi}_1 / \hat{\Xi}_2 > 0$.

\subsection{Coefficients of the vector perturbations in the general model}
\label{ssec:DefCoefsVec}

In this Appendix, we will define the coefficients entering the quadratic action for the vector field in the general model \eqref{eq:SV2}.
\begin{align}
	q_V &= 1 - \frac{1}{2\mu} \alpha_1 + \frac{1}{\mu} \left( 1 - 2 \frac{H \phi}{\Lambda^2} \right) \alpha_{2,X} \nonumber \\
	&\qquad \qquad + \frac{1}{2\mu^2} \left\lbrace  \frac{3H^2 \mu - \left( 1 + H \phi / \Lambda^2 \right) \dot{H}}{\Lambda^2} \alpha_3 - \mu  \frac{H \phi}{\Lambda^2} \left( 2 - \frac{H \phi}{\Lambda^2} \right) \alpha_{3,X} \right\rbrace  \,, \nonumber \\
	\mathcal{C}_1 &= 1 + \frac{1}{2 \left( 1 + H \phi / \Lambda^2 \right)} \left[ - \alpha_1 + 2 \left( 1 - \frac{H \phi + \dot{\phi}}{\Lambda^2} \right) \alpha_{2,X} + \frac{3H^2 + 2 \dot{H}}{\Lambda^2} \alpha_3 \right. \nonumber \\
	& \qquad \qquad \qquad \qquad \qquad \qquad \left. - \left( \frac{H \phi}{\Lambda^2} + \left( 1 - \frac{H \phi}{\Lambda^2} \right) \frac{\dot{\phi}}{\Lambda^2} \right) \alpha_{3,X} \right] \,, \nonumber \\
	\mathcal{C}_2 &= \left( \alpha_{1,X} + d_{1,X} \right) \frac{\dot{\phi}}{H^2} + \frac{\dot{H}}{\Lambda^2 \mu} \alpha_3 + \frac{1}{H^2} \p_t \left( H \left[ 4 (\alpha_{2,X} + d_{2,X} ) + \frac{\dot{H}}{\Lambda^2 \mu} \alpha_3 - \frac{H \phi}{\Lambda^2} \left( \alpha_{3,X} + d_{3,X} \right) \right] \right) \nonumber \\
	& \qquad + 2 q_V + \frac{\p_t (q_V H)}{H^2} \,, \nonumber \\
	\mathcal{C}_3 &= 2 \frac{\phi}{M_{\text{Pl}}} \left( \alpha_{2,X} + d_{2,X} \right) - \frac{H}{M_{\text{Pl}}} \left( \frac{H \phi - \dot{\phi}}{2 \Lambda^2 \mu} \alpha_3 + \frac{\phi^2}{\Lambda^2} \left( \alpha_{3,X} + d_{3,X} \right) \right) \,, \nonumber \\
	\mathcal{C}_4 &= \frac{\dot{\phi} - H \phi}{2 \Lambda^2 M_{\text{Pl}} \mu} \alpha_3 \,,
\end{align}
where we have used
\begin{equation}
	\mu = 1 + \frac{H \phi + \dot{\phi}}{2\Lambda^2} \,.
\end{equation}

\subsection{Coefficients of the scalar perturbations in the general model}
\label{ssec:DefCoefsScal}

Finally, we define the coefficients entering the quadratic action for the scalar sector of the general model \eqref{eq:SS2}.
\begin{align}
	\omega_1 &= -2 M_{\text{Pl}}^2 H - \phi^3 ( \alpha_{1,X} + d_{1,X} )- 4 \Lambda^2 H \left[ (\alpha_2 + d_2 ) + \frac{\phi^4}{\Lambda^4} ( \alpha_{2,XX} + d_{2,XX} ) \right]  \\
	&\qquad + \frac{H^2 \phi^3}{\Lambda^2} \left[ \left(\alpha_{3,X} + d_{3,X} \right) + \frac{\phi^2}{\Lambda^2} \left(\alpha_{3,XX} + d_{3,XX} \right) \right] \,, \nonumber \\
	\omega_2 &= \omega_1 + 2 M_{\text{Pl}}^2 H q_T \,, \nonumber \\
	\omega_3 &= -2 \phi^2 q_V \,, \nonumber \\
	\omega_4 &= - 3 M_{\text{Pl}}^2 H^2 + \frac12 \phi^4 \alpha_{0,XX} - \frac32 H \phi^3 \left[ \left(\alpha_{1,X} + d_{1,X} \right) - \frac{\phi^2}{\Lambda^2} \left(\alpha_{1,XX} + d_{1,XX} \right) \right] \nonumber \\
	&\qquad - 3 \Lambda^2 H^2 \left[ 2 ( \alpha_2 + d_2 ) + 2 \frac{\phi^2}{\Lambda^2} (\alpha_{2,X} + d_{2,X}) + \frac{\phi^4}{\Lambda^4} (\alpha_{2,XX} + d_{2,XX}) - \frac{\phi^6}{\Lambda^6} (\alpha_{2,XXX} + d_{2,XXX}) \right] \nonumber \\
	&\qquad + \frac12 \frac{H^3 \phi^3}{\Lambda^2} \left[ 9 \left(\alpha_{3,X} + d_{3,X} \right) - \frac{\phi^4}{\Lambda^4} \left(\alpha_{3,XXX} + d_{3,XXX} \right) \right] \,, \nonumber \\
	\omega_5 &= \omega_4 - \frac32 H (\omega_1 + \omega_2) \,, \nonumber
\end{align}
\begin{align}
	\omega_6 &= - \phi^2 ( \alpha_{1,X} + d_{1,X} ) + 4 H \phi \left[ (\alpha_{2,X} + d_{2,X} ) - \frac{\phi^2}{\Lambda^2} ( \alpha_{2,XX} + d_{2,XX} ) \right]  \\
	&\qquad - \frac{H^2 \phi^2 }{\Lambda^2} \left[ \left(\alpha_{3,X} + d_{3,X} \right) - \frac{\phi^2}{\Lambda^2} \left(\alpha_{3,XX} + d_{3,XX} \right) \right] + 2 \frac{\dot{H} H \phi}{\mu \Lambda^2} \alpha_3 \,, \nonumber \\
	\omega_7 &= - \dot{\phi} ( \alpha_{1,X} + d_{1,X} ) - 4 \left[ \dot{H}(\alpha_{2,X} + d_{2,X} ) + \frac{H \phi \dot{\phi}}{\Lambda^2} ( \alpha_{2,XX} + d_{2,XX} ) \right] \nonumber \\
	&\qquad + \frac{H( 2 \dot{H} \phi + H \dot{\phi})}{\Lambda^2} \left(\alpha_{3,X} + d_{3,X} \right) + \frac{H^2 \phi^2 \dot{\phi}}{\Lambda^4} \left(\alpha_{3,XX} + d_{3,XX} \right) - \frac{\dot{H} H^2}{\mu \Lambda^2} \alpha_3 - \p_t \left( \frac{\dot{H} H}{\mu \Lambda^2} \alpha_3 \right) \,, \nonumber \\
	\omega_8 &= \frac{\dot{H} \phi^2}{\mu \Lambda^2} \alpha_3 \,, \nonumber \\
	\omega_9 &= \frac{H \phi (H \phi - \dot{\phi})}{\mu \Lambda^2} \alpha_3 \,.\nonumber
\end{align}
On a side note, these coefficients do not reproduce the ones of the special model depicted in \eqref{eq:coefsomegaspec} in the limit $\alpha_2 = - d_2$ and $\alpha_3 = - d_3$, showing the non-trivial relation between the special and the general models. This highlights the importance of either the tuning in the special model or the non-minimal couplings in the general model to construct a healthy theory.

\section{Operator relevance about the de Sitter point}
\label{sec:OperatordS}

We have seen in Section \ref{sssec:PertSpec} that the scalar and vector velocities vanish about the dS point. This could potentially suggest the presence of strong coupling issues associated with the breakdown of perturbative unitarity as we reach those fix points. We will show that this is not the case. Schematically, the action for the  mode $\psi$ reads
\begin{equation}
	S^{(2)}_{S,\psi} = \int \d t \, \d^3 x \, a^3 \left( - \frac12 Z\mnup \p_{\mu} \psi \p_{\nu} \psi \right) = \int \d t \, \d^3 x \, \frac{a^3}{2} \hat{Q}(t) \left( \dot{\psi}^2 - \frac{\hat{c}(t)^2}{a^2} (\p_i \psi)^2 \right) \,,
\end{equation}
defining an effective metric $Z\mnup$.  To determine the strong coupling scale, or the scale at which perturbative unitarity gets broken, we first normalize the field and the spacetime coordinates as performed for instance in \cite{PhysRevD.95.123523},
\ba
\tilde t=\int \hat c(t)\, \d t\,, \quad \tilde x^i=x^i\,, \quad {\rm and}\quad \psi =(\hat Q \hat c)^{-1/2}\tilde \psi\,,
\ea
so that in terms of these variables the field is canonically normalized
\ba
S^{(2)}_{S,\psi} = \int \d \tilde t \, \d^3 \tilde x \, \frac{a^3}{2} \left( \(\p_{\tilde t}\tilde \psi\)^2 - \frac{1}{a^2} \(\p_{\tilde x^i} \tilde\psi\)^2 \right) \,.
\ea
Treated as an EFT, we would expect the theory to include an infinite number of irrelevant operators of the form $a^{3-2L} \psi^N \dot{\psi}^M (\p_i \psi)^{2L}$ entering  at the respective  scale $\Lambda_{NML}$, where $\Lambda_{NML}$ is expected to be at least of order  $\Lambda$ and where $N$, $M$ and $L$ are positive integers respecting $N + 2M + 4L > 4$.
In terms of the normalized fields and coordinates, this operator enters at the physical scale $\tilde{\mu}_{NML}$ given by
\begin{equation}
	\tilde{\mu}_{NML}
= \hat{Q}^{\frac{N + M + 2L}{2(N+2M+4L-4)}} \left( \hat{c}^2 \right)^{\frac{N - M + 2L + 2}{4(N+2M+4L-4)}} \Lambda_{NML}\,.
\end{equation}
Now, the kinetic term $\hat{Q}$ and the velocity $\hat{c}^2$ are power laws of the scale factor $a(t)$ when approaching the dS point, for both the vector and the scalar mode. Let us define
\begin{equation}
	p_Q = \frac{\text{ln}(\hat{Q})}{\text{ln}(a)}, \qquad p_c = \frac{\text{ln}(\hat{c}^2)}{\text{ln}(a)} \,,
\end{equation}
so that
\begin{equation}
p_{NML}	\equiv \frac{\text{ln} ( \tilde{\mu}_{NML} / \Lambda_{NML} )}{\text{ln} (a)} = \frac{(N+2L)(2p_Q+p_c) + M(2p_Q-p_c) +2p_c}{4(N+2M+4L-4)} \,.
\end{equation}
The validity of the EFT is preserved as long as no scale $\tilde{\mu}_{NML}$ has a much smaller value than the corresponding scale $\Lambda_{NML}$, for any non-negative integers $N$, $M$, $L$ such that $N + 2M +4L > 4$. This translates into
\begin{equation}
	p_{NML} > 0, \quad \text{for all } N, M, L \ge 0, \quad N + 2M +4L > 4 \,.
\end{equation}
The scaling of the kinetic terms and velocities of the scalar and vector modes in terms of the scale factor $a(t)$ can be deduced from Figs.~\ref{fig:velocities} and \ref{fig:kin} and is summarized in the Table \ref{tab:apowers} below.
\begin{table}[h!]
\begin{center}
\begin{tabular}{ c | c | c }
  & vector & scalar $\psi$ \\ \hline
  $\hat{Q}$ & $a^3$ & $a^6$ \\
  $\hat{c}^2$ & $a^{-3}$ & $a^{-3}$
\end{tabular}
\caption[Scaling of kinetic terms and velocities about the dS point.]{Power-law behaviour of the vector and scalar kinetic terms and velocities as functions of the scale factor $a(t)$ about the dS point.}
\label{tab:apowers}
\end{center}
\end{table}

\noindent It follows that the value of $p_{NML}$ for the vector and scalar $\psi$ modes about the dS point read
\begin{equation}
	\begin{cases}
		(p_{NML})_{V} &= \frac{3}{4} \frac{N + 3M + 2L - 2}{N+2M+4L-4} \,, \\
		(p_{NML})_{S,\psi} &= \frac{3}{4} \frac{3N + 5M + 6L - 2}{N+2M+4L-4} \,.
	\end{cases}
\end{equation}
From here, it is easy to prove that $p_{NML}$ is always strictly positive for any of the allowed values of $N$, $M$ and $L$, for both the scalar and the vector. This concludes the proof that our EFT does not suffer any strong coupling issue due to the vanishing of the velocities about the de Sitter point (in link with the divergence of the associated kinetic terms).

\chapter{Causal scalar effective field theory}

\section{Causal time advances and Lorentz invariant UV completions}
\label{timeadvance}

As noted by Wigner and Eisenbud \cite{Eisenbud,Wigner:1955zz}, for scattering in a potential of finite range $a$, it is natural to obtain a scattering time advance of $2a/v$ for spherical wave scattering since this reflects the time advance of a wave which scatters directly off the hard boundary at $r=a$, relative to a wave which makes it to $r=0$. Clearly this does not violate causality, and so the causality condition of Wigner-Eisenbud for monopole ($\ell=0$) scattering is
\be\label{WEbound}
\Delta T > - \frac{2a}{v}   -\frac{{\cal O}(1)}{\omega} \, ,
\ee
with $v$ the group velocity of the wave. Given this, one may wonder whether we have been too strict in our consideration of monopole scattering by not allowing any time advance. The key difference is that we are interested in the scattering of essentially massless particles in the relativistic limit for which $\omega $ is large in comparison to the potential $V$, and the scale of variations of the potential $r_0$. More precisely we assume ${\rm Max}[V^{(n)}(r)]\ll \omega^{n+1}$ for all $n \ge 0$.
In this limit, no resolvable time advance is consistent with Lorentz invariant causality. 

To understand why this is the case, let us consider the case of relativistic scattering off of a \mbox{(quasi-)hard} sphere. To make comparison with the non-relativistic problem, consider a complex massive scalar field $\Phi$ of mass $m$, which is charged under a $U(1)$ gauge field whose Coulomb potential $q A_0 = V(r)$ takes the form
\be
V(r) = V_0 \theta(a-r) \, .
\ee
The equation of motion for the complex scalar is
\be
m^2 \Phi-\nabla^2 \Phi + D_t^2 \Phi =0  \, ,
\ee
where $D_t =\partial_t + i V$. For a given frequency and multipole we have
\be
(\omega-V(r))^2 \Phi = m^2 \Phi-\frac{1}{r^2} \frac{\partial }{\partial r} \( r^2 \frac{\partial \Phi}{\partial r} \) +\frac{\ell(\ell+1)}{r^2} \Phi \, .
\ee
The non-relativistic problem is obtained as usual by replacing $\omega = m+\omega_{\rm NR}$ and neglecting $\omega_{\rm NR}^2$ and $V^2$ terms.
Focussing on the monopole case $\ell=0$ for simplicity, the solution for $r<a$ which is regular at $r=0$ is
\be
\Phi(r) = \frac{A}{r} \sin \(\kappa_0 \,  r\) \, ,
\ee
with $\kappa_0= \sqrt{(\omega-V_0)^2-m^2}$. Denoting $k = \sqrt{\omega^2-m^2}$,
the solution for $r>a$ can be parametrized as
\be
\Phi(r) = \frac{A'}{2i r} \(e^{2i \delta} e^{i k r} -e^{-ik r}  \)   \, .
\ee
Matching at $r=a$ determines the relativistic phase shift to be
\be
e^{2 i \delta } = e^{-2i a k} \frac{\kappa_0 \cos(a \kappa_0)+ i k \sin(a \kappa_0)}{\kappa_0 \cos(a \kappa_0)- i k \sin(a \kappa_0)} \, .
\ee
Now in the true hard sphere limit $|V_0| \rightarrow \infty$ for which the field vanishes for $r<a$
the phase shift reduces to
\be
e^{2 i \delta } = e^{-2i a k}\,,
\ee
and as expected this gives the relativistic version of the time advance noted by Wigner and Eisenbud
\be
\Delta T =2 \frac{\p \delta}{\p \omega}= -\frac{2 a}{v}\,,
\ee
with $v= \frac{\d \omega}{\d k}=k/\omega$, and a similar behaviour occurs even at finite $V_0$  consistent with the bound \eqref{WEbound}.

Crucially however this effect occurs because the potential is sharper than the frequencies being considered. If we consider rather the situation where the frequencies are large in comparison to the typical scale of variation of the potential, we may use the WKB approximation for which the phase shift will take the approximate form
\be
\delta = \int_0^{\infty} \d r \( \kappa(r)- \sqrt{\omega^2-m^2} \) \, ,
\ee
where now
\be
\kappa(r) = \sqrt{(\omega-V(r))^2-m^2} \, .
\ee
For $\omega>{\rm Max}\left( |V(r)|\right)$ in the massless case $m=0$, the leading WKB correction to the time delay vanishes for $m=0$ since the leading contribution to the phase shift is frequency-independent. The first order correction to the WKB phase shift gives a frequency-dependent term which gives rise to a time-delay
\be
\Delta T \sim \frac{V'(0)}{\omega^3}+\dots
\ee
In the high frequency limit we are working in where $\omega^2\gg |V'(r)|$ this time delay/advance is unresolvable $|\omega \Delta T|\ll 1$ and higher order WKB corrections are similarly negligible. 

The massive case is slightly more subtle. The leading WKB term gives a correction
\be
\Delta T \approx  \int_0^{\infty} \d r \frac{m^2 (2 \omega-V(r)) V(r)}{\omega^2 (\omega-V(r))^2} \approx \int_0^{\infty} \d r \frac{2 m^2 V(r)}{\omega^3} \, ,
\ee
where in the first step we assumed $m^2\ll((\omega-V(r))^2,\omega^2)$ and in the last step we assumed $\omega \gg {\rm Max}\left( |V(r)|\right)$. At first sight, it looks like we can easily obtain a time advance from a region of negative potential.  However, for the situations considered in the main text, any background configuration can be parametrized by an overall amplitude and scale in terms of a dimensionless function. Similarly consider a potential of the form $V(r)= V_0 f\left(r/r_0\right)\,,$ where $f(x)$ is a dimensionless function. The maximum time advance relative to a freely propagating massive particle we can create in this region is then of order
\be
|\Delta T| \sim \frac{ m^2 V_0 r_0}{\omega^3}  \, .
\ee
By assumption, for the WKB approximation to be valid we need $\omega \gg r_0^{-1}$. Furthermore we have assumed $V_0 \ll \omega$. Thus we have the bound
\be
\omega |\Delta T| \ll m^2r_0^2\,.
\ee
For the theories considered in the Chapter, we assume the fundamental field is massless and any effective mass generated for fluctuations around a given background solution will be bounded in the sense $m^2 \lesssim{\cal O}(1) r_0^{-2}$, and hence these potential time advances are unresolvable $\omega |\Delta T| \ll 1 $. Thus provided we consider the region $\omega \gg (r_0^{-1}, {\rm Max}\left( |V(r)|\right))$ we do not expect to obtain any resolvable time advance.

In summary, although time advances for monopole scattering are allowed in the non-relativistic and low frequency region without contradicting causality, for the scattering of massless or light (in the scale of the background) high frequency scattering is not expected to lead to any resolvable time advance and this is implicit in our use of this criterion in the main text.

\subsection*{Positivity of Lorentz invariant UV completions}

The previous example was particularly trivial since it does not lead to any interesting time delay at high frequencies. To make it more interesting, and to generate a resolvable time delay, consider now a UV theory of two charged scalars, whose fluctuations may be described by one light field $\Phi$ and one heavy field $H$ with mass $M$. Integrating out the heavy scalar will give EFT corrections to the previously considered theory which describe the scattering and will give rise to a time delay.  Focussing on monopole fluctuations, it is natural to rescale $\Phi = \varphi/r$ and $H = h/r$. We will assume the quadratic action for the monopole fluctuations in the UV completion takes the $U(1)$ invariant form
\ba
S &=& \int \d t \int_0^{\infty} \d r  \int \d \Omega \,  \( |D_t \phi|^2 - | \partial_r \phi |^2-m^2 |\phi|^2  +|D_t h|^2 - | \partial_r h |^2- M^2 |h|^2    \right. \\
 &+&\left. \alpha h^*\partial_r \phi + \beta h^* D_t \phi + \alpha^* h \partial_r \phi^* + \beta^* h (D_t\phi)^* \) \, ,\nn
\ea
where we have dropped any mass mixing terms which can be traded for derivative interactions by a field redefinition.
This is manifestly relativistically causal by virtue of the Lorentz invariant two derivative terms which dominate the dynamics at high energy and determine the causal support of the retarded propagators. Integrating out the heavy field gives a low energy effective theory whose cutoff is $\Lambda = M$ and whose full effective action is
\begin{equation}
	S=\int \d t \int_0^{\infty} \d r  \int \d \Omega \,  \Big( |D_t \phi|^2 - | \partial_r \phi |^2-m^2 |\phi|^2 
+  (\alpha \partial_r \phi + \beta D_t  \phi)^* \frac{1}{M^2+D_t^2-\partial_r^2} (\alpha \partial_r \phi+\beta D_t \phi) \Big) \, .\nn
\end{equation}
The effective dispersion relation is
\be
((\omega-V)^2-k_r^2-m^2) ((\omega-V)^2-k_r^2-M^2)- |\alpha  k_r - \beta (\omega-V)|^2 =0 \, .
\ee
Due to the presence of odd powers of $k_r$ in the dispersion relation, the outgoing and ingoing waves have different magnitudes for their momenta $k_r^{\pm}$ and the WKB scattered wave may be parametrized as
\be
\phi =A(r) \( e^{i \int_0^{r} k_r^+ \d r  }- e^{i \int_0^{r}  k_r^- \d r } \) \, .
\ee
which is matched against the asymptotics
\be
\phi = A'  \( e^{2 i \delta} e^{i \sqrt{\omega^2-m^2} r}- e^{-i  \sqrt{\omega^2-m^2} r} \)\,,
\ee
to give the WKB phase shift
\be
\delta =  \int_0^{\infty} \d r \[\frac{1}{2} (k_r^+(r)+k_r^-(r))- \sqrt{\omega^2-m^2}  \] \, .
\ee
In the regime of validity of the low energy EFT, the leading two derivative terms in the effective action are
\be\label{toyEFT}
 S= \int \d t \int_0^{\infty} \d r  \int \d \Omega \,  \( |D_t \phi|^2 - | \partial_r \phi |^2-m^2 |\phi|^2+\frac{1}{M^2} |\alpha \partial_r \phi+\beta D_t\phi|^2 + \dots  \) \, ,
\ee
and the time delay takes the form
\be
\Delta T = \Delta T_{M=\infty}+ \Delta T_{\rm EFT} \, ,
\ee
where $\Delta T_{M=\infty}$ is the delay obtained previously and the leading EFT correction is
\ba  \Delta T_{\rm EFT} &=& 2 \frac{\p}{\p \omega } \int_0^{\infty} \d r \[\frac{1}{2} (k_r^+(r)+k_r^-(r))-\kappa(r) \] \, , \nn \\
&=& \frac{1}{M^2} \frac{\p}{\p \omega }\int_0^{\infty} \d r \[\frac{1}{2\kappa(r)}\left| \alpha \kappa(r)-(\omega-V) \beta\right|^2 +\frac{1}{2\kappa(r)}\left| \alpha \kappa(r)+(\omega-V) \beta\right|^2   \] + \dots \nn \\
 &=& \frac{1}{M^2}\frac{\p}{\p \omega }\ \int_0^{\infty} \d r \[ |\alpha|^2 \kappa+|\beta|^2 \frac{(\omega-V)^2}{\kappa} \] +\dots \nn \\
 &=&  \frac{1}{M^2} \int_0^{\infty} \d r \[ |\alpha|^2 \frac{(\omega-V)}{\kappa}+|\beta|^2 \frac{(\omega-V)}{\kappa}\(1-\frac{m^2}{\kappa^2}\) \] + \dots  \, .
\ea
In the WKB region considered, $\kappa \gg m$, and $\omega \gg {\rm Max}[V(r)]$ and so both terms are manifestly positive.
Since in this example we know the UV completion, we can directly infer the cutoff in $\omega$ of the low energy EFT by asking at what energy scale does the dispersion relation depart from that implied by the two derivative action \eqref{toyEFT}. This is when $(\omega-V) \sim M^2/(|\alpha|+|\beta|)$ and so we infer that the largest time delay calculable within the low energy EFT we could create is bounded by
\be
|\omega \Delta T_{\rm EFT}| \lesssim (|\alpha|+|\beta|) r_0 \, .
\ee
Since the RHS can be made arbitrarily large by increasing $r_0$, remaining in the region of validity of the low energy EFT, this positive time delay can be made resolvable.
Thus as anticipated, a consistent unitary Lorentz invariant UV completion of an EFT for a massless or light field gives rise to a positive, generally resolvable, time delay $\Delta T>0$ in the WKB region, and the EFT contribution itself is by itself positive $ \Delta T_{\rm EFT}>0$.

\section{Conventions} \label{ap:DefDimLess}
In this Appendix, we summarize some our relations and conventions.
For completeness, we consider the EFT including up to dimension-14 operators and work with the following form of the Lagrangian,
\begin{align}
	\L =& - \frac12 (\p \phi)^2 - \frac12 m^2 \phi^2+ \frac{g_8}{\Lambda^4} (\p \phi)^4 \nonumber \\
	&  + \frac{g_{10}}{\Lambda^6} (\p \phi)^2 \Big[ (\phi_{, \mu \nu})^2  -  (\Box \phi)^2 \Big] + \frac{g_{12}}{\Lambda^8} (( \phi_{, \mu \nu} )^2 )^2 + \frac{g_{14}}{\Lambda^{10}} ( \phi_{, \mu \nu} )^2 ( \phi_{, \alpha \beta \gamma} )^2 \, .
	\label{eq:Lhigh}
\end{align}
The dimension-14 operator is constrained by the following positivity bounds
\begin{equation}
	 -2 g_{12} < g_{14} < \frac{27}{5} (2g_8 - g_{10}) \ .
\end{equation}
The relations between the parameters considered here and those included in \cite{Tolley:2020gtv} and \cite{Caron-Huot:2020cmc} are given in the  Table~\ref{tab:dictionary} below.

\begin{table}[!h]
	\begin{center}
	\begin{tabular}{ | c | c | c | c | } \hline
		 EFT & Tolley \emph{et al.},  \cite{Tolley:2020gtv} & Caron-Huot \emph{et al.}, \cite{Caron-Huot:2020cmc} \\ \hline
		 $g_8$ & $\frac{1}{4}\tilde{a}_{1,0}$ & $\frac{1}{2} \Lambda^4 g_2$ \\
		 $g_{10}$ & $-\frac{1}{3} \tilde{a}_{0,1}$ & $\frac{1}{3} \Lambda^6 g_3$ \\
		 $g_{12}$ & $\tilde{a}_{2,0}$ & $4 \Lambda^8 g_4$ \\
		 $g_{14}$ & $\frac{4}{5}\tilde{a}_{1,1}$ & $-\frac{8}{5}\Lambda^{10} g_5$ \\ \hline
	\end{tabular}
	\caption[Dictionary for various conventions of the shift-symmetric EFT coefficients.]{Parameters dictionary relating the conventions used in this work, defined in Eq.~\eqref{eq:Lhigh} and others presented in the literature.}
	\label{tab:dictionary}
	\end{center}
\end{table}

In order to extremize the causality bounds, it is convenient to work with dimensionless parameters.
The relations between the dimensionless parameters and their dimensionfull counterparts is provided  in Table \ref{tab:dictionaryDimensionless} below.
\begin{table}[!h]
	\begin{center}
	\begin{tabular}{ | c | c | c | c | c | c | c | c | } \hline
		 Dimensionless parameter & $f(r)$ & $R$ & $\epsilon_1$ & $\epsilon_2$ & $\Omega$ & $B$ & $R_t$ \\ \hline
 &  & &  & &  &  &  \\[-6pt]
		 Definition & $\frac{\bar{\phi}(r)}{\bar{\Phi}_0}$ & $\frac{r}{r_0}$ & $\frac{\bar{\Phi}_0}{r_0 \Lambda^2}$ & $\frac{1}{r_0 \Lambda}$ & $\frac{\omega}{\Lambda}$ & $\frac{b}{r_0}$ & $\frac{r_t}{r_0}$ \\[5pt] \hline
	\end{tabular}
	\caption[Parameters dictionary between dimensionless and dimensionfull parameters.]{Parameters dictionary relating the dimensionless and dimensionfull ones.}
	\label{tab:dictionaryDimensionless}
	\end{center}
\end{table}
It is worth noting that $\bar{\Phi}_0$ carries the scale of the background field $\bar{\phi}$, $r_0$ is its typical scale of variation, whereas $\omega$ is the frequency of the scattered perturbation. The cutoff of the scalar EFT in Eq.~\eqref{eq:L} is given by $\Lambda$ if the dimensionless couplings $g_i$ are all considered to be at most of order 1. Finally, $b$ and $r_t$ are respectively the impact parameter of the free theory and the turning point of the higher-multipole scattering events.

\section{NLO corrections to the time delay at $\ell=0$} \label{ap:NLO}
In this Appendix, we provide the explicit expressions required for computing the time delay at the next order in the EFT, which we refer to as next-to-leading order (NLO).
At NLO, the equation of motion for the monopole $\ell=0$ is given by,
\begin{align}
	&\left. \hat{W}_0(R) \right|_{\rm NLO}= 1152 g_8^3 \epsilon _1^6 f'(R)^6 +224 g_8 g_{12} \Omega ^2 \epsilon _1^4 \epsilon _2^2 f'(R)^2 f''(R)^2  \\
	&+144 \frac{g_8^2}{\Omega ^2} \epsilon _1^4 \epsilon _2^2 \left(2 \frac{f'(R)^3 f''(R)}{R}+2 f'(R)^2 f''(R)^2+f^{(3)}(R) f'(R)^3\right) \nonumber \\
	&-96 g_8 g_{10} \epsilon _1^4 \epsilon _2^2 \left(\frac{f'(R)^4}{R^2}-3\frac{f'(R)^3 f''(R)}{R}\right) \nonumber \\
	&-8 g_{12} \epsilon _1^2 \epsilon _2^4 \left(2\frac{f'(R)^2}{R^4}+\p_R\(4 \frac{f'(R)f''(R)}{R^2} -2 \frac{f''(R)^2}{R} + f^{(3)}(R) f''(R)\)\right) \nonumber \\
	&-4 g_{14} \Omega ^2 \epsilon _1^2 \epsilon _2^4 \left(12\frac{f'(R) f''(R)}{R^3}+4\frac{f^{(3)}(R) f'(R)-3 f''(R)^2}{R^2}+2\frac{f^{(3)}(R) f''(R)}{R}+\p_R\(f^{(3)}(R) f''(R)\)\right) \nonumber \\
	&- 12 \frac{g_{10}}{\Omega ^2} \epsilon _1^2 \epsilon _2^4 \left(\frac{f'(R)^2}{R^4}+\p_R\(\frac{f'(R)f''(R)}{R^2}\)\right) \, .\nonumber
\end{align}
The sound speed square and effective potential are given by
\begin{align}
	&\left. c_s^2(\omega^2,R) \right|_{\rm NLO}= -128 g_8^3 \epsilon_1^6 f'(R)^6 - 96 g_8 g_{12} \epsilon_1^4 \epsilon_2^2 \frac{\omega^2}{\Lambda^2} f'(R)^2 f''(R)^2  \\
	& + 96 g_8 g_{10} \epsilon_1^4 \epsilon_2^2 \left( \frac{f'(R)^4}{R^2} + \frac{f'(R)^3 f''(R)}{R} \right) \nonumber \\
	&+ 8 g_{12} \epsilon_1^2 \epsilon_2^4 \left( 2 \frac{f'(R)^2}{R^4} + \p_R\(4 \frac{f'(R)f''(R)}{R^2} -2 \frac{f''(R)^2}{R} + f^{(3)}(R) f''(R)\) \right) \nonumber \\
	&+ 4 g_{14} \epsilon_1^2 \epsilon_2^4 \frac{\omega^2}{\Lambda^2} \left( 12 \frac{f'(R) f''(R)}{R^3} +4 \frac{-3 f''(R)^2 + f'(R) f^{(3)}(R)}{R^2} +2 \frac{f''(R) f^{(3)}(R)}{R} + \p_R\(f^{(3)}(R) f''(R)\) \right) \ \, , \nonumber
\end{align}
and
\begin{align}
	&\left. V_{\text{eff}}(R) \right|_{\rm NLO}= -48 g_8^2 \epsilon _1^4 \left(2\frac{f'(R)^3 f''(R)}{R}+4 f'(R)^2 f''(R)^2+f^{(3)}(R) f'(R)^3 \right)  \\
	&+12 g_{10} \epsilon _2^2 \epsilon _1^2 \left(\frac{f'(R)^2}{R^4}+\p_R\(\frac{f'(R)f''(R)}{R^2}\)\right) \, . \nonumber
\end{align}
The integrand of the time delay at NLO is given by
\begin{align}
	\left. \mathcal{I}_0(\omega^2,R) \right|_{\rm NLO}= &8 (\omega r_0) \epsilon_1^2 \left[ 104 g_8^3 \epsilon _1^4 f'(R)^6 -2 \frac{g_8^2}{\Omega ^2} \epsilon _1^2 \epsilon _2^2 \left(3 \frac{f'(R)^4}{R^2}-4 f'(R)^2 f''(R)^2\right) - 6 g_8 g_{10} \epsilon _1^2 \epsilon _2^2 \frac{f'(R)^4}{R^2} \right. \nonumber \\
	&\left. +72 g_8 g_{12} \Omega ^2 \epsilon _1^2 \epsilon _2^2 f'(R)^2 f''(R)^2 -\frac{45}{4} g_{14} \Omega ^2 \epsilon _2^4 \left(\frac{f'(R)^2}{R^4} + \frac{f^{(3)}(R) f'(R)-f''(R)^2}{R^2}\right) \right. \nonumber \\
	&\left. + \left(g_{12}-\frac34 \frac{g_{10}}{\Omega ^2}\right) \epsilon _2^4 \left(\frac{f'(R)^2}{R^4}-\frac{f^{(3)}(R)f'(R)+f''(R)^2}{R^2}\right) \right] \nonumber \\
	&+ \text{total derivatives}  \, .
	\label{eq:I0NLO}
\end{align}
We do not write the total derivative terms explicitly since they vanish upon integration in the $\ell=0$ case considered here. Note that the total derivatives include terms like $f'(R)^2/R^3, f'(R)f''(R)/R^2$ and $f''(R)^2/R$ that diverge when evaluated at the origin. In this analysis, we have been careful to cancel the divergences so that the total derivatives in the last line of Eq.~\eqref{eq:I0NLO} actually vanish upon integration from $0$ to $\infty$.

\section{Higher-order multipoles} \label{ap:NLOmultipole}

In this Appendix, we provide the leading-order expressions to the various functions entering the computation of the time delay for $\ell>0$, as defined in Section \ref{sec:multipole}. Note that since we are focusing on a regime where $\omega r_0 \gg 1$ in order to safely ignore all WKB corrections, we will ignore all $1/\Omega$ corrections for consistency. Furthermore, such a regime also allows us to forget about NLO corrections, hence they will be omitted here. The function $W_{\ell}(R)$ reads, at leading order,
\begin{align}
	\left. W_{\ell}(R) \right|_{\rm LO} = &\left(1-\frac{B^2}{R^2}\right) \left(1+ 8 g_8 \epsilon _1^2 f'(R)^2 + 96 g_8^2 \epsilon _1^4 f'(R)^4 \right)  \\
	&+ 8 g_{12} \Omega ^2 \epsilon _1^2 \epsilon _2^2 \left\lbrace \left(1-\frac{B^2}{R^2}\right) \left(f''(R)-\frac{f'(R)}{R}\right)+\frac{f'(R)}{R}\right\rbrace^2 \nonumber \\
	& +12 g_{10} \epsilon _1^2 \epsilon _2^2 \frac{B^2}{R^2} \left(\frac{f'(R)^2}{R^2}-\frac{f'(R)f''(R)}{R}\right) \nonumber \,.
\end{align}
This means that the square sound velocity and the effective potential at leading order read
\begin{align}
	\left. c_s^2(\omega^2,R) \right|_{\rm LO}=& 1 -8 g_8 \epsilon _1^2 f'(R)^2 -32 g_8^2 \epsilon _1^4 f'(R)^4  \\
	& -8 g_{12} \Omega ^2 \epsilon _1^2 \epsilon _2^2 \left(2\frac{B^2}{R^2} \left(\frac{f'(R)f''(R)}{R}-f''(R)^2\right)+f''(R)^2\right) -24 g_{10} \epsilon _1^2 \epsilon_2^2 \frac{f'(R) f''(R)}{R} \, ,\nonumber \\
	\label{eq:velocityHighL}
	\left. V_{\text{eff}}(R) \right|_{\rm LO}=& \frac{L^2}{R^2} \left[ \vphantom{\frac12} 1 -8 g_{12} \Omega ^2 \epsilon_1^2 \epsilon_2^2 f''(R)^2 -12 g_{10} \epsilon_1^2 \epsilon_2^2  \left(\frac{f'(R)^2}{R^2}+\frac{f'(R) f''(R)}{R}\right) \right]  \\
	&-8\frac{L^4}{R^4} g_{12} \epsilon _1^2 \epsilon _2^4 \left(\frac{f'(R)^2}{R^2}-f''(R)^2\right) \, .\nonumber
\end{align}
We have decided to express the effective potential in terms of the orbital number $L$ rather than the reduced effective impact parameter $B$ in order to make contact with the free theory where $V_{\rm eff, free}= L^2/R^2$. It is worth mentioning once again that the effective potential term is suppressed by $(\omega r_0)^{-2}=\epsilon_2^2 / \Omega^2$ with respect to the speed of sound term. Hence, the leading-order effective potential should include terms up to $\mathcal{O}(\epsilon_1^2)$. Note that the terms $L^2 \epsilon_1^2 \epsilon_2^2$ and $L^4 \epsilon_1^2 \epsilon_2^4$ seem to be higher-order and appear to be unnecessarily taken into account. However, recalling that $L=\Omega B/\epsilon_2$ implies $\epsilon_2 L \sim \mathcal{O}(\epsilon^0)$. This means that $L^2 \epsilon_1^2 \epsilon_2^2 \sim L^4 \epsilon_1^2 \epsilon2_4 \sim \mathcal{O}(\epsilon_1^2)$, so all terms considered are indeed leading order in the effective potential. To show that this is indeed the correct functional form for the potential, one could rewrite the term of interest, \ie $V_{\rm eff}/(\omega r_0)^2$ rather than just the effective potential, in terms of the variable $B$ that does not hide any dependence on $\epsilon_i$,
\begin{align}
	\left. \frac{V_{\text{eff}}(R)}{(\omega r_0)^2} \right|_{\rm LO}=& \frac{B^2}{R^2} \left[ \vphantom{\frac12} 1 -8 g_{12} \Omega ^2 \epsilon_1^2 \epsilon_2^2 f''(R)^2 -12 g_{10} \epsilon_1^2 \epsilon_2^2  \left(\frac{f'(R)^2}{R^2}+\frac{f'(R) f''(R)}{R}\right) \right] \\
	& - 8 \frac{B^4}{R^4} g_{12} \Omega ^2 \epsilon _1^2 \epsilon _2^2\left(\frac{f'(R)^2}{R^2}-f''(R)^2\right)  \, . \nonumber
\end{align}
In this set up, the corresponding turning point is now $R_t$, which is given by
\begin{equation}
	\left. R_t \right|_{\rm LO} = B \left[ 1 - 4 g_{12} \Omega^2 \epsilon_1^2 \epsilon_2^2 \frac{f'(B)^2}{B^2} - 6 g_{10} \epsilon_1^2 \epsilon_2^2 \left(\frac{f'(B)^2}{B^2}+\frac{f'(B) f''(B)}{B}\right) \right] \,.
\end{equation}
Moreover, we have
\begin{align}
	\left. U_{\ell}(R) \right|_{\rm LO} =& 4 \left(1-\frac{B^2}{R^2}\right) \left( g_8 \epsilon _1^2 f'(R)^2 + 10 g_8^2 \epsilon _1^4 f'(R)^4 \right)  \\
	& -6 g_{10} \epsilon _1^2 \epsilon _2^2 \left\lbrace \left(1-\frac{B^2}{R^2}\right) \left(\frac{f'(R)^2}{R^2}-\frac{f'(R) f''(R)}{R}\right)- \left( \frac{f'(R)^2}{R^2} + \frac{f'(R) f''(R)}{R} \right) \right. \nonumber \\
	&\qquad \left. + \frac{B^2}{R^2} \left( \frac{f'(B)^2}{B^2} + \frac{f'(B) f''(B)}{B} \right) \right\rbrace \nonumber \\
	& +4 g_{12} \Omega^2 \epsilon _1^2 \epsilon _2^2 \left\lbrace \left[\left(1-\frac{B^2}{R^2}\right) \(\frac{f'(R)}{R}-f''(R)\)-\frac{f'(R)}{R}\right]^2 - \frac{B^2}{R^2}\frac{f'(B)^2}{B^2} \right\rbrace \,,\nonumber
\end{align}
with $U_{\ell}(R_t)=0$. Having all the ingredients, the dimensionless time delay can now be expressed in the following form
\ba
\omega \Delta T_{b}(\omega) = (\omega r_0) \left[ \int_B^{\infty} \left( \frac{\Upsilon^{(0)}_{\ell}(R)}{\sqrt{ 1 - \frac{B^2}{R^2} }} + \Upsilon^{(1)}_{\ell}(R) \sqrt{ 1 - \frac{B^2}{R^2} } + \Upsilon_{\ell}^{(2)} \left( 1 - \frac{B^2}{R^2} \right)^{3/2} \right) \mathrm{d}R + \Upsilon_{\ell}^{(3)} \right],\
\ea
where
\begin{align}
	\Upsilon_{\ell}^{(0)}(R) =& 12 g_{10} \epsilon _1^2 \epsilon _2^2 \left\lbrace \frac{f'(R)^2}{R^2}+\frac{f'(R)f''(R)}{R} -\frac{B^2}{R^2} \left(\frac{f'(B)^2}{B^2}+\frac{f'(B) f''(B)}{B}\right) \right\rbrace  \\
	& +24  g_{12} \Omega ^2 \epsilon _1^2 \epsilon _2^2 \left(\frac{f'(R)^2}{R^2}- \frac{B^2}{R^2}\frac{f'(B)^2}{B^2}\right) \,,\nonumber \\
	\Upsilon_{\ell}^{(1)}(R) =& 8 g_8 \epsilon _1^2 f'(R)^2+80 g_8^2 \epsilon _1^4f'(R)^4 - 48 g_{12}\Omega ^2 \epsilon _1^2 \epsilon _2^2 \left(\frac{f'(R)^2}{R^2} - \frac{f'(R) f''(R)}{R} \right)  \\
	&-12 g_{10} \epsilon _1^2 \epsilon _2^2 \left(\frac{f'(R)^2}{R^2}-\frac{f'(R) f''(R)}{R}\right) \,, \nonumber\\
	\Upsilon_{\ell}^{(2)}(R) =& 24 g_{12} \Omega ^2 \epsilon _1^2 \epsilon _2^2 \left(\frac{f'(R)^2}{R^2}-\frac{2 f'(R) f''(R)}{R}+f''(R)^2\right) \,, \\
	\Upsilon_{\ell}^{(3)}(R) =& 12 g_{12} \pi B \Omega ^2 \epsilon _1^2 \epsilon _2^2 \frac{f'(B)^2}{B^2} \,.
\end{align}
Note that $\Upsilon_{\ell}^{(0)}(R) = \mathcal{F}[f(R)] - B^2/R^2 \mathcal{F}[f(B)]$, where $\mathcal{F}$ is a functional of the function $f$. This immediately shows that $\Upsilon_{\ell}^{(0)}(R=B)=0$, hence avoiding any divergence around the lower bound of the integral.

\section{Extremization method} \label{ap:procedure}

The method used to extremize the causality bounds for the simple profile considered in \eqref{eq:profilefR2} (with $p=3$) is summarized below. In principle the same method could be applied to more generic profiles and in less symmetric situations. The dimensionless time delay is given as a function of
\begin{equation}
	(\omega \Delta T) = (\omega \Delta T)(g_{10}, g_{12}, \mathcal{P}) \,
\end{equation}
where the parameters are listed in the vector
\begin{equation}
	\mathcal{P} = \left\{ g_8, a_0, a_2, a_4, a_6, \epsilon_1, \epsilon_2, \Omega, B \right\} \,.
\end{equation}
In our analysis $g_8$ will be fixed to be either $0$ or $1$ but we include it for completeness. In order to remain within the regime of validity of the EFT we only consider $-5<a_i<5$ so that $f(R)$ is $\mathcal{O}(1)$. More importantly, during the extremization procedure we constrain the parameters in $\mathcal{P}$ such that the analysis remains in the regime of validity of the EFT as given in Eq.~\eqref{eq:eps} (Eq.~\eqref{eq:eftGal} for the Galileons) by replacing $\ll1$ by $<1/2$. Since the suppression of higher-order EFT corrections always comes as the square of these parameters, this ensures that the terms that we neglect are suppressed by at least a factor of $0.25$. Furthermore, we also need to ensure that the WKB formula is valid up to the order that we compute it. To do so, we explicitly compute corrections to the WKB formula in the monopole case and check that they are negligible. For higher multipoles however, we rely instead on dimensional analysis to compute the order of magnitude of the corrections that are being neglected. This requires enforcing Eq.~\eqref{eq:WKBreq}. For a more detailed discussion on the validity of the EFT and WKB approximation we refer to the analysis in Sections \ref{sec:monopole} and \ref{sec:multipole}. Note that in our analysis, we separated the case $\ell=0$ and $\ell>0$, and also $g_8=0$ and $g_8=1$, which gave four separate sets of causal regions. However, the method used in each of them was identical and will be detailed below.

The boundary of the causal region for a given set of parameters is defined by $(\omega \Delta T) = -1$, which can be solved for $g_{12}$ to give the equation of a line in the $(g_{10},g_{12})$ plane
\begin{equation}
	g_{12} = m(\mathcal{P}) g_{10} + p(\mathcal{P}) \equiv \mathcal{Y}_{\mathcal{P}}(g_{10}) \,.
\end{equation}
Now, the extremization process differentiates between lower and upper bounds. In both cases, let us define a vector $\mathcal{G}$ corresponding to set of discrete points in the interval $[0,2.5]$. The parameter $g_{12}$ will take values drawn from $\mathcal{G}$, \ie $g_{12} \in \mathcal{G}$.

The tightest lower bound for $g_{10}$ for a given value of $g_{12}=\mathcal{G}_i$ is achieved by finding the optimal set of parameters $\mathcal{P}_i^{(\rm lower)}$ such that the negative value of $g_{10}$ at the intersection between the two lines defined by $g_{12}=\mathcal{Y}_{\mathcal{P}_i^{(\rm lower)}}(g_{10})$ and $g_{12}=\mathcal{G}_i$ is maximal. It can be defined as
\begin{equation}
	\mathcal{P}_i^{(\rm lower)}= {\rm Max} \left\{ g_{10} < 0 \left| g_{12}=\mathcal{Y}_{\mathcal{P}}(g_{10})\, \& \, g_{12}=\mathcal{G}_i \right. \right\} \,,
\end{equation}
and the `causal' region\footnote{This method does not `prove' causality, it simply indicates the absence of obvious acausality.} $\mathcal{R}_i^{(\rm lower)}$ would consist of all points in the $(g_{10}, g_{12})$ plane that are ``above" this line, meaning
\begin{equation}
	\mathcal{R}_i^{(\rm lower)} = \left\{ (g_{10},g_{12}) \left| g_{10} \in \mathbb{R}, g_{12} > \mathcal{Y}_{\mathcal{P}_i^{(\rm lower)}}(g_{10}) \right. \right\} \,.
\end{equation}
Equivalently, the tightest upper bound for a given $i$ is given by
\begin{equation}
	\mathcal{P}_i^{(\rm upper)}= {\rm Min} \left\{ g_{10} > 0 \left| g_{12}=\mathcal{Y}(\mathcal{P})\, \& \, g_{12}=\mathcal{G}_i \right. \right\} \,,
\end{equation}
and the associated `causal' region
\begin{equation}
	\mathcal{R}_i^{(\rm upper)} = \left\{ (g_{10},g_{12}) \left| g_{10} \in \mathbb{R}, g_{12} < \mathcal{Y}_{\mathcal{P}_i^{(\rm upper)}}(g_{10}) \right. \right\} \,.
\end{equation}
Note that in the case where $\ell=0$, the method does not identify any upper bound, as described previously. This process is iterated for all values of $i$ (and it could be optimized further by exploring more values in the range $[0,2.5]$ or by extending this range) and the final causal region $\mathcal{R}_{\rm causal}$ is obtained by taking the union of all lower and upper regions labelled by $i$,
\begin{equation}
	\mathcal{R}_{\rm causal} = \cup_{i} \cup_{j=\text{lower, upper}} \mathcal{R}_i^{(j)} \,.
\end{equation}

\section{Gravitationally-coupled Galileons}
\label{app:Matter_coupling}
In most of this work, we have considered the scalar field EFT to describe a single low energy degree of freedom in its own right in flat spacetime and in the absence of any other light degrees of freedom. For such low energy EFTs, one can in principle consider an arbitrary external source $J$ that would spontaneously generate an arbitrary (Lorentz-violating) background profile for the scalar field.

We now explore a `Galileon' field which, in some contexts, can be thought of as describing a degree of freedom reminiscent of an infrared modification of gravity (as is the case from instance in the Dvali-Porrati-Gabadadze model \cite{Dvali:2000hr} or massive gravity \cite{deRham:2010ik,deRham:2010kj}). In this case the EFT is not precisely a low energy description, and the presence of other light degrees of freedom may not always be safely ignored. Generating a non-trivial profile for the field typically comes at the price of introducing a non-trivial stress-energy tensor which would also be expected to ever-so-slightly affect the geometry. The subtle issue of backreaction on the geometry can be put aside for now, but in this Appendix, we establish which source would be required to generate the spherically-symmetric background profile we have considered so far. In particular we explore whether there are any physical requirements to be imposed on that source, and whether the source satisfies the null or weak energy condition.  In the present case, we consider the coupling of the Galileon to matter through the trace of the stress-energy tensor $T^{\mu}_{\phantom{\mu} \mu}$ which generically arises in massive gravity theories. Thus the source in Eq.~\eqref{eq:lag} is now given by
\begin{equation}
	J=\frac{1}{M_{\rm Pl}}  T^{\mu}_{\phantom{\mu}\mu} \ ,
\end{equation}
The Galileon interactions (and possible mass term) are small corrections compared to the kinetic term and thus the equation of motion for the source reads
\begin{equation}
	\Box \phi = - \frac{g_{\rm matter}}{M_{\rm Pl}} T^{\mu}_{\phantom{\mu}\mu} \,.
	\label{eq:boxphiT}
\end{equation}
The stress-energy tensor needs to respect the spherical symmetry and hence can be written in the following form
\begin{equation}
	T^{\mu}_{\phantom{\mu}\nu} = \text{diag}(- \rho(r), p_r(r), p_{\Omega}(r), p_{\Omega}(r)) \,,
\end{equation}
where $p_r$ and $\rho$ are respectively the radial pressure and energy density of the fluid, and $p_{\Omega}$ is the angular pressure. For simplicity, we write $p_{\Omega} = A p_r$, where $A$ is a constant that will be constrained by requiring asymptotic flatness of the spacetime. The trace of the stress-energy tensor is then simply given by $T^{\mu}_{\phantom{\mu}\mu} = p_r (1+2A) - \rho$.   Energy-momentum conservation implies
\begin{equation}
	p'_r + 2(1-A) \frac{p_r}{r} = 0 \,.
\end{equation}
This first-order differential equation for the radial pressure $p_r$ is solved by,
\begin{equation}
	p_r(r) = \bar p_r\  r^{-2(1-A)} \,, \qquad \rho(r) = \bar p_r\  (1+2A) r^{-2(1-A)} - T^{\mu}_{\phantom{\mu}\mu}(r) \,.
\end{equation}
Asymptotic flatness (or `vacuum') demands that at large radius $p_r, \rho \sim r^n$ with $n<-3$, which effectively provides the bound $A<-1/2$. Furthermore, for the source to be physical, we should at the very least demand the weak energy condition which requires
\begin{equation}
	\rho > 0 \,, \qquad  \rho + p_r > 0 \,, \qquad {\rm and }\qquad \rho+A p_r>0 \,. \label{eq:WEC}
\end{equation}
Defining $T_{\rm max} = {\rm Max}_{r>0} \left\{ r^{2(1-A)} \left| T^{\mu}_{\phantom{\mu}\mu}(r) \right| \right\}$, then if one were to choose
\begin{equation}
	A<-1 \,, \qquad \bar p_r < \frac{T_{\rm max}}{2(1+A)}<0\,,
	\label{eq:solWEC}
\end{equation}
and as long as $\left| T^{\mu}_{\phantom{\mu}\mu}(r) \right|$ is bounded and $r^{2(1-A)} T^{\mu}_{\phantom{\mu}\mu}(r) \rightarrow 0$ when $r \rightarrow \infty$, which is ensured for exponentially suppressed background profiles as the one considered in Eq.~\eqref{eq:profilefR2}, then the weak energy condition is respected. Note that if one is only interested in the null energy condition, then $\rho$ is unconstrained, but to satisfy the other two conditions in Eq.~\eqref{eq:WEC} we still require that Eq.~\eqref{eq:solWEC} holds. We have thus proven that some fluids with negative pressure along some direction (and positive pressure along others) can represent a physical source generating an asymptotically flat spacetime, satisfying  the weak energy condition and leading to any bounded profile $\bar \phi(r)$. Note that this stress-energy tensor diverges at the origin, indicating that the source ought to be regularized but since  the scalar field remains finite, one would not expect the regularization to impact the outcome of this study.

\chapter{Causal vector effective field theory}

\section{Equations of Motion} \label{ap:eom}
The equation of motion for the vector field $A^{\mu}$ with Lagrangian given by Eq.~\eqref{eq:Lagrangian} and sourced by an arbitrary current $J^\nu$ is
\begin{equation}
	\partial^\mu \mathcal{E}^{(F)}\mn =g J_\nu \ , 	\label{eq:eomA}
\end{equation}
where
\begin{equation}
	\mathcal{E}^{(F)}\mn \equiv F\mn + \sum_{i=1}^8 c_i \mathcal{E}^{(i)}\mn \,,
	\label{eq:eomF}
\end{equation}
where
\begin{align}
\mathcal{E}^{(1)}\mn =& - \frac{8}{\Lambda^4} F\mn F^{\alpha \beta} F_{\alpha \beta} \\
\mathcal{E}^{(2)}\mn =& - \frac{8}{\Lambda^4} F_{\mu \alpha} F_{\nu \beta} F^{\alpha \beta} \\
\mathcal{E}^{(3)}\mn =& \frac{2}{\Lambda^6} \left[ F_{\mu\alpha} \left( \p^{\alpha} F_{\nu\beta} \p_{\gamma} F^{\beta\gamma} + F^{\beta \gamma} \p_{\gamma} \p_{\alpha} F_{\nu \beta} \right) - F^{\alpha\beta} \p_{\alpha} F\du{\mu}{\gamma} \left( \p_{\gamma} F_{\nu \beta} + \p_{\nu} F_{\beta\gamma} \right) \right] - \left( \mu \leftrightarrow \nu \right) \\
\mathcal{E}^{(4)}\mn =& \frac{1}{\Lambda^6} \left[ \frac12 F_{\mu\alpha} F_{\nu\beta} \left( \p_{\gamma} \p^{\beta} F^{\alpha\gamma} - \p_{\gamma} \p^{\alpha} F^{\beta\gamma} \right) - F\du{\mu}{\alpha} \left( \left( \p_{\alpha} F_{\beta\gamma} - \p_{\beta} F_{\alpha\gamma} \right) \p^{\gamma} F\du{\nu}{\beta} + \p_{\gamma} F\du{\beta}{\gamma} \p_{\nu} F\du{\alpha}{\beta} \right. \right. \nonumber \\
	&\left. \left. \qquad  \vphantom{+ \p_{\gamma} F\du{\beta}{\gamma} \p_{\nu} F\du{\alpha}{\beta}} + F^{\beta\gamma} \p_{\nu} \p_{\gamma} F_{\alpha\beta} \right) - F^{\alpha\beta} \left( \p^{\gamma} F_{\mu\alpha} \left( \p_{\beta} F_{\nu \gamma} - \p_{\nu} F_{\beta\gamma} \right) + \p_{\alpha} F\du{\mu}{\gamma} \p_{\nu} F_{\beta\gamma} \right) \vphantom{\frac12} \right]- \left( \mu \leftrightarrow \nu \right) 
\end{align}
\begin{align}
\mathcal{E}^{(5)}\mn =& \frac{1}{\Lambda^6} \left[ F\du{\mu}{\alpha} \left( \p_{\alpha} F_{\beta\gamma} \p_{\nu} F^{\beta\gamma} + F^{\beta\gamma}\p_{\nu}\p_{\alpha}F_{\beta\gamma} \right) + F^{\alpha\beta} \left( \p_{\gamma} F_{\alpha\beta} \p_{\mu} F\du{\nu}{\gamma} + \p_{\gamma} F\du{\mu}{\gamma} \p_{\nu} F_{\alpha\beta} \right) \right. \nonumber \\
	&\left. \qquad + F\mn \left( F^{\alpha\beta} \p_{\gamma} \p_{\beta} F\du{\alpha}{\gamma} + \p_{\beta} F_{\alpha\gamma} \p^{\gamma} F^{\alpha\beta} \right) \right] - \left( \mu \leftrightarrow \nu \right)  \\
\mathcal{E}^{(6)}\mn =& \frac{1}{2\Lambda^8} \left[ - F\du{\mu}{\alpha} \p_{\nu} \p_{\alpha} \left( \p_{\rho} F_{\beta \gamma} \p^{\rho} F^{\beta \gamma} \right) - \p_{\mu} F\du{\nu}{\alpha} \p_{\alpha} \left( \p_{\rho} F_{\beta \gamma} \p^{\rho} F^{\beta \gamma} \right) - \p_{\alpha} F\du{\mu}{\alpha} \p_{\nu} \left( \p_{\rho} F_{\beta \gamma} \p^{\rho} F^{\beta \gamma} \right)  \right. \nonumber \\
	&\qquad + 2 \p^{\alpha} F\mn \left( 2 \p^{\rho} F^{\beta\gamma} \p_{\gamma}\p_{\alpha} F_{\beta\rho} + \p_{\alpha} F^{\beta\gamma} \p_{\gamma}\p^{\rho} F_{\beta\rho} + F^{\beta\gamma} \p_{\alpha} \p_{\gamma} \p^{\rho} F_{\beta\rho} \right) \nonumber \\
	&\qquad \left. + 4 \Box F\mn \p_{\beta} \left( F_{\alpha\gamma} \p^{\gamma} F^{\alpha\beta} \right) \right] - \left( \mu \leftrightarrow \nu \right) \\
\mathcal{E}^{(7)}\mn =& \frac{1}{\Lambda^8} \left[ \p_{\mu} F^{\alpha\beta} \left( \p_{\gamma} F^{\gamma\rho} \p_{\nu}\p_{\rho} F_{\alpha\beta} - \p_{\alpha} F^{\gamma\rho} \p_{\nu} \p_{\beta} F_{\gamma\rho} \right)  + \p^{\alpha} F\mn \p^{\rho} \left( \p_{\beta} F^{\beta\gamma} \p_{\gamma} F_{\alpha\rho} \right) \right. \nonumber \\
	&\qquad \left. +2 \p_{\alpha}\p_{\gamma} F\mn \p_{\rho} \left( F^{\alpha\beta} \p_{\beta} F^{\gamma\rho} \right) + F^{\alpha\beta} \p_{\mu} \p_{\alpha} F^{\gamma \rho} \p_{\nu} \p_{\beta} F_{\gamma \rho} \right] - \left( \mu \leftrightarrow \nu \right) \\
\mathcal{E}^{(8)}\mn =& \frac{1}{\Lambda^8} \left[ -F_{\mu\alpha} \p_{\gamma} \p^{\rho} \left( \p^{\alpha} F_{\beta\rho} \p_{\nu} F^{\beta\gamma} \right) +2 \p_{\mu} F^{\alpha\beta} \p^{\rho} F_{\alpha\gamma} \p^{\gamma} \p_{\beta} F_{\nu \rho} + 2F^{\alpha\beta} \p_{\mu}\p^{\gamma} F_{\beta\rho} \p^{\rho} \p_{\alpha} F_{\nu\gamma} \right. \nonumber \\
	&\qquad \left. + \p_{\beta} F_{\mu\alpha} \p^{\rho} \left(  \p^{\alpha} F_{\gamma\rho} \p_{\nu} F^{\beta\gamma} + \p^{\alpha} F^{\beta\gamma} \p_{\nu} F_{\gamma\rho} - F_{\gamma\rho} \p^{\alpha} \p_{\nu} F^{\beta\gamma} - F^{\beta\gamma} \p^{\alpha} \p_{\nu} F_{\gamma\rho} \right) \right]  - \left( \mu \leftrightarrow \nu \right) \,.
\end{align}

In Section \ref{sec:spherical}, we focus on the perturbations around a spherically symmetric background by perturbing the field as in Eq.~\eqref{eq:Aspherical}. In this setting, one can find that one of Maxwell's equations lead to a constraint and fixes one of the unphysical modes. To find this constraint, we perform a field-redefinition of $u_1^{\ell}$ and write
\begin{align}
	v_1^{\ell} \equiv& \frac{i\omega  u_1^{\ell}}{r}-\left(u_2^{\ell}\right)' -\frac{16}{\Lambda ^4} \left(2 c_1+c_2\right) A_0' A_0'' u_2^{\ell} \nonumber \\
	&- \frac{2}{\Lambda^6} \frac{1}{r} \left\lbrace \vphantom{\frac{A_0'}{r}} c_3 A_0' (A_0' +4r A_0'') \left(u_2^{\ell}\right)'' + 2c_3 \left( 2r(A_0'')^2 + A_0' \left( 3 A_0'' + 2r A_0^{(3)} \right) \right) \left(u_2^{\ell}\right)' \right. \nonumber \\
	& \qquad + \left[ c_3 \left( \omega^2 - \frac{2}{r^2} \right) (A_0')^2 + A_0'' \left\lbrace 2(3c_3-c_4-2c_5)A_0'' +3(c_3-c_4-2c_5)r A_0^{(3)} \right\rbrace \right. \nonumber \\
	& \qquad \left. \left. + \frac{A_0'}{r} \left\lbrace 2 \left( 2c_3(r^2 \omega^2-1) + c_4+2c_5 \right) A_0'' + 2 (3c_3 -c_4 -2c_5) r A_0^{(3)} + (c_3 -c_4 -2c_5) r^2 A_0^{(4)} \right\rbrace \right] \right\rbrace u_2^{\ell} \nonumber \\
	&+ \frac{1}{\Lambda^8} \sum_{n=0}^4 d_n (u_2^{\ell})^{(n)} \,, \label{eq:constraint}
\end{align}
where
\begin{align}
	d_0 =& 8\left(-c_8\left(L^2-3\right)+2c_6+c_7\right) \frac{(A_0')^2}{r^5} + 4\left(\left(3c_6+c_8\right)L^2-11c_6-4c_7-9c_8\right) \frac{A_0'A_0''}{r^4} \nonumber \\
	& + \left(\left(2c_6+c_8\right)L^2+2c_7+10c_8\right) \frac{\omega^2 (A_0')^2}{r^3} + 4\left(7c_6+3c_7+2c_8\right) \frac{A_0' A_0^{(3)}}{r^3} +4\left(7c_6+2c_7+3c_8\right) \frac{(A_0'')^2}{r^3} \nonumber \\
	& + \left(-c_8\left(L^2+12\right)-16c_6-20c_7\right) \frac{\omega^2 A_0' A_0''}{r^2} - 2 \left(c_6\left(L^2+3\right)+2c_7-c_8\right) \frac{A_0'A_0^{(4)}}{r^2} \nonumber \\
	&- 2 \left(c_6\left(3L^2+7\right)+6c_7-c_8\right)\frac{A_0'' A_0^{(3)}}{r^2} - \left(2c_6+c_8\right) \frac{\omega^4 (A_0')^2}{r} \nonumber \\
	& +2\left(-c_6+7c_7+5c_8\right) \frac{\omega^2 A_0' A_0^{(3)}}{r} +2 \left(9c_7+5c_8\right) \frac{\omega^2 (A_0'')^2}{r} \nonumber \\
	& -2\left( \left(c_6+c_8\right) \frac{A_0' A_0^{(5)}}{r} + 2\left(3c_6-c_7+2 c_8\right) \frac{ A_0'' A_0^{(4)}}{r} + \left(5c_6-2c_7+3c_8\right) \frac{(A_0^{(3)})^2}{r} \right) \nonumber \\
	& +c_8\omega^4A_0'A_0'' + \omega^2\left(\left(2c_6+8c_7+5c_8\right)A_0^{(4)}A_0'+2\left(3
	c_6+8c_7+4c_8\right)A_0^{(3)}A_0''\right) \nonumber \\
	& - 2 \left(c_6+c_8\right)\left(A_0'' A_0^{(5)} + 3A_0^{(3)}A_0^{(4)}\right)+128(2c_1+ c_2)^2(A_0')^3 A_0'' \\
	d_1 =& 4\left(c_7-c_8\left(L^2+2\right)\right) \frac{(A_0')^2}{r^4} + 4\left(c_8\left(L^2+4\right)+2c_6-c_7\right) \frac{A_0'A_0''}{r^3} \nonumber \\
	& - \left(c_8 \left(2L^2+5 \right)+8c_6-2c_7\right) \frac{A_0' A_0^{(3)}}{r^2} - 2\left(c_8\left(L^2+4\right)+4c_6\right) \frac{(A_0'')^2}{r^2}+\left(-4c_6+2c_7+c_8\right) \frac{\omega^2 (A_0')^2}{r^2} \nonumber \\
	& -\frac{A_0'}{r}\left(\left(4c_6+2c_7+c_8\right)\omega^2
	A_0''+\left(-4c_6+4c_7+3c_8\right)A_0^{(4)}\right) +\frac{3\left(4c_6-3c_8\right)A_0^{(3)}A_0''}{r} \nonumber \\
	&+2\left(\left(c_6+c_8\right) A_0' A_0^{(5)}+\left(4c_6-c_8\right) A_0''A_0^{(4)} +\left(3c_6-2c_8\right) (A_0^{(3)})^2\right) \nonumber \\
	& +\omega^2\left(\left(2c_7+3c_8\right) A_0'A_0^{(3)} +c_8 (A_0'')^2\right) \vphantom{\frac{(A_0')^2}{r^4}} \\
	d_2 =& \left(\left(2 c_6 + c_8  \right) L^2 - 6 c_7 + 6 c_8 \right) \frac{(A_0')^2}{r^3} - \left(c_8\left(L^2 + 2\right) + 16 c_6 -8 c_7\right) \frac{A_0' A_0''}{r^2} \nonumber \\
	& - \frac{2}{r}\left(\left(2 c_6 + c_8 \right)\omega^2 (A_0')^2 + \left(c_6 + 3 c_7 -2 c_8 \right) A_0' A_0^{(3)} + \left(c_7 - 2 c_8 \right) (A_0'')^2 \right) \nonumber \\
	& + A_0'\left(2 c_8\omega^2 A_0'' + \left(2 c_6 + 4 c_7 +7 c_8 \right) A_0^{(4)} \right) + 2\left(3 c_6 + 2 c_7 + 3 c_8 \right) A_0'' A_0^{(3)} \vphantom{\frac{(A_0')^2}{r^3}} \\
	d_3 =& \left(-4 c_6 + 2 c_7 +  c_8 \right) \frac {(A_0')^2} {r^2} - \left(4 c_6 + 2 c_7 + c_8 \right) \frac {A_0' A_0''} {r} + 2 \left(2 c_7 + 3 c_8 \right) A_0' A_0^{(3)} + 2 c_8 (A_0'')^2 \\
	d_4 =& c_8 A_0' A_0'' - \left(2 c_6+c_8\right) \frac{\left(A_0'\right){}^2}{r} \,.
\end{align}
  Note that we only need to go up to order $1/\Lambda^8$ for the terms $c_1^2, c_1 c_2$ and $c_2^2$ in order to get the $\epsilon_1^4$ contributions at leading order. In terms of this new variable we find that the constraint is given by
 \begin{equation}
 	v_1^{\ell} = 0 \ .
 \end{equation}
\section{Explicit expressions for propagation around spherically symmetric backgrounds}\label{ap:SphericalExpr}
The equations of motion for the propagating modes are given by Eq.~\eqref{eq:ModesEOM} with $W_{I,\ell}$ given by
\begin{align}
	W_{2,\ell} =& 1 -\frac{B^2}{R^2} + 16 \epsilon_1^2 \frac{B^2}{R^2} \left(2c_1+c_2\right) f'(R)^2 \nonumber \\
	& + 2 \epsilon _1^2 \epsilon _2^2 \left\lbrace (-5c_3+c_4+2c_5) \frac{B^2}{R^2}\frac{f'(R)^2}{R^2} + \left( \frac{B^2}{R^2}(7c_3+c_4+2c_5)-4c_3 \right) f''(R)^2 \right. \nonumber \\
	&\qquad -2 \left. \left( \left( \frac{B^2}{R^2}(c_3+c_4+2c_5) + 3c_3 \right) f''(R) +2c_3 \left(1-2\frac{B^2}{R^2}\right)R f^{(3)}(R) \right) \frac{f'(R)}{R} \right\rbrace \nonumber \\
	&+2\epsilon_1^2 \epsilon_2^2 \Omega^2 (2c_6-c_8) \frac{B^2}{R^2} \left\lbrace \left(2\frac{B^2}{R^2}-1\right)\frac{f'(R)^2}{R^2} +\left(1-\frac{B^2}{R^2}\right)f''(R)^2 \right. \nonumber \\
	&\qquad - \left. \left( \frac{B^2}{R^2} f''(R) + \left(1-\frac{B^2}{R^2}\right)R f^{(3)} \right) \frac{f'(R)}{R} \right\rbrace \,, \label{eq:W2}
\end{align}
and
\begin{align}
	W_{4,\ell} =& 1 -\frac{B^2}{R^2} + 8 \epsilon_1^2 \frac{B^2}{R^2} c_2 f'(R)^2 \nonumber \\
	& + 2 \epsilon _1^2 \epsilon _2^2 \left\lbrace (-3c_3+c_4) \frac{B^2}{R^2}\frac{f'(R)^2}{R^2} + \left( \frac{B^2}{R^2}(c_3-c_4)-4c_3 \right) f''(R)^2 \right. \nonumber \\
	&\qquad + \left. \left( 2 c_3 \left( \frac{B^2}{R^2}-3 \right) f''(R) + \left(\frac{B^2}{R^2}(c_3-c_4)-4c_3\right)R f^{(3)}(R) \right) \frac{f'(R)}{R} \right\rbrace \nonumber \\
	&-2\epsilon_1^2 \epsilon_2^2 \Omega^2 c_8 \frac{B^2}{R^2} \left\lbrace \left(2\frac{B^2}{R^2}-1\right)\frac{f'(R)^2}{R^2} +\left(1-\frac{B^2}{R^2}\right)f''(R)^2 \right. \nonumber \\
	&\qquad - \left. \left( \frac{B^2}{R^2} f''(R) + \left(1-\frac{B^2}{R^2}\right)R f^{(3)} \right) \frac{f'(R)}{R} \right\rbrace \,. \label{eq:W4} 
\end{align}

Let us now write down the turning point for both physical modes
\begin{equation}
	R^{t}_{I,\ell} = B \left[ 1 - \epsilon_1^2 \Psi_{I,\ell}^{(1)}(B) - \epsilon_1^2 \epsilon_2^2 \Psi_{I,\ell}^{(2)}(B) - \epsilon_1^2 \epsilon_2^2 \Omega^2 \Psi_{I,\ell}^{(3)}(B) \right] \,,
\end{equation}
where
\begin{align}
	\Psi_{2,\ell}^{(1)}(B) =& 8 (2c_1+c_2) f'(B)^2 \nonumber \\
	=& (f_2+g_2) f'(B)^2 \,, \nonumber \\
	\Psi_{2,\ell}^{(2)}(B) =& -(5c_3-c_4-2c_5) \frac{f'(B)^2}{B^2} -2(4c_3+c_4+2c_5) \frac{f'(B)f''(B)}{B} +(3c_3+c_4+2c_5) f''(B)^2 \nonumber \\
	& +4c_3 f'(B) f^{(3)}(B) \nonumber \\
	=& - \frac13 \left\lbrace (f_3+3(g_3-4h_3)) \frac{f'(B)^2}{B^2} -2(f_3+3(g_3+2h_3)) \frac{f'(B)f''(B)}{B} \right. \nonumber \\
	&+\left. (f_3+3g_3+4h_3) f''(B)^2 +8h_3 f'(B) f^{(3)}(B) \vphantom{\frac{f'(B)^2}{B^2}} \right\rbrace \,, \nonumber \\
	\Psi_{2,\ell}^{(3)}(B) =& (2c_6 -c_8) \left( \frac{f'(B)^2}{B^2} - \frac{f'(B)f''(B)}{B} \right) \nonumber \\
	=& 2(2f_4+ g_{4}) \left( \frac{f'(B)^2}{B^2} - \frac{f'(B)f''(B)}{B} \right) \,,
\end{align}
and
\begin{align}
	\Psi_{4,\ell}^{(1)}(B) =& 4c_2 f'(B)^2 \nonumber \\
	=& -(f_2-g_2) f'(B)^2 \,, \nonumber \\
	\Psi_{4,\ell}^{(2)}(B) =& -(3c_3-c_4) \frac{f'(B)^2}{B^2} -4c_3 \frac{f'(B)f''(B)}{B} -(3c_3+c_4) \left( f''(B)^2 + f'(B) f^{(3)}(B) \right) \nonumber \\
	=& \frac13 \left\lbrace -(f_3-3g_3-8h_3) \frac{f'(B)^2}{B^2} +8h_3 \frac{f'(B)f''(B)}{B} + (f_3-3g_3+4h_3) \left( f''(B)^2 + f'(B) f^{(3)}(B) \right) \right\rbrace \,, \nonumber \\
	\Psi_{4,\ell}^{(3)}(B) =& - c_8 \left( \frac{f'(B)^2}{B^2} - \frac{f'(B)f''(B)}{B} \right) \nonumber \\
	=& - 2(2f_4 - g_{4}) \left( \frac{f'(B)^2}{B^2} - \frac{f'(B)f''(B)}{B} \right) \,.
\end{align}
Finally, the functions $U_{I,\ell}$ appearing in the phase shift and time delay expressions are given by
\begin{align}
	U_{I,\ell} =& \frac{B^2}{R^2} \left[ \epsilon_1^2 \left( \Psi_{I,\ell}^{(1)}(R) - \Psi_{I,\ell}^{(1)}(B) \right) + \epsilon_1^2 \epsilon_2^2 \left( \Psi_{I,\ell}^{(2)}(R) - \Psi_{I,\ell}^{(2)}(B) \right) + \epsilon_1^2 \epsilon_2^2 \Omega^2 \left( \Psi_{I,\ell}^{(3)}(R) - \Psi_{I,\ell}^{(3)}(B) \right) \right] \nonumber \\
	&+ \left( 1 - \frac{B^2}{R^2} \right) \left( \epsilon_1^2 \epsilon_2^2 \Upsilon_{I,\ell}^{(1)}(R) + \epsilon_1^2 \epsilon_2^2 \Omega^2 \Upsilon_{I,\ell}^{(2)}(R) \right) \,,
	\label{eq:defUIell}
\end{align}
where it is clear that the first term in square bracket vanishes when $R \rightarrow B$ which ensures the convergence of the time delay. The analytical expressions for the functions entering Eq.~\eqref{eq:defUIell} above are given by
\begin{align}
	\Upsilon_{2,\ell}^{(1)}(R) =& -2c_3 \left( 3 \frac{f'(R)f''(R)}{R} +2 \left( f''(R)^2 + f'(R)f^{(3)}(R) \right) \right) \nonumber \\
	=& \frac43 h_3 \left( 3 \frac{f'(R)f''(R)}{R} +2 \left( f''(R)^2 + f'(R)f^{(3)}(R) \right) \right) \,, \nonumber \\
	\Upsilon_{2,\ell}^{(2)}(R) =& -(2c_6-c_8) \left\lbrace 2 \frac{f'(R)^2}{R^2} - \frac{f'(R)f''(R)}{R} - f''(R)^2 + f'(R)f^{(3)}(R) \right\rbrace \nonumber \\
	=& -2(2f_4+g_{4}) \left\lbrace 2 \frac{f'(R)^2}{R^2} - \frac{f'(R)f''(R)}{R} - f''(R)^2 + f'(R)f^{(3)}(R) \right\rbrace \,,
\end{align}
and
\begin{align}
	\Upsilon_{4,\ell}^{(1)}(R) =& \Upsilon_{2,\ell}^{(1)}(R) \,, \nonumber \\
	\Upsilon_{4,\ell}^{(2)}(R) =& c_8 \left\lbrace 2 \frac{f'(R)^2}{R^2} - \frac{f'(R)f''(R)}{R} - f''(R)^2 + f'(R)f^{(3)}(R) \right\rbrace \nonumber \\
	=& 2(2f_4-g_{4})) \left\lbrace 2 \frac{f'(R)^2}{R^2} - \frac{f'(R)f''(R)}{R} - f''(R)^2 + f'(R)f^{(3)}(R) \right\rbrace \,.
\end{align}
Finally, the expression for the time delay is
\begin{align}
	&(\omega \Delta T_{b,I,\ell}(\omega)) \nonumber \\
	&= 2(\omega r_0) \left[ \int_{B}^{\infty} \frac{\p_{\omega} \left( \omega U_{I,\ell}(R) \right)}{\sqrt{1- \frac{B^2}{R^2}}} dR + \frac{\pi}{2} \left( B - \p_{\omega} \left( \omega R^{t}_{I,\ell} \right) \right) \right] \nonumber \\
	&= 2(\omega r_0) \left[ \int_{B}^{\infty} \frac{B^2}{R^2} \epsilon_1^2 \frac{\left[ \left( \Psi_{I,\ell}^{(1)}(R) - \Psi_{I,\ell}^{(1)}(B) \right) + \epsilon_2^2 \left( \Psi_{I,\ell}^{(2)}(R) - \Psi_{I,\ell}^{(2)}(B) \right) + 3\epsilon_2^2 \Omega^2 \left( \Psi_{I,\ell}^{(3)}(R) - \Psi_{I,\ell}^{(3)}(B) \right) \right]}{\sqrt{1- \frac{B^2}{R^2}}} dR \right. \nonumber \\
	& \qquad \qquad \quad + \int_{B}^{\infty} \sqrt{1- \frac{B^2}{R^2}} \left( \epsilon_1^2 \epsilon_2^2 \Upsilon_{I,\ell}^{(1)}(R) + 3 \epsilon_1^2 \epsilon_2^2 \Omega^2 \frac{B^2}{R^2} \Upsilon_{I,\ell}^{(2)}(R) \right) dR \nonumber \\
	& \qquad \qquad \quad \left. + \frac{\pi}{2} B \left( \epsilon_1^2 \Psi_{I,\ell}^{(1)} + \epsilon_1^2 \epsilon_2^2 \Psi_{I,\ell}^{(2)} + 3 \epsilon_1^2 \epsilon_2^2 \Omega^2 \Psi_{I,\ell}^{(3)} \right) \vphantom{\frac{\left( \epsilon_1^2 \Phi_{I,\ell}^{(1)} + \epsilon_1^2 \epsilon_2^2 \Phi_{I,\ell}^{(2)} + 3 \epsilon_1^2 \epsilon_2^2 \Omega^2 \Phi_{I,\ell}^{(3)} \right)}{\sqrt{1- \frac{B^2}{R^2}}}} \right] \nonumber \\
	&= 2(\omega r_0) \left[ \int_{B}^{\infty} \frac{B^2}{R^2} \frac{\left[ \epsilon_1^2 \Psi_{I,\ell}^{(1)}(R) + \epsilon_1^2 \epsilon_2^2 \Psi_{I,\ell}^{(2)}(R) + 3 \epsilon_1^2 \epsilon_2^2 \Omega^2 \Psi_{I,\ell}^{(3)}(R) \right]}{\sqrt{1- \frac{B^2}{R^2}}} dR \right. \nonumber \\
	& \qquad \qquad \quad \left. + \int_{B}^{\infty} \sqrt{1- \frac{B^2}{R^2}} \left( \epsilon_1^2 \epsilon_2^2 \Upsilon_{I,\ell}^{(1)}(R) + 3 \epsilon_1^2 \epsilon_2^2 \Omega^2 \frac{B^2}{R^2} \Upsilon_{I,\ell}^{(2)}(R) \right) dR \right] \,. 
\end{align}
Note that the integral over the constant $\Psi_{I,\ell}^{(j)}(B)$ can be performed and exactly cancels out the extra term that is free of any integration.
\section{Optimization method for the vector case}
\label{ap:method}

The algorithm used to optimize the causality bounds will be described in this Section. For any $2d$ plot in the $(\mathcal{W}_J, \mathcal{W}_K)$ plane, we start by fixing all the remaining coefficients collectively denoted by $\{ \mathcal{W}_L \}$ to constants corresponding to a particular UV completion or another interesting case. This is necessary in order to reduce the complexity of the optimization by only allowing $2$ of the $8$ coefficients to vary \footnote{Note however that the causality bounds derived in this paper are insensitive to $g_4'$ and that $g_2$ is set to either $1$ or $0$ without loss of generality.}.

The vector causality constraints reduce to $2$ constraints in the even and odd sector and the extremization procedure can be done in each independently. When presenting the results, we can choose to show the causality bounds of each sector individually or to show the final result which is achieved by taking their union. In the following, we forget about the two sectors and assume we are specialising to a given one, only to take the union of both at the end.

The boundary of the causality constraint reads $(\omega \Delta T) = -1$ and can be solved for $\mathcal{W}_J$ as a function of $\mathcal{W}_K$ and all the other parameters of the problem. Next, we discretize the direction $\mathcal{W}_K$ by letting this coefficient take values in the interval $\left[ \mathcal{W}_K^{\rm (min)} ,\mathcal{W}_K^{\rm (max)} \right]$ divided in $n_K$ equal steps of length $\Delta \mathcal{W}_K = (\mathcal{W}_K^{\rm (max)} - \mathcal{W}_K^{\rm (min)})/n_K$,

\begin{equation}
    \mathcal{W}_K = \left\{ \mathcal{W}_K^{\rm (min)}, \mathcal{W}_K^{\rm (min)} + \Delta \mathcal{W}_K , \cdots , \mathcal{W}_K^{\rm (max)} \right\} \,.
\end{equation}
For each such value of $\mathcal{W}_K$, we find the maximal (if it exists) value that $\mathcal{W}_J$ can take by extremizing over the parameters of the background profile and the ones controlling the EFT expansion, under the constraint that one remains in the regime of validity of the EFT and the WKB approximations simultaneously. This way, we get a set of extremal parameters per discretized value of $\mathcal{W}_K$ that we plug back into the equation $(\omega \Delta T) = - 1$, which in turn gives an optimized straight line in the $(\mathcal{W}_J, \mathcal{W}_K)$ plane separating the causality-violating region from the allowed one and is read as an upper bound. One gets such a line for each discretized value of $\mathcal{W}_K$ and hence can form an envelope by imposing all upper constraints derived in this way.

The next step is to repeat this operation for lower bounds. If both upper and lower bounds exist, their union gives the sought-after compact causality bounds. Finally, we refine our bounds by performing the same procedure once again after swapping the role of $\mathcal{W}_J$ and $\mathcal{W}_K$, i.e. discretizing the other axis and getting left/right bounds rather than upper/lower.

As mentioned earlier, this is done independently in the even and odd sectors of the vector. The theory is only causal if neither mode propagate in a causality-violating way, hence the final causality bounds are obtained by imposing causality on both sectors simultaneously.

\section{Partial UV completions}
 Together with the positivity and causality bounds, we will show explicitly the values for the coefficients of known partial UV completions. We will focus on tree level, partial UV completions given by the interactions of the photon with a scalar and and axion. In some cases, we will also show the values for the partial UV completions involving a graviton that were analyzed in \cite{Henriksson:2021ymi,Henriksson:2022oeu,Haring:2022sdp}. Lastly, we will also analyze some of these bounds that can be compared to one-loop, partial UV completions from QED like theories. We will consider the standard (spinor) QED \cite{Euler:1935qgl,Euler:1935zz,Costantini:1971cj,Karplus:1950zz}, scalar QED \cite{Weisskopf:1936hya,Yang:1994nu}, and vector QED \cite{Yang:1994nu,Vanyashin:1965ple}. The coefficients for all these partial UV completions are shown in Table \ref{tab:UVcomp}.

\begin{table}[h!]
	\centering
	\begin{tabular}{ | c || c | c | c | c | c | c | c | c || c | c | }
	\hline
	UV completion & $g_2$ & $f_2$ & $f_3$ & $g_3$ & $h_3$ & $f_4$ & $g_4$ & $g_4'$ & $g_{4,1}$ & $g_{4,2}$ \\ \hline
	scalar & $1$ & $1$ & $3$ & $1$ & $0$ & $\frac12$ & $1$ & $0$ & $1$ & $0$  \\
	axion & $1$ & $-1$ & $-3$ & $1$ & $0$ & $-\frac12$ & $1$ & $0$ & $1$ & $0$ \\ \hline \hline
	scalar QED & $1$ & $\frac34$ & $\frac{5}{14}$ & $\frac{3}{28}$ & $\frac{1}{28}$ & $\frac{1}{84}$ & $\frac{41}{420}$ & $-\frac{1}{168}$ & $\frac{17}{840}$ & $\frac{1}{336}$ \\
	spinor QED & $1$ & $-\frac{3}{11}$ & $-\frac{10}{77}$ & $\frac{4}{77}$ & $-\frac{1}{77}$ & $-\frac{1}{231}$ & $\frac{13}{660}$ & $-\frac{5}{462}$ &$\frac{41}{4620}$ & $\frac{5}{924}$ \\
	vector QED & $1$ & $\frac{1}{28}$ & $\frac{5}{294}$ & $-\frac{47}{1764}$ & $\frac{1}{588}$ & $\frac{1}{1764}$ & $\frac{131}{8820}$ & $-\frac{23}{1176}$ & $-\frac{83}{17640}$ & $\frac{23}{2352}$ \\ \hline \hline
	spin-2 even I & $1$ & $1$ & $0$ & $1$ & $0$ & $\frac{1}{2}$ & $1$ & $-6$ & $-5$ & $3$ \\
	spin-2 even II & $1$ & $0$ & $0$ & $-1$ & $0$ & $0$ & $1$ & $-2$ & $-1$ & $1$ \\
	spin-2 odd & $1$ & $-1$ & $0$ & $1$ & $0$ & $-\frac12$ & $1$ & $-6$ & $-5$ & $3$ \\
	minimally-coupled spin-2 & $1$ & $0$ & $0$ & $-\frac12$ & $0$ & $0$ & $\frac12$ & $-1$ & $-\frac12$ & $\frac12$ \\ \hline
\end{tabular}
	\caption[Values of the Wilson coefficients for known partial UV completions (for $g_2=1$).]{Values of the Wilson coefficients for known partial UV completions, where $g_2$ is normalized to unity.}
	\label{tab:UVcomp}
\end{table}
One should note that the time delay for the even sector of the axion vanishes as it should. Similarly, the time delay for the odd sector vanishes for the scalar partial UV completions. 

\section{Comparison with other works}

In this Section, we provide a conversion chart that enables the reader to go from our conventions to the ones of HMRV \cite{Henriksson:2021ymi,Henriksson:2022oeu} and HHKMP \cite{Haring:2022sdp}. 

\begin{table}[h!]
	\centering
	\begin{tabular}{ | c | c | c | c | }
		\hline
		EFT dimension & This Paper & HMRV & HHKMP \\
		\hline
		\multirow{2}{*}{$8$} & $c_1$ & $f_2 = 8c_1 +2c_2$ & $f_2 = 8c_1 +2c_2$ \\ 
		& $c_2$ & $g_2 = 8c_1 +6c_2$ & $g_2 = 8c_1 +6c_2$ \\
		\hline
		\multirow{3}{*}{$10$} & $c_3$ & $f_3 = -3(c_3+c_4+c_5)$ & $f_3 = -3(c_3+c_4+c_5)$ \\
		& $c_4$ & $g_3 = -c_5$ & $g_3 = -c_5$ \\
		& $c_5$ & $h_3 = -\frac32 c_3$ & $h_3 = - \frac32 c_3$ \\
		\hline
		\multirow{3}{*}{$12$} & $c_6$ & $f_4 = \frac14 c_6$ & $f_4 = \frac14 c_6$ \\
		& $c_7$ & $g_{4,1} = \frac12(c_6+c_8)+c_7$ & $g_4 = \frac12 (c_6-c_8)$ \\
		& $c_8$ & $g_{4,2} = -\frac12 (c_7+c_8)$ & $g_4' = c_7+c_8$ \\ \hline
	\end{tabular}
	\caption{Conversion table for the massless photon EFT coefficients.}
	\label{tab:conversion}
\end{table}

\pagebreak

\end{appendices}

\end{document}